\documentclass{ws-rv9x6}
\usepackage{subfigure}   
\usepackage{ws-rv-thm}   
\usepackage{ws-rv-van}   
\usepackage{amsmath,amssymb,amsfonts,xfrac,mathrsfs}
\usepackage{hyperref}
\makeindex

\usepackage{bm}
\newcommand{\p}{\bm{p}}
\newcommand \f {\not\!}

\begin{document}

\chapter[Jet MC]{Jet Quenching: From Theory to Simulation.\label{jet1}}

\author[S. Cao, A. Majumder, R. Modarresi-Yazdi, I. Soudi, Y. Tachibana ]{
Shanshan Cao,$^1$\footnote{shanshan.cao@sdu.edu.cn} 
Abhijit Majumder,$^{2,}$\footnote{majumder@wayne.edu} 
Rouzbeh Modarresi-Yazdi,$^3$\footnote{rouzbeh.modarresi-yazdi@mail.mcgill.ca} 
Ismail Soudi$^{2,4,5}$\footnote{ismail.i.soudi@jyu.fi}
and Yasuki Tachibana$^6$\footnote{ytachibana@aiu.ac.jp}}

\address{ 
$^1$Institute of Frontier and Interdisciplinary Science,
Shandong University, Qingdao, Shandong, China 266237\\
$^2$Department of Physics and Astronomy, Wayne State University, Detroit, Michigan, USA 48201\\
$^3$Department of Physics, McGill University, 3600 University street, Montreal, QC, Canada H3A 2T8\\
$^4$University of Jyväskylä, Department of Physics, P.O. Box 35, FI-40014 University of Jyväskylä, Finland\\
$^5$Helsinki Institute of Physics, P.O. Box 64, FI-00014 University of Helsinki, Finland\\
$^6$Akita International University, Yuwa, Akita-city, Japan 010-1292
}

\begin{abstract}
With the explosion of data on jet based observables in relativistic heavy-ion collisions at the Large Hadron Collider and the Relativistic Heavy-Ion Collider, perturbative QCD based simulations of these processes, often interacting with an expanding viscous fluid dynamical background, have taken center stage. This review is meant to bridge the gap between theory, simulation and phenomenology of jet modification in a dense medium. We will demonstrate how the existence of such end-to-end event generators with semi-realistic or even fully realistic final states allows for the most rigorous comparisons between pQCD based jet modification theory and experiment. State-of-the-art calculations of several jet based observables are presented. Extensions of this theory to jets in the small systems of $p$-$A$ and $e$-$A$ collisions is discussed. 
\end{abstract}


\body

\date{May 2023}

\tableofcontents

\section{Introduction} 
    
On page 292 of his 1989 text~\cite{Field:1989uq}, R.~D.~Field exclaims ``Present day QCD Monte Carlo models for hadron-hadron collisions contain many approximations and should not be taken too seriously." While we encourage any serious student of pQCD to read this textbook, we urge them to lend no credence to this sentence. 
Over the last several decades, there have been several end-to-end Monte Carlo event generators developed to address proton-proton ($p$-$p$) collisions, e.g., PYTHIA~\cite{Sjostrand:2014zea}, HERWIG~\cite{Bahr:2008pv}, SHERPA~\cite{Sherpa:2019gpd}, etc. These generators have been successful in addressing many aspects of $p$-$p$ collisions (with the exception of phenomena in high multiplicity events).

Unlike the case for $p$-$p$ collisions, for jets in heavy-ion ($A$-$A$) collisions, and indeed for almost all aspects of $A$-$A$ collisions, Monte Carlo event generators have become the leading tool to compare between theory and experimental data. There are many reasons for this. Perhaps the most important reason is that there is no one \emph{effective}~\footnote{While everything is in principle governed by QCD, the description of phenomena at different energy and density regimes, even within the same heavy-ion collision, require a patchwork of effective theories and phenomenology.} theory that can encompass hadron production in a heavy-ion collision. Hadrons with transverse momenta $\lesssim 2$~GeV are predominantly described by a fluid dynamical simulation~\cite{Bjorken:1982qr, Kolb:2000sd, Kolb:2003dz,Teaney:2000cw, Gale:2013da} followed by a Cooper-Frye process where boosted thermal distributions are used to convert the viscous fluid into hadrons~\cite{Cooper:1974mv}.
Hard jets, formed almost immediately at the time of nuclear overlap, propagate through the fluid and become modified by interaction with the fluid~\cite{Majumder:2010qh,Cao:2020wlm}. These are typically described by pQCD based \emph{sub}-event-generators, followed by hadronization via a combination of Lund string breaking~\cite{Andersson:1983ia} and parton recombination~\cite{Fries:2003vb,Molnar:2003ff,Greco:2003xt,Hwa:2002tu} processes. The initial stage of the collision which seeds both the soft bulk and hard jet simulations, can itself be simulated either by QCD motivated approaches~\cite{Schenke:2012wb} or straightforward parametrizations~\cite{Moreland:2014oya}.

Other chapters in this volume will describe many of the non-jet related portions of a heavy-ion collision. This chapter will focus on the description of jet propagation in a dense medium, in particular, the focus will remain on the construction of multi-stage Monte Carlo jet quenching event generators~\cite{Putschke:2019yrg,Cao:2017zih}. We will demonstrate how generators in each stage can be derived from first principles pQCD. Our hope is that after reading this chapter, the reader will not only have a better grasp of the theoretical background of various jet quenching generators, but will be able to design generators of their own and incorporate these within multi-stage frameworks~\cite{Putschke:2019yrg}.

The remainder of this chapter is organized as follows: In Sec.~\ref{sec:Theory}, we recapitulate the basic pQCD based theories that have been developed to describe the propagation of a hard parton in a dense medium. In Sec.~\ref{sec:Theory-2-EventGeneration} we highlight how theoretical expressions derived to describe distributions of partons, are modified to simulate real events, first applied to jets in vacuum in $p$-$p$ collisions, and then for jets in the dense matter created in $A$-$A$ collisions. Section~\ref{sec:Pheno} will cover phenomenological comparisons with experimental data, where the utility of multi-stage frameworks  will be elucidated. New and upcoming developments will be briefly discussed in Sec.~\ref{sec:new-stuff}.

\section{Theory} \label{sec:Theory} 
    
Theoretical simulation of jets in a $p$-$p$ or in an $A$-$A$ collision starts with the concept of factorization~\cite{Collins:1985ue,Collins:1981uw,Collins:1988ig,Collins:1989gx}, which allows one to separate the cross section or yield of a process, with a large exchanged transverse momentum ($p_T$), into a product of probability distributions. In the case of $p$-$p$ collisions, one may express the cross section as:
\begin{eqnarray}
    \frac{d\sigma_{pp}}{dy dp_T^2 } = \int d x_a  d x_b G(x_a,Q^2) G(x_b,Q^2) 
    \frac{d \hat{\sigma}}{d \hat{t}} \frac{D(z,Q^2)}{\pi z} \,\, 
    + \mathcal{O}\left(\frac{\Lambda}{Q}\right).
\label{eq:pp-factorized-formula}
\end{eqnarray}
The equation above expresses the cross section for a process with a detectable large $p_T$, taking place in a small rapidity window $dy$. The neglect of terms suppressed by powers of the hard scale $\mathcal{O} (\Lambda/Q)$, allows one to express the cross section as the product of the parton distribution functions within a nucleon $G(x_{a/b})$ (carrying fraction $x_{a/b}$ of the nucleon momentum), with the hard partonic scattering cross section ($d\hat{\sigma}/d \hat{t}$), and the final observable (fragmentation, jet etc.) function $D(z)$ (carrying fraction $z$ of the outgoing parton's momentum).

As one moves to the case of a heavy-ion collision, there is an overall multiplicative increase in the cross section due to the fact that each heavy-ion collision, depending on the range of centrality considered, engenders several binary nucleon-nucleon collisions ($N_{\rm bin}$). Ignoring this overall multiplicative $N_{\rm bin}$ enhancement, the ``per binary collision" cross section in an $A$-$A$ collision can be expressed, rather schematically as, 
\begin{eqnarray}
  \mbox{}\!\!\!\!  \frac{d \sigma_{AA}}{dy dp_T^2} &=& \!\!\!\int \!\! dx_a dx_b \Tilde{G}(x_a,Q^2) \Tilde{G}(x_b, Q^2) 
  \frac{d\hat{\sigma}}{d \hat{t}}
  \mathscr{P}(\delta p, \delta \mu^2, \rho) \Tilde{D}(z,\delta p, Q^2,\delta \mu^2). 
  \label{eq:AA-factorized-formula}
\end{eqnarray}
The tilde on the parton distribution functions indicates that these nucleons are not free and are bound within a nucleus, affecting their parton distribution functions. The tilde on the $D$ indicates that the fragmentation or jet function is also modified due to the presence of a medium. The $\mathscr{P}$, which strictly speaking is also a function of the forward momentum $p$ and the virtuality $\mu^2$ of the parton [i.e. $\mathscr{P} (p,\delta p, \mu^2 , \delta \mu^2 , \rho )$] indicates that both the momentum $p$ and the virtuality $\mu$ of the hard parton(s) are modified due to interaction with the medium which has a density $\rho$. The main point of Eq.~\eqref{eq:AA-factorized-formula} is that each term in the convoluted product is a probability distribution, with no interference between terms. While widely held to be true, and extensively used by all researchers in the field, there is no general proof of factorization in the case of $A$-$A$ collisions [as indicated by Eq.~\eqref{eq:AA-factorized-formula}]. Limited proofs of factorization in $p$-$p$ collisions, as indicated by Eq.~\eqref{eq:pp-factorized-formula} exist~\cite{Collins:1985ue}.
In many calculations~\cite{Salgado:2003gb,He:2015pra,Schenke:2009gb,Qin:2009bk} $\mathscr{P}$ is simplified to only a probability for energy or forward momentum loss. In many prior discussions (e.g., Ref.~\cite{Majumder:2010qh,Majumder:2007iu} and references therein), the $\mathscr{P}$ factor is absorbed within the $\Tilde{D}$. We keep these separate to indicate that there may be medium modifications to the fragmentation (e.g. from shower thermal recombination~\cite{Hwa:2004ng,Majumder:2005jy}) beyond the loss of energy and virtuality of the propagating hard partons. 

\begin{figure}
    \centering
    \includegraphics[width=0.9\textwidth,height=0.25\textwidth]{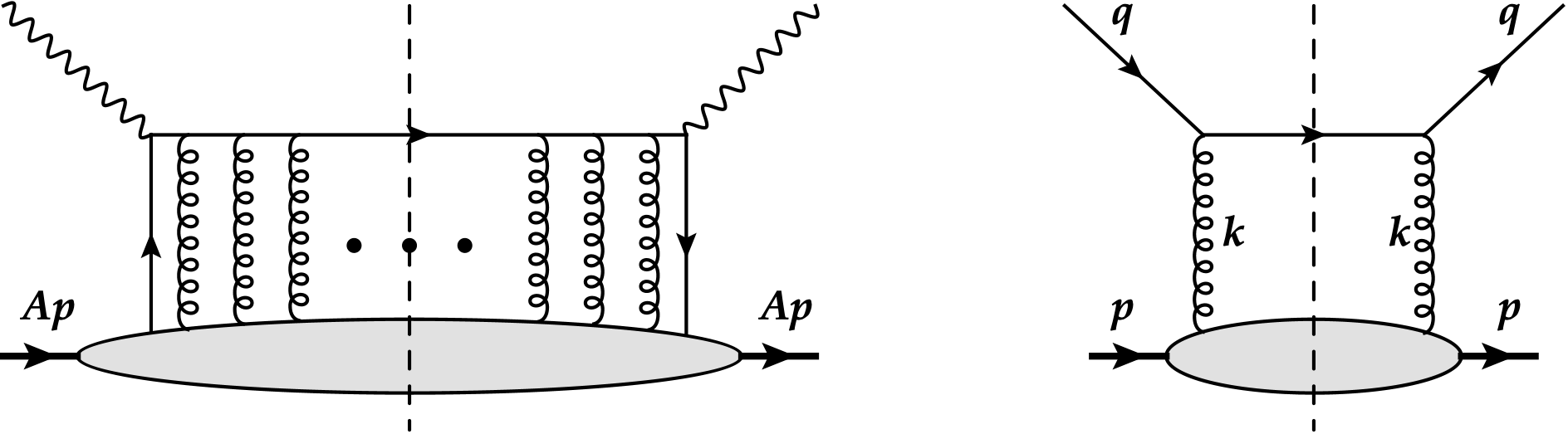}
    \vspace{-0.3cm}
    \caption{Left: A multiple scattering contribution to deep-inelastic scattering on a large nucleus. Right: One scattering of the propagating quark. }
    \label{fig:A-DIS-mult-scat}
\end{figure}

The probability of modification $\mathscr{P}$ may be further factorized into its hard partonic part and soft medium portion. The simplest example of the use of a factorized form within a medium is in the calculation of final state transverse momentum from multiple scattering in deep-inelastic scattering on a large nucleus (A-DIS). In this case, the double differential cross section for the outgoing lepton $L_2$ and the outgoing quark $l$, can be factorized as~\cite{Majumder:2007hx,Qin:2012fua},
\begin{eqnarray}
    \frac{E_{L_2} d \sigma}{d^3L_2 d^3 l } =  \Tilde{G}(x_B) \frac{ E_{L_2} d \hat{\sigma}} {d^3 L_2}  \mathscr{P}(\vec{l}).
\end{eqnarray}
The probability distribution for outgoing parton modification is given as, 
\begin{eqnarray}
    \mathscr{P} (l^+,l_\perp) \propto \exp\left[ - \frac{\left( l^+ - q^+ - \hat{e}_+\tau^+ \right)^2}{\hat{e}_{2+} \tau^+} \right] \exp\left[- \frac{l_\perp^2}{\hat{q}_+\tau^+}\right].
\end{eqnarray}
In the equation above, $\tau^+$ is the light-cone distance traveled by the single hard parton. The transport coefficients $\hat{e}_+, \hat{e}_{2+}$~\cite{Majumder:2008zg} quantify the light-cone drag and diffusion per unit light-cone length, while $\hat{q}_+$~\cite{Baier:2002tc, Majumder:2012sh, Kumar:2020pdl} quantifies the transverse diffusion per unit light-cone length experienced by the hard parton which starts with light-cone momentum $l^+ = q^+, l_\perp = 0$ and $l^- = {l_\perp^2}/{(2l^+)} = 0$. The equations above are calculated from diagrams with an arbitrary number of scatterings, and no emissions in the final state, as shown in the Fig.~\ref{fig:A-DIS-mult-scat}. Throughout this chapter we will move between the light-cone and Cartesian versions of the transport coefficients, where, for a light-like parton (traveling a distance $L$ in the $z$-direction $L^0 = L$), 
\begin{eqnarray}
    \hat{q}_+ = \sfrac{\langle k_\perp^2 \rangle_L}{L^+} =  \sfrac{ \langle k_\perp^2 \rangle_L}{(\sqrt{2}L)} = \sfrac{\hat{q}}{\sqrt{2}}.
\end{eqnarray}

Within these diagrams, the hard parton travels on-shell along light-like trajectories, and thus each of the successive scatterings in the medium is independent. One could study the effect of each separately as in the right panel of Fig.~\ref{fig:A-DIS-mult-scat}, where an incoming on-shell quark scatters off the glue field of the medium and continues to propagate on mass shell, but with changed momentum. Studying the momentum modes of the distribution of the scattered parton, yields the various transport coefficients.

\subsection{Transport coefficients}

To relate these transport coefficients such as $\hat{q}, \hat{e}, \hat{e}_2$ with properties of the medium, these have to be calculated for a given medium, and related to some intrinsic property of that medium. Differing assumptions regarding the media used, lead to a variety of expressions regarding these transport coefficients, which are used in the literature. 

We illustrate this with the case of the transverse momentum broadening coefficient $\hat{q}$, calculated from the diagram in the right panel of Fig.~\ref{fig:A-DIS-mult-scat}. Imagine a quark in a well defined momentum state $ | q \rangle \equiv |q^+, 0 ,0_\perp \rangle$ 
impinging on a medium $| M \rangle $ and then exiting 
in the state 
\[
| q + k \rangle \equiv \left| q^+ + k^+ , \sfrac{ \left( k_\perp^2 \right)}{[2 (q^+ + k^+) ] } ,  \vec{k}_\perp \right\rangle.
\]
The medium state absorbs this change in momentum and becomes $| X \rangle$. At lowest order, both incoming and outgoing quarks are assumed on-shell.

Consider the reaction in the rest frame of the medium. In this frame $q^0>0$, and we have defined the $z$-axis such 
that $q_z > 0$. We will assume that $k_\perp \ll q^+$ and  $q^- \rightarrow 0 $. We can now invoke the scaling with $\lambda$, and assume that $q^+ \sim Q$ is the hard scale. The exchanged transverse momentum $k_\perp \sim \lambda Q$, where $\lambda \ll 1$.
To keep the outgoing quark on-shell we will require 
$ k^- = {k_\perp^2}/{(2 q^+)} \,\sim \lambda^2 Q $. 
We have further assumed that $k^+ \sim \lambda^2 Q$ and is thus ignored in the above equation.


The spin-color-averaged differential transition probability (matrix element) for this 
process is given to leading order as (interaction picture), 
\begin{eqnarray}
\frac{ d W(k,q)}{d^3 k}  &=&  \frac{g^2}{2 N_c}\langle q ; M | \int d^4 x d^4 y \Bar{\psi}(y) \f\! A(y) \psi(y) \nonumber \\ 
&\times& | q + k; X \rangle 
\langle q + k ; X | \Bar{\psi}(x) \f \!A(x) \psi(x) | q; M \rangle, 
\end{eqnarray}
where, $A_\mu = t^a A^a_\mu$.  
%
%
%
%
%
The mean $k^2_\perp$ 
which yields $\hat{q}$ has the obvious definition, 
\begin{eqnarray}
\hat{q} = \int \frac{d^3 k}{(2\pi)^3}  k_\perp^2  \frac{{d W(k,q)}/{d^3 k}}{t} = \int \frac{d^3 k}{(2\pi)^3} k_\perp^2 \frac{ d\Gamma (k,q)}{d^3 k}.  \label{eq:q-hat-from-rate}
\end{eqnarray}
In the equation above $t$ represents the time spent by the 
hard quark propagating through the dense medium. Thus, $\Gamma$ is the transition probability per unit time, or the rate for a parton to scatter off the gluon field. The rate depends on the assumptions made regarding the medium.
In later sections, the rate will be calculated in a medium describable by the Hard-Thermal-Loop (HTL) effective theory~\cite{Braaten:1989kk,Braaten:1989mz,Frenkel:1989br}.

For expressions beyond using the rate, one can simplify the Dirac trace:
\begin{align}
\hspace{-.25em}
\langle M | {\rm Tr}[\f q  \f A(y) (\f q + \f k) \f A(x)] |M \rangle 
&\simeq
8 ( q^+ )^2 {\rm Tr}[t^a t^b] \langle M | A_a^- (y) A_b^- (x) | M \rangle.
\end{align}
The $k_\perp^2$ may be combined 
with the vector potentials to yield, $\nabla_\perp A^- \simeq {F_{\perp}}^{-}$. Absorbing both factors of $k_\perp$, 
we obtain an expression containing only field strength tensors.
Substituting the above simplifications, one obtains, 
\begin{eqnarray}
\mbox{}\!\!\!\!\!\hat{q}\!\! &=&\!\! \frac{4 \pi^2 \alpha_s}{ N_c } \!\! \int \!\!\frac{dy^+ d^{2} y_{\perp} d^{2} k_{\perp} }{(2 \pi)^{3}} 
e^{ -i \frac{k_{\perp}^{2}}{2q^{+} } \cdot y^{+} +  i\vec{k}_{\perp} \cdot \vec{y}_{\perp}  } 
%
\langle M | {F^{-}}_{\perp} (y^-,y_{\perp}) {F_{\perp}}^{-} (0)  | M \rangle. \label{eq:qhat-FF}
\end{eqnarray}
This is the leading order definition of $\hat{q}$. It is leading order in terms of coupling with the hard parton, but may be of arbitrary order in terms of coupling in the medium. That is to say, this expression does not constrain the source of the gluon field. Note that nothing is specified about $|M\rangle$, it may indeed be an 
arbitrary medium. 

In the subsequent subsections, we will develop the theory of jet propagation, first in vacuum and then in a dense medium. All expressions will assume \emph{maximal} factorization as discussed above. The theory will be developed with the goal of straightforward transition to event generation.

\subsection{Jets in Vacuum}

In the preceding subsections, we discussed the application of factorized pQCD in hard scattering in both $p$-$p$ and $A$-$A$ collisions. While there are typically 2 factorizations in $p$-$p$, the initial distribution from the hard scattering cross section from the final jet/fragmentation function, $A$-$A$ collisions engender an additional factorization of the energy loss distribution $\mathscr{P}$ from the hard scattering and the final fragmentation. 

\begin{figure}
\centerline{\includegraphics[width=11cm]{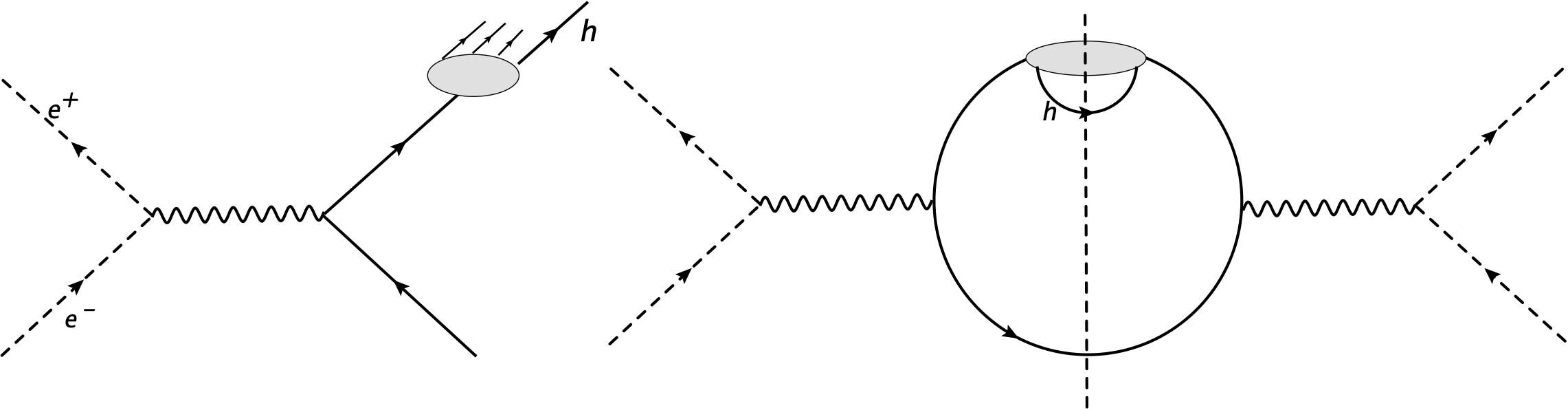}}
\vspace{-0.3cm}
\caption{ $e^+e^-$ annihilation to a hadronic state with one identified hadron or jet. Left panel shows the Feynman diagram for the amplitude, while the right panel shows the total cross section as portion of the imaginary part of the forward scattering process. }
\label{fig:epem_basic}
\end{figure}

Consider a back-to-back quark anti-quark state created in the annihilation of an electron and a positron, where the outgoing quark defines the positive $z$-axis. Let us also consider a collinear hadronic state $h$ (with momentum $p_h$) formed in the fragmentation of the outgoing quark:
\begin{equation}
    e^+ + e^-  \rightarrow \gamma^* \rightarrow \bar{q} + q \rightarrow h + X. \label{eq:chemical equation}
\end{equation}
Assuming that the outgoing quark and anti-quark will occupy opposing hemispheres, the arbitrary state $X$ will represent the entire final state without $h$.
The collinear hadronic state can refer to a single or several hadrons (either enumerated or clustered into a jet) and can either be described by a fragmentation function, written for a single hadron as, 
\begin{align}
   \mbox{}\!\!\! D_h (z) \!= & \frac{z^3}{2} \!\!\int\!\! \frac{d^4 p d^4 x}{(2\pi)^4} e^{-i p \cdot x} \delta 
    \!\left( \!z\! - \!\frac{p_h^+}{p^+} \!\right) \!\sum_X\! \langle h X | \bar{\psi}(x) | 0 \rangle 
    \!\frac{\gamma^+}{2 p_h^+}\!
    \left\langle 0 | \psi(0) | h X \right\rangle\!, \label{eq:Dz}
\end{align}
or by a jet function $J(z)$, obtained by replacing $h \rightarrow J$ in the equation above. We invoke collinear dynamics, i.e., the fragmentation function only depends on the momentum fraction $z$ of the large light cone momentum $p^+ = (p^0 + p_z)/\sqrt{2}$,  and all other dependencies have been ignored. In this review, we will ignore any transverse momentum dependence in the initial parton distribution and  final fragmentation functions, while the in-medium scattering matrix elements which yield transport coefficients like $\hat{q}$ will be dominated by their transverse momentum exchange with the projectiles.

\begin{figure}
\centerline{\includegraphics[width=11cm]{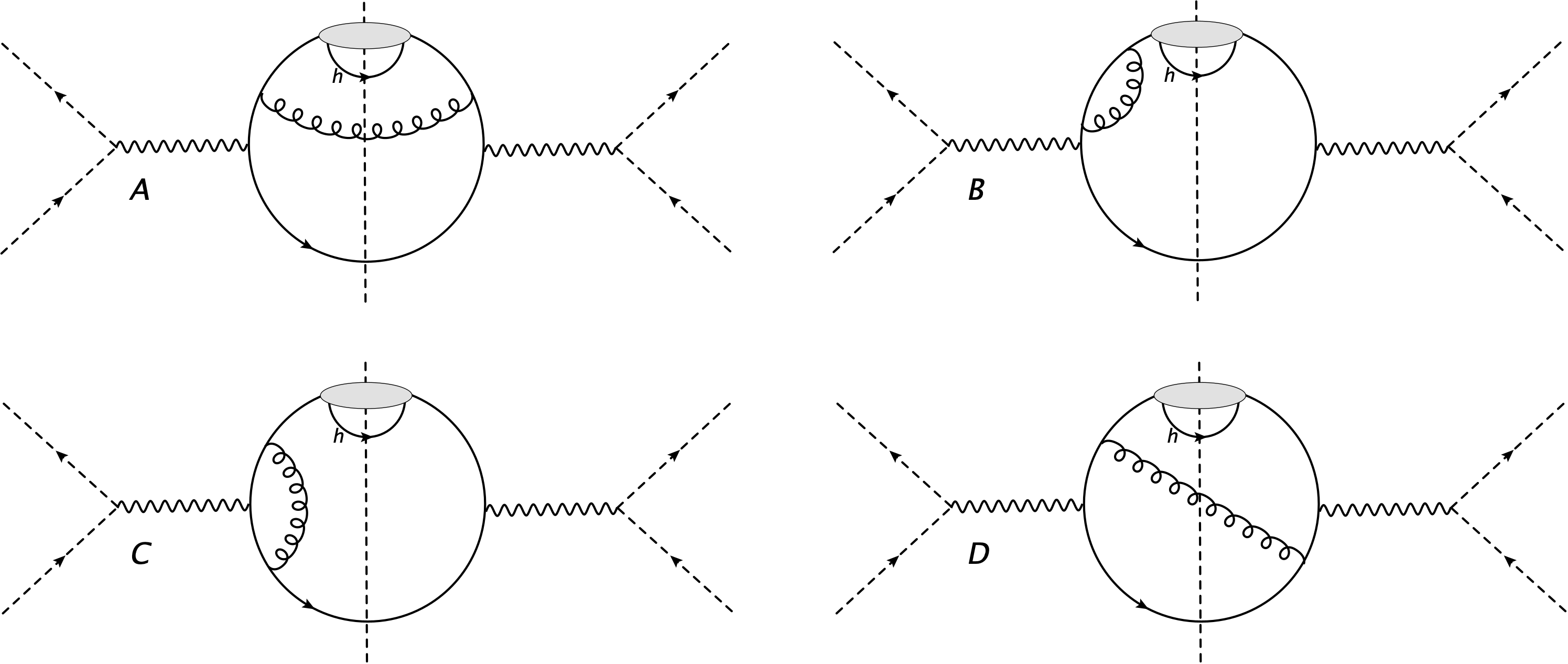}}
\vspace{-0.3cm}
\caption{ $e^+e^-$ annihilation to a hadronic state with one identified hadron or jet at NLO. One also needs to include the mirror images, across the cut line, of diagrams B, C and D to obtain the full cross section at NLO (one also needs an LSZ factor of $\frac{1}{2}$ for B). }
\label{fig:epem_NLO}
\end{figure}

The cross section for the process can be calculated using the imaginary part of the forward scattering cross-section as shown in the right panel of Fig.~\ref{fig:epem_basic}. The differential cross section as a function of $z$ can be expressed as, 
\begin{equation}
    \frac{d \sigma }{d z} = \sigma_0 D(z) ,
\end{equation}
where $\sigma_0$ is the full leading order cross section, and $D(z)$ is given in Eq.~\eqref{eq:Dz}. 
In order to develop an event generator for energy loss, one needs to know the space-time location of the quark that will undergo energy loss. At a minimum, one needs to know the region over which the quark can be treated as a parton, and at what point fragmentation will set in. 
At this order of calculation ($\alpha_S^0$), $\sigma_0$ is calculated assuming on-shell leptons and, more importantly, on-shell quarks and anti-quarks. There is no notion of virtuality, and thus no notion of where in space one could consider the fragmentation to start.

At the next-to-leading order (NLO), the diagrams in Fig.~\ref{fig:epem_NLO} have at least one gluon either emitted or absorbed from the quark line that leads to the fragmentation function, and thus we expect these diagrams to affect the fragmentation function. One typically chooses a forward light cone gauge, i.e., $A^+ = 0 $, which moves the leading log singularity into diagrams A and B, and suppresses the contributions from C and D. Summing these diagrams, we obtain the cross section at (LO + NLO) as, 
\begin{eqnarray}
\frac{d \sigma^q}{dz} &=& \sigma_0 \left[ D( z) 
\left\{1 -    \int\limits_0^1 dy \int\limits_0^{Q^2y(1-y)} 
\frac{dl_\perp^2}{l_\perp^2} \frac{\alpha_s (l_\perp^2) }{2\pi} \left[ P(y) + g(y,l_\perp^2) \right]  \right\} \right. \nonumber \\
&+&\left.   \int\limits_z^1 \frac{dy}{y} 
\int\limits_0^{Q^2y(1-y)} \frac{d l_\perp^2}{l_\perp^2} \frac{\alpha_s (l_\perp^2) }{2\pi} \left\{ P(y) + g(y,l_\perp^2) \right\}
D \left( \frac{z}{y} \right) 
\right] \label{eq:NLO_dsigma_dz} \\
&=& \sigma_0 \left[  D(z) 
+  \int_0^{Q^2} \frac{d \mu^2}{\mu^2} \frac{\alpha_S(\mu^2)}{2\pi}\int_z^1 \frac{dy}{y} \left\{ P_{+}(y) + f_{+}(y,\mu^2) \right\}
D \left( \frac{z}{y} \right) \right].  \nonumber
\end{eqnarray}
In the first line of the above equation, the virtual correction, which describes the interference between the Born term and the emission and absorption of a gluon, as in diagram B in Fig.~\ref{fig:epem_NLO},  is included as a negative correction with the LO term. It includes a 3 dimensional integration over the transverse momentum of the virtual gluon ($\vec{l}_\perp$) and the forward momentum fraction of the gluon ($y$). 
The virtual correction does not actually terminate at $l_\perp^2 = Q^2y(1-y)$, where $Q$ is the mass of the virtual photon. The portion from $Q^2 y(1-y)$ to $\infty$ is absorbed in a UV renormalization of the coupling. 
Also note that the upper limit of $l_\perp^2$ even in the real emission contribution is actually $Q^2y (1-y)$. As a result, we change variables to the virtuality of the radiating parton $\mu^2 = l_\perp^2 /[y(1-y)]$ in the last line.

The second line of Eq.~\eqref{eq:NLO_dsigma_dz} contains the real correction, the emission of a real gluon in the final state. 
In the 3rd line, the virtual function is included with the real correction using the $(+)$ prescription for the splitting function,
\begin{eqnarray}
    P_+ (y) = \lim_{\epsilon \rightarrow 0} \left[ P (y) - \delta(1 - \epsilon - y ) \int_\epsilon^{1-\epsilon} dx P(x)  \right] .  \label{eq:plusfunc}
\end{eqnarray}
While the combination of real and virtual terms removes the infrared divergence, it does not remove the collinear divergence at $\mu \rightarrow 0$.
The functions $ g(y,l_\perp^2), f(y,\mu^2)$ represent terms that contain further dependence on the transverse momentum $l_\perp$ or the scale $\mu$ and as a result will not yield a leading log contribution (in $\mu^2$). These are often called coefficient functions, i.e., terms that do not possess a collinear singularity.

The resolution of this singularity is to identify that it takes place when $l_\perp$ or 
$\mu \rightarrow 0$, i.e., at a very soft scale, at a time $\tau \sim 2E/\mu^2 $ far away from the hard scattering process. One may subtract, and include this singularity within a re-normalized definition of the fragmentation (or jet) function:
\begin{eqnarray}
    \mbox{} \!\!\!\!\!\! D(z,\mu_F^2) &=& D(z) + \!\!\int\limits_0^{\mu_F^2} \frac{d\mu^2}{ \mu^2} \frac{\alpha_S(\mu^2) }{2\pi} \int\limits_z^1 \frac{dy}{y} \left\{ P_+(y) + f_D(y,\mu^2) \right\} D \left( \frac{z}{y}\right)\!.  \label{eq:D_NLO_vac}
\end{eqnarray}
In the above equation, terms involving $f_D(y,\mu^2)$ represent parts of the coefficient function that are included within the re-normalized fragmentation function. The choice of $f_D$ defines a scheme. One could choose $f_D = 0$, or take $f_D = f$. While not explicitly written, $f_D \equiv {f_D}_+$ [see Eq.~\eqref{eq:plusfunc}]. 

At the next-to-leading order, we obtain:
\begin{eqnarray}
    \delta D &=& \int_0^{\mu_F^2} \frac{d\mu^2}{\mu^2} \frac{\alpha_S (\mu^2)}{2 \pi}
    \int_z^1 \frac{dy}{y} \left\{ P_+(y) + f_D(y,\mu^2) \right\} \int_0^{\mu^2} \frac{d 
    \Bar{\mu}^2 }{ \Bar{\mu}^2 } \frac{\alpha_S (\Bar{\mu}^2)}{2 \pi} \nonumber
    \\
   &\times& \int_z^{1} \frac{d\Bar{y}}{\Bar{y}} \left\{ P_+(\Bar{y}) + f_D(\Bar{y},
   \Bar{\mu}^2) \right\}  D \left( \frac{z}{y\Bar{y}} \right) \label{eq:delta_D_NLO} \\
    &=& \frac{1}{2!} \left[ \int_0^{\mu_F^2} \frac{d\mu^2}{\mu^2} \frac{\alpha_S(\mu^2)}{2 \pi}  %
    \int_z^1 \frac{dy}{y} \left\{ P_+(y) + f_D(y, \mu^2) \right\}
    \right]^2 \otimes D(z), \nonumber 
\end{eqnarray}
where, the convolution notation $\otimes D(z) \implies D(\sfrac{z}{y\Bar{y}} )$,  once the $y$ and $\Bar{y}$ integrations are written out. One can now easily generalize to the full resummed form of the re-normalized fragmentation function, 
\begin{eqnarray}
    D(z,\mu_F^2) = e^{  \int\limits_0^{\mu_F^2} \frac{d\mu^2}{\mu^2} \frac{\alpha_S(\mu^2)}{2 \pi}  %
    \int_z^1 \frac{dy}{y} \left\{ P_+(y) + f_D(y, \mu^2) \right\}     } \otimes D(z). \label{eq:exponential_D_+}
\end{eqnarray}
Taking $f_D = 0$, and differentiating with $\mu_F^2$ (and writing $\mu_F$ as just $\mu$ for simplicity), we obtain the well known Leading Order (LO) Dokshitzer-Gribov-Lipatov-Altarelli-Parisi (DGLAP)~\cite{Dokshitzer:1977sg,Gribov:1972ri,Gribov:1972rt,Altarelli:1977zs} equation for a non-singlet~\cite{Field:1989uq} fragmentation function, 
\begin{eqnarray}\label{eq:DGLAP}
    \mu^2 \frac{\partial D(z,\mu^2) }{ \partial \mu^2} = \frac{\alpha_S(\mu^2) }{2 \pi } \int_z^1 \frac{dy}{y} P_+(y) D \left( \frac{z}{y}, \mu^2 \right). 
\end{eqnarray}

While the above equation is very useful for calculating the evolution of the fragmentation function (and also the PDF when used for the initial state), the presence of the ($+$)-function makes event generation not straightforward. This can be remedied by expanding the ($+$)-function in the exponential in Eq.~\eqref{eq:exponential_D_+} (with $f_D=0$), yielding 
\begin{eqnarray}
     D(z,\mu^2) = \lim_{\epsilon \rightarrow 0}  e^{  \int\limits_0^{\mu^2} \frac{d\Bar{\mu}^2}{\Bar{\mu}^2} \frac{\alpha_S(\Bar{\mu}^2)}{2 \pi}  %
    \int\limits_z^1 \frac{dy}{y} \left[ P(y) - \delta(1 - \epsilon - y ) \int\limits_\epsilon^{1-\epsilon} dx P(x)  \right]     } \otimes D(z),
\end{eqnarray}
and then separating and absorbing the singular (in $\mu$) part of the integral in the exponential within a renormalized fragmentation function as,
\begin{eqnarray}
     D(z,\mu^2) = \lim_{\epsilon \rightarrow 0}  e^{  \int\limits_{\mu_0^2}^{\mu^2} \frac{d\Bar{\mu}^2}{\Bar{\mu}^2} \frac{\alpha_S(\Bar{\mu}^2)}{2 \pi}  %
    \int\limits_z^1 \frac{dy}{y} \left[ P(y) - \delta(1 - \epsilon - y ) \int\limits_\epsilon^{1-\epsilon} dx P(x)  \right]     } \otimes D(z,\mu_0^2).
\end{eqnarray}
The scale $\mu_0$ is intended to be the lower limit at which one can apply pQCD, typically taken to be $\mu_0 \simeq 1$~GeV. 
One then notices that the second term in the exponent does not participate in the convolution ($\otimes$), because of the $\delta(1-\epsilon-y)$ factor, and can thus be separated out as, 
\begin{eqnarray}
    D(z,\mu^2) &=& \Delta( \mu^2, \mu_0^2) e^{ \int\limits_{\mu_0^2}^{\mu^2} \frac{d\Bar{\mu}^2}{\Bar{\mu}^2} \frac{\alpha_S(\Bar{\mu}^2)}{2 \pi}  %
    \int\limits_z^1 \frac{dy}{y}  P(y)  } \otimes D(z,\mu_0^2).
\end{eqnarray}
Where, $\Delta(\mu^2 , \mu_0^2 )$ is called the Sudakov form factor, and includes all order contributions to the case of no resolvable emission. 

In the absence of the portion of the expression above that deals with real emissions, the meaning of the Sudakov form factor becomes clear:
$D(z,\mu^2)  =  \Delta(\mu^2,\mu_0^2) D(z, \mu_0^2) $; 
it represents the probability that a parton with a virtuality at or below $\mu^2$ has transitioned via unresolvable emission to (or only contained) a parton with a virtuality at or below $\mu_0^2$, prior to the fragmentation process, that produces a state carrying away a fraction $z$ of the original parton's light-cone momentum. Considering the decay of a parton with virtuality $\mu^2$ into two other partons with virtuality $\mu_0^2$, the limits of the $y$ integration can be easily computed to be, 
\begin{eqnarray}
    y_{min} \equiv \epsilon = \frac{1}{2} - \frac{\sqrt{ 1 - 4 \mu_0^2/\mu^2  }  }{2} \simeq \frac{\mu_0^2}{\mu^2}, \;\;\; y_{max} = 1 - y_{min}.
\end{eqnarray}
These are physical limits that are more constraining than the $\epsilon \rightarrow 0$ limits placed [in Eq~\eqref{eq:plusfunc}]. One thus obtains the final and finite form of the Sudakov form factor as, 
\begin{eqnarray}
    \Delta(\mu^2 , \mu_0^2 ) =  \exp\left[ -\int\limits_{\mu_0^2}^{\mu^2} \frac{d \Bar{\mu}^2}{\bar{\mu}^2} 
\frac{\alpha_S(\Bar{\mu}^2)}{2 \pi}
\int\limits_{\sfrac{\mu_0^2}{\Bar{\mu}^2}}^{1 - \sfrac{\mu_0^2}{\Bar{\mu}^2}} dy P(y)  \right] . \label{eq:vac_sudakov}
\end{eqnarray}
Sampling the Sudakov factor allows for the determination of the parent parton's virtuality. Sampling the splitting function then allows for a determination of the splitting momentum fraction $y$. The forward light-cone variable $p^+$ is split between the offspring partons as $yp^+$ and $(1-y)p^+$, and this allows a determination of the transverse momentum between the two outgoing offspring partons, 
\begin{eqnarray}
    l_\perp^2 = \mu^2 y(1-y) - \mu_1^2 (1-y) - \mu_2^2 y.
\end{eqnarray}
Where $\mu_1^2$ is the virtuality of the parton with a fraction $y$ of the parent's light-cone momentum (limited by a maximum value of $\mu_1^2 \leq y^2 \mu^2$), and $\mu_2^2$ is that for the parton with a fraction $1-y$ [limited by a maximum value of $ \mu_2^2 \leq (1-y)^2 \mu^2$]. 

The determination of both $l_\perp$ and $y$ immediately allows for a computation of the mean light-cone formation time of the split, 
\begin{eqnarray}
    \tau^+ = \frac{2p^+ y (1-y)}{l_\perp^2}. 
\end{eqnarray}
For a quark splitting into a quark and gluon, the light-cone formation time represents the light-cone time required for the parent quark to complete one oscillation ($p^- \tau^+ \sim 1$), which establishes its wave number. The formalism developed so far, since Eq.~\eqref{eq:Dz} has no notion of position or time. This is because, all position-time variables have been integrated out. To accurately incorporate position-time information with momentum information, one has to use a Wigner-Function formalism, which was first set up in Ref.~\cite{Majumder:2013re}. For the purpose of this review we will posit that for each parton with a forward light-cone momentum $p^+$ and an off-shellness $\mu^2$, the mean lifetime is given as $2p^+/\mu^2$. This assumption is sufficient to understand that a parton that starts out with a virtuality $\mu^2 \sim Q^2$ (the hard scale), will take some time (and as a consequence, some distance) to drop down to a virtuality of $\mu^2 \sim \mu_{med}^2 $, a scale where it is strongly affected by the medium, by the process of repeated splittings.

\subsection{Jets in a medium: High virtuality} 
\label{sec:jets_in_a_medium_high_virtuality}

Partons produced in a hard scattering tend to be considerably off their mass shell. This off-shellness or virtuality is lost  in a series of emissions. As argued above, these emissions engender a formation time, and as such lead to a propagation of the parton within this time. As a result, when the hard partons propagate through a medium there is a delay before the virtuality of the parton reaches a value where medium effects, i.e., multiple scattering, dominantly influence the process of splitting. The effect of scattering on the hard parton and its splitting is codified by the transport coefficient $\hat{q}$, defined as the mean square transverse momentum exchanged per unit length (defined here using the large light-cone length):
\begin{eqnarray}
    \hat{q}_+ = \frac{1}{N_{Events}} \sum_{i}^{N_{Events}} \frac{\langle {k^i_\perp}^2 \rangle }{L_i^+}.
\end{eqnarray}

This later stage of parton evolution is described by the condition that the virtuality of the parton is equal to that caused by multiple scattering within its formation time, 
\begin{eqnarray}
    \mu^2 = \mu_{med}^2 = \hat{q}_+ \tau^+ = \hat{q}_+ \frac{2 p^+}{ \mu_{med}^2} \;\; \implies \mu_{med}^2 = \sqrt{ 2p^+ \hat{q}_+ },
\end{eqnarray}
which further yields $\tau_{med}^+ = \sqrt{{2p^+}/{\hat{q}_+}}$. For a parton with a large energy (or $p^+$), there is a considerable range of time during which the parton remains at a virtuality $\mu^2 \gtrsim \mu_{med}^2$. This stage is referred to as the high virtuality sector of parton evolution.

\begin{figure}
\centerline{\includegraphics[width=9cm]{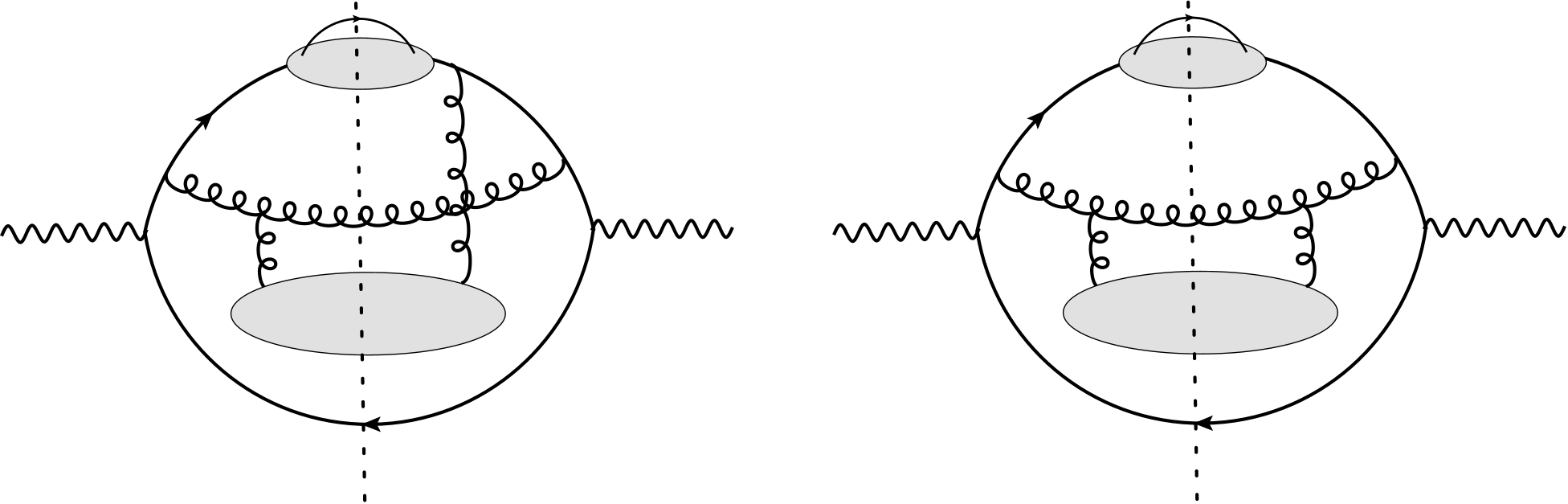}}
\caption{The first scattering correction to the processes of Fig.~\ref{fig:epem_NLO}. The lower blob, represents a QCD medium penetrated by the outgoing $q , \Bar{q}$, and radiated gluon.   }
\label{fig:single_scatter_NLO}
\end{figure}

Consider the process of Eq.~\eqref{eq:chemical equation}, taking place within an evolving medium. The simplest extension of the diagrams of Fig.~\ref{fig:epem_NLO} is shown in Fig.\ref{fig:single_scatter_NLO}. This represents one of several diagrams, which include the first perturbative correction from scattering in the medium to the process of splitting. Counting couplings that lead to the formation of a radiated gluon, (and absorbing couplings related to scattering within the eventual definition of the in-medium transport coefficient), we write down the medium modified correction to the NLO contribution to the fragmentation function as, 

\begin{eqnarray}
     D(z,\mu_F^2) &=& D(z) + \!\!\int\limits_0^{\mu_F^2} \frac{d\mu^2}{ \mu^2} \frac{\alpha_S(\mu^2) }{2\pi} \left[ \int\limits_z^1 \frac{dy}{y} \left\{ P_+(y) + f_D(y,\mu^2) \right\} D \left( \frac{z}{y}\right)\!  \right. \nonumber \\
    &+& \left. \int\limits_z^1 \frac{dy}{y}  P_+(y) \int\limits_0^{\#\tau^+} d \xi^+ \hat{q}(\xi^+) \frac{f_M (y, \xi^+, \mu^2)}{\mu^2} D\left(\frac{z}{y} \right) \right]. \label{eq:D_NLO_1_scat}
 \end{eqnarray}

In the equation above, $\tau^+$ is the formation time and the $\#$ preceding it indicates that the integration is taken up to a multiple of the formation time (typically 3). The factor $f_M$ represents an $\mathcal{O}(1)$ function that correlates the location of the scattering with the momentum fraction and virtuality of the parent parton in the split. The full form of this expression is derived in Ref.~\cite{Sirimanna:2021sqx}. An earlier simplified form which yields results in close agreement with the full expression and with data is:
\begin{eqnarray}
    f_M(y, \xi^+, \mu^2) = \frac{2 - 2 \cos\left( \frac{\xi^+}{\tau^+}\right)}{y(1-y)} = \frac{2 - 2 \cos\left( \frac{\mu^2 \xi^+}{2p^+ } \right)}{y(1-y)}. \label{eq:fM}
\end{eqnarray}
One immediately notes that the size of the in-medium contribution is modulated by the factor ${\int d\xi^+ \hat{q}}/{\mu^2}$. When this factor is small compared to unity, one re-scattering per emission is sufficient and the perturbation in scattering can terminate at Eq.~\eqref{eq:D_NLO_1_scat}. If one were to add additional scatterings, each additional scattering would engender a factor of $\mu_{med}^2/\mu^2$. 

Following the steps from Eq.~\eqref{eq:D_NLO_vac} to~\eqref{eq:vac_sudakov}, we can derive the medium modified Sudakov form factor as, 
\begin{eqnarray}
    \Delta_M (\mu^2, \mu_0^2) &=& \exp \left[ -\int\limits_{\mu_0^2}^{\mu^2}  \frac{d \Bar{\mu}^2}{\bar{\mu}^2} 
\frac{\alpha_S(\Bar{\mu}^2)}{2 \pi}
\int\limits_{\sfrac{\mu_0^2}{\Bar{\mu}^2}}^{1 - \sfrac{\mu_0^2}{\Bar{\mu}^2}} dy P(y) 
\Bigg\{ 1 \right.  \nonumber \\
&+&  \left.  \frac{ \int_{0}^{\# \tau^+}  d \xi^+  \hat{q} (\vec{r} + \hat{n}\xi^+)
f_M(y,\Bar{\mu}, \xi^+) }{\Bar{\mu}^2} \Bigg\} \right] . \label{eq:in-medium-sudakov}
\end{eqnarray}
One notices a clear enhancement of the Sudakov factor due to re-scattering, implying a decrease in the probability of no emission. 
Thus scattering in the medium increases the probability of emission. Once the virtuality of the radiating parton is determined by sampling the Sudakov, the splitting fraction is determined by sampling the medium modified splitting function: $P_M(y) = P(y)\left[1 + \int_0^{\# \tau^+} d\xi^+ \hat{q} f_M(y,\mu,\xi^+) /\mu^2 \right]$. 

\begin{figure}
    \centering
      \includegraphics[width=0.4\textwidth,height=0.375\textwidth]{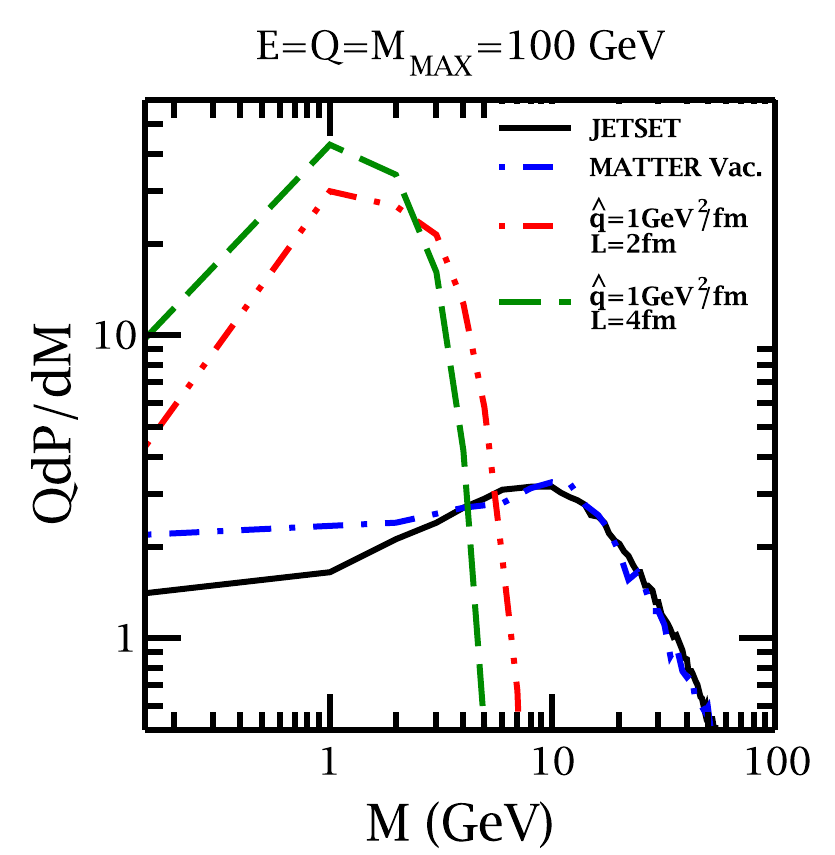}
      \hspace{1cm}
    \includegraphics[width=0.4\textwidth,height=0.375\textwidth]{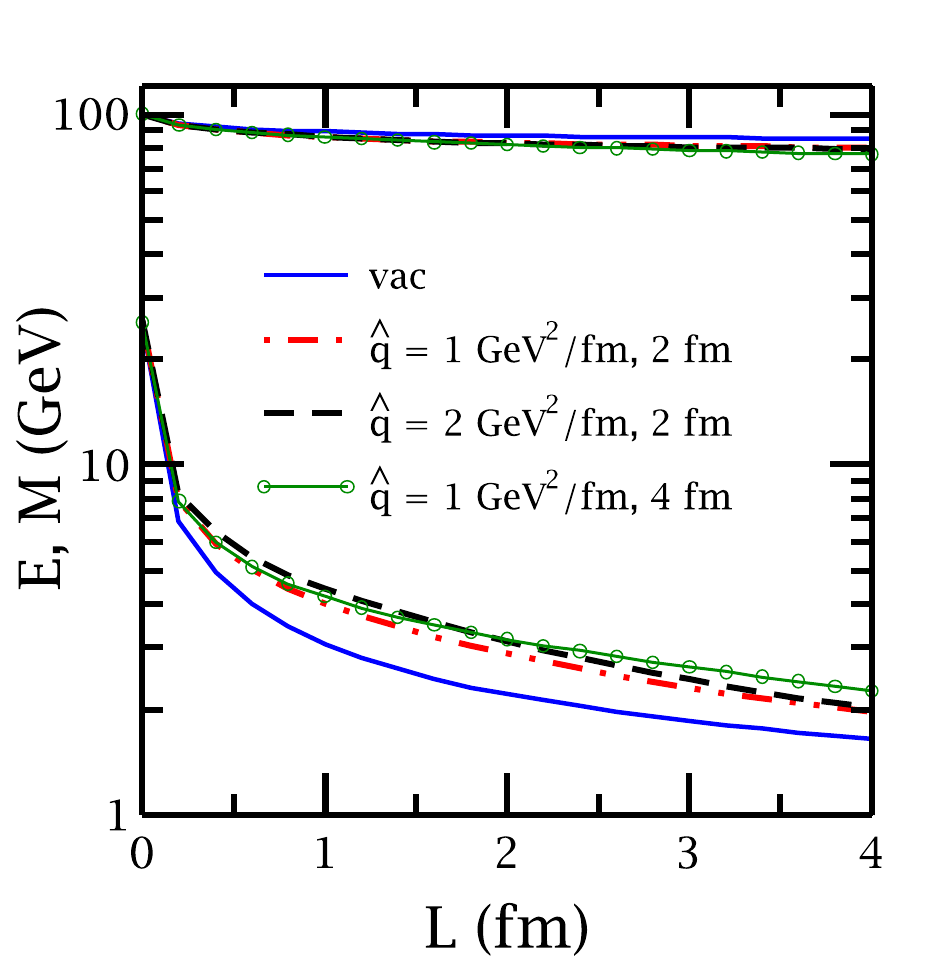}
    \vspace{-0.3cm}
    \caption{Left: The virtuality distribution of the hardest parton in a shower at $t=0$~fm/$c$ in both JETSET and MATTER. Originating parton has $E=100$~GeV. Also plotted are the virtuality distributions at $2$~fm/$c$ and $4$~fm/$c$ for the hardest parton in the shower. Right: The mean of these distributions is plotted as a function of length traversed in a static medium. One notes that the virtuality of the hardest parton in a medium is slightly larger than its virtuality in a vacuum. Plots are taken from Ref.~\cite{Majumder:2014gda}}
    \label{fig:parton_virtuality_distribution}
\end{figure}

Assuming that $f_M$ is given by Eq.~\eqref{eq:fM}, one can plot the distribution of initial virtualities and the virtuality of the leading parton that traverses some distance in a static medium in Fig.~\ref{fig:parton_virtuality_distribution}. The plots indicate that the virtuality of the leading parton in a dense medium is actually larger (though barely) than the virtuality in vacuum. 
This is to be expected: once the parton transitions to the low virtuality stage, the average virtuality will be held at $\mu_{med}^2 = \hat{q} \tau$, which will eventually be higher than the virtuality at that point in a vacuum shower.

This method of calculating ``vacuum-like" emissions with small corrections to the probability is only valid as long as the factor ${\int d\xi^+ \hat{q}}/{\mu^2}$, which emanates from single re-scattering is $\lesssim 1$. At the start of the shower, when $\mu^2$ is large, and as a result $\tau^+ = {2p^+}/{\mu^2}$ is small, the factor is $\ll 1$. However, this grows with each successive split. As the factor approaches unity, the probability of multiple scattering also grows and will quickly subsume the pure vacuum contribution. 

While the transport coefficient $\hat{q}$ can be calculated in many different ways, as highlighted in the upcoming sections, it will be shown in the section on event generation that a very important aspect of jet quenching is the interaction with the medium. This involves both the perturbative aspect of recoil as well as the non-perturbative portion of energy deposition and thermalization. The first of these, has to be constructed in such a way that it reproduces the $\hat{q}$ used in both high and low virtuality stages. The distribution of transverse momentum exchange in the scattering between jet and in-medium partons, should yield $\hat{q}$ as its second moment.

  
    \subsection{Jets in a medium: Low virtuality} 
            As partons continue to propagate through a medium, their virtuality continues to drop, albeit slower than in the vacuum. At some point, the virtuality generated by multiple scattering ($\mu^2 \simeq 
\hat{q}_+ \tau^+$), becomes comparable to or larger than that of a parton in vacuum undergoing multiple splits ($\mu^2 = 2p^+/\tau^+$). At this stage, multiple scattering induced energy loss begins to dominate over vacuum emission. This stage starts at a different point for each parton, depending on its  momentum $p^+$, and the density of the ambient medium ($\hat{q}$):
\begin{equation}
    \tau^+_{\rm Low Virt.} \gtrsim \sqrt{ {2p^+}/{\hat{q}} } \,\,. 
\end{equation}

Most models of jet energy loss in QGP medium fall in this lower virtuality category.
Due to the larger interaction rate with the medium, the jet undergoes multiple scatterings per emission leading to the largest fraction of the jet energy loss taking place at this stage.

In this section, we describe two major approaches to low-virtuality energy loss calculations: opacity expansion and finite-temperature field theory. The two formalisms are related, and indeed, it has been shown that they can be arrived at from the BDMPS-Z formalism~\cite{Arnold:2008iy}.  

\subsubsection{BDMPS-Z}
    The BDMPS-Z formalism seeks to compute the radiated gluon spectrum, and thus the radiative energy loss of fast partons traveling through a QGP medium. For a process where a parton of momentum $\mathbf{P}=\mathbf{p}+\mathbf{k}$ splits to a gluon of momentum $\mathbf{k}$ and a quark of momentum $\mathbf{p}$, the radiated gluon spectrum is given by~\cite{Arnold:2008iy}
    \begin{align}
        \omega \frac{d}{d\omega} &\left(I - I_{\mathrm{vac.}}\right) =\; \frac{\alpha_s\; z\; P_{a\to bc}(z)}{[z(1-z)E]^2} \mathrm{Re} \int_{0}^{\infty} dt_1 \int_{t_1}^{\infty} dt_2 \nonumber \\
        &\left[\mathbf{\nabla}_{\mathbf{x}}\cdot\mathbf{\nabla}_{\mathbf{y}}\{G(\mathbf{y},t_2;\mathbf{x},t_1)-G_{\mathrm{vac}}(\mathbf{y},t_2;\mathbf{x},t_1)\} \right]_{x=y=0}
        \label{eq:BDMPSZ_gluon_spectrum_config_space}
    \end{align}
    where $z\equiv k/P$ is the momentum fraction taken by the gluon. The function $G(\mathbf{x},t;\mathbf{x}',t_1)$ is the Green's function solution to the Schr\"odinger equation with the Hamiltonian describing the dynamics of the radiating system in the transverse plane. It is given by
    \begin{equation}
        H(t) = \delta E(\mathbf{h},t) - i\Gamma_3(\mathbf{x},t).
        \label{eq:bdmpsz.hamiltonian}
    \end{equation}
    The first term of the Hamiltonian is the energy difference between an initial state with a high-energy parton of momentum $\mathbf{p}+\mathbf{k}$ and a final state with a parton of momentum $\mathbf{p}$ and a radiated gluon of momentum $\mathbf{k}$. It can be written as
    \begin{equation}
        \delta E(\mathbf{h},t) = \frac{h^2}{2z(1-z)P}+\frac{m^2_q(t)}{2(1-z)P} + \frac{m^2_g(t)}{2zP} - \frac{m^2_q}{2P}
        \label{eq:energy.diff.bremsstrahlung}
    \end{equation}
    where 
    \begin{equation}
        \mathbf{h}\equiv \frac{p\mathbf{k}_{\perp} - k\mathbf{p}_{\perp}}{p+k}.
    \end{equation}
    The second term in Eq.~(\ref{eq:bdmpsz.hamiltonian}) describes the interaction of the system with the medium
    \begin{equation}
        \Gamma_3(\mathbf{x},t) = \sum_{\beta \in \{1, z, (1-z)\}} C_{\beta} \bar{\Gamma}_2(\beta\;\mathbf{x},t) 
    \end{equation}
    with $z\equiv k/P$. 
    The factors $C_{\beta}$, are combinations of quadratic Casimirs of the partons involved in the process,
    \begin{align}
        C_{(1)} =\;& \frac{1}{2}\left(-C^{1}_{R}+C^{z}_{R}+C^{(1-z)}_{R}\right)\nonumber\\
        C_{(z)} =\;& \frac{1}{2}\left(\;\;C^{1}_{R}\;-C^{z}_{R}+C^{(1-z)}_{R}\right)\nonumber\\
        C_{(1-z)} =\;& \frac{1}{2}\left(\;\;C^{1}_{R}\;+C^{z}_{R}-C^{(1-z)}_{R}\right)
        \label{eq:casimir.factors}
    \end{align}
    with $C_R$ equalling $C_A=4/3$ for gluons and $C_F=4$ for quarks. $\bar{\Gamma}_2$ is associated with the Fourier transform of the elastic collision rate with a Casimir factor removed. It is given by
    \begin{equation}
        \bar{\Gamma}_2(\mathbf{x},t) = \int d^2 q_{\perp} \frac{d\bar{\Gamma}_{\mathrm{elas.}}(t)}{d^2q_{\perp}} \left(1-e^{i\mathbf{q}_{\perp}\cdot\mathbf{x}}\right). 
    \end{equation}
    An important assumption in BDMPS-Z is the nature of the soft scatterings with the medium. The medium is thought of as a series of well-separated, static sources, located at $\mathbf{x}_{i}$, which source a colour-screened Yukawa potential (GW model~\cite{Gyulassy:1993hr})
    \begin{equation}
        v_{i}(\mathbf{q}) = 4\pi \;\alpha_s\;T^a(R)\otimes T^{a}(i) \;\frac{e^{-i\mathbf{q}\cdot\mathbf{x}_i}}{\mathbf{q}^2+m^2_D}
        \label{eq:gw.potential}
    \end{equation}
    where $g$ is the strong coupling constant, $T^a(R)$ and $T^{a}(i)$ are the generator of $SU(3)$ for the incident and target partons, respectively. Finally, $m_D$ is the Debye mass
    \begin{equation}
        m^2_D = \frac{2\pi\;\alpha_{s}}{3}\left(N_{\mathrm{c}} + \frac{N_{\mathrm{f}}}{2}\right) \;T^2.
    \end{equation}
\label{sec:BDMPSZ}
        \subsubsection{Opacity expansion} \label{sec:OE}
                The opacity expansion approach to jet energy loss stems from the principal assumption that the scattering centers in a QGP medium are well separated. Stated more directly, the assumption is that the distance between the scattering centers is much larger than the length-scale set by the Debye mass: $\Delta x = x_{i}-x_{i-1} \gg m^{-1}_D$~\cite{Gyulassy:1993hr}. This is a reasonable assumption in the extremely high temperature limit of QGP where the strong coupling $g$ is small, effectively assuming a thin QGP composed of a collection of well-separated charges. 
    
    The initial attempts at the opacity expansion by Gyulassy, Levai and Vitev (GLV)~\cite{Gyulassy:1999zd,Gyulassy:2000er} and independently by Wiedemann~\cite{Wiedemann:2000za}, then, considered a static source approximation. This meant that these static sources were modeled with the GW potential of Eq.~\eqref{eq:gw.potential}. 
    Considering a ``thin'' QGP medium leads naturally to the idea of opacity expansion, where the radiated gluon spectrum is expanded in powers of opacity $L/\lambda_{\mathrm{mfp}}$~\cite{Gyulassy:2000er},
    \begin{equation}
        \frac{L}{\lambda_{\mathrm{mfp}}} = \frac{L}{(\rho \sigma_{\mathrm{el}})^{-1}} = \int_0^{L} dz \int d^2\mathbf{q} \frac{d\sigma_{\mathrm{el}}}{d^2\mathbf{q}}\rho(z, \tau=z) ,
    \end{equation}
    where $\rho(z)$ is the number density of scattering centers in the medium. The GLV model now assumes that the jet is produced inside the medium, with production amplitude given by~\cite{Gyulassy:1999zd}
    \begin{equation}
        \mathcal{M}_J = J(P) e^{iPx_0},
    \end{equation}
    where the jet parton is produced by source $J$, a slowly varying function and with momentum $P_0$ at space-time location $x_0$. The amplitude for the production of a hard parton which undergoes $n_s$ elastic scatterings, exchanging $q_i$ with scatterings centers located at $x_i$, with an associated gluon bremsstrahlung, is given by~\cite{Gyulassy:1999zd}
    %
    %
    %
    %
    
    \begin{equation}
       \begin{split}
           \mathcal{M}^{(n_s)}_J\otimes \mathcal{M}_{n_s} =& \int d^4P\; J(P) e^{iPx_0}\int \prod_{i=1}^{n_s} \Big(-2iP^{0}V(\mathbf{q}_i)e^{-i\mathbf{q}_i\cdot\mathbf{x}_i}T_{a_i}\Big)\\
           &\times \mathcal{M}_{n_s}(k,p;q_1,\cdots,q_{n_s})\delta^{4}(P-k-p+\sum q_i).
       \end{split}
    \end{equation}
    In the above, $\mathcal{M}_{n_s}$  is the sum over all time-ordered diagrams with $n_s$ scatterings where a gluon of momentum $k$ is emitted. The potential $V(\mathbf{q}_i)$ is related to Eq.~\eqref{eq:gw.potential}
    \begin{equation}
        V(\mathbf{q}) = \frac{4\pi\alpha_s}{\mathbf{q}^2 + m^2_D}.
    \end{equation}
    Note that the elastic scatterings can also be experienced by the emitted gluon. To arrive at an expression for the inclusive gluon spectrum, the above amplitude is squared and averaged over initial and summed over all final state colors. Further taking an eikonal limit and assuming static scattering centers allows for writing the general formula for the induced gluon emission to the $n$-th order in opacity is given by~\cite{Gyulassy:2000er}

    %
    \begin{align}
        z \frac{dN^{(n)}}{dzd^2\mathbf{k}_{\perp}} =\;&\frac{C_R\;\alpha_s}{\pi^2}\frac{1}{n!}\left(\frac{L}{\lambda_g(1)}\right)^n\int\prod_{i=1}^{n}\left(d^2\mathbf{q}_i\left(\frac{\lambda_g(1)}{\lambda_g(i)}\right)\right)\left[\bar{v}^2_i(\mathbf{q}_i)-\delta(\mathbf{q}_i)\right]\nonumber\\
        &\times\Bigg(-2\mathbf{C}_{(1 \cdots n)}\cdot\sum_{m=1}^{n}\mathbf{B}_{\left(m+1\;\cdots n\right)\left(m \cdots n\right)}\nonumber \\
        &\times \left[\cos{\left(\sum^{m}_{k=2}\omega_{\left(k \cdots n\right)}\Delta x_{k}\right)}-\cos{\left(\sum^{m}_{k=1}\omega_{\left(k \cdots n\right)}\Delta x_{k}\right)}\right]\Bigg),
        \label{eq:glv.nth.OPE.gluon.dist}
    \end{align}
    where $\mathbf{k}_{\perp}$ is the transverse momentum of the gluon relative to the incoming hard parton. The mean-free-path of the gluon, $\lambda_g(i)$ and the number density of the scattering centers, $\rho$, can vary along the path of the hard parton and the notation $\lambda_g(i)$ is used to signify this fact. The terms $\mathbf{C}$ and $\mathbf{B}$, called the cascade and Gunion-Bertsch terms, respectively, are given by
    \begin{align}
       \mathbf{C}_{(1\; 2\cdots m)} =\;&\frac{\left(\mathbf{k}_{\perp}-\mathbf{q}_{1}-\mathbf{q}_{2}\cdots-\mathbf{q}_{m}\right)}{\left(\mathbf{k}_{\perp}-\mathbf{q}_{1}-\mathbf{q}_{2}\cdots-\mathbf{q}_{m}\right)^2}\nonumber\\
       \mathbf{B}_{(i_1 i_2 \cdots i_m)(j_1 j_2 \cdots j_n)} =\;& \mathbf{C}_{(i_1 i_2\cdots i_m)} - \mathbf{C}_{(j_1 j_2 \cdots j_n)} \nonumber \\
       \mathbf{B}_{(i)} =\;& \frac{\mathbf{k}_{\perp}}{\mathbf{k}^2_{\perp}}- \mathbf{C}_{(i)}\nonumber\\
       \mathbf{B}_{(n+1 \cdots n)(n)} \equiv\;& \mathbf{B}_{(n)}.
       \label{eq:cascade.gunion.bertsch.terms}
    \end{align}

    The phase factor in the cosine terms, $\omega_{(k\cdots n)}\Delta x_{k}$, is the LPM phase factor that controls the destructive interference. Finally, the inverse formation time is given by
    \begin{equation}
        \omega_{(1\cdots n)} = \frac{\left(\mathbf{k}_{\perp} - \mathbf{q}_{1}\cdots-\mathbf{q}_{n}\right)^2}{2zE}.
        \label{eq:inv.form.time.GLV}
    \end{equation}

    The DGLV opacity expansion~\cite{Djordjevic:2003zk,Djordjevic:2007at,Djordjevic:2008iz,Djordjevic:2009cr} introduced thermal masses for quarks to the GLV framework and replaced the static scattering sources in the medium with dynamical ones by using the Hard Thermal Loop (HTL) gluon propagator. The move to dynamic scattering centers results in a change in the mean free path and the replacement of the scattering potential to an HTL potential~\cite{Djordjevic:2007at,Djordjevic:2008iz}
    \begin{equation}
        \lambda_{\mathrm{static}} \Rightarrow\;\lambda_{\mathrm{dynamic}},\;
        \frac{1}{\mathbf{q}^2 + m^2_D} \Rightarrow\; \frac{1}{\mathbf{q}^2 \left(\mathbf{q}^2+m^2_D\right)}.
    \end{equation}
    The effective dynamic mean free path is given by $\lambda_{\mathrm{dynamic}} = (3\alpha_s T)^{-1}$ and is related to the static mean free path via~\cite{Djordjevic:2007at,Djordjevic:2008iz} {$\lambda_{\mathrm{dynamic}} = 9 \frac{1.202}{\pi^2}\frac{4 + N_{f}}{6 + N_{f}}\lambda_{\mathrm{static}}$}.     
    %
    The inclusion of the quark mass and accounting for the thermal gluon mass also results in modifications to the denominators of cascade and Gunion-Bertsch terms of Eq$.$~\ref{eq:cascade.gunion.bertsch.terms} as well as the inverse formation time of Eq$.$~\ref{eq:inv.form.time.GLV}~\cite{Djordjevic:2007at,Djordjevic:2008iz}
    \begin{align}
        \mathbf{C}_{(1\; 2\cdots m)} =\;&\frac{\left(\mathbf{k}_{\perp}-\mathbf{q}_{1}-\mathbf{q}_{2}\cdots-\mathbf{q}_{m}\right)}{\left(\mathbf{k}_{\perp}-\mathbf{q}_{1}-\mathbf{q}_{2}\cdots-\mathbf{q}_{m}\right)^2 + \chi^2}\nonumber\\
        \mathbf{B}_{(i)} =\;& \frac{\mathbf{k}_{\perp}}{\mathbf{k}^2_{\perp}+\chi^2}- \mathbf{C}_{(i)} \nonumber\\
        \omega_{(1\cdots n)} =\;& \frac{\left(\mathbf{k}_{\perp} - \mathbf{q}_{1}\cdots-\mathbf{q}_{n}\right)^2 + \chi^2}{2zE},
    \end{align}
    where $\chi^2 = M^2 z^2_+  + m^2_g (1-z_+)$ where $M$ is the quark mass and $z_+ \equiv k^+/p^+$ is the fraction of the plus-momentum of the radiated gluon to that of the jet. Finally, $m_g = m_D/\sqrt{2}$ is the gluon plasmon mass. 
    For completeness, we present the first order ($n=1$) gluon bremsstrahlung spectrum in the DGLV opacity expansion for a fluid at rest, with a running coupling~\cite{Xu:2014ica, Xu:2014wua, Xu:2015bbz, Xu:2014tda}
    \begin{equation}
        \begin{split}
            &z_E\frac{\mathrm{d}N^{(1)}_{i \to g i}}{\mathrm{d}z_E}=\;\frac{18 C_R}{\pi} \frac{4+N_f}{16+9N_f}\int d\tau \rho(\mathbf{x}) \int d^2k_{\perp} \alpha_s\left(\frac{\mathbf{k}^2_{\perp}}{z_+ (1-z_+)}\right)\\
            &\times \int d^2q_{\perp} \frac{\alpha^2_s(\mathbf{q}^2_{\perp})}{\mathbf{q}^2_{\perp}(\mathbf{q}^2_{\perp}+m^2_D(\mathbf{x}))} \frac{-2(\mathbf{k}_{\perp}-\mathbf{q}_{\perp})}{(\mathbf{k}_{\perp}-\mathbf{q}_{\perp})^2 + \chi^2(\mathbf{x})}\\
            &\times\left[\frac{k_{\mathrm{\perp}}}{\mathrm{k}^2_{\perp}+\chi^2(\mathbf{x})} - \frac{(\mathbf{k}_{\perp}-\mathbf{q}_{\perp})}{(\mathbf{k}_{\perp}-\mathbf{q}_{\perp})^2 + \chi^2(\mathbf{x})}\right]\\
            &\times\left[1-\cos{\left(\frac{(\mathbf{k}_{\perp}-\mathbf{q}_{\perp})^2+\chi^2(\mathbf{x})}{2z_{+} E}\tau\right)}\right]\left(\frac{z_E}{z_+}\right)\left|\frac{dz_+}{dz_E}\right|,
        \end{split}
        \label{eq:lo_dglv_gluon_brem_spec}
    \end{equation}
    where $\tau$ replaces $\Delta x$ in the argument of the cosine (see Eq.~\eqref{eq:glv.nth.OPE.gluon.dist}) and $z_E \equiv k/p$ is the fractional energy and is related to the fractional plus-momentum via $z_+ = z_E \left[1 + \sqrt{1-\left(k_{\perp}/(z_E E)\right)^2}\right]$. The Debye mass and the density of scatterers are also shown as explicit functions of $\mathbf{x}$, the space-time position of the hard parton at proper time $\tau$.  
    \begin{figure}
        \centering
        \includegraphics[width=\textwidth]{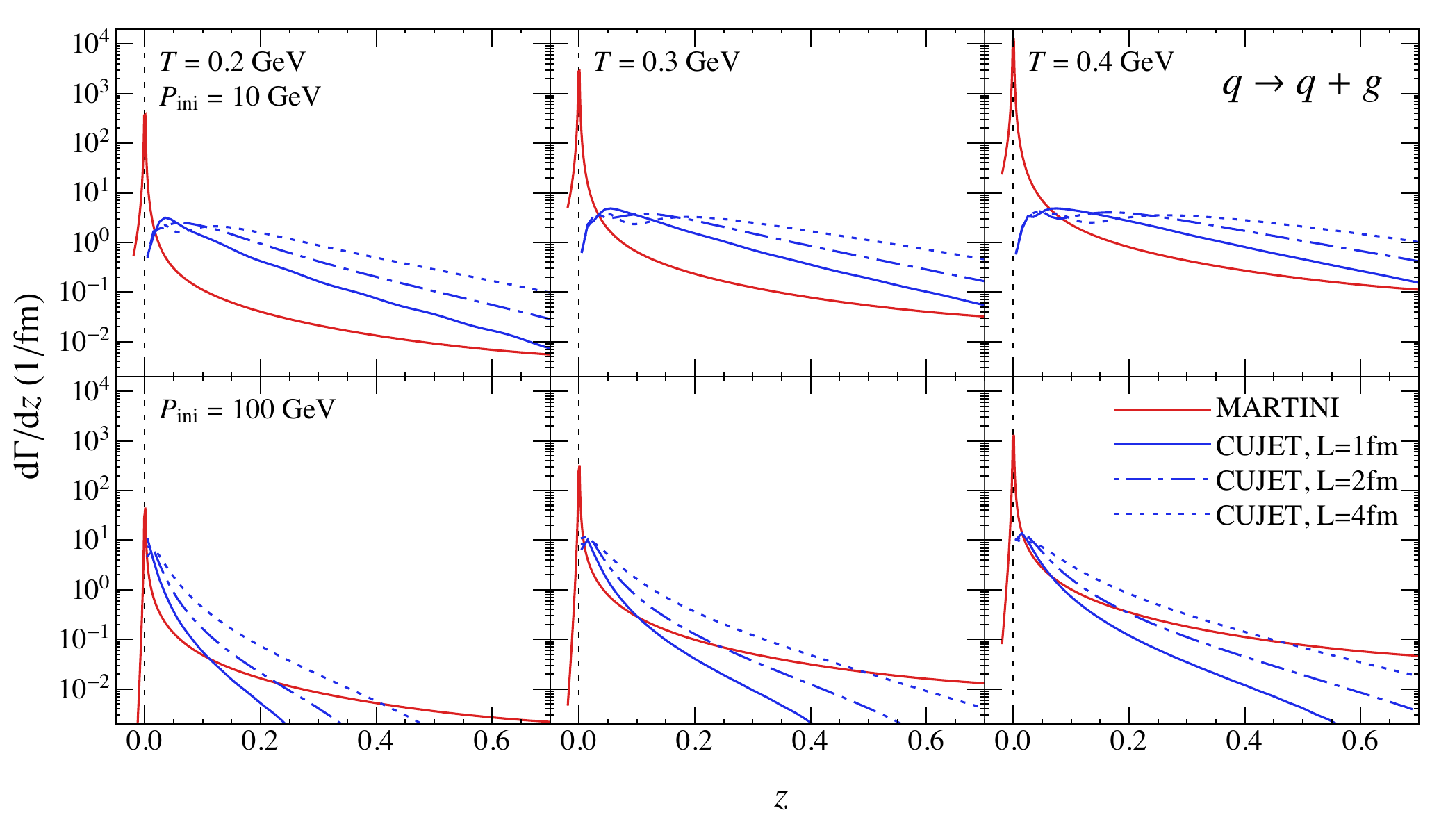}
        \vspace{-0.4cm}
        \caption{Comparison between the gluon bremsstrahlung rates of CUJET and MARTINI which use LO-DGLV (Eq.~\eqref{eq:lo_dglv_gluon_brem_rate}) and AMY (Eq.~\eqref{eq:splitting_rate_amy}), respectively. The rates are labeled by the Monte Carlo model which uses them. The top row shows the rates for a hard parton of momentum $p=10$~GeV and the bottom row presents the same for $p=100$~GeV, both for three representative temperatures available in heavy-ion collisions. In these figures, $z=p_g/p$ is the momentum fraction of the radiated gluon to that of the incoming hard parton. The AMY rates (MARTINI) are path-length independent, while the DGLV rates (CUJET) have explicit time/length dependence (Plot from Ref.~\cite{Shi:2022rja}).}
        \label{fig:rates_dglv_vs_amy}
    \end{figure}
        \subsubsection{Finite Temperature Field Theory} \label{sec:FTFT} 
            The calculation of jet energy loss within finite-temperature field theory is most famously associated with the work done by Arnold, Moore and Yaffe~\cite{Arnold:2002ja,Arnold:2001ba,Arnold:2001ms} and Jeon and Moore~\cite{Jeon:2003gi}, collectively known as the AMY framework. The assumption of the framework is that of a thermalized, infinite slab of QGP medium that is at asymptotically high temperature such that the strong coupling, $g$, is taken to be small and perturbative techniques can be applied. Furthermore, the hard parton is assumed to have been created in the distant past, and effects of the creation of the jet at $t=0$ (interference diagrams with vacuum emission) are not considered.  

The AMY framework then proceeds by carefully considering the separation of scales that is induced by high temperatures $g^2T\ll gT\ll T$. 
At the leading order, the important range of momentum transfers is $\mathcal{O}(gT)$ as $\mathcal{O}(g^2T)$ and $\mathcal{O}(T)$ exchanges are subdominant. The time scale of $\mathcal{O}(gT)$ collisions is, approximately, of order $m^{-1}_D\approx (gT)^{-1}$. Given the assumption of the smallness of the strong coupling $(g\ll 1)$, the elastic collision time scale is much smaller than the formation time of the radiative process, $\mathcal{O}((g^2T)^{-1})$. Thus, in principle, the radiating system of parton and gluon can undergo an arbitrary number of elastic collisions during the emission process which will destructively interfere with each other, a process called the LPM effect and accounted for in the AMY rates. 

The splitting rate for inelastic process $a\to bc$ where $b$ carries $z\equiv k/p$ momentum fraction of the incoming parton is given by
    \begin{align}
        \frac{\mathrm{d} \Gamma^{a}_{bc}}{\mathrm{d} z} (p,z) =\;\frac{\alpha_s P_{a\rightarrow bc}(z)}{[2p\,z(1{-}z)]^2}
        \int \! \frac{\mathrm{d}^2 \mathbf{h}_{\perp}}{(2\pi)^2} ~\text{Re} \left[ 2\mathbf{h}_{\perp} \cdot \mathbf{g}_{(z,p)}(\mathbf{h}_{\perp}) \right]\;.
        \label{eq:splitting_rate_amy}
    \end{align}
    $P_{a \rightarrow bc}(z)$ is the DGLAP splitting kernel for $a\to bc$ process, given by
    \begin{equation}
        P_{a \to bc}(z) =\begin{cases} 
        2 C_\mathrm{A} \frac{[1-z(1{-}z)]^2}{z(1{-}z)}\;& g\to gg\\ 
        C_\mathrm{F} \frac{1+(1{-}z)^2}{z}\;& q\to qg\\
        \frac{1}{2} \left(z^2+(1{-}z)^2\right)\;& g\to q\bar{q}
        \end{cases}.
    \end{equation}
    Finally, the function $\mathbf{g}_{(z,p)}(\mathbf{h}_{\perp})$ is related to the amplitude of a hard parton of momentum $p$ radiating a parton of momentum $k$ and encodes the transverse dynamics of the process. It is the solution to the following integral equation, which represents the re-summation of the infinite set of ladder diagrams
     \begin{align}
            2\mathbf{h}_{\perp} =\;& i \delta E(z,p,\mathbf{h}_{\perp}) \mathbf{g}_{(z,p)}(\mathbf{h}_{\perp}) 
            + \int \frac{\mathrm{d}^2\mathbf{q}_{\perp}}{(2\pi)^2}~C(q_\perp)\nonumber\\
            &\times\sum_{\beta \in \left\{1, z, (1-z)\right\}} C_{\beta} \left( \mathbf{g}_{(z,p)}(\mathbf{h}_{\perp}) - \mathbf{g}_{(z,p)}(\mathbf{h}_{\perp} -\beta\;\mathbf{q}_{\perp}) \right),
        \label{eq:AMY.linear.integral.equation}
    \end{align}
    where the factors $C_{1,z,1-z}$ are the same as Eq$.$~\ref{eq:casimir.factors} and $\delta E(z,p,\mathbf{h}_{\perp})$, the energy difference between the initial and final states of the radiation process is given by Eq.~\eqref{eq:energy.diff.bremsstrahlung}. The two dimensional vector $\mathbf{h}_{\perp}\equiv (\mathbf{k}\times\mathbf{p})\times e_{\parallel}$, the argument of $\mathbf{g}_{(z,p)}(\mathbf{h}_{\perp})$ measures the collinearity of the outgoing partons and its magnitude is parametric of $\mathcal{O}(gT^2)$.  
    
    Finally, $C(q_{\perp})$ is the transverse momentum broadening kernel, the rate of exchange of transverse momentum $q_{\perp}$ between the hard parton and the medium. Its value, stripped of the Casimir factor of the jet $\bar{C}(\mathbf{q}_{\perp})=C^{-1}_{s}C(\mathbf{q}_{\perp})$, and at the leading order in $\alpha_s$ is given by~\cite{Arnold:2008vd}
    \begin{align}
        \bar{C}_{\mathrm{LO}}\left(\mathbf{q}_{\perp}\right) &= \frac{g^2 T^3}{q^2_{\perp}\left(q^2_{\perp} + m^2_D\right)}\int \frac{\mathrm{d}^3p}{\left(2\pi\right)^3} \frac{p-p_z}{p}\times\nonumber
        \\& 
        [2 C_A n_B(p)(1+n_B(p'))+ 4N_f T_f n_F(p)(1-n_F(p'))],
        \label{eq:transverse.momentum.broadening.kernel.LO}
    \end{align}
    where $n_{B/F}$ are the Bose-Einstein and Fermi-Dirac distributions, respectively, and $p'= p + {(q^2_{\perp}+2\mathbf{q}_{\perp}\cdot \mathbf{p})}/{[2(p-p_z)]}$. 
    
    Figure~\ref{fig:rates_dglv_vs_amy} shows a comparison of the AMY rates against those of LO-DGLV, labeled as MARTINI and CUJET, respectively. A visible difference between the two is the inclusion of energy gain from the medium in AMY (region where $z<0$ in Fig.~\ref{fig:rates_dglv_vs_amy}). There are also other differences visible, namely the dependence on the momentum of the incoming hard parton, where LO-DGLV shows a more steep fall off, as a function of the momentum fraction, for larger jet momenta.  

    The rates described above are for an infinite medium. They are the rate of inelastic splitting for large times and neglect the effect of the formation time of the radiation. The inclusion of the finite-size effect was addressed by Caron-Huot \& Gale in Ref.~\cite{Caron-Huot:2010qjx}. The starting point is to Fourier transform the BDMPS-Z  formula for radiated spectrum in Eq$.$~\ref{eq:BDMPSZ_gluon_spectrum_config_space} to transverse momentum space and differentiating with respect to time to arrive at the rate
    \begin{align}
        \frac{d\Gamma^{a}_{bc}}{dz}(p,z,t) =&\; \frac{\alpha_s P^{(0)}_{a\to bc}(z)}{p^2 z^2(1-z)^2} \mathrm{Re\;} \int_{t}^{\infty} dt_1 \int \frac{d^2\mathbf{q}_1\;d^2\mathbf{q}_2}{(2\pi)^4} \mathbf{q}_1\cdot \mathbf{q}_2\nonumber \\ 
        &\times\left[K(t,\mathbf{q}_2;t_1,\mathbf{q}_1)-K_0(t,\mathbf{q}_2;t_1,\mathbf{q}_1) \right],
        \label{eq:caron_huot_intermediate}
    \end{align}
    with $P^{(0)}_{a\to bc}(z)$ the usual DGLAP kernel. The above is the rate for the process $a\to bc$,  where the incoming hard parton $a$  carries momentum $\mathbf{p}$ and is produced at time $t=0$. This parton then splits to partons $b$ and $c$ with momentum fractions $z$ and $1-z$, respectively, with $z\equiv k/p$.  $K(t,\mathbf{q}_2;t_1,\mathbf{q}_1)$ is the Fourier transform of the Green's function of the light-cone Hamiltonian of Eq.~\eqref{eq:bdmpsz.hamiltonian}
    \begin{equation}
        K(t_2,\mathbf{q}_2;t_1,\mathbf{q}_1) = \int \frac{d^2\mathbf{y}\;d^2\mathbf{x}}{(2\pi)^4} e^{i\mathbf{q}_2\cdot \mathbf{y}} e^{-i\mathbf{q_1}\cdot \mathbf{x}} G(t_2,\mathbf{y};t_1,\mathbf{x}), 
    \end{equation}
    and $K_{0}(t,\mathbf{q}_2;t_1,\mathbf{q}_1)$ is its counterpart in vacuum. By integrating the time integral of Eq.~\eqref{eq:caron_huot_intermediate} by parts and some rearrangements, we get
    \begin{equation}
        \frac{d\Gamma^{a}_{bc}}{dz}(p,z,t) = \frac{\alpha_s P^{(0)}_{a\to bc}(z)}{p^2 z^2(1-z)^2} \mathrm{Re\;} \int_{0}^{t} dt_1 \int_{\mathbf{q}_1,\mathbf{q}_2} \frac{i\mathbf{q}_1\cdot \mathbf{q}_2}{\delta E(\mathbf{q_2})} \mathcal{C}(t) K(t, \mathbf{q}_2;t_1,\mathbf{q}_1).
        \label{eq:ftft.with.finite.size}
    \end{equation}
    where $\mathcal{C}$ is the Fourier transform of the interaction term of the Hamiltonian. It is given by~\cite{Caron-Huot:2010qjx}
    \begin{equation}
        \mathcal{C} \psi(\mathbf{p}) = \int_{\mathbf{q}} \bar{C}(\mathbf{q}) \sum_{\beta\in(1,z,(1-z))} C_{\beta} \psi(\mathbf{p}-\beta \mathbf{q}).
    \end{equation}
    $\bar{C}(\mathbf{q})$ in the above is the Casimir-stripped elastic collision rate whose leading order expressions was provided in Eq.~\eqref{eq:transverse.momentum.broadening.kernel.LO}.  The radiation rate of Eq.~\eqref{eq:ftft.with.finite.size} is equivalent to the BDMPS-Z radiation spectrum and it is straightforward to recover the AMY rate of Eq.~\eqref{eq:splitting_rate_amy} as its $t\to\infty$ limit. 
    
    As is evident from the above discussion, an important ingredient of the bremsstrahlung rate in a QGP medium is the transverse momentum broadening kernel, $\bar{C}(q_{\perp})$.  Recently, techniques of electrostatic QCD (EQCD) have also allowed for higher order evaluations of this kernel. Its next-to-leading order (NLO) expression was first computed by Caron-Huot~\cite{Caron-Huot:2008zna}
    \begin{align}
        \frac{C^{\mathrm{NLO}}_{\mathrm{EQCD}}(q_{\perp})}{g^4 T^2 C_s C_A}=&\frac{7}{32q^3_{\perp}} + \frac{-m_D - 2 \frac{q^2_{\perp}-m^2_D}{q_{\perp}} \arctan{(\frac{q_{\perp}}{m_D})}}{4\pi(q^2_{\perp}+m^2_D)^2} \nonumber \\ 
        &+ \frac{m_D-\frac{q^2_{\perp}+4m^2_D}{2q_{\perp}}\arctan{(\frac{q_{\perp}}{2m_D})}}{8\pi q^4_{\perp}} - \frac{\arctan{(\frac{q_{\perp}}{m_D})}}{2\pi q_{\perp} (q^2_{\perp} + m^2_D)} \nonumber\\
        + \frac{\arctan{(\frac{q_{\perp}}{2m_D})}}{2\pi q^3_{\perp}} 
        &+ \frac{m_D}{4\pi(q^2_{\perp} + m^2_D)}\Big[\frac{3}{q^2_{\perp}+4m^2_D}-\frac{2}{q^2_{\perp}+m^2_D}-\frac{1}{q^2_{\perp}}\Big].
    \end{align}
    
   The NLO evaluation of the broadening kernel was then used in NLO calculation of  $\hat{q}$ via
    \begin{equation}
        \hat{q} = C_s \int_0^{q_{\mathrm{max}}} \frac{d^2q_{\perp}}{(2\pi)^2} q^2_{\perp} \bar{C}(q_{\perp}).
    \end{equation}
    The resulting  $\hat{q}$ is given by~\cite{Caron-Huot:2008zna}
    \begin{eqnarray}
            \frac{\hat{q}}{g^4 C_s T^3} &=&\; \frac{C_A}{6\pi} \Big[\log{(\frac{T}{m_D})}+\frac{\zeta(3)}{\zeta(2)} \log{(\frac{q_{\mathrm{max}}}{T})}-0.068854926766592\cdots)\Big] \nonumber \\
            &+&\frac{N_f T_f}{6\pi}\Big[\log{(\frac{T}{m_D})} + \frac{3}{2}\frac{\zeta(3)}{\zeta(2)}\log{(\frac{q_{\mathrm{max}}}{T})} - 0.072856349715786\cdots\Big] \nonumber \\
            &+& 2.9185\frac{C_A}{6\pi}\frac{m_D}{T} ,
    \end{eqnarray}
    where $q_{\mathrm{max}}$ is an ultraviolet cutoff. The NLO collision kernel has also been used in NLO determination of thermal photon production rate in Ref.~\cite{Ghiglieri:2013gia} as well as calculations of jet energy loss rates~\cite{Ghiglieri:2015ala}, though these rates have not yet been implemented in simulations of jet energy loss. Using the same EQCD techniques, a non-perturbative (NP) evaluation of the collision kernel has also become available~\cite{Panero:2013pla,Moore:2019lua} and used in the computation of AMY rates~\cite{Schlichting:2021idr}. This is done by inserting the NLO and NP evaluations of the collision kernel into Eq.~\eqref{eq:AMY.linear.integral.equation} and allows for a study of the effect of the broadening kernel, without complications of evaluating the AMY rates at higher order. 
    \begin{figure}
        \centering
        \includegraphics[width=\linewidth]{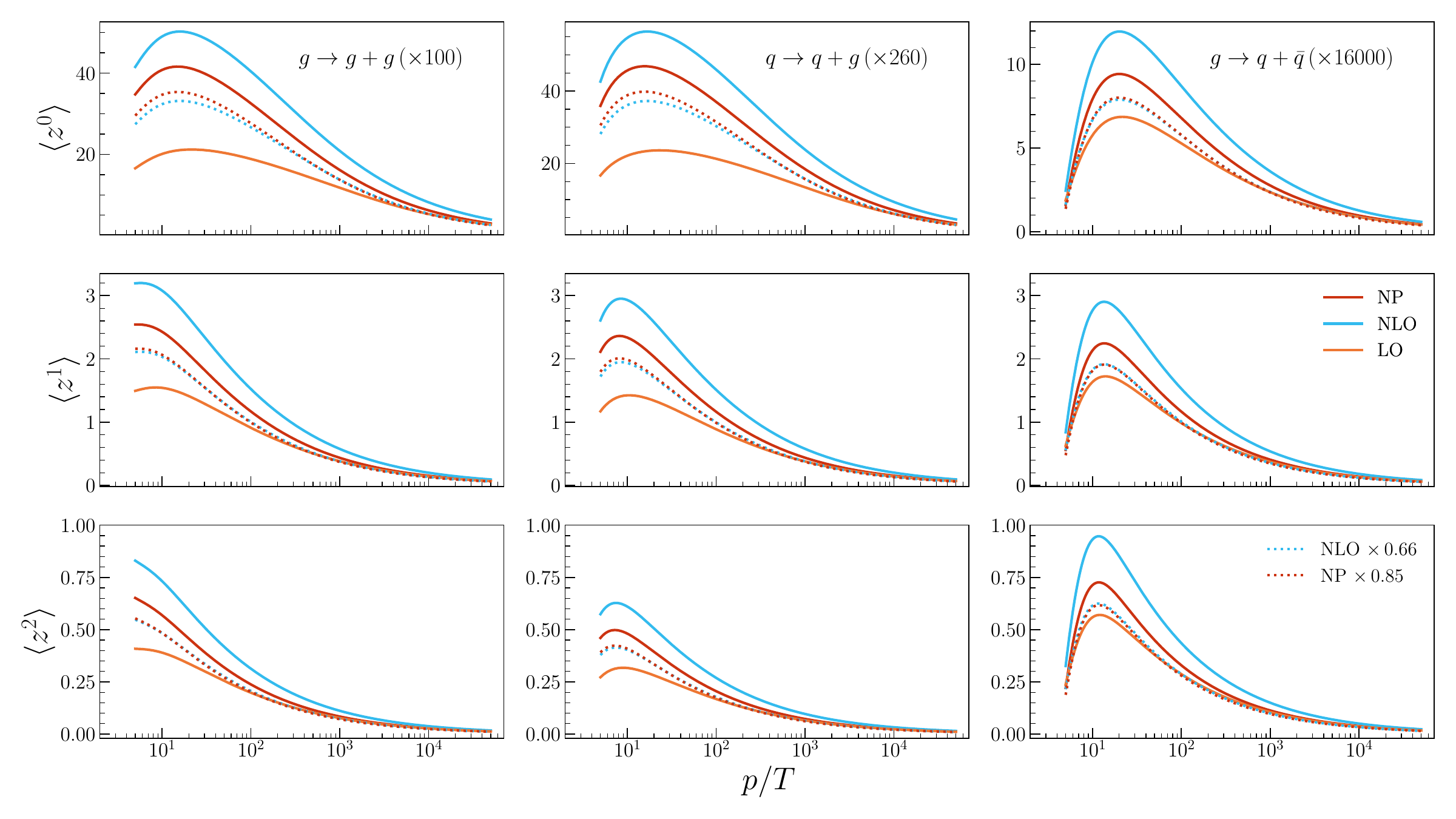}
        \vspace{-0.5cm}
        \caption{Moments of AMY rates as given by Eq.~\eqref{eq:moments_of_amy_rates}. The rates are labeled with the collision kernel that was used to generate them. The horizontal axis is the temperature scaled energy of the incoming hard parton. The vertical axes, from top to bottom, are the zeroth, first and second moment of the rates, respectively. Each column represents an inelastic splitting channel and is multiplied by the number indicated at the top so that each row shares the same vertical axis. The solid lines are direct evaluations of Eq.~\eqref{eq:moments_of_amy_rates} while the dotted lines are the NLO and NP rates, multiplied by constant factors for a visual match with the LO curve. Figure from Ref.~\cite{Yazdi:2022bru}.}
        \label{fig:amy_rates_moments}
    \end{figure}
    
The first three moments of the AMY rates evaluated using the LO, NLO and NP momentum broadening kernels are provided in Fig.~\ref{fig:amy_rates_moments}. Each rate set is labeled by the kernel used to compute it: the LO kernel generates the LO rate, and so on. The moments are defined by
    \begin{equation}
        \langle z^i \rangle = g^{-4}T^{-1} \left(\int_{0^+}^{1} |z|^i \frac{d\Gamma}{dz} dz - \int_{-\infty}^{0^{-}} |z|^i \frac{d\Gamma}{dz} dz\right).
        \label{eq:moments_of_amy_rates}
    \end{equation}
Thus, when scaled by $g^4 T \Delta t$,  $\langle z^0 \rangle$ and $\langle z^1 \rangle$  give
the difference between the number of emitted, absorbed particles and the net energy loss fraction by the incoming hard parton in time $\Delta t$, respectively. $\langle z^2 \rangle$ is related to the variance of energy loss per event via ${\langle z^2 \rangle}/{\langle z^0 \rangle} - ({\langle z^1 \rangle}/{\langle z^0 \rangle})^2$.  

The solid lines in the figure are the directly computed moments and demonstrate the $\mathrm{LO} < \mathrm{NP} < \mathrm{NLO}$ ordering. The dotted lines are the NLO and NP curves, scaled by constant factors to match the high (temperature-scaled) momentum tail of the LO rate with the scaling factors found to be, approximately, $\alpha^{\mathrm{NLO}}_{s} = \sqrt{0.66}\;\alpha^{\mathrm{LO}}_{s}$  and $\alpha^{\mathrm{NP}}_{s} = \sqrt{0.85}\;\alpha^{\mathrm{LO}}_{s}$. This matching of the large $p/T$ tail of the moments also flips the ordering of the resulting curves relative to the LO rate which can signal the slow convergence in the perturbative expansion for the momentum broadening kernel. Finally, the reduced but remaining difference for $p/T \lesssim 100$  indicates that while a redefinition of the strong coupling can remove the difference between the rate sets for $p/T > 100$, for momenta $p \lesssim 100 T$ this difference may still be observable. 
        
\subsection{Energy and virtuality dependence of $\hat{q}$ }\label{sec:Theory-2-q2-dependence}
In the preceding subsections, we have introduced the description of hard jet partons in a dense medium, from a high virtuality stage where multiple emissions are punctuated by rare scattering, to a low virtuality stage where each emission is stimulated via multiple scattering. In all cases, the transverse momentum exchanged with the medium, quantified by the transport coefficient $\hat{q}$ plays a leading role. 

\begin{figure}[!h]
    \centering
    \includegraphics[width=0.8\textwidth]{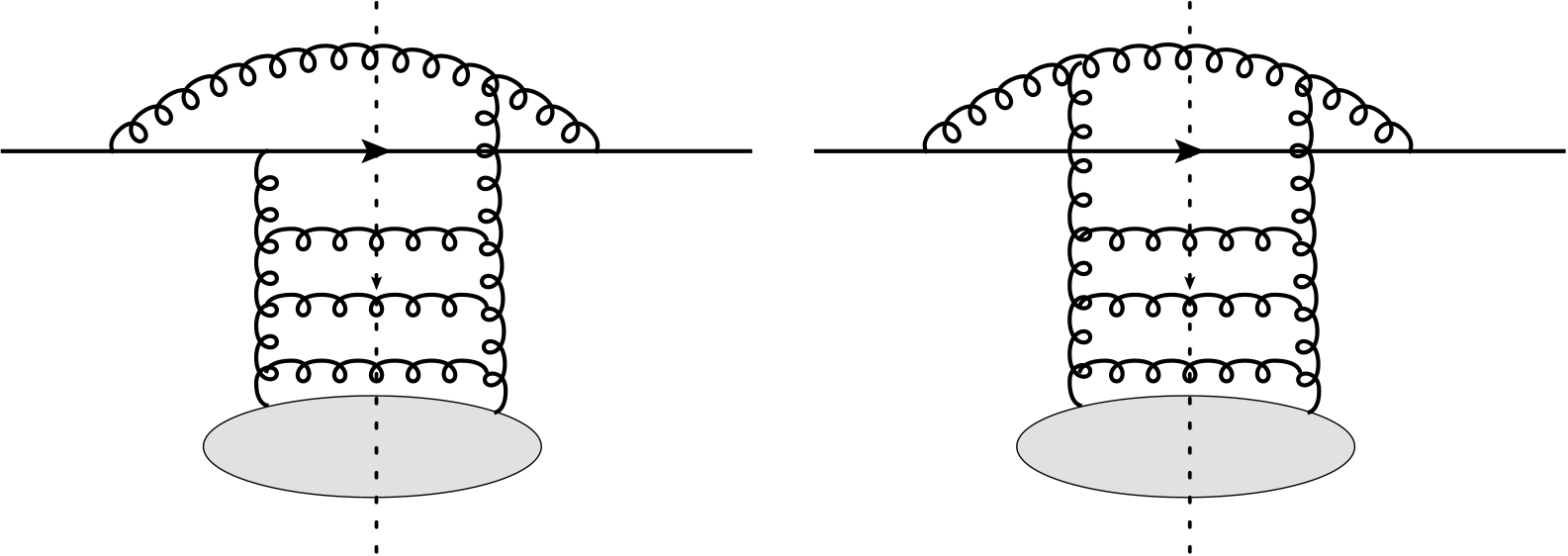}
    \vspace{-0.4cm}
    \caption{Diagrams that lead to evolution of $\hat{q}$ at high virtuality.}
    \label{fig:q-hat-evolution}
\end{figure}

While $\hat{q}$ obviously depends on the medium density, it also depends on the energy and virtuality of the hard parton. As a result, the value of $\hat{q}$ can change with energy and be different in the high and low virtuality phase. To calculate the dependence on the energy of the parton, further assumptions will have to be made regarding the medium constituent. This will be done in the subsequent sections. The expression for $\hat{q}$ derived in Eq.~\eqref{eq:qhat-FF} is the expression for a high energy and low virtuality parton, one that does not split in the process of scattering. At higher virtuality, parton splitting in the process of scattering influences the effective value of $\hat{q}$ in these processes.

To understand this, consider the process of scattering induced emission of a hard quark off the gluon field within a medium as indicated in Fig.~\ref{fig:q-hat-evolution}. These diagrams, when considered in the Breit frame consist of multiple splits in the target gluon prior to interaction with the quark, and radiative emission from the quark in the process of scattering. The first of these increases the value of $\hat{q}$, while the latter reduces the effective $\hat{q}$. There is no closed form for the final result, which results in a slow reduction of $\hat{q}(\mu^2)$ with increasing off-shellness of the quark $\mu^2$, as worked out in Ref.~\cite{Kumar:2019uvu}. In the later section on phenomenology this will be referred to as \emph{modified coherence}, which distinguishes it from the closely related coherence based arguments that also lead to a reduction of the effective $\hat{q}$~\cite{MehtarTani:2011tz,CasalderreySolana:2012ef}.

\section{From theory to event generation}~\label{sec:Theory-2-EventGeneration}
    
In the preceding sections, we discussed the theory of modification of hard partons in a dense medium. Directly applying this formalism to calculate observables that have been fashioned from Eqs.~\eqref{eq:pp-factorized-formula} and \eqref{eq:AA-factorized-formula}, yields insights on how the formulation works. However, these methods are often limited to a few observables at a time. The first subsection below on inclusive semi-analytics serves as an illustration of how far this method can be pushed. 

With the explosion of jet-based observables in the last decade, semi-analytic approaches have faced challenges in simultaneously describing multiple observables from the same calculation. This has necessitated the transition to simulation or event generation, where experiment-like events are generated. These simulated events can then be used to calculate any observable of interest, similar to what is done in experiment. As we build the components of these event generators, we will build the argument for event generator frameworks: simulation environments that allow multiple different sub-simulators to be combined into an end-to-end simulator. 

    \subsection{Inclusive Semi-analytics} 
        
Initially when J.D. Bjorken \cite{Bjorken:1982tu} posited that the presence of QGP will lead to jet “quenching”, the energy loss of high energy partons was estimated by considering elastic scatterings only.
However, it was soon realized that multiple scatterings with the medium will induce radiation from the hard partons.
It turns out that medium induced radiation is the dominant mechanism of jet quenching \cite{Baier:1998yf,Zakharov:1998sv}.
Several studies have worked on analytical~\cite{Baier:2001yt,Armesto:2011ir,Ghiglieri:2015ala,Mehtar-Tani:2017web,Mehtar-Tani:2014yea} and semi-analytical~\cite{Jeon:2003gi,Bass:2008rv,Majumder:2011uk,Mehtar-Tani:2018zba,Schlichting:2020lef,Mehtar-Tani:2022zwf,Isaksen:2022pkj,Qin:2007rn,Qin:2009gw} methods to investigate the energy loss cascade due to elastic and radiative interactions with the medium.

The spectrum obtained in Sec.~\ref{sec:FTFT} describes the distribution of partons emitted from an initial parton.
The radiated partons can themselves undergo further interactions with the medium constituents leading to subsequent radiation.
By resumming the subsequent radiation, one can obtain the distribution of energy lost by the initial hard parton.
The medium spectrum can be written as a convolution between the distribution of energy $D(\epsilon)$ that a hard parton loses due to medium induced radiation and the vacuum spectrum \cite{Baier:2001yt,Mehtar-Tani:2014yea,Arleo:2017ntr}, as follows
\begin{align}
    \frac{d\sigma^{\rm medium}}{dp_\bot^2} = \int d\epsilon~D(\epsilon) \frac{d\sigma^{\rm vacuum}(p_{\bot} + \epsilon)}{dp_\bot^2}\;.
\end{align}
An expansion of this expression in powers of the energy $\epsilon$ leads to the form
\begin{align}
    \frac{d\sigma^{\rm medium}}{dp_\bot^2}
    \simeq& \frac{d\sigma^{\rm vacuum}}{dp_\bot^2} + \int d\epsilon~\epsilon D(\epsilon) \frac{d\sigma^{\rm vacuum}}{dp_\bot^2}\;,\\
    \simeq& \frac{d\sigma^{\rm vacuum}}{dp_\bot^2} + \Delta E \frac{d}{dp_\bot}\frac{d\sigma^{\rm vacuum}}{dp_\bot^2}\;,
\end{align}
which can be resummed to
\begin{align}
    \frac{d\sigma^{\rm medium}}{dp_\bot^2}
    \simeq \frac{d\sigma(p_\bot + \Delta E)}{dp_\bot^2}
    \;.
\end{align}

The energy loss distribution due to subsequent medium induced radiation is given by a Poisson distribution that resums $n$ independent radiations (c.f. Sec.~\ref{sec:FTFT}) carrying total energy $\epsilon = \sum_{i=1}^n \omega_i$
\begin{align}\label{eq:EnergyLossDist}
    D(\epsilon) 
    = \sum_{n=0}^{\infty} \frac{1}{n !} \left[\prod_{i=1}^{n} \int d\omega_i \frac{dI}{d\omega_i}\right] \delta\left(\epsilon - \sum_{i=1}^{n} \omega_i\right) \exp\left[-\int d\omega~\frac{dI}{d\omega}\right]\;.
\end{align}
By computing the first moment of the energy loss distribution, one can obtain the mean energy loss, which shifts the vacuum spectrum.
While this approach is useful to obtain an estimate of the suppression of jets in the medium, it misses the dynamics of fluctuations.
In the following section, we will discuss dynamical approaches to obtain the energy loss distribution.

\subsubsection{Effective kinetic theory} \label{sec:eff-kin-theory}
The energy loss distribution in Eq.~(\ref{eq:EnergyLossDist}) gives an estimate of the energy loss but does not capture the full dynamics of the cascade.
Moreover, as the radiated partons lose energy, their energy becomes comparable to medium scales and elastic scattering contribution becomes important \cite{Blaizot:2014jna,Ghiglieri:2015ala,Schlichting:2020lef,Mehtar-Tani:2022zwf}.

In order to understand the dynamics of the cascade, one can write an effective kinetic theory (EKT) for the evolution of the phase-space distribution of the partons.
At the leading order of the QCD coupling constant, the evolution of the phase-space distribution is governed by the following Boltzmann equation\cite{Arnold:2000dr}
\begin{align}\label{eq:Boltzmann}
    \left(\partial_{t} + \frac{\bm{p}}{|\bm{p}|}\bm{\nabla}_x\right) f_a(\bm{x}, \bm{p}, t) = C^{2\to2}_a[\{f_i\}] + C^{1\rightleftarrows 2}_a[\{f_i\}]\;.
\end{align}
The collision integral $C^{2\to2}_a[\{f_i\}]$ describes the leading order elastic scattering where the number of particles is conserved.
Medium induced radiation is obtained in the collision integral $C^{1\rightleftarrows 2}_a[\{f_i\}]$, where both $1\to2$ radiations and $2\to1$ mergings are included.

While the position dependence of the phase-space distribution is important to understand the interplay between the medium dynamics and jet suppression, unfortunately, a numerical discretization of both position and momentum is prohibitively expensive.
Thus, we simplify Eq.~(\ref{eq:Boltzmann}) by integrating out the position dependence $f_a(\p,t) = \int d^3x~ f_a(\bm{x}, \bm{p}, t)$ and considering the evolution of the phase-space distribution $f_a(\p,t)$.
Conversely, Monte Carlo methods, described in the following sections, circumvent this problem by following the evolution of each parton using stochastic methods, which are averaged to describe the evolution of the phase-space distribution.

The phase-space distribution can be decomposed into a medium distribution $n_a(p)$ and the hard parton distribution $\delta f_a(p)$,
\begin{align}
    f_a(\bm{p})
    =\;& n_a(p) + \delta f_a(\bm{p})\;.
\end{align}
Since hard partons are dilute compared to medium constituents, the Boltzmann equation can be linearized in terms of the hard distribution $\delta f$, neglecting contributions from the interactions of the hard particles with each other.
We define the energy distribution
\begin{align}
    D(z,t)
    \equiv z\frac{dN}{dz} = \int d^3 p~ \frac{p}{E} \delta\left(\frac{|\bm{p}|}{E} - z\right) f(\bm{p}, t)\;,
\end{align}
where $E$ is the energy of the initial parton and $z=p/E$ is the momentum fraction carried by each parton in the medium cascade.

\subsubsection{Medium induced radiation} \label{sec:medium-induced-radiation}
Medium induced radiations described by the spectrum from Sec.~\ref{sec:FTFT} can be incorporated as an effective $1\leftrightarrow 2$ collision integral.
Including both $1\to2$ splittings and $2\to1$ mergings, the collision integral is written as \cite{Arnold:2003rq}
\begin{align}
    &C^{1\to2}_a[\{f_i\}]
    = \sum_{bc} \int_0^1 dz~ \left\{ -\frac{1}{2} \frac{d\Gamma^a_{bc}(\p,z)}{dz}
    \right. \nonumber\\
    &
    \times\Big[f_a(\p)(1\pm f_b(z\p)) (1\pm f_c((1-z)\p)) 
    - f_b(z\p) f_c((1-z)\p) (1\pm f_a(\p))\Big] \nonumber\\
    &+ \frac{\nu_b}{\nu_a} \frac{d\Gamma^b_{ac}\left(\frac{\p}{z},1\right)}{dz}
    \left[f_b\left(\frac{\p}{z}\right)(1\pm f_a(\p)) \left(1\pm f_c\left(\frac{(1-z)\p}{z}\right)\right)\right. \nonumber\\
    &\left.\left.- f_a(\p) f_c\left(\frac{(1-z)\p}{z}\right) \left(1\pm f_b\left(\frac{\p}{z}\right)\right)\right]\right\} \;.
\end{align}
The splitting rates ${d\Gamma^a_{bc}(\p,z)}/{dz}$ are given by Eq.~(\ref{eq:splitting_rate_amy}), which considers the medium length to be infinite such that the splitting rates are time-independent.  
These evolution equations can be straightforwardly promoted to the case for a time-dependent splitting rate when considering finite medium length \cite{Isaksen:2022pkj} or expanding medium \cite{Adhya:2019qse,Adhya:2021kws}.

To understand the medium cascade, let us consider gluon initiated processes only.
For hard gluons $(z\gg T/E)$, the Bose enhancement is exponentially suppressed and can be neglected to obtain a DGLAP-like equation \cite{Blaizot:2012fh}
\begin{equation}
    \label{eq:medium-induced-radiation}
    \begin{split}
    \partial_{t} D_g(z,t) 
        =&\; \frac{\alpha_s}{\pi} \sqrt{\frac{\hat{q}}{E}} \int_0^1 dx~ \mathcal{K}_{gg}(x) \\
        &\times\left[\Theta(x-z)\sqrt{\frac{x}{z}} D_g\left(\frac{z}{x},t\right) - \frac{x}{\sqrt{z}}D_g(z)\right]\;,
    \end{split}
\end{equation}
where the shorthand kernel is related to the rate in Eq.~(\ref{eq:splitting_rate_amy}) as follows
\begin{align}
    \frac{\mathrm{d} \Gamma_{g\to gg}}{\mathrm{d} z} (z E,z) 
    =& \frac{1}{\sqrt{z}} \mathcal{K}_{gg}(x)\;.
\end{align}

Apart from the change from the vacuum splitting function to the medium-induced splitting rates, the main difference between Eq.~(\ref{eq:medium-induced-radiation}) and the DGLAP equation in Eq.~(\ref{eq:DGLAP}) is the dependence on the parent parton's scale ($z$), which give rise to a turbulent cascade reminiscent of wave turbulence \cite{nazarenko_2011,zakharov2012kolmogorov}.
Wave turbulence is a well studied phenomenon that emerges in a wide range of physical systems, from plasma physics to hydrodynamics.
These systems are characterized by the transport of a scalar quantity (here energy) between a large separation of scales from a source to a sink.
When the dynamics of the non-equilibrium evolution are governed by interactions that are local in the energy domain, the inertial range far away from the source and sink of the system develops scale independent universal behavior known as the Kolmogorov spectrum.
The transport of the energy injected at the source is mediated by a cascade of interactions between the different scales, which transfer energy from one scale to the next, all the way to the sink, without any deposition in the inertial range.

In the case of the gluon cascade described by Eq.~(\ref{eq:medium-induced-radiation}), the source is the hard parton and the sink is the thermal bath.
Considering the energy flux through an arbitrary scale $\mu/T$, given by the integral of Eq. (\ref{eq:medium-induced-radiation}) as follows, \cite{Blaizot:2015jea,Mehtar-Tani:2018zba,Schlichting:2020lef}
\begin{align}
    \frac{d E}{dt} = -  \frac{\alpha_s}{\pi} \sqrt{\frac{\hat{q}}{E}} \int_{\mu/E}^1 \!\! dx \int_0^1 \! dz\; \mathcal{K}_{gg}(z) \left[\Theta(z-x)\sqrt{\frac{z}{x}} D_g\left(\frac{x}{z},t\right) - \frac{z}{\sqrt{x}}D_g(x)\right].
\end{align}
The first integral can be rewritten using a change of variable $x \to x/z$ leading to cancellation with the second term, one obtains
\begin{align}
    \frac{d E}{dt} = -  \frac{\alpha_s}{\pi} \sqrt{\frac{\hat{q}}{E}} \int_0^1 \! dz \; z\mathcal{K}_{gg}(z) \int_{\mu/E}^{\frac{\mu}{zE}} \!\! dx \; \frac{1}{\sqrt{x}}D_g(x)\;.
\end{align}
By plugging the distribution $D_g(x) \propto {1}/{\sqrt{x}}$, the dependence of the energy flux on the scale $\mu$ cancels out, leading to a scale invariant energy flux.

The BDMPS-Z spectrum introduced in Sec.~\ref{sec:FTFT} can lead to complex dependence on the momentum fraction $(\mathcal{K}_{gg}(z))$ which, depending on the elastic scattering kernel used, may not be solvable  analytically.
However, the Kolmogorov spectrum we obtained is only determined by the overall factor ${1}/{\sqrt{x}}$ which is the same for the different rates obtained in Sec.~\ref{sec:FTFT}.

\subsubsection{Elastic scatterings}\label{sec:elastic-scatterings}
The elastic scatterings of the hard parton with the medium constituents are described by the collision integral
\begin{align}\label{eq:elastic-scattering}
    C^{2\to2}[\{f_i\}]
    =& \frac{1}{4|\bm{p}|\nu_a} \sum_{bcd} \int_{\bm{kp'k'}} |\mathcal{M}^{ab}_{cd}(\bm{p}, \bm{k}, \bm{p'}, \bm{k'})|^2 (2\pi)^4 \delta^{(4)}(p+k-p'-k')\nonumber\\
    &\times\left[f^a(\bm{p'}) f^b(\bm{k'}) (1\pm f^c(\bm{p})) (1\pm f^d(\bm{k})) \right. \nonumber\\
    &-\left. f^c(\bm{p}) f^d(\bm{k}) (1\pm f^a(\bm{p'})) (1\pm f^b(\bm{k'}))\right]\;.
\end{align}

Due to the thermal bath, the matrix element $|\mathcal{M}^{ab}_{cd}|^2$ are regulated by self-energies computed using Hard Thermal Loop (HTL) theory \cite{Braaten:1989kk,Braaten:1989mz, Frenkel:1989br}.
However, thermal effects are only important for soft momentum exchange $q \sim gT$ when the $t$- or $u$-channels are of the order of the thermal mass squared and modifications of other terms can be neglected \cite{Arnold:2002ja}.

Another method of regulating the matrix elements is to introduce a cut-off $\mu$ on the momentum exchange, and separate the collision integral into large and small angle scatterings, as follows,
\begin{align}
    C^{2\to2}[\{f_i\}]
    =& C^{2\to2}_{\rm large}[\{f_i\}] + C^{2\to2}_{\rm small}[\{f_i\}]\;.
\end{align}
The small angle scatterings can be further simplified using an expansion in the momentum exchange $q$, transforming the collision integral into a differential Fokker-Planck equation \cite{Blaizot:2014jna,Ghiglieri:2015ala,Schlichting:2020lef}.
After linearizing the Fokker-Planck equation, one obtains the following equation for the energy distribution
\begin{align}
    C^{2\to2}_{\rm small}[\{f_i\}]
    =& -\nabla_p \mathcal{J}_a + S_a\;,
\end{align}
where the current $\mathcal{J}_a$ describes momentum drag and diffusion, which are characterized by the drag coefficient $\eta_D$ and the momentum broadening $\hat{q}$, respectively.
In the small angle approximation, quark exchange processes lead to negligible momentum exchange but give rise to conversion between partonic spices from quarks to gluons and vice versa.

Advances in EQCD theory, as mentioned in Sec.~\ref{sec:FTFT}, have enabled the computation of the momentum broadening and drag coefficient at NLO in the coupling constant.
The advantage of this reorganization is that it is still valid at NLO, allowing the use of the NLO transport coefficients to extend the EKT equations to NLO \cite{Ghiglieri:2015ala}.

\subsubsection{Medium response}
When the Boltzmann equation is linearized, the momentum of the initiating parton is considered to be larger than the medium scale $p\gg T$ and terms that evaluate the medium distribution at this momentum $n_a(p)$ are neglected.
However, when the radiated partons' energy becomes comparable to the medium scale $T$, these terms can become important, leading to modifications of the medium distribution.

In Refs \cite{Schlichting:2020lef,Mehtar-Tani:2022zwf}, the medium distribution is taken to be static, and the medium response is manifest in the parton distribution $\delta f$ itself.
At asymptotically late times, the fragments thermalize inside the medium and the system is fully described by the conserved quantities: energy $E$, momentum $\bm{P}$ and valence charge $N_f$, which lead to a change of the thermodynamic quantities: temperature $T$, flow velocity $\bm{u}$ and chemical potential $\mu_f$.

Due to linearization, the equilibrium distribution is given by
\begin{align}
    f^{\rm eq}_a(\bm{p}) 
    = n_a(p) + \delta f^{\rm eq}_a(\bm{p})\;,
\end{align}
where the distribution $\delta f^{\rm eq}_a(\bm{p})$ is a perturbation of the thermal distribution due to the energy deposited by the hard parton.
The perturbation can be obtained by computing the modification of the equilibrium distribution due to a small change in the thermodynamic quantities.
For example, for energy conservation, we have
\begin{align}\label{eq:AssymptoticDist}
    \delta f^{\rm eq}_a(\bm{p})
    = \delta T \partial_T n_a(p)\;,
\end{align}
where $\delta T$ is obtained by matching the energy of $\delta f^{\rm eq}_a(\bm{p})$ with the initial energy of the hard parton.

The distribution obtained in Eq.~(\ref{eq:AssymptoticDist}) is the late time response of the medium to the energy deposited by the hard parton.
However, this does not capture the dynamics due to the propagation of medium response in space.
In Sec. ~\ref{sec:medium_response}, we will discuss more realistic approaches coupling medium fragmentation and hydrodynamical medium evolution to understand the space-time propagations of the medium perturbation.

\subsubsection{Energy cascade}\label{sec:energy-cascade}
\begin{figure}
    \begin{center}
        \includegraphics[width=0.5\textwidth]{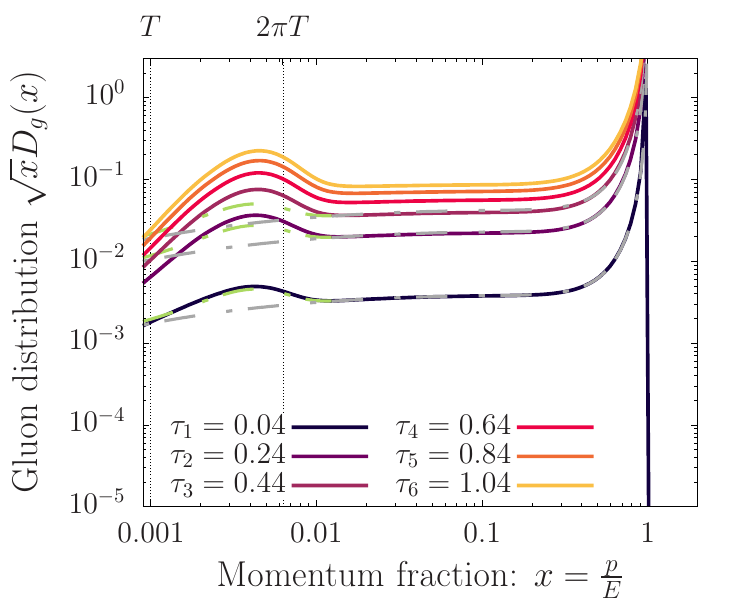}\includegraphics[width=0.5\textwidth]{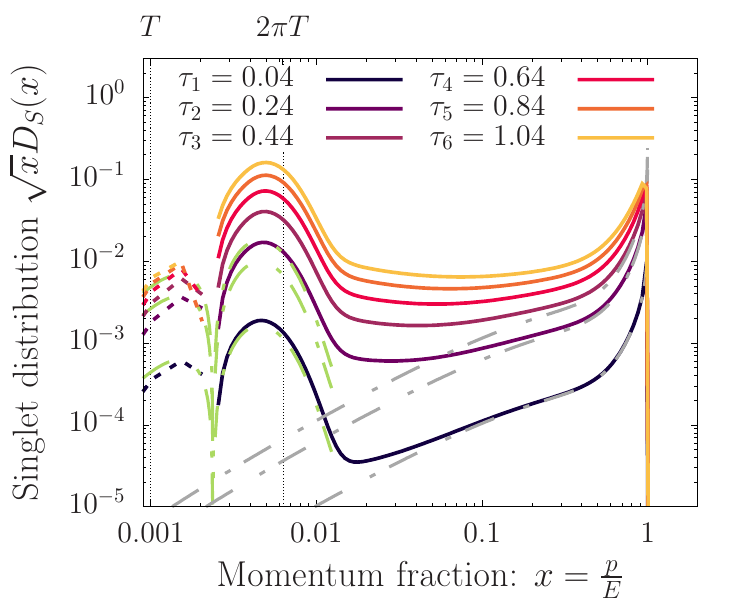}
        \vspace{-0.7cm}
    \end{center}
        \vspace{-0.25cm}
    \caption{Early time evolution of the energy distribution for gluon (left) and singlet $\left(\frac{q+\bar{q}}{2} \right)$ (right).
        The dashed lines correspond to single emission spectrum scaled with time. Figures from Ref.~\cite{Schlichting:2020lef}.}
    \label{fig:LinearBehaviour}
\end{figure}

\begin{figure}
    \begin{center}
        \includegraphics[width=0.5\textwidth]{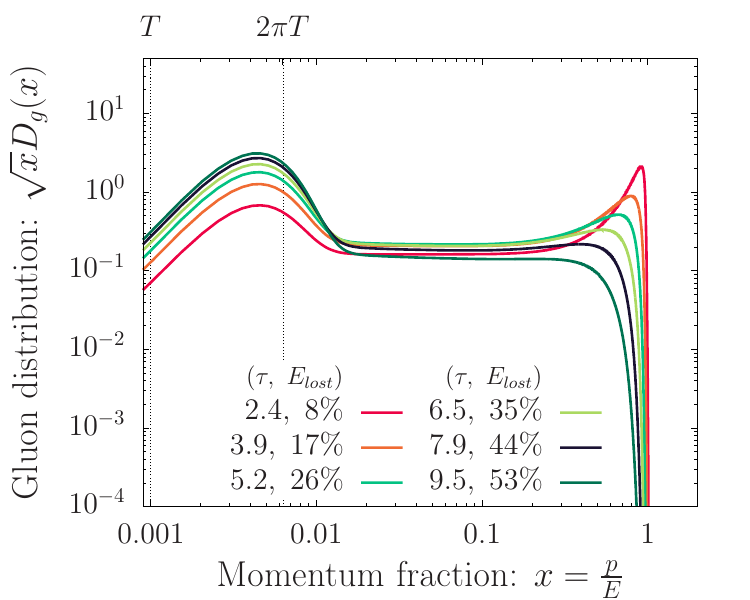}\includegraphics[trim= 0 20.3cm 0 0, clip,width=0.5\textwidth]{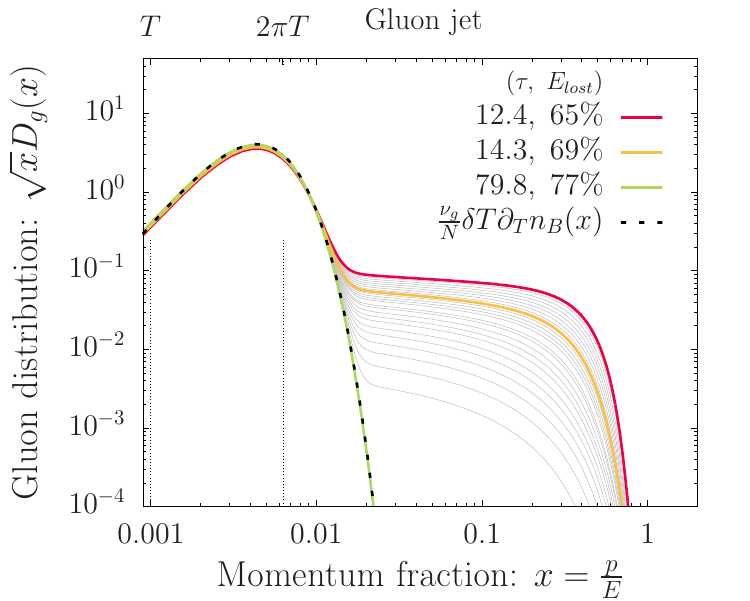}
        \vspace{-0.8cm}
    \end{center}
    \vspace{-0.25cm}
    \caption{Intermediate (left) and late (right) time behavior of the gluon energy distribution.
        $E_{\rm lost}$ is the percentage of energy in the low-lying modes with $x\leq 2\pi T/E$. Figures from Ref.~\cite{Schlichting:2020lef}.}\label{fig:Turbulence}
\end{figure}

The evolution of the energy distribution is characterized by three regimes \cite{Blaizot:2012fh,Mehtar-Tani:2018zba,Schlichting:2020lef}:
\begin{itemize}
    \item Early time evolution due to single emission.
    \item Turbulent cascade due to multiple successive splittings.
    \item Late time decay and equilibration of the soft modes.
\end{itemize}

In the following, we will discuss the evolution of a gluon initiated jet described by an initial normalized Gaussian distribution $D_g(z,0) \propto e^{-{(1-z)^2}/{\sigma^2}}$ with initial energy $E=1000T$ \cite{Schlichting:2020lef}.
In the early stages of the evolution, the energy distribution is dominated by single emissions, which leads to a linear growth of the energy distribution with time (c.f. Fig.~\ref{fig:LinearBehaviour}).
These single emissions populate the intermediate region of the energy distribution $ T/E < z < 1$ between the initial gluon and the medium scales.
Energy is accumulated in the medium scales where the response of the medium is important.
For the singlet distribution, a negative value of the energy distribution is obtained at medium scales.
This is understood as medium partons gaining energy, leading to a hardening of the distribution.

Once the single emission populates the intermediate region between the initial gluon and the medium scales, an energy cascade develops.
The wave turbulence cascade transports energy from the initial hard gluon $z\sim 1$ to the medium scales $z\sim T/E$ without any deposition in the intermediate region.
The characteristic Kolmogorov spectrum is obtained as shown in the left of panel Fig.~\ref{fig:Turbulence}, where $\sqrt{z}D_g(z)$ is virtually flat between $0.02 \lesssim z \lesssim 0.3$.

At later times, the energy of the hard parton thermalizes with the medium, leading to a perturbation on top of the medium as defined in Eq.~(\ref{eq:AssymptoticDist}).
This distribution is shown in the right panel of Fig.~\ref{fig:Turbulence} where the energy is concentrated in the low-lying modes $z\lesssim 2\pi T/E$.

    \subsection{Event generation with Jets: $p$-$p$ collisions}~\label{p-p-jets}
        In the preceding section we discussed an advanced calculation of the longitudinal momentum distribution of partons emanating from the scattering and radiation off a hard onshell parton, propagating through a static medium. This represents, in some ways, the limit of what can be calculated with semi-analytical methods. A real jet is seldom an on-shell parton; it starts far off mass shell and undergoes multiple splits prior to approaching the case of the near on-shell parton described above. 

While there have been several semi-analytical calculations of the fragmentation, jet and parton distribution functions for vacuum and medium modified showers within a DGLAP like formalism~\cite{Majumder:2009zu,Kang:2014xsa,Chien:2015vja}, the evolution of a parton with virtuality has never been combined with the multiple scattering and radiation calculations of on-shell partons within a semi-analytical calculation. Such efforts are very computationally involved as semi-analytical calculations typically integrate over the entire phase space. Secondly, once carried out, semi-analytical calculations cannot be used to calculate a different observable, say the soft drop prong distribution within a jet. These constraints have necessitated the focus on event generation, or the theoretical calculation of experiment like events.

In this section, we highlight the steps involved to carry out the simplest case of parton evolution in vacuum in a $p$-$p$ collision. Full events are obtained by simulating each factorized piece in Eq.~\eqref{eq:pp-factorized-formula}. In this Chapter, we will focus on the simulation of the final state [leading to the fragmentation or jet function in Eq.~\eqref{eq:pp-factorized-formula}], pointing the reader to Refs.~\cite{Sjostrand:1984,Sjostrand:1985xi} for the simulation of the initial state and hard scattering. 

In vacuum, the parton/hadron fragmentation function $D_a (z, Q^2)$ at a given virtuality scale $Q^2$ is typically described by the DGLAP evolution equation~\cite{Gribov:1972ri,Lipatov:1974qm,Dokshitzer:1977sg,Altarelli:1977zs}, where $z$ is the fractional momentum of the daughter parton/hadron taken from the initial parton with flavor $a$. Based on the DGLAP equation, one may construct the Sudakov form factor as~\cite{Hoche:2014rga}
\begin{equation}
\label{eq:sudakov}
\Delta_{ai} (Q_\mathrm{max}^2, Q^2_a)
=\exp \left[-\int\limits_{Q^2_a}^{Q_\mathrm{max}^2}\frac{d{Q}^2}{{Q}^2}\frac{\alpha_\mathrm{s}({Q}^2)}{2\pi}\int\limits_{z_\mathrm{min}}^{z_\mathrm{max}}dzP_{ai}(z,{Q}^2)\right],
\end{equation}
which represents the probability of no splitting between a maximum possible scale $Q^2_\mathrm{max}$ and a given parton scale $Q^2_a$. In the equation above, $P_{ai}$ is the parton splitting function of a particular channel $i$, $z_\mathrm{min}$ and $z_\mathrm{max}$ are the lower and upper kinematic limits of $z$. After we combine all splitting channels, the Sudakov form factor reads $\Delta _a(Q_\mathrm{max}^2, Q^2_a) = \prod_i \Delta_{ai} (Q_\mathrm{max}^2, Q^2_a)$.

Based on this Sudakov form factor, event generators are developed to simulate parton showers. For a splitting process $a\rightarrow bc$, if a random number $r\in (0,1)$ is smaller than $\Delta_a(Q_\mathrm{max}^2, Q_\mathrm{min}^2)$, parton $a$ is considered stable with its virtuality set as the lower limit $Q_\mathrm{min}^2$, which is usually taken as the hadronization scale ($\sim 1$~GeV$^2$) in vacuum parton showers. Otherwise, one solves the equation $r = \Delta_a(Q_\mathrm{max}^2, Q_a^2)$ to find the scale $Q_a^2$ at which $a$ splits. The specific splitting channel $i$ is determined from the branching ratios:
\begin{equation}
\label{eq:branching}
\mathrm{BR}_{ai}(Q_a^2)=\int_{z_\mathrm{min}}^{z_\mathrm{max}} dz\; P_{ai} (z, Q_a^2).
\end{equation}
The longitudinal momentum fractions of the two daughter partons are then sampled according to the splitting function $P_{ai} (z, Q_a^2)$, and $z^2Q_a^2$ and $(1-z)^2Q_a^2$ are used as the new upper limits of virtualities ($Q^2_\mathrm{max}$) for the two daughters respectively with which their virtualities ($Q_b^2$ and $Q_c^2$) can be calculated again using Eq.~(\ref{eq:sudakov}). In the end, the transverse momentum of the daughters relative to the mother is given by 
\begin{equation}
\label{eq:transverse}
k_\perp^2=z(1-z)Q_a^2-(1-z)Q_b^2-zQ_c^2.
\end{equation}
This completes one splitting process, which can be iterated until all daughter partons reach the preset value of $Q^2_\mathrm{min}$.
Here, we have neglected the rest mass of partons, which can be taken into account by replacing $Q^2_a$ with $M^2_a=Q^2_a+m^2_a$ in the discussion above, with $m_a$ being the rest mass of parton $a$. 

This routine has been implemented in event generators that simulate parton showers in $p+p$ collision. The detailed implementation, e.g., the definition of the kinematic variables and their limits, may vary between different models. For instance, in PYTHIA 6~\cite{Sjostrand:2006za}, $z$ is defined as the fractional energy $E_b/E_a$, and thus the kinematic cuts of $z$ are given by
\begin{equation}
\label{eq:cutWithE}
z_\mathrm{max/min}=\frac{1}{2}\Bigg[1+\frac{M_b^2-M_c^2}{M_a^2}
\pm \frac{|\vec{p}_a|}{E_a}\frac{\sqrt{(M_a^2-M_b^2-M_c^2)^2-4M_b^2M_c^2}}{M_a^2}\Bigg].
\end{equation}
This can be obtained by solving the momentum $p$ of parton $b/c$ in the rest frame of parton $a$ using $M_a=\sqrt{p^2+M_b^2}+\sqrt{p^2+M_c^2}$ and then implementing a collinear boost with $\pm |\vec{p}_a|/E_a$. Before knowing the daughter parton species $b$ and $c$, one may also temporarily assume zero masses for them compared to $M_a$, and use  
\begin{equation}
\label{eq:cutWithES}
z_\mathrm{max/min}=\frac{1}{2}\left(1\pm \frac{|\vec{p}_a|}{E_a}\right)
\end{equation}
in Eqs.~(\ref{eq:sudakov}) and (\ref{eq:branching}).

In another event generator, MATTER~\cite{Majumder:2013re,Cao:2017qpx}, $z$ is defined as fraction of $p^+$ the light cone coordinate. The lower and upper limits of $z$ are obtained when parton $b$ is collinear with parton $a$, or by setting $k_\perp^2=0$ in Eq.~(\ref{eq:transverse}), as 
\begin{equation}
\label{eq:cutWithP}
z_\mathrm{max/min}=\frac{1}{2}\Bigg[1+\frac{M_b^2-M_c^2}{M_a^2}
\pm \frac{\sqrt{(M_a^2-M_b^2-M_c^2)^2-4M_b^2M_c^2}}{M_a^2}\Bigg].
\end{equation}
Compared to Eq.~(\ref{eq:cutWithE}), the factor $|\vec{p_a}|/E_a$ is absent here. This forbids our further simplification using $M_b=M_c=0$, which leads  to $z_\mathrm{max}=1$ and $z_\mathrm{min}=0$ where the splitting function may be divergent. An alternative approximation is setting $M_b^2=M_c^2=Q_\mathrm{min}^2$, which yields
\begin{equation}
\label{eq:cutWithPS}
z_\mathrm{max/min}=\frac{1}{2}\left[1 \pm \sqrt{1-\frac{4Q_\mathrm{min}^2}{M_a^2}}\right]
\approx \frac{1}{2}\left[1 \pm \left(1-2Q_\mathrm{min}^2/M_a^2\right)\right],
\end{equation}
or $z_\mathrm{max}=1-Q_\mathrm{min}^2/M_a^2$ and $z_\mathrm{min}=Q_\mathrm{min}^2/M_a^2$. Again, the rest masses of daughter partons have been neglected in Eqs.~(\ref{eq:cutWithES}) and~(\ref{eq:cutWithPS}), which can be introduced via $M^2_{b/c} = Q_\mathrm{min}^2+m^2_{b/c}$ if needed, e.g., for heavy quarks. 

There is no unique choice of the maximum virtuality $Q^2_\mathrm{max}$. It can be treated as a model parameter to fit the hadron/jet spectra observed in $p+p$ collisions. In PYTHIA 6, this is set as $Q_\mathrm{max}^2 = 4Q_\mathrm{hard}^2$ by default, with $Q_\mathrm{hard}^2$ being the transverse momentum exchange square of the initial hard scattering. The pre-factor of 4 here can be changed via the ``PARP(67)" parameter in the PYTHIA 6 package. In contrast, MATTER initiates parton showers from each individual parton instead of the entire $p+p$ collision. Without the knowledge of the initial hard scattering scale, it uses the initial parton energy square $E^2$ as its $Q_\mathrm{max}^2$ at the beginning~\cite{Cao:2017qpx}. Within the JETSCAPE framework~\cite{Putschke:2019yrg} where hard partons are first generated by PYTHIA hard scatterings and then evolve through the MATTER shower,  $Q_\mathrm{max}^2=p_\mathrm{T}^2/4$ is set by default in MATTER to describe the hadron/jet spectra in $p+p$ collisions, which can be changed by users via the ``vir\_factor" parameter in JETSCAPE.

    \subsection{Event generation with Jets: $A$-$A$ collisions}~\label{A-A-jets}
        As for the case of event generation of jets in $p$-$p$, for jets in $A$-$A$ collisions, we start with Eq.~\eqref{eq:AA-factorized-formula}. We will ignore the simulation of the hard scattering cross section and focus on the physics of the nuclear parton distribution function $\Tilde{G}$, and the simulation of the parton evolution function $\mathscr{P}$ and the medium modified fragmentation function $\Tilde{D}$ (or the medium modified jet function $\Tilde{J}$). In each case, the reader will be systematically led to and through the working of exact modules within the JETSCAPE framework. 
        \subsubsection{NPDFs} 
            
Nuclei are bound states of nucleons. As a result, the distribution of partons within a bound nucleon (within a nucleus) will be different from that of a free nucleon~\cite{Bodek:1983qn,EuropeanMuon:1988tpw}. 
This is typically understood in terms of the ratio of the parton distribution function, per nucleon, within a nucleus divided by that of a free nucleon (typically to that per nucleon in Deuterium), 
\begin{equation}
    R_{i/A}(x, Q^2) = { f_{i/A} (x,Q^2) }\Big/ { f_{i/D} (x, Q^2) }.
\end{equation}
This effect, often generically (and erroneously) referred to as nuclear shadowing, consists of 4 separate effects: a shadowing or suppression of $R_{i/A} (x,Q^2)$ ($\lesssim 1$), for $x\lesssim 0.1 $, an anti-shadowing or enhancement ($R_{i/A}(x,Q^2) \gtrapprox 1$), for $0.1 \leq x \leq 0.3$, a milder suppression, the EMC effect at $0.3 \leq x \leq 0.7$ and an enhancement as $x$ approaches $1$ and remains non-zero even beyond $1$ (Note that beyond $x\gtrapprox 1$, $f_{i/p}(x,Q^2)$, the PDF in a proton goes to zero, while $f_{i/A}(x,Q^2)$ is non-zero, hence $R_{i/A}(x) \rightarrow \infty$) due to Fermi motion of nucleons within the nucleus. 

At this time, there exist several formulations for $A$-dependent nuclear shadowing, the most popular being the impact parameter independent versions from Refs.~\cite{Eskola:2016oht,Eskola:2009uj}. Other versions that include impact parameter dependence are also available~\cite{Li:2001xa}. 

Incorporation of shadowing functions within both semi-analytic efforts and within event generators is now rather straightforward. Pre-generated subroutines can be called that include these effects. Generators such as PYTHIA can directly draw initial state distributions that include $A,x,Q^2$ dependent shadowing. 

        \subsubsection{Medium Modified Parton Shower} 
            How the medium modified parton shower is modeled in AA collisions depends on the virtuality scale of the jet partons. When the parton virtuality is much higher than the medium temperature scale, the parton evolution is still dominated by successive splittings, similar to its shower in vacuum. This is known as the ``rare-scattering-multiple-emission region". To the contrary, when its virtuality approaches the medium scale, scatterings inside the medium prevent its virtuality from further decreasing, and we enter the ``multiple-scattering-few-emission region". 

The high virtuality region can be simulated using the MATTER event generator, described above in Sec.~\ref{sec:jets_in_a_medium_high_virtuality}. One samples a medium modified Sudakov factor [see Eq.~\eqref{eq:in-medium-sudakov}] to determine the virtuality of the hard parton. Medium modified splitting functions are sampled to determine the forward momentum fraction shared between partons after the split. 
This is followed by determination of the virtualities of the outgoing partons and their relative transverse momentum. Single scattering is simulated on one of the out-going partons to generate an outgoing  recoil parton and a hole in the medium from where it arose. 

Successive splits lead to a reduction in the virtuality of the hard partons until they approach the medium induced scale $\mu_{med}^2 \approx \hat{q} \tau$, at which point they are transitioned to a low virtuality energy loss module. So far three separate energy loss modules have been combined with MATTER to produce a multi-scale generator: LBT, MARTINI and CUJET. These are described in the following.

            \subsubsection{LBT}
                The linear Boltzmann transport (LBT) model~\cite{Luo:2023nsi} simulates elastic and inelastic scatterings of low virtuality jet partons inside a thermal medium based on the Boltzmann equation
\begin{eqnarray}
  \label{eq:boltzmann1}
  p_a\cdot\partial f_a(x_a,p_a)=E_a(\mathcal{C}_\mathrm{el}+\mathcal{C}_\mathrm{inel}),
\end{eqnarray}
where $f_a(x_a,p_a)$ is the phase space distribution of jet parton $a$, and $E_a$ denotes the parton energy. 

From the elastic part of the collisional integral $\mathcal{C}_\mathrm{el}$, one may extract the elastic scattering rate as 
\begin{eqnarray}
 \label{eq:rate2}
 \Gamma_\mathrm{el}^{ab\rightarrow cd} &=&\frac{\gamma_b}{2E_a}\int \frac{d^3 p_b}{(2\pi)^3 2E_b}\int\frac{d^3 p_c}{(2\pi)^3 2E_c}\int\frac{d^3 p_d}{(2\pi)^3 2E_d}\nonumber\\
&\times& f_b(E_b)\left[1\pm f_c(E_c) \right]\left[1\pm f_d(E_d)\right] S_2(\hat s,\hat t, \hat u)\nonumber\\
&\times& (2\pi)^4\delta^{(4)}(p_a+p_b-p_c-p_d)|\mathcal{M}_{ab\rightarrow cd}|^2,
\end{eqnarray}
where $b$ represents a thermal parton with spin-color degeneracy $\gamma_b$, $c$ and $d$ are the final state particles of a $2\rightarrow 2$ scattering. By convention, $c$ is used to denote the final state of the jet parton. It keeps the flavor of $a$ if it is a heavy quark; otherwise, $E_c>E_d$ is assumed for a light flavor jet. In the rest frame of the medium, the thermal distribution function $f_i(E_i)=1/(e^{E_i/T}\mp 1)$ is assumed for $i=b,c,d$, with $T$ the local temperature of the medium. Here and in Eq.~(\ref{eq:rate2}), the upper sign in $\mp$ ($\pm$) is for gluons and the lower sign for quarks. The initial-state-averaged and final-state-summed (over spin and color degeneracies) scattering matrices $|\mathcal{M}_{ab\rightarrow cb }|^2$ at the leading order~\cite{Auvinen:2009qm} are applied, whose collinear divergence is regulated using a double-$\theta$ function $S_2(\hat s,\hat t, \hat u)=\theta(\hat s\ge2\mu_\mathrm{D}^2)\theta(-\hat s+\mu_\mathrm{D}^2\le \hat t\le -\mu_\mathrm{D}^2)$, with $\hat{s},\hat{t},\hat{u}$ the Mandelstam variables, $\mu_\mathrm{D}=6\pi\alpha_\mathrm{s}T^2$ the Debye screening mass. The total rate of elastic scattering of $a$ then reads $\Gamma^a_{\rm el}= \sum_{b,c,d} \Gamma_{\rm el}^{ab\rightarrow cd}$.

The inelastic scattering rate is related to the rate of medium-induced gluon emission in LBT as
\begin{equation}
\label{eq:rateInel}
\Gamma^a_\mathrm{inel}=\int dzdk_\perp^2 \frac{1}{1+\delta^{ag}}\frac{dN^a_g}{dzdk_\perp^2 dt},
\end{equation}
with the gluon spectrum taken from the higher-twist energy loss calculation~\cite{Guo:2000nz,Zhang:2003wk,Majumder:2009ge,Majumder:2007ae},
\begin{eqnarray}
\label{eq:gluondistribution}
\frac{dN_g^a}{dz dk_\perp^2 dt}=\frac{2\alpha_\mathrm{s} C_A \hat{q}_a(x) P_a(z)k_\perp^4}{\pi \left({k_\perp^2+z^2 m_a^2}\right)^4} \, {\sin}^2\left(\frac{t-t_i}{2\tau_f}\right),
\end{eqnarray}
in which $z$ and $k_\perp$ are the fractional energy and the transverse momentum of the emitted gluon with respect to jet parton $a$, $m_a$ is the parton mass, $t_i$ is the production time of $a$ or the last time of its splitting, $\tau_f={2Ez(1-z)}/{(k_\perp^2+z^2m_a^2)}$ is formation time of the emitted gluon, and $P_a(z)$ is the splitting function. To avoid divergence at $z\rightarrow 0$, a lower cut-off $z_\mathrm{min}=\mu_\mathrm{D}/E$ is applied. The Kronecker-delta function $\delta^{ag}$ in Eq.~(\ref{eq:rateInel}) is imposed to prevent double counting on the gluon emission rate from the $g\rightarrow gg$ process. The medium information is all absorbed in the jet quenching parameter $\hat{q}_a$, which is evaluated using Eq.~(\ref{eq:rate2}) with a weight of $q_\perp^2=[\,\vec{p}_c-(\vec{p}_c\cdot\hat{\vec p}_a)\hat{\vec p}_a]^2$ on its right hand side. 

With the scattering rates above, the elastic and inelastic scattering probabilities of parton $a$ during a time step $\Delta t$ is then given by 
\begin{align}
&P^a_\mathrm{el} =1-e^{-\Gamma^a_\mathrm{el} \Delta t},\\
&P^a_\mathrm{inel} =1-e^{-\Gamma^a_\mathrm{inel} \Delta t}.
\end{align}
The total probability is then 
\begin{equation}
P^a =1-e^{-(\Gamma^a_\mathrm{el}+\Gamma^a_\mathrm{el}) \Delta t} = P^a_\mathrm{el} + P^a_\mathrm{inel} - P^a_\mathrm{el}P^a_\mathrm{inel},
\end{equation}
which can be understood as the combination of pure elastic scattering without gluon emission $P^a_\mathrm{el}(1-P^a_\mathrm{inel})$ and inelastic scattering $P^a_\mathrm{inel}$. Based on these probabilities, the Monte Carlo method is used at each time step to determine whether a scattering happens, and if so, whether it is pure elastic or inelastic. For pure elastic, an $ab\!\rightarrow \!cd$ process is then sampled according to the differential rate Eq.~(\ref{eq:rate2}). For inelastic processes, this $ab\!\rightarrow\! cd$ process is also sampled first since gluon emission is induced by elastic scattering in LBT. The gluon momentum is then sampled from parton $c$ according to Eq.~(\ref{eq:gluondistribution}). Finally, the kinematics of $c$ and $d$ need to be adjusted~\cite{Luo:2023nsi} to ensure energy-momentum conservation of the $ab\!\rightarrow\! cd\!+\!g$ process. 

In the LBT model, both the final state of the jet parton ($c$) and the final state of the medium parton ($d$) are tracked in simulation, the latter of which is named as ``recoil" parton. Additionally, the initial state of the thermal parton ($b$) is also recorded, which represents the energy-momentum depletion from the medium when a recoil parton is generated. This is called ``negative" parton, or ``back-reaction", and is essential in energy-momentum conservation in each scattering process. Recoil and ``negative" partons constitute the ``jet-induced medium excitation" in LBT. Recoil partons and radiated gluons are allowed to rescatter with the medium in the same way as jet partons do. 

In phenomenological studies, the strong coupling strength $\alpha_\mathrm{s}$ is treated as a model parameter in LBT. In most studies on jets~\cite{Luo:2018pto,He:2018xjv,He:2022evt}, constant $\alpha_\mathrm{s}$ is used, which is adjusted to describe the nuclear modification factor of jets, although running $\alpha_\mathrm{s}$ with respect to the medium temperature and parton energy is also assumed in some studies on heavy quarks~\cite{Cao:2017hhk,Xing:2019xae}, especially for a simultaneous description of their observables from low to high $p_\mathrm{T}$. Non-perturbative interactions have also been introduced for studying heavy quarks at low $p_\mathrm{T}$~\cite{Xing:2021xwc}. In addition, while most LBT calculations assume zero masses for light flavor partons and gluons, their thermal masses have been introduced in recent work~\cite{Liu:2021dpm,Liu:2023rfi} for describing heavy quark interactions with the medium.

\label{sec:lbt}
            \subsubsection{MARTINI}\label{sec:martini}
                The Modular Algorithm for Relativistic Treatment of Heavy Ion Interactions~\cite{Schenke:2009gb} or MARTINI is a realization of the AMY energy loss formalism. MARTINI can be used as a stand-alone Monte Carlo generator in a single-stage energy loss calculation (combined with an initial state, hard scattering and fluid simulation) or as a component of a multi-stage model, in particular as a low-virtuality energy loss model within the JETSCAPE framework~\cite{JETSCAPE:2019udz}. 

The model assumes that the incoming hard partons coming from hard scattering are massless and on mass shell. It uses the AMY inelastic splitting rates of Eq.~\eqref{eq:splitting_rate_amy} and gluon-mediated elastic scattering processes of Eq.~\eqref{eq:elastic-scattering}. Quark-mediated processes where the identity of the incoming hard parton is changed are also included but only in the conversion limit. In this approximation, the energy loss during the process is considered minimal and the outgoing parton inherits the full momentum of the incoming hard parton (see Sec.~\ref{sec:elastic-scatterings}). Included channels are $q\to g$, $g\to q$ and $q\to \gamma$ where $q$ signals both quarks and anti-quarks. MARTINI, then, solves the following coupled Fokker-Planck type equation for the evolving parton distribution in the medium
\begin{align}
    \frac{df_{q}}{dt}(p) =&\; \int_{-\infty}^{\infty} dk\; f_{q}(p+k)\frac{d\Gamma(p+k)}{dk} - f_{q}(p)\frac{d\Gamma(p,k)}{dk},
\end{align}
where $d\Gamma(p,k)/dk$ denotes the rate for a hard parton of momentum $p$ to radiate a parton of momentum $k$. The elastic and conversion processes are also implemented in a similar manner.  

The rates are computed and tabulated beforehand and read into MARTINI for interpolation at run time. For a given time interval $\Delta t$ and an incoming parton with energy $p$, MARTINI calculates the total probability of interacting with the medium at local temperature $T$ as
\begin{equation}
    P_{\mathrm{int}}(p) = \Delta t \left[\Gamma_{\mathrm{rad}}(p,T) + \Gamma_{\mathrm{elas}}(p,T) + \Gamma_{\mathrm{conv}}(p,T)\right].
\end{equation}
In the above, $\Gamma_{i}$ is the total rate for channel $i$. It should be noted that the radiative channels in MARTINI are $1\to 2$ channels and no recoil parton is generated in the process. The elastic and conversion channels, however, do include the recoil/hole prescription as described in Sec.~\ref{sec:medium_response}. 

The main parameters of MARTINI are those governing the running of the strong coupling. The coupling is evaluated using the LO pQCD expression, 
\begin{equation}
    \alpha_s(\mu^2) = \frac{4\pi}{\left(11 -\frac{2}{3} N_f\right)\log{\left(\frac{\mu^2}{\Lambda^2_{\mathrm{QCD}}}\right)}},
\end{equation}
with $\Lambda_{\mathrm{QCD}} = 200\mathrm{\;MeV}$ QCD scale parameter. The renormalization scale at which the coupling is evaluated is the mean transferred transverse momentum and is treated separately for each process
\begin{equation}
    \mu = \sqrt{\langle p^2_T\rangle}=\begin{cases}
        \kappa_e \sqrt{\hat{q} \lambda_{\mathrm{mfp}}} & \mathrm{elastic\;\&\; conversion \;channels}\\
        \kappa_r \left(\hat{q}p\right)^{1/4} & \mathrm{radiative\;channels}
    \end{cases}.
\end{equation}
In the above, MARTINI uses $\hat{q}$, the mean squared exchanged transverse momentum per unit length, as the second moment of the HTL re-summed elastic scattering rate in the $q_{\perp}\ll T$ limit~\cite{Arnold:2008vd} 
    \begin{equation}
        \frac{d\Gamma_{\mathrm{elas.}}}{d^2\mathbf{q}_{\perp}} = \frac{C_{R}}{(2\pi)^2}\frac{4\pi \alpha_{s,0} m^2_D T}{\mathbf{q}^2_{\perp}(\mathbf{q}^2_{\perp} + m^2_D)},
    \end{equation}
where $m^2_D$ is the squared Debye mass and $C_{R}$ the Casimir factor of the jet parton. Thus, using Eq.~\eqref{eq:q-hat-from-rate}, $\hat{q}$ is given by
\begin{align}
    \hat{q} =\;& \int^{q_{\mathrm{max}}} d^{2}\mathbf{q}_{\perp} \mathbf{q}^2_{\perp} \frac{d\Gamma_{\mathrm{elas.}}}{d^2\mathbf{q}_{\perp}}\nonumber\\
    =\;&  C_{R}\alpha_{s,0}m^2_D\log{\left(1+\frac{q^2_{\mathrm{max}}}{m^2_D}\right)}.
\end{align}
where $q_{\mathrm{max}}=\sqrt{6pT}$ with $p$ denoting the momentum of the incoming hard parton. The mean free path, $\lambda_{\mathrm{mfp}} = \Gamma^{-1}_{\mathrm{elas.}}$ is calculated as the inverse of the total elastic scattering rate
\begin{equation}
    (\lambda_{\mathrm{mfp}})^{-1} = \int_{q_{\mathrm{min}}}^{q_{\mathrm{max}}} d^{2}\mathbf{q}_{\perp}\frac{d\Gamma_{\mathrm{elas.}}}{d^2\mathbf{q}_{\perp}},
\end{equation}
where $q_{\mathrm{max}}$ is the same as before and $q_{\mathrm{min}}=0.05T$ to be consistent with the elastic scattering rate tables. The free parameters of MARTINI with respect to running coupling, then, are $\alpha_{s,0}, \kappa_r$ and $\kappa_e$, which are fixed by fits to data. The remaining parameter of the model is the momentum cut that is applied to energy loss, radiated partons and recoil partons (see Sec.~\ref{sec:medium_response}). 

These cuts can be considered as the boundary between what is considered as a hard parton versus what should be viewed as a particle of the medium. In the case of energy loss, if the energy of a given hard parton (in the rest frame of the fluid) falls below this momentum cut, it is no longer allowed to interact with the medium. Similarly, in a radiative event, if the radiated parton is determined to have total momentum below this cutoff, its energy is subtracted from the jet parton (i.e., the energy loss is affected), but the radiated parton is not included in the event record. The same thought process is applied to the recoil partons involved in the elastic scattering processes. The value of this momentum cut can be changed separately for each of the cases described above. That is, one can choose $p^{\mathrm{eloss}}_{\mathrm{cut}} \neq p^{\mathrm{rad}}_{\mathrm{cut}} \neq p^{\mathrm{recoil}}_{\mathrm{cut}} $ though mostly these are chosen to be the same, with typical choices being a fixed value of $2\mathrm{\;GeV}$  or a multiple of the local temperature. 

In stand-alone or single-stage simulations, MARTINI also allows for the inclusion of the finite-size effect in the LO AMY rates. The numerical cost of solving the time-dependent AMY rates of Eq.~\eqref{eq:ftft.with.finite.size} makes their usage in dynamic jet energy loss calculations computationally prohibitive. Thus MARTINI uses a separation condition~\cite{Park:2016jap, Park:2018acg}
\begin{equation}
    \Delta r_{\perp} \Delta p_{\perp} > U(k, T),
    \label{eq:rndm_walk_finite_size}
\end{equation}
where $U(k,T)$ is the minimum allowed uncertainty and $k$ the momentum of the radiated parton. $U(k,T)$ is then parametrized by 
\begin{equation}
    U(k,T) = 0.25 \left(\frac{k}{T}\right)^{0.11}.
\end{equation}
where the numerical factors are chosen to modify the LO AMY rates [Eq.~\eqref{eq:splitting_rate_amy}] in an infinite medium to reproduce the rates which would result from Eq.~\eqref{eq:ftft.with.finite.size}~\cite{Park:2021yck}. After a radiative event, the radiated particle travels collinearly with the parent parton. At each subsequent time step, the condition of Eq.~\eqref{eq:rndm_walk_finite_size} is checked and if the partons are deemed to be coherent, only elastic scatterings with the thermal medium are allowed. Over time, the two particles will receive enough kicks from the medium particles to satisfy the condition, at which point the partons are once again allowed to have bremsstrahlung events.
            \subsubsection{CUJET}\label{sec:cujet}
                
Columbia University Jet flavor tomography program or CUJET~\cite{Buzzatti:2011vt, Buzzatti:2013scw, Xu:2014ica} is an implementation of the LO-DGLV parton energy loss model. As a stand-alone program for calculating energy loss, it uses the pQCD spectrum of un-quenched partons (including a $K$-factor to account for higher order corrections) and computes the energy loss by integrating Eq.~\eqref{eq:lo_dglv_gluon_brem_spec}. Elastic scatterings, in this application of CUJET, are calculated using the Thoma-Gyulassy model~\cite{Thoma:1990fm}
    \begin{equation}
    \begin{split}
        \frac{d[\Lambda(\mathbf{x})E(\mathbf{x})]}{d\tau} =&\;-C_R \pi \alpha_s\left(m^2_D\right)\alpha_s\left(6\Lambda E T\right) T^2 (1+\frac{N_f}{6})\\
        &\; \times\Theta(\Lambda E-M)\\
        &\; \times \log{\left(\frac{6T \sqrt{\Lambda^2 E^2-M^2}}{\Lambda E-\sqrt{\Gamma^2 E^2-M^2}+6T}\frac{1}{m_D}\right)}
    \end{split}.
    \label{eq:TG_model_elastic_in_cujet}
    \end{equation}
$E$, $T$ and $\Lambda$ are the energy of the hard parton, the local temperature and the relativistic correction accounting for the boost to the lab frame, respectively and are functions of the transverse spacetime location of the hard parton
\begin{equation}
    \mathbf{x}=(x_{0,\perp} +\tau \cos{\phi}, y_{0,\perp} + \tau \sin{\phi})
\end{equation} 
i.e. $E = E(\mathbf{x})$. In the above, $\mathbf{x}_0 = (x_{0,\perp},y_{0,\perp})$ denotes the production point of the jet that has traveled for the proper time $\tau$ in direction $\phi$. The relativistic correction factor is given by $\Lambda(\mathbf{x}) = u^{\mu}_{\mathrm{fluid}}v_{\mu,\mathrm{jet}}$, where $u^{\mu}$ is the fluid four-velocity. 

The un-quenched pQCD spectrum is then taken as the initial production probability. For each sampled hard parton transverse momentum from this distribution, energy loss is calculated using the radiative and elastic energy loss of Eqs.~\eqref{eq:lo_dglv_gluon_brem_spec} and~\eqref{eq:TG_model_elastic_in_cujet}, including the effect of fluctuations in the number of radiated gluons and elastic collisions as emphasized in Ref.~\cite{Wicks:2005gt}. This results in a probability distribution, which can then be folded with the original un-quenched pQCD spectrum, including the distribution of binary collisions in a given centrality. Finally, this spectrum is folded with a fragmentation function to arrive at a spectrum for the observable hadronic or leptonic spectrum.

The more recent implementation of CUJET in Ref.~\cite{Shi:2022rja}\;, as a Monte Carlo generator of low virtuality energy loss in a JETSCAPE framework, also uses the LO-DGLV energy loss formalism but applies it as a rate
    \begin{equation}
        \begin{split}
            &\frac{\mathrm{d}\Gamma^{\mathrm{DGLV}}_{i \to g i}}{\mathrm{d}z}(p,z,\tau, T) =\; \frac{18 C^R_{i}}{\pi^2} \frac{4+N_f}{16+9N_f} \rho(T)
    	\int{\mathrm{d}^2\mathbf{k}_{\perp}} \Bigg\{
    	\frac{1}{z_+} \left| \frac{\mathrm{d}z_+}{\mathrm{d}z}\right|\alpha_s
        \Big( \frac{\mathbf{k}^2_{\perp}}{z_+ - z_+^2} \Big)\\
        &\times \int \mathrm{d}^2\mathbf{q}_{\perp} \Bigg[
            	\frac{ \alpha_s^2(\mathbf{q}^2_{\perp})}{\mathbf{q}^2_{\perp}(\mathbf{q}^2_{\perp} + m_D(T)^2)}\\
        &\times\frac{-2}{(\mathbf{k}_{\perp}-\mathbf{q}_{\perp})^2+\chi^2(T)}
        \bigg(\frac{\mathbf{k}_{\perp}\cdot(\mathbf{k}_{\perp}-\mathbf{q}_{\perp})}{\mathbf{k}_{\perp}^2+\chi^2(T)} - \frac{(\mathbf{k}_{\perp}-\mathbf{q}_{\perp})^2}{(\mathbf{k}_{\perp}-\mathbf{q}_{\perp})^2+\chi^2(T)} \bigg)\\
        & \times \bigg(1-\cos\bigg(\frac{(\mathbf{k}_{\perp}-\mathbf{q}_{\perp})^2+\chi^2(T)}{2 z_+ p} \tau\bigg)\bigg) \Bigg] \Bigg\}.
        \end{split}
        \label{eq:lo_dglv_gluon_brem_rate}
    \end{equation}

In the above equation, $\rho(T)$ is the density of the dynamical scatterers in the medium, which is related to the entropy density via $\rho(T)=s(T)/4$. The entropy density is connected to temperature by the QCD equation of state.  
In both implementations, the running coupling is evaluated using its LO pQCD expression,
\begin{equation}
    \alpha_s(\mu^2) = \begin{cases}
        \alpha_{s,\mathrm{max}} & \mu \leq \Lambda_{\mathrm{QCD}}e^{2\pi/\alpha_{s,\mathrm{max}}}\\
        \frac{4\pi}{9\log{(\mu^2/\Lambda^2_{\mathrm{QCD}})}} & \mu > \Lambda_{\mathrm{QCD}}e^{2\pi/\alpha_{s,\mathrm{max}}}
    \end{cases}
    \quad,
\end{equation}
where $\alpha_{s,\mathrm{max}}$ is the only free parameter of the model, to be fixed by comparisons to data. Finally, the Debye mass, $m^2_D(T)$, is solved for using the self-consistent equation
\begin{equation}
    m^2_D(T) = 4\pi \alpha_s(m^2_D) T^2 (1 + {N_f}/{6}).
\end{equation}
            \subsubsection{From static to dynamical medium}
                For most event generators, routines of Monte Carlo simulations are first developed in a static medium and then extended to a dynamical one for realistic heavy-ion collisions. 

The QGP medium is generated by hydrodynamic simulation, which is now acknowledged as the ``standard model" of the bulk evolution in heavy-ion collisions. The hydrodynamic simulation provides a data file that includes the space-time distribution of the medium flow velocity, temperature, and other thermodynamic quantities requested by jet energy loss models. At each time step, with the knowledge of the position of a given jet parton, one may obtain the local flow velocity and thermodynamic quantities of the medium from the hydrodynamic data. Using the velocity, this hard parton is boosted into the local rest frame of the expanding medium, in which it can scatter with the medium based on algorithms developed in the previous subsections. The scattering rate here depends on both the medium temperature or other thermodynamic quantities, as well as the parton momentum in the local rest frame of the medium. After scattering, the final state of the jet partons, together with other newly produced partons, is boosted back to the global frame, where they propagate to locations of the next time step. This procedure is iterated for each parton until it exits the QGP, i.e., when the local temperature of its surrounding medium is below the hadronization temperature. 

            \subsubsection{Medium Response}\label{sec:medium_response} 
When a jet traverses a medium, the medium becomes excited due to the interaction with the jet. The excitation propagates through the medium, consequently influencing the distribution of final-state hadrons originating from the medium constituents. Since the jet induces the excitation and its propagation~\cite{Qin:2009uh,Neufeld:2009ep}, these hadron distribution alterations are correlated with the jet. Hence, this phenomenon, generally referred to as the \emph{medium response}, causes sizable impacts on jet observables~\cite{Casalderrey-Solana:2016jvj,Tachibana:2017syd,KunnawalkamElayavalli:2017hxo,Chen:2017zte,Chang:2019sae,Cao:2022odi}
\footnote{Conversely, there are studies that have discussed the impact of medium response to multiple minijet propagations on soft observables, such as $v_{2}$.~\cite{Schulc:2013kra,Floerchinger:2014yqa,Schulc:2014jma,Okai:2017ofp,Pablos:2022piv}} 
In this subsection, we show how the medium response is modeled in Monte Carlo event generators. 

In several MC generators~\cite{Zapp:2012ak,Zapp:2013vla,Wang:2013cia,He:2015pra,Yazdi:2022bru,Shi:2022rja}, interaction with the medium in some momentum transfer regions is modeled as scatterings with medium constituent partons, as previously mentioned. 
For each scattering, these generators simulate a collision between a jet parton and a parton sampled from the medium. 
The outgoing medium parton of the scattering is called the recoil parton. 
The recoils are treated the same as jet particles and involve successive scatterings and medium-induced radiations. 
In the two-stage models, recoil partons, supposed to have small virtuality, are passed to a module that handles low-virtuality jet partons.

On the other hand, the energy and momentum of the incoming medium parton of the scattering becomes a deficiency in the medium, called a hole. 
To track the energy-momentum balance of jets, the generators also keep records of holes. 
The evolution of recoils and the generation of holes are considered a representation of the semi-hard medium response. 
Recoils and holes are generated not only by elastic processes but also by scatterings of inelastic processes involving radiations.
In inelastic processes, to ensure energy and momentum conservation, as described in the reference~\cite{Luo:2023nsi}, the energy and momentum of the outgoing partons of the process are adjusted.

Incidentally, if the contribution of the hole is not converted to the depletion in the medium fluid through the source term mentioned later, its contribution must be appropriately subtracted to ensure energy-momentum conservation. 
There are various techniques for subtraction in jet analysis: simple methods include subtracting the four-momenta of holes that fall within a jet cone~\cite{JETSCAPE:2022jer}, and other approaches involve performing subtraction during the merging of subjets in jet reconstruction~\cite{Luo:2023nsi,JETSCAPE:2023hqn}, with some methods considering the use of a $\phi$-$\eta$ plane grid, similar to that in a detector~\cite{KunnawalkamElayavalli:2017hxo}.

Approximations relying on particle picture-based descriptions of jet constituents in MC generators, including recoils and holes, break down as the energies of those partons approach a comparable scale to the ambient medium temperature. 
Such soft components of jets are supposed to thermalize while diffusing their energy and momentum in the medium~\cite{Tachibana:2020mtb}. 
Once the thermalization is complete, they fully integrate with the medium, and all their energy and momentum are transported hydrodynamically via the bulk medium flow ~\cite{Stoecker:2004qu,Casalderrey-Solana:2004fdk}.

The bulk medium evolution, influenced by the energy-momentum injection from the thermalized portion of jets, can be modeled by solving the hydrodynamic equation with a source term~\cite{Chaudhuri:2005vc,Betz:2008ka}, 
\begin{align}
\partial_{\mu} T^{\mu \nu}_{\mathrm{fluid}} (x)= J^{\nu} (x), 
\label{eq:hydro-source}
\end{align}
where $T^{\mu \nu}_{\mathrm{fluid}} (x)$ is the energy-momentum tensor of the QGP fluid. The source term $J^{\nu} (x)$ accounts for four-momentum density incoming through the thermalization process. 

\begin{figure*}
  \centering
  \includegraphics[width=.75\textwidth]{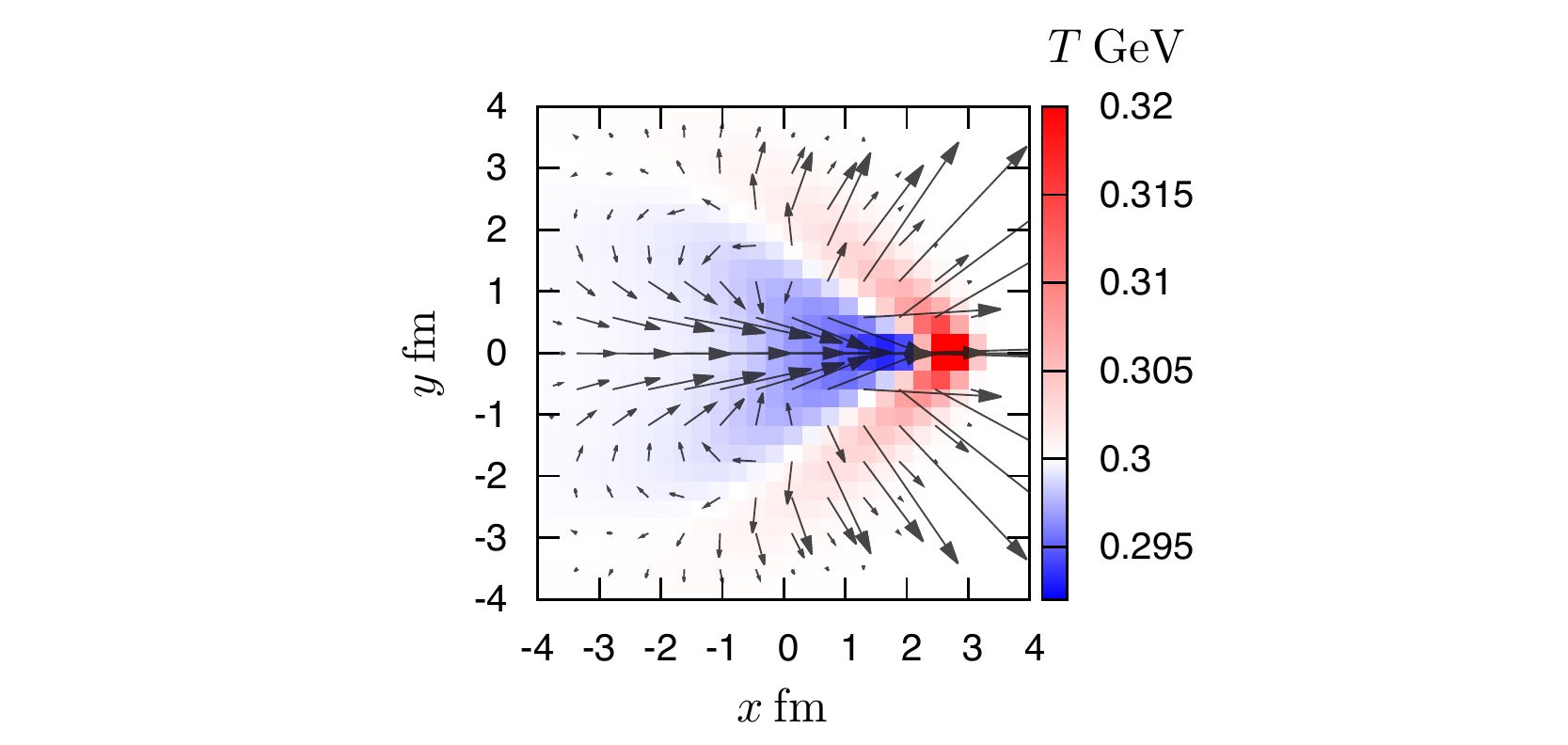}
  \vspace{-0.25cm}
  \caption{Temperature (heat map) and flow velocity (arrows) distributions in an ideal QGP fluid. 
  An energetic parton, which is currently at $\left(x,y,z\right)=\left(3~\mathrm{fm},0,0\right)$, travels in $x$-direction while depositing its energy and momentum into the three-dimensional ideal QGP fluid with uniform and static initial conditions at a temperature of 300 MeV~\cite{Tachibana:2012sa}. Figure from~\cite{Tachibana:2014yai}.
  }
  \label{fig:machcone2}
\end{figure*} 
Figure~\ref{fig:machcone2} shows the distribution of temperature change and flow velocity induced by a single high-energy particle passing through a three-dimensional medium fluid as a solution to Eq.~\eqref{eq:hydro-source}. 
In this example, energy and an equal amount of momentum in the travel direction (along the $x$-axis) are constantly deposited into the fluid cell containing the particle. 

One can observe a V-shaped wavefront, with the particle's position at its vertex (indicated by the region with a temperature rise). 
In three-dimensional space, this wavefront takes the form of a conical shape known as a Mach Cone ~\cite{Stoecker:2004qu,Casalderrey-Solana:2004fdk}. 
The Mach Cone is a type of shock wave formed by the interference of sound waves when an object moves through a fluid at a speed exceeding that of sound.

Inside the Mach Cone behind the particle, there is an area where the temperature decreases. 
Additionally, a strong forward flow following the particle is known as a diffusion wake~\cite{Casalderrey-Solana:2004fdk,Casalderrey-Solana:2006lmc,Gubser:2007ga,Betz:2008ka}. 
This diffusion wake is caused by the influx of momentum in the direction of the particle motion.

The source term $J^{\nu} (x)$ in Eq.~\eqref{eq:hydro-source} provides a profile of deposited energy and momentum, which can be modeled in a variety of ways. 
When integrating with MC jet shower generators, the source term becomes a superposition of four-momentum deposition from each jet parton: 
\begin{align}
  J^{\nu} (x) = \sum_{i} j_i^{\mathrm{\nu}}(x), 
\end{align}
where $i$ runs over all jet partons. 
For practical reasons for computations, the source term is often modeled with a Gaussian in the $\tau$-$\eta_{\mathrm{s}}$ coordinate system, as in Refs.~\cite{Chen:2017zte,Chen:2017zte,Pablos:2022piv,Casalderrey-Solana:2020rsj}:
\begin{equation}
j_i^{\mathrm{\nu}}(x) = 
\frac{-{d p^{\nu}_{i}}/{d\tau}}{\left(2\pi\right)^{3/2} \tau \sigma_{r}^2 \sigma_{\eta_{\mathrm{s}}}} 
 e^{-\frac{\left(x-x_i\right)^2+\left(y-y_i\right)^2}{2\sigma_{r}^2}}
 e^{-\frac{\left(\eta_{\mathrm{s}}-\eta_{\mathrm{s},i}\right)^2}{2\sigma_{\eta_{\mathrm{s}}}^2}}, 
\label{eq:source-term-gauss}
\end{equation}
where $p^{\mu}_{i}$ and $(x_i,y_i,\eta_{\mathrm{s},i})$ are four momentum and the position in the $\tau$-$\eta_{\mathrm{s}}$ coordinate system of the $i$-th jet parton at the proper time $\tau$, respectively. 
Here the instantaneous thermalization of the deposited four-momentum is assumed. 
The parameters $\sigma_{r}$ and $\sigma_{\eta_{\mathrm{s}}}$ controls the widths of the Gaussian smearing in the transverse and longitudinal directions, respectively.

For the MC generators, which simulate scatterings with sampled medium partons for each energy-momentum transfer within a medium (e.g., LBT), a lower cutoff energy limit ($E^{\mathrm{dep}}_{\mathrm{cut}}$) is introduced for the jet shower partons described based on the particle picture. 
Then, if soft partons with energy (in the local rest frame of the fluid) below the cutoff or holes are created, they are assumed to be thermalized and a source term is generated for each of them. For the case with the source term in Eq.~\eqref{eq:source-term-gauss}, one sets 
\begin{align}
  -\frac{d p^{\nu}_{i}}{d\tau} = -\Delta p^{\nu}_{i} \delta\left(\tau - \tau_i\right) =
  \begin{cases}
  p_i^{\mathrm{\nu}}\delta\left(\tau - \tau_i\right) 
  &\!\!\!\text{for partons with }p_i^{\mu}u_{\mu}\!<\! E^{\mathrm{dep}}_{\mathrm{cut}}\\
  - p_i^{\mathrm{\nu}} \delta\left(\tau - \tau_i\right)
  &\!\!\!\text{for hole partons}\\
  0 &\!\!\!\text{otherwise}
  \end{cases}\!,
\end{align}
where $\tau_i$ is the creation proper time of the $i$-th jet parton and $u^{\mu}$ is the flow velocity of the fluid at the parton's position.

As a dynamic and relativistic extension of the source term in Eq.~\eqref{eq:source-term-gauss}, using the causal diffusion equation~\cite{Aziz:2004qu}
\begin{align}
\left[
\frac{\partial}{\partial t}+t_\mathrm{relax}\frac{\partial^2}{\partial t^2}-D_\mathrm{diff}\nabla^2
\right]
\mathcal{J}_i^{\nu}(x)
&=
0,
\label{eq:causal_diff}
\end{align}
the source term can be modeled as~\cite{Tachibana:2020mtb}  
\begin{align}
j^{\nu}_i(x) = \mathcal{J}_i^{\nu}(x) \delta\left(\tau - \tau_i - \tau_{\mathrm{th}} \right). 
\label{eq:causal_diff_source}
\end{align}
This corresponds to a phenomenological description of the spacetime evolution of energy and momentum during the thermalization process, from when the particle picture for the jet parton no longer works effectively until it becomes part of the fluid, using~Eq.~\ref{eq:causal_diff}. 
Here $\tau_{\mathrm{th}}$ is the proper time taken for this thermalization process. 
The parameters $t_\mathrm{relax}$ and $D_\mathrm{diff}$ represent the relaxation time and the diffusion coefficient, respectively. Their values are selected to satisfy the causality: $v_\mathrm{sig}=(D_\mathrm{diff}/t_\mathrm{relax})^{1/2}\leq 1$, where $v_\mathrm{sig}$ is the speed of signal propagation. 
This approach is adopted in the Causal Liquefier module of the JETSCAPE package~\cite{JETSCAPE:2020uew}, and the source term of Eq.~\ref{eq:causal_diff_source} is generated by using the solution of  Eq.~\eqref{eq:causal_diff} 
for the initial condition $\mathcal{J}_i^{\nu} = \Delta p^{\nu}_{i}\delta^{(3)}(\vec{x}-\vec{x}_i)$ and 
${\partial \mathcal{J}_i^\nu}/{\partial t} =0$ at $\tau=\tau_i$. 

Hydrodynamic flow induced by jet energy-momentum deposition leads to correlations to jet propagation in hadron production in the bulk medium fluid. 
Therefore, for jet reconstruction or hadronic jet-correlated observables, 
one needs to take into account the contributions of the medium hadrons in a model that incorporates the hydrodynamic medium response by Eq.~\eqref{eq:hydro-source}. 
Simply, as in the conventional hydrodynamic models, medium hadrons can be obtained using the Cooper-Frye formula~\cite{Cooper:1974mv}. 
This approach requires an appropriate background to construct observables comparable with experimental data~\footnote{To estimate the pure contribution from jets, one should subtract the same hydrodynamic event without jet production.}.

            \subsubsection{Hadronization of jets}\label{sec:hadronization} 
                
    Due to the confinement property of QCD, the partons coming out of a high energy collision get dressed into hadrons. Therefore, in order to study jets and jet energy loss in QGP and compare them to experimental results, one also has to convert the partons coming out of calculations or simulations of high energy collisions to hadrons. Here, we briefly discuss the various models employed in the hadronization of jets and high-$p_T$ partons. \\

\emph{Fragmentation Functions:}
    An example of a fragmentation function was shown in Eq$.$~\ref{eq:Dz}. The fragmentation function, $D(z,\mu^2_F)$ encodes the non-perturbative physics of hadronization and can be thought of as providing the probability, when evaluated at a factorization scale $\mu_F$, of a given parton of momentum $P$ to fragment to a specified hadron carrying momentum $z_{h}P$. These functions are process-independent and can be fitted to experimental data to be used in theoretical calculations.
    Usage of fragmentation functions in heavy-ion collision is in the mode of Eq.~\ref{eq:AA-factorized-formula} where $\tilde{D}$ is taken as the $p$-$p$ or vacuum fragmentation function and the medium modifications are absorbed in the $\mathscr{P}$ term, computed using the various models and theories introduced in the previous sections. 
    As for the limitations of fragmentation functions, an important one is that they provide hadrons that are collinear to the incoming parton. Therefore, while useful for calculations of inclusive spectra or correlations, one cannot rely on them for jet-substructure studies. There is interest in transverse momentum-dependent fragmentation functions, though these have yet to be used in jet energy loss calculations. Other limitations, at present, are the lack of more recent experimental data and fits for fragmentation functions (particularly for parton to photon fragmentation). This issue was studied in Ref.~\cite{Klasen:2017dsy} where it was found that using NLO matrix elements for the hard process, followed by a parton-shower approach, had more success at matching experimental data for hard photons in heavy-ion collisions. \\

\emph{Lund String Model:}
    The Lund string model~\cite{Andersson:1983ia,andersson_1998,Ferreres-Sole:2018vgo} is the hadronization model employed by the Pythia Monte Carlo generator. From Lattice QCD studies, it is known that the color potential between a static pair of quark and anti-quark has the following shape as a function of the separation between the two fermions~\cite{Bali:1994de}
    \begin{equation}
        V_{q,\bar{q}}(r) \sim -\frac{4}{3}\frac{\alpha_s}{r} + \kappa r,
    \end{equation}
    where $\kappa \approx 1\;\mathrm{GeV}/\mathrm{fm}$ is the string tension. 
    
    The Lund model focuses on the linear, confining part of the potential as hadronization is a long-distance phenomenon. In this picture, the quark and anti-quark are connected by a massless color string and the hadronization process occurs via splittings of the string. Gluons are incorporated as excitations or kinks on the string, thus a string starts from a $q(\bar{q})$ and can pass through a number of gluons before terminating on a $\bar{q}(q)$.   

    In the fragmentation process, if the length of the string is long enough, and string breaking is energetically favorable, a new pair of $q_{1}\bar{q}_1$ is sampled from the vacuum. The process is taken to be symmetric: it should not matter from which end of the system ($q$ or $\bar{q}$ the fragmentation process starts). This leads to the Lund symmetric fragmentation function for the distribution of the light cone momentum fraction, $z$, taken by the hadron~\cite{Sjostrand:1984}
    \begin{equation}
        f(z) = \frac{(1-z)^a}{z} e^{-bm^2_{\perp}/z},
    \end{equation}
    where $a$ and $b$ are parameters of the model. The transverse mass of the hadron, $m_{\perp}$, is given by
    \begin{equation}
        m^2_{\perp} = m^2 + p^2_{\perp}, 
    \end{equation}
    and is used if the quarks have transverse momenta. If this is not so, then $m_{\perp}$ reduces to $m$. 
    
    If the produced $q\bar{q}$ pair are without mass or transverse momenta, they are produced as real particles in the same vertex. However, if they possess either mass or transverse momentum, then they cannot be produced in the same vertex as real particles. Instead, they are deemed to have been produced as virtual particles that have to tunnel out of a distance of $m_{q,\perp}/\kappa$. This results in a Gaussian suppression factor of $e^{\left(-\pi m^2_{q,\perp}/\kappa\right)}$ for string breaking, suppressing strange or heavy quark production~\cite{Andersson:1983ia}. Given the large suppression implied by this effect for charm and bottom quarks, it is assumed that these quarks are produced in hard events only and are not generated in string breaking.

    The model is also capable of producing baryons. In this case, a diquark-anti-diquark pair is generated without breaking the string. Then, by generating a meson between the two baryons, the string is broken, and the hadrons are created. 

    The Lund model can be used for hadronization in Monte Carlo generators, for both $p$-$p$ and heavy-ion collisions. The model can be used in two ways. In the typical method, the color of each partons is tracked through the shower (referred to as the colored method in JETSCAPE). For cases with multiple scattering in the medium, the colors of the surviving partons are assigned at exit (referred to as colorless hadronization in JETSCAPE)~\cite{JETSCAPE:2019udz}.
    The difference between the two methods lies in string formation and color assignment. In the first method, the colored hadronization scheme, the assigned colors during the hard process generation are respected and carried through to the fragmentation stage. On the other hand, in the colorless hadronization method, the color information of the event is wiped at the end of the evolution and strings are connected (colors assigned) by minimizing the parton distances in the $\eta$-$\phi$ plane. The distance measure used in this process, for two partons $i$ and $j$, is given by
    \begin{equation}
        \Delta R_{i,j} = \sqrt{\Delta\eta_{i,j}^2 + \Delta\phi_{i,j}^2},
    \end{equation}
    where $\Delta \eta_{i,j}$ and $\Delta \phi_{i,j}$ denote the difference between the pseudorapidity and azimuthal angles of the partons, respectively. Thus, the colored hadronization scheme is more easily associated with $p$-$p$ collisions while colorless hadronization is most suitable for heavy-ion collision events, where the propagation of the hard parton through the QGP can result in the random changes in the color information of the parton shower.  \\

\emph{Cluster Model:}
    Cluster model~\cite{Webber:1983if} is available as the default hadronization model of HERWIG~\cite{Bahr:2008pv,Bellm:2015jjp} and SHERPA~\cite{Sherpa:2019gpd}. In this model, the final state gluons in the event record are assigned a constituent gluon mass ($m_g\approx \mathcal{O}(1)$ GeV) and allowed to decay to quarks and anti-quarks. Thus, the event record becomes a list of color-connected quark-antiquark and di-quarks and anti-diquarks. The formed color singlet clusters are then treated as excited hadron resonances and decayed to the final state, stable hadrons, which are observed experimentally.  
    This model of hadronization has fewer free parameters relative to the Lund model though it does not perform as well as in reproducing experimental results. It is also mostly applied to $p$-$p$ event simulations. \\

\emph{Hybrid Hadronization:}
As the name suggests, {Hybrid Hadronization}~\cite{Han:2016uhh,Kordell:2021prk} is a mixture of string based hadronization and a somewhat modified version of cluster hadronization, referred to as the Recombination or Coalescence Model~\cite{Hwa:2004ng,Hwa:2002tu,Fries:2003vb,Fries:2003kq,Kolb:2004gi,Molnar:2003ff}. In the Recombination model, one uses Cooper-Frye hadronization to freeze out the QGP fluid into an assembly of constituent quarks and anti-quarks. These quarks and anti-quarks are then recombined into hadronic resonances using a space-time and momentum space probability distribution, which is enhanced for constituent quarks and anti-quarks that are close in the phase space. In Hybrid Hadronization, recombination is a probabilistic process that varies from event-to-event. Quarks and anti-quarks that remain unrecombined are then attached to the ends of the string and then fragmented to hadrons.

    \section{Phenomenology}~\label{sec:Pheno}   
        Given the range of components in a typical simulation of a heavy-ion collision, given the uncertainties of transitions from one sub-simulator to the next, the ability to vary sub-simulators within an end-to-end simulation, and not just the parameters, is an essential requirement to systematically compare with entire data sets in heavy-ion collisions. 
Thus, it is our recommendation that serious attempts at phenomenology should be carried out using an event generator (or simulator) framework. At the time of this writing, the only such complete framework is the Jet Energy-loss Tomography with a Statistically and Computationally Advanced Program Envelope (JETSCAPE) model factory. 

        \subsection{Non-Framework Efforts} 
            In the remainder of this Chapter, we will discuss wide ranging comparisons with experimental data using JETSCAPE based simulations. However, in the interest of completeness, we will highlight other simulators which are also currently, widely used:

\begin{itemize}

    \item HIJING: The Heavy-Ion Jet Interaction Generator was by far the first end-to-end generator~\cite{Wang:1991hta,Gyulassy:1994ew}. It would simulate a heavy-ion $A$-$A$ and $p$-$A$ collision as an assembly of $p$-$p$ collisions without the production of a Quark Gluon Plasma. It is still used today to simulate baseline events in a variety of experiments. 

    \item JEWEL: The Jet Evolution With Energy Loss stands apart from most other approaches to jet modification in that it is not directly derived from any particular model of energy loss. Scatterings off a PYTHIA shower are sampled using $2$-to-$2$ scatterings, with radiative corrections and recoils added similar to the first hard scattering~\cite{Zapp:2008gi,Zapp:2008gi,Zapp:2012ak,KunnawalkamElayavalli:2017hxo}. While not incorporated within a fluid dynamical simulation, JEWEL is widely used in comparisons with data. 

    \item EPOS: While not a framework, the EPOS model~\cite{Pierog:2013ria,Pierog:2009zt,Porteboeuf:2010um} provides another end-to-end simulator for $p$-$p$, $p$-$A$ and $A$-$A$ collisions. It is also one of the early adopters of the hydro+hard scattering approach in small systems.
    
    \item HYBRID: The hybrid Strong weak approach to jet modification takes PYTHIA hard scattering and parton shower events, and induces strong coupling energy loss on each of the partons in the shower~\cite{Casalderrey-Solana:2014bpa,Casalderrey-Solana:2015vaa, Casalderrey-Solana:2016jvj}. In this sense, it is also a multi-stage simulator, with earlier versions incorporated in JETSCAPE~\cite{Park:2019sdn}.

    \item CoLBT-Hydro: While the LBT model of scattering, recoil and radiation~\cite{Luo:2023nsi} is one of the standard low virtuality generators in JETSCAPE, CoLBT-hydro is the enhancement of this approach with soft partons and holes thermalized within a fluid dynamic source term, modifying the future evolution of the fluid. Thus, it reproduces exact medium response from jets, without a putative high virtuality stage~\cite{Chen:2017zte,Chen:2020tbl}.

\end{itemize}
        
        \subsection{Frameworks - Model Factory}\label{sec:frameworks_modelfac} 
            \begin{figure}
    \centering
    \includegraphics[width=\linewidth]{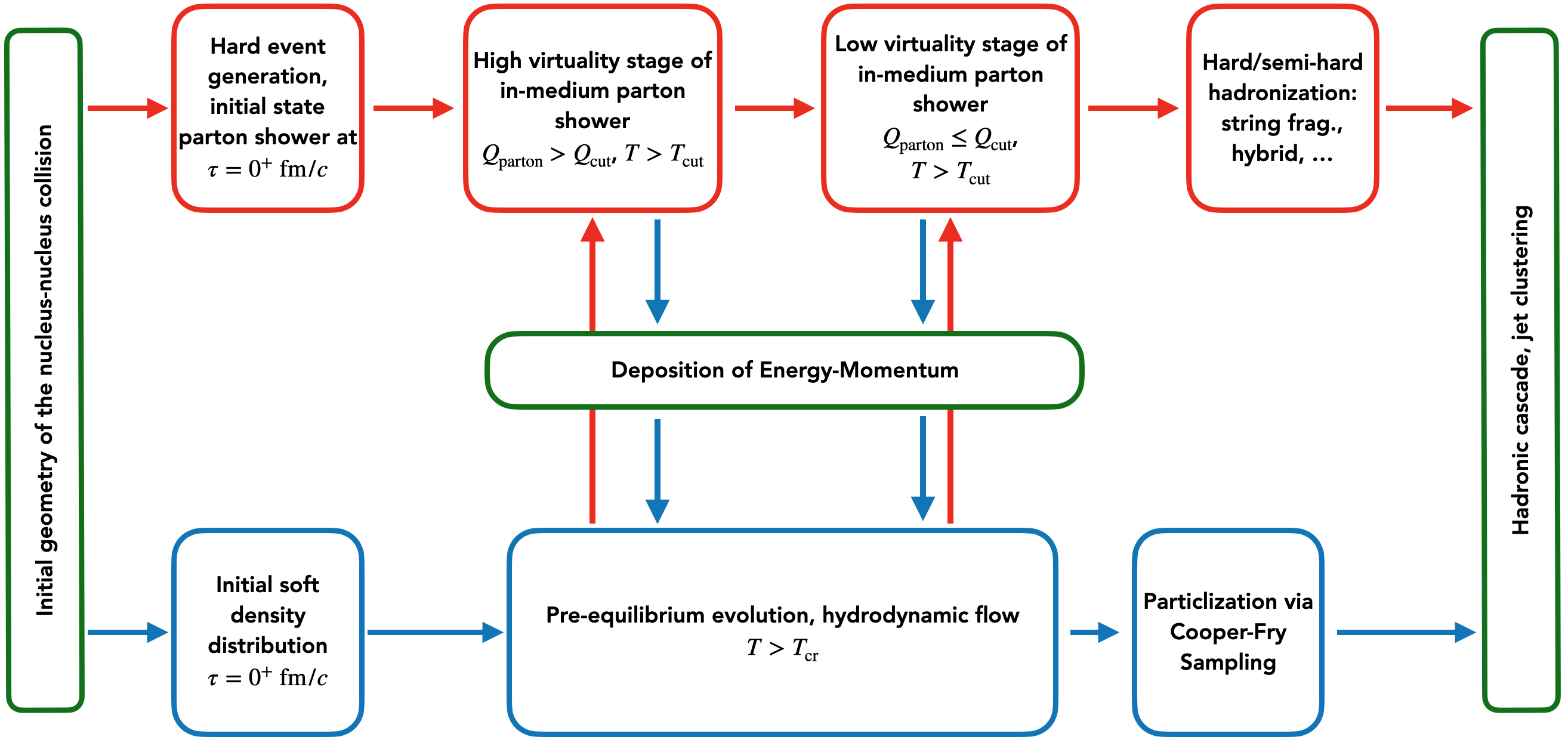}
    \caption{Flow of information in a JETSCAPE framework.}
    \label{fig:jetscape_framework_flowchart}
\end{figure}

For a complete description of different stages of jet-QGP interactions, the JETSCAPE framework~\cite{JETSCAPE:2017eso,Putschke:2019yrg} has been developed as a platform on which jet models designed for different kinematic regions are allowed to work together within a unified framework. As illustrated in Fig.~\ref{fig:jetscape_framework_flowchart}, this framework consists of modules for initialization, bulk evolution, jet evolution, hadronization and hadronic scattering. 

To start with, an initialization model (e.g., Glauber, Trento) is applied to determine the spatial distribution of both the energy density of the bulk matter and the production vertices of jet partons. The bulk part then evolves through the pre-equilibrium stage model (e.g., free streaming) and the hydrodynamic model (e.g., MUSIC), providing the spacetime profile of the medium temperature and flow velocity. 

Meanwhile, a hard event generator (e.g., PYTHIA) is used to produce jet partons from hard scatterings between nucleons. These initial partons can include contributions from multi-parton interactions and initial-state showers, and are placed at the locations drawn from the initialization model. These hard partons then interact with the bulk matter through several stages. For example, when the virtuality scale of a hard parton is higher than a preset medium scale ($Q_\mathrm{cut}$), this parton belongs to a rare-scattering-induced-multiple-emission stage, and thus evolves based on the Sudakov formalism with medium-modified splitting functions (e.g., MATTER). During sequential splittings, the parton virtuality gradually drops towards $Q_\mathrm{cut}$. If it reaches $Q_\mathrm{cut}$ before exiting the QGP, or with the local temperature above $T_\mathrm{cut}$, it is transferred to a transport model (e.g., LBT, MARTINI, CUJET) where it undergoes multiple scatterings through the QGP with gluon emissions induced by these scatterings. 

During jet-medium interactions, the jet modules take the medium information from the bulk modules, based on which the former solves evolution equations for the jet partons. At the same time, jet modules can convert soft partons produced from these interactions into energy-momentum deposition and feed it back into the bulk modules. This deposition affects the subsequent evolution of the bulk and constitutes the jet-induced medium excitation. 

A parton stops interacting with the QGP when its surrounding medium temperature is below $T_\mathrm{cut}$. 
Below this temperature threshold, jet partons cease interacting with the medium and  
are converted to hadrons through their own hadronization models (e.g., string fragmentation, hard-soft recombination, see Sec.~\ref{sec:hadronization}).  
Note that if a hard parton leaves the plasma with its virtuality higher than the hadronization scale in vacuum (usually taken as $Q_\mathrm{vac}=1$~GeV), it is evolved by MATTER vacuum shower to $Q_\mathrm{vac}$ first and then hadronized. This is crucial for ensuring that the observables smoothly approach their baselines in $p$-$p$ collisions when medium effects vanish. In the end, hadrons from hard and soft parts can evolve through hadronic scatterings together (e.g., SMASH) until the system is sufficiently dilute. One can analyze the final state particles from this framework in the same way as experimentalists treat signals from detectors, and compare the extracted observables with experimental data. The JETSCAPE framework is designed such that its modules can be easily modified, replaced or added, as long as one clearly defines the criteria that send a parton between different modules.

        
        \subsection{Data Comparison} \label{sec:data-comparison}
            In this section, we present comparisons of jet simulations and energy loss calculations to experimental data. In particular, we will demonstrate the benefit of a framework approach to Monte Carlo simulations, as emphasized in Sec.~\ref{sec:frameworks_modelfac}. 
The jet energy loss results shown here are results of jet interaction with simulations of the soft sector. The parameters of these soft-sector calculations were fixed in a major Bayesian analysis of heavy-ion collisions~\cite{Bernhard:2019bmu}, using Trento + Free-streaming + VISHNU(2+1)D with UrQMD hadronic.

\subsubsection{Two-stage simulation comparison of CUJET \& MARTINI}
    Here we consider CUJET and MARTINI, discussed previously in Sec.~\ref{sec:cujet} and Sec.~\ref{sec:martini}, respectively. These models are used as low-virtuality models in a two stage simulation, the high-virtuality stage of which is handled by MATTER. 
    
    \begin{figure}
        \centering
        \includegraphics[width=\linewidth]{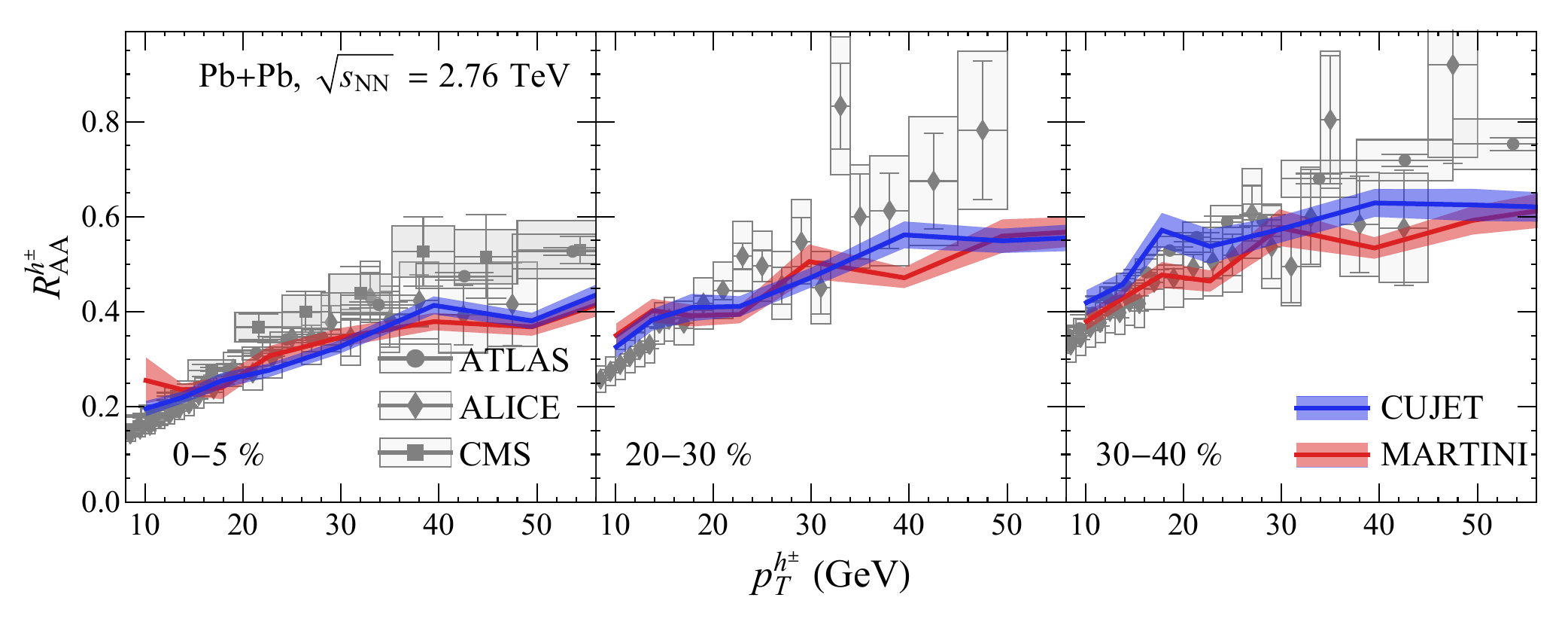}
        \vspace{-1cm}
        \caption{Comparison of the calculation of charged hadron nuclear modification factor as a function of the charged hadron transverse momentum ($p_T$) for pseudo-rapidity $|\eta|<1.0$ using two-stage simulations of CUJET or MARTINI for three centrality classes. Data points are experimental measurements from ATLAS~\cite{ATLAS:2015qmb}, ALICE~\cite{ALICE:2012aqc} and CMS~\cite{CMS:2012aa}. Figure taken from Ref.~\cite{Shi:2022rja}.}
        \label{fig:martini_vs_cujet_charged_hadron_raa}
    \end{figure}
    As a pictorial representation of the evolution of the simulation, consider Fig.~\ref{fig:jetscape_framework_flowchart}. The event flow begins by the generating of the hard process along with its location in the transverse plane at midrapidity. This is done using Pythia and includes multi-parton interactions as well as initial state parton shower. The evolution of the hard partons then proceeds on a parton-by-parton basis. If a parton is highly virtual ($Q>Q_{\mathrm{cut}} $), its evolution is handled by MATTER. Partons with virtuality lower than this value are passed to the low-virtuality model, CUJET or MARTINI. Hard partons are further evolved in the medium with interactions governed by CUJET or MARTINI, experiencing elastic or radiative events, until they are frozen out of evolution. This condition is met when either local temperature falls below $T_{\mathrm{cut}}=160$ MeV or parton momentum falls below the momentum cut $p_{\mathrm{cut}}=2$ GeV. Once all partons are frozen out, the event is prepared for hadronization which is done via the colorless hadronization mechanism described in Sec.~\ref{sec:hadronization}.
   \begin{figure}
        \centering
        \includegraphics[width=\linewidth]{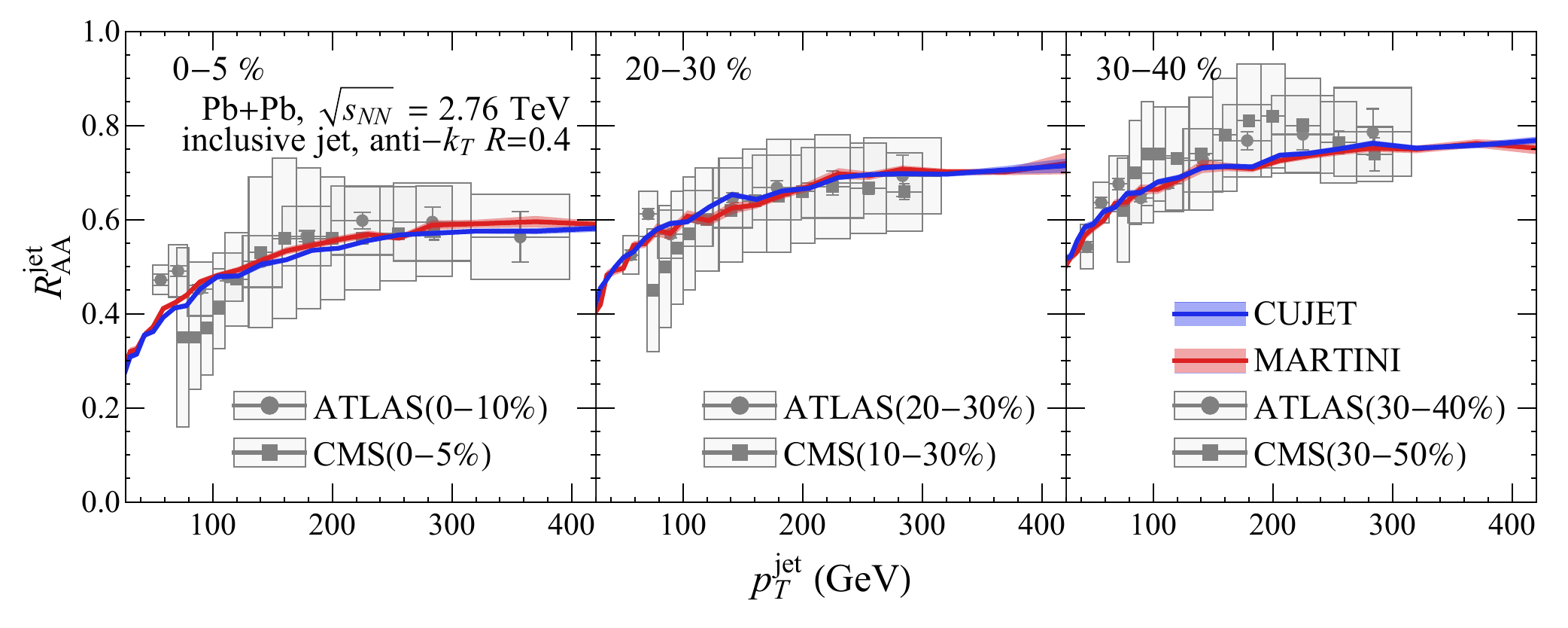}
        \vspace{-1cm}
        \caption{Nuclear modification factor of inclusive jets, in three centrality classes, for Pb$-$Pb collisions at $\sqrt{s}=2.76$ ATeV. Experimental data are from ATLAS~\cite{ATLAS:2014ipv} and CMS~\cite{CMS:2016uxf} Collaborations, with $|\eta_{\mathrm{jet}}|<2.1$ and $|\eta_{\mathrm{jet}}|<2.0$, respectively. The theory curves are computed for $|\eta_{\mathrm{jet}}|<2.0$. Jet cone radius for theory curves and experimental data is $R=0.4$. Figure taken from Ref.~\cite{Shi:2022rja}}.
        \label{fig:martini_vs_cujet_jet_raa}
    \end{figure}
    All parameters that do not pertain to CUJET or MARTINI are fixed to values provided by the JETSCAPE Collaboration~\cite{JETSCAPE:2019udz}. The remainder is the parameters governing the running of the strong coupling in CUJET or MARTINI, which are fixed by fits to charged hadron nuclear modification factor data for Pb$-$Pb collisions at $\sqrt{s}=2.76$ ATeV and $0$-$5\%$ centrality. Charged hadron nuclear modification factor or charged hadron $R_{\mathrm{AA}}$ is defined as
    \begin{equation}
        R^{h^{\pm}}_{\mathrm{AA}} = \frac{\frac{d\sigma^{h^{\pm}}_{AA}}{dp_T d\eta}}{\langle N_{\mathrm{bin}}\rangle \frac{d\sigma^{h^{\pm}}_{pp}}{dp_T d\eta}},
        \label{eq:charged_RAA}
    \end{equation}
    where $\langle N_{\mathrm{bin}}\rangle$ is the average number of binary collisions provided by the Monte Carlo Glauber model. Figure~\ref{fig:martini_vs_cujet_charged_hadron_raa} shows the data used in the fit (left) and two other centralities. The agreement between the two multi-stage models and the data is very good, as is the agreement between the models themselves. The latter can be due to the inclusive nature of charged hadron $R_{\mathrm{AA}}$. Thus, we consider more discriminatory observables: jets and their substructure. Given the three dimensional nature of jets, they are more likely to show sensitivity to the differences underlying our models. 

    Figure~\ref{fig:martini_vs_cujet_jet_raa} shows the results of a calculation of jet nuclear modification factor, defined analogously to charged hadron $R_{AA}$ of Eq.~\ref{eq:charged_RAA},
    \begin{equation}
        R^{\mathrm{jet}}_{\mathrm{AA}} = \frac{d\sigma^{\mathrm{AA}}/dp_T d\eta}{\langle N_{\mathrm{bin}\rangle} d\sigma^{\mathrm{pp}}/dp_T d\eta}.
        \label{eq:jet_raa}
    \end{equation}
    The inclusive jet $R_{\mathrm{AA}}$ results of Fig.~\ref{fig:martini_vs_cujet_jet_raa} are calculations of this quantity using the same events which generated Fig.~\ref{fig:martini_vs_cujet_charged_hadron_raa}. The anti-$k_T$ algorithm~\cite{Cacciari:2005hq, Cacciari:2011ma} is used to cluster the jets for a jet cone size of $R=0.4$. 

    Much like the charged hadron $R_{\mathrm{AA}}$, results for inclusive jets are in good agreement with the data. The two composite models of MATTER + CUJET and MATTER + MARTINI are also in agreement with each other. Two-stage simulations, therefore, are readily capable of reproducing experimental observations of charged hadron and jet nuclear modification factors. Nuclear modification factor of inclusive jets can also be considered for different jet cone radii. 
    
    \begin{figure*}[!h]
        
        \includegraphics[width=\linewidth]{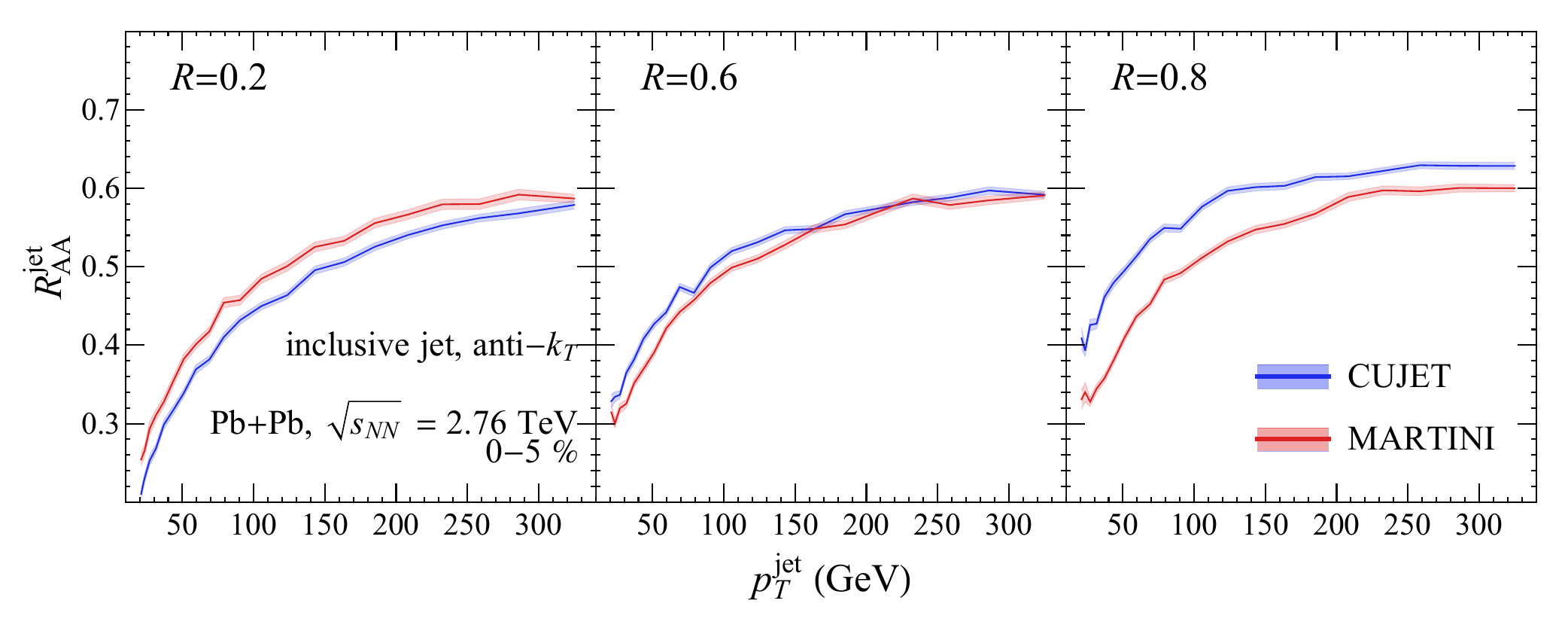}
        \vspace{-1cm}
        \caption{Inclusive jet nuclear modification factor for Pb-Pb collisions at  $\sqrt{s}=2.76$ ATeV and $0$-$5\%$ centrality. The figure shows the results for jet cone radii $R=0.2, 0.6$ and $0.8$ for left, center and right figures, respectively.  The jets are clustered at midrapidity, with $|\eta_{\mathrm{jet}}|<2.0$. The relative movement of the results MATTER+CUJET model relative to MATTER+MARTINI model is clearly visible. Figure from Ref.~\cite{Shi:2022rja}.}
        \label{fig:martini_vs_cujet_jet_raa_func_R}

        \includegraphics[width=\linewidth]{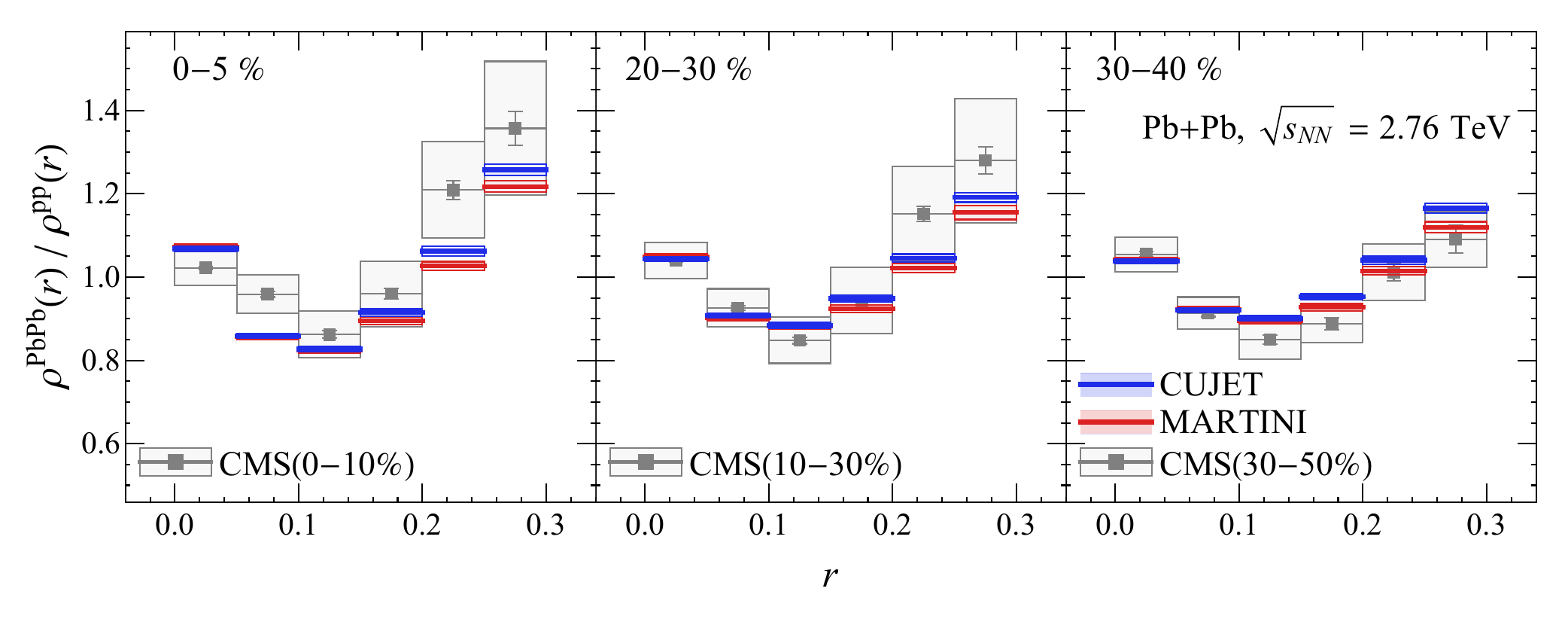}
        \vspace{-1cm}
        \caption{Jet shape ratio for the Pb-Pb system at $\sqrt{s}=2.76$ ATeV, for three centrality classes. The jets are clustered for $0.3 < |\eta_{\mathrm{jet}}|<2.0$ for jets with $p^{\mathrm{jet}}_{T} > 100$ GeV and a cut placed on the charged hadron tracks $p^{\mathrm{trk}}_T > 1.0$ GeV, to match the cuts placed on the data. Experimental results are from the CMS Collaboration~\cite{CMS:2013lhm}. The centrality classes of the data are, left to right, $0$-$10\%$, $10$-$30\%$ and $30$-$50\%$ while the theory curves are computed for $0$-$5\%$, $20$-$30\%$ and $30$-$40\%$. Figure from Ref.~\cite{Shi:2022rja}.}
    \label{fig:martini_vs_cujet_jet_shape_ratio}
    \end{figure*}
    By considering different jet cone radii at midrapidity, we can see in Fig.~\ref{fig:martini_vs_cujet_jet_raa_func_R}  a relative movement between the models where increasing the jet cone radius causes the CUJET model results to move above the MARTINI results. This is indicative of an interplay between the radiative and elastic channels. From the rates themselves, in Fig.~\ref{fig:rates_dglv_vs_amy}, it is evident that for a given hard parton, MARTINI, using AMY rates, is more likely to radiate soft gluons. These gluons are more readily deflected by subsequent elastic scatterings with the medium as compared to harder gluons which are relatively more likely from the LO-DGLV rates of CUJET. Thus, as the jet cone radius is increased, CUJET will recover some of the previously radiated, softer gluons that had been scattered out of the jet cone. On the other hand, it will take a larger jet cone radius for MARTINI to recover the radiated soft gluons. We can finally consider a jet-substructure observable that complements this logic. Jet shape is defined as 
    \begin{equation}
        \rho(r) \equiv \frac{N_{\mathrm{norm}}}{N_{\mathrm{jet}}} \sum_{\mathrm{jets}} \sum_{r\in [r_{\mathrm{min}}, r_{\mathrm{max}})} \frac{p^{\mathrm{trk}}_T/p^{\mathrm{jet}}_T}{r_{\mathrm{max}}-r_{\mathrm{min}}},
        \label{eq:jet_shape}
    \end{equation}
    which looks at the shape of the jet in the $\phi-\eta$ plane, as a function of $r=\sqrt{\Delta \phi^2 + \Delta \eta^2}$, using charged particles. Calculation of the ratio of this quantity in $A$-$A$ and $p$-$p$ collisions is presented in Fig.~\ref{fig:martini_vs_cujet_jet_shape_ratio}, constructed using the same events as the previous observables, for three centrality classes. It is clear that close to the jet axis, the two models are in good agreement and systematic differences appear only as we move towards the outer annuli of the jets (bins of $r>0.05$).  

    To conclude, what is presented here is the ability of a Monte Carlo framework to simultaneously reproduce observations of the nuclear modification of charged hadrons and jets, as well as substructures of jets. Furthermore, the framework also allows us to faithfully compare different models of jet energy loss, within an evolving QGP, in a significantly more controlled fashion than previously available. 

\subsubsection{Two-stage simulations of MATTER+LBT}
Here, compared to experimental data, we demonstrate the results from the two-stage jet MC simulations by MATTER+LBT employing the effective $\hat{q}$ with the explicit virtuality dependence discussed in Sec.~\ref{sec:Theory-2-q2-dependence} within the JETSCAPE framework. 
All results presented here incorporate the medium response by the recoil prescription in both MATTER and LBT and do not include the hydrodynamic medium response via Eq.~\eqref{eq:hydro-source}

\begin{figure*}[!h]
  \includegraphics[width=.495\textwidth]{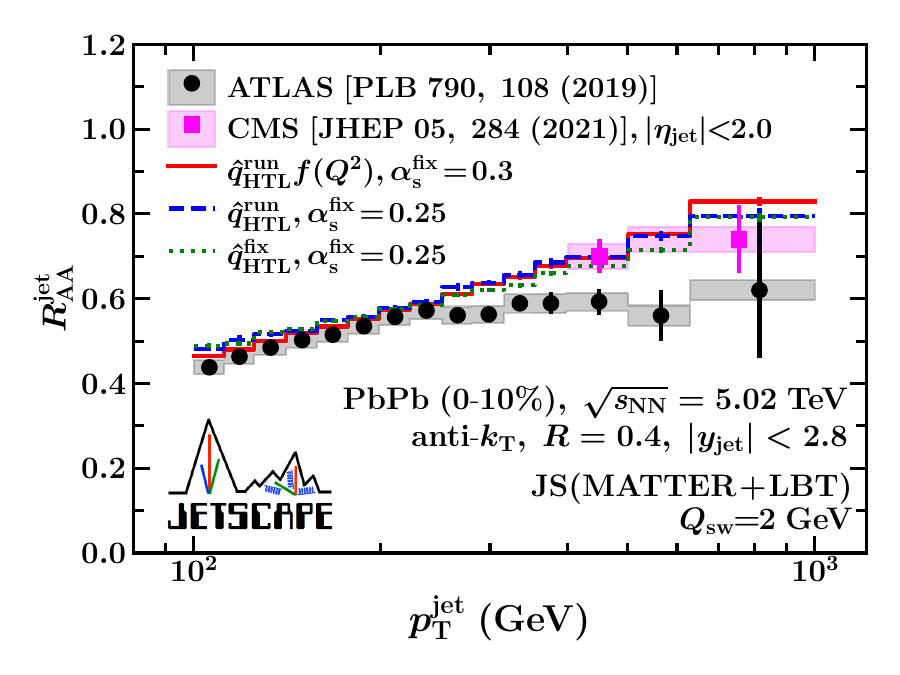}
  \includegraphics[width=.495\textwidth]{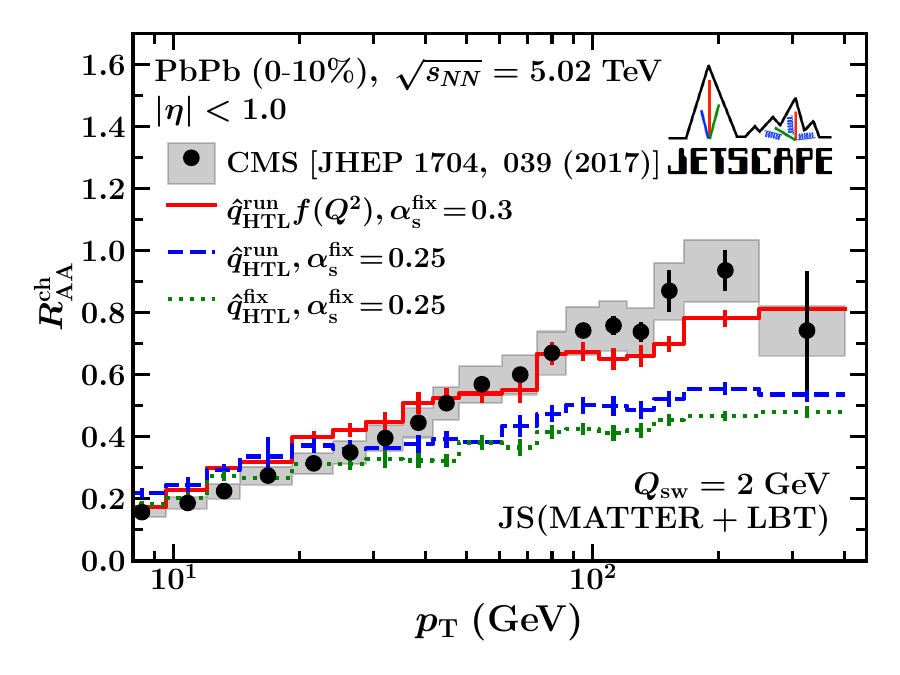}  

\vspace{-1.2em}
  
  \caption{Nuclear modification factor of inclusive jets of cone radii $R = 0.4$ (left) and charged particle (right) for $0$-$10\%$ Pb-Pb collisions at $\sqrt{s}=5.02$ ATeV. 
  Here, three different setups for the effective $\hat{q}$ are compared: with virtuality dependence and running $\alpha_s$ (solid red), no virtuality dependence with running $\alpha_s$ (dashed blue), and no virtuality dependence with fixed $\alpha_s$ (dotted green). 
  Experimental data are from ATLAS~\cite{ATLAS:2018gwx} and CMS~\cite{CMS:2021vui} for jets and from CMS~\cite{CMS:2016xef} for charged particles. 
  Figures from Ref.~\cite{JETSCAPE:2022jer}. 
  }
  \label{fig:raa_5020TeV}
\end{figure*}   
Figure~\ref{fig:raa_5020TeV} displays the effect of the virtuality dependence in the inclusive jet and charged particle $R_{\mathrm{AA}}$. 
Our full results, incorporating virtuality dependence, effectively match the experimental data for both the suppression of reconstructed jets and charged particles.
There is no noticeable disparity between the three presented setups in the jet $R_{\mathrm{AA}}$ throughout the currently available $p_{T}$ range.
However, distinct differences are apparent in the charged particle $R_{\mathrm{AA}}$. The behavior of weakened suppression at very large $p_{T}$ ($>$ 300~GeV) observed in the experimental data is accurately captured only by including the virtuality dependence.

\begin{figure*}[!h]
  \includegraphics[width=.495\textwidth]{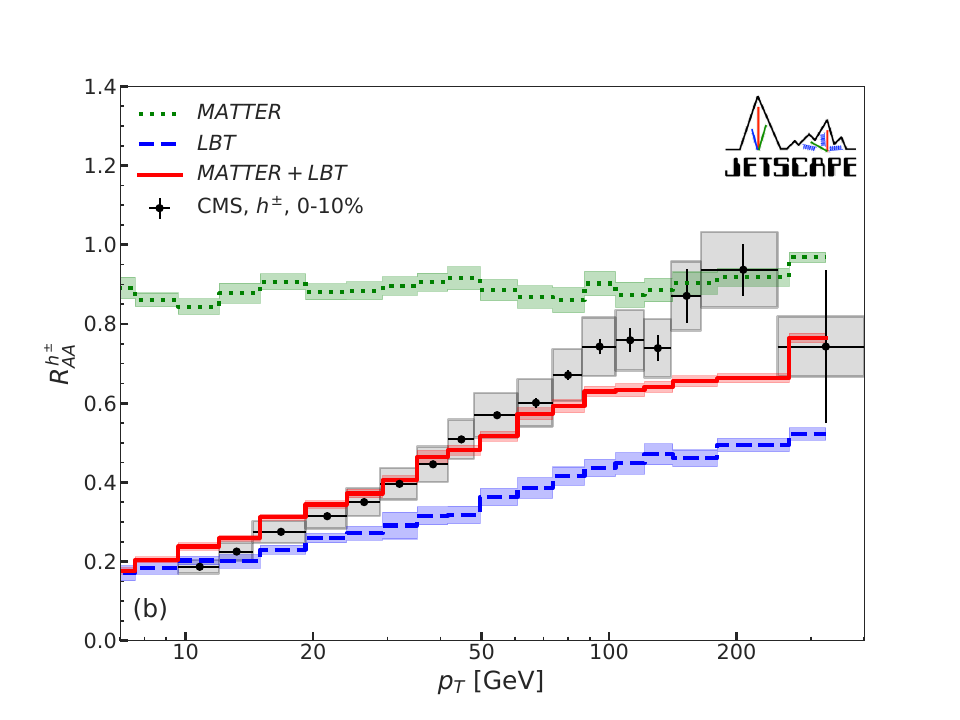}
  \includegraphics[width=.495\textwidth]{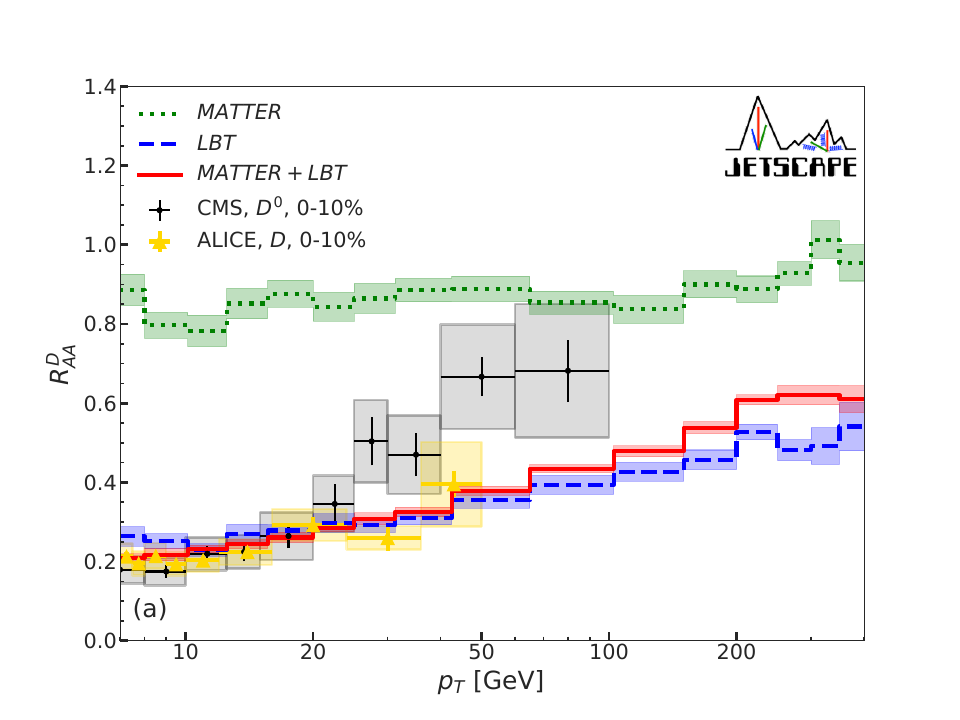}     

\vspace{-1.em}
  
  \caption{Nuclear modification factor of inclusive charged particle (left) and D-meson with (right) for $0$-$10\%$ Pb-Pb collisions at $\sqrt{s}=5.02$ ATeV. 
  Here, results from the two-stage simulation of MATTER+LBT (solid red) and single-stage simulations of MATTER (dotted green) and LBT (dashed blue) are compared. 
  In both cases, the virtuality dependence in the effective $\hat{q}$ is on in MATTER. 
  Experimental data are from CMS~\cite{CMS:2016xef} for charged particles and from CMS~\cite{CMS:2017qjw} and ALICE~\cite{ALICE:2018lyv} for D-meson. 
  Figures from Ref.~\cite{JETSCAPE:2022hcb}. }
  \label{fig:single_vs_twostage}
\end{figure*}

In Fig.~\ref{fig:single_vs_twostage}, 
the results from the two-stage simulation of MATTER+LBT and single-stage simulations of MATTER and LBT are compared for $R_{\mathrm{AA}}$ of inclusive charged particles and $D$ meson at $\sqrt{s_{\mathrm{NN}}} = 5.02$~TeV. 
The energy loss of inclusive charged particles becomes dominated by the contribution in the high virtuality phase by MATTER as we move toward the large $p_{T}$ region. 
In contrast, the energy loss in the low-virtuality LBT phase remains predominant in the $D$ meson spectrum, even at high $p_{T}$.

\begin{figure*}[!h]
  \includegraphics[width=.495\textwidth]{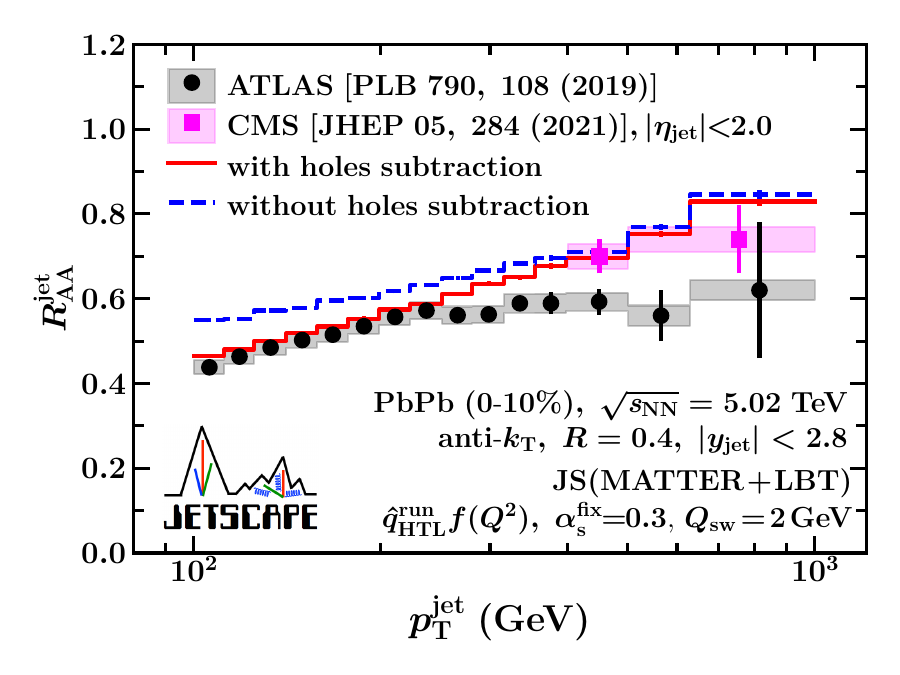}
  \includegraphics[width=.495\textwidth]{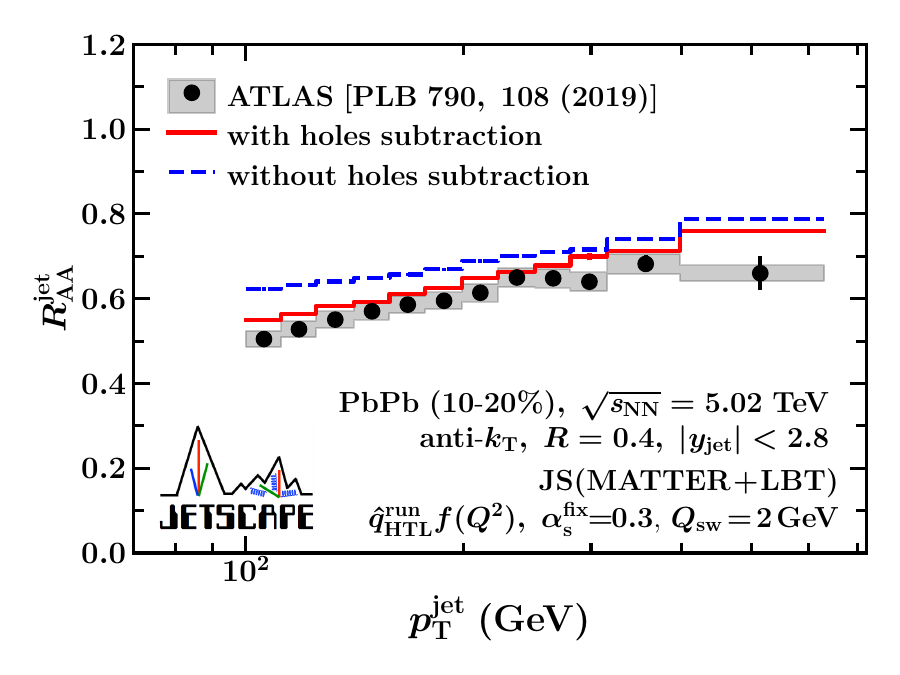}

\vspace{-.6em}
  
  \includegraphics[width=.495\textwidth]{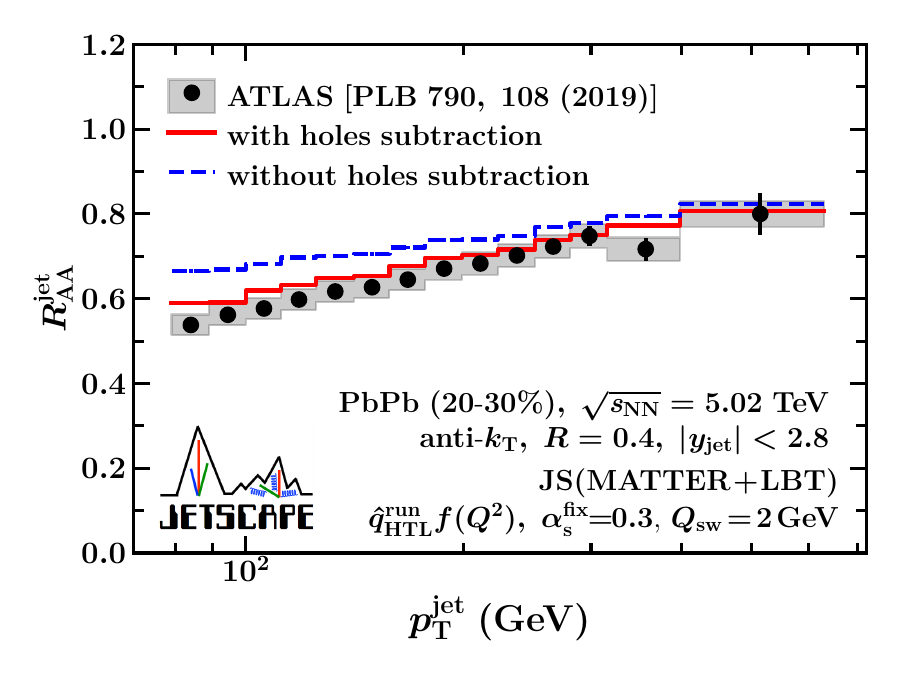}
  \includegraphics[width=.495\textwidth]{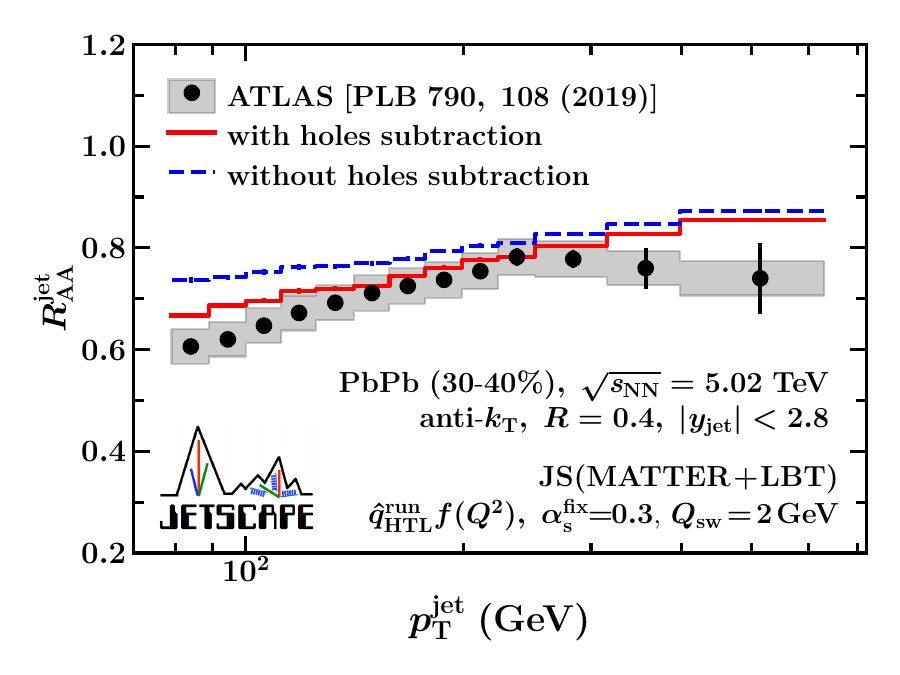}

  \vspace{-.6em}
  
  \includegraphics[width=.495\textwidth]{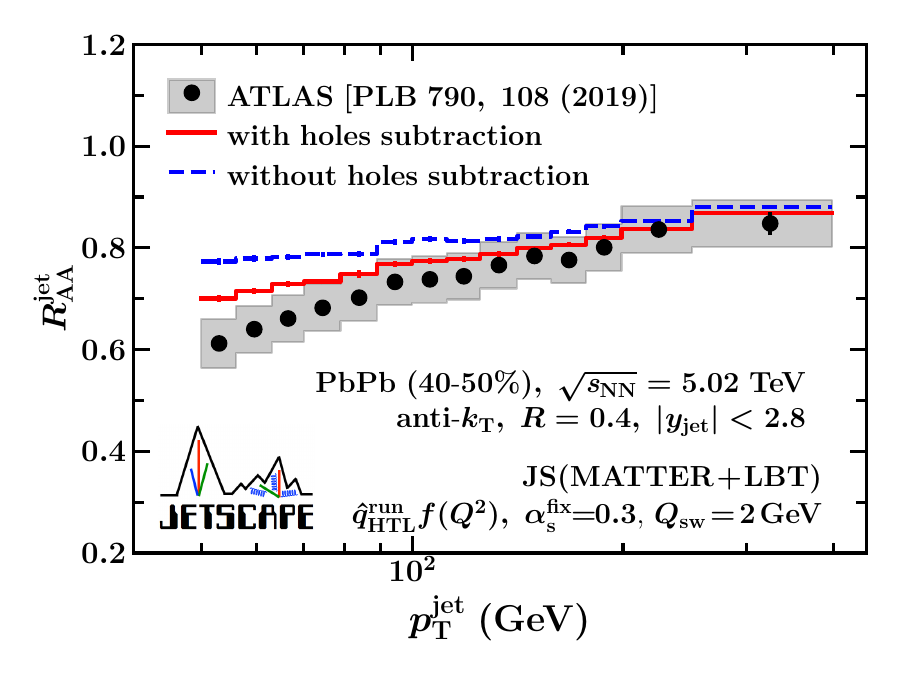}
  \includegraphics[width=.495\textwidth]{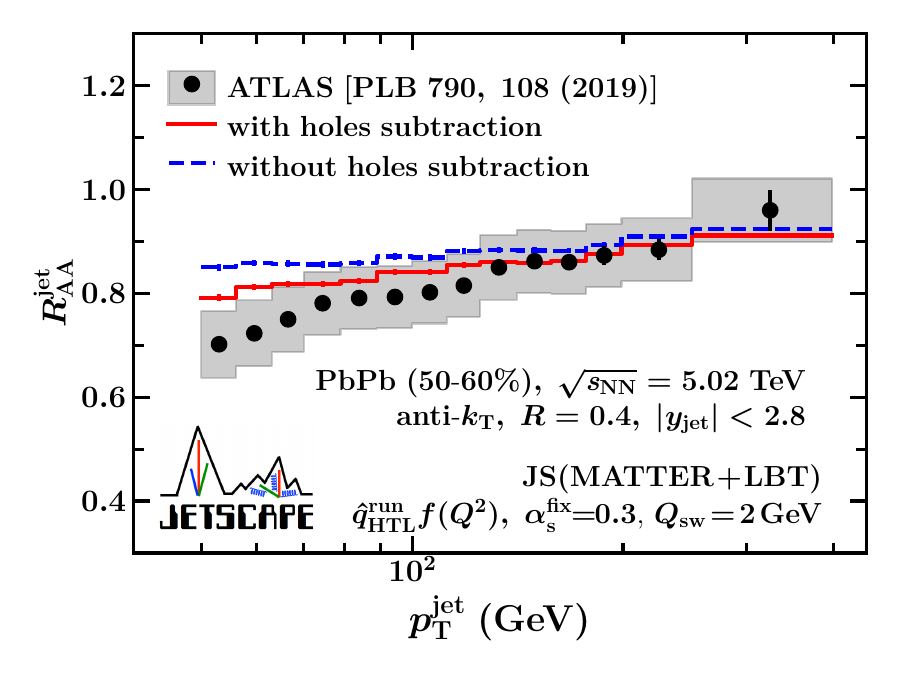}

\vspace{-1.em}
  
  \caption{Inclusive jet nuclear modification factor in Pb-Pb collisions at $\sqrt{s}=5.02$ ATeV for different centralities. 
  Here, results from the MATTER+LBT with the virtuality-dependent effective $\hat{q}$ are shown with and without the holes subtraction. 
  Experimental data are from ATLAS~\cite{ATLAS:2018gwx} and CMS~\cite{CMS:2021vui}. 
  Figures from Ref.~\cite{JETSCAPE:2022jer}. 
  }
  \label{fig:jet_raa_5020TeV_cent}
\end{figure*}  
\begin{figure*}[!h]
  \includegraphics[width=.495\textwidth]{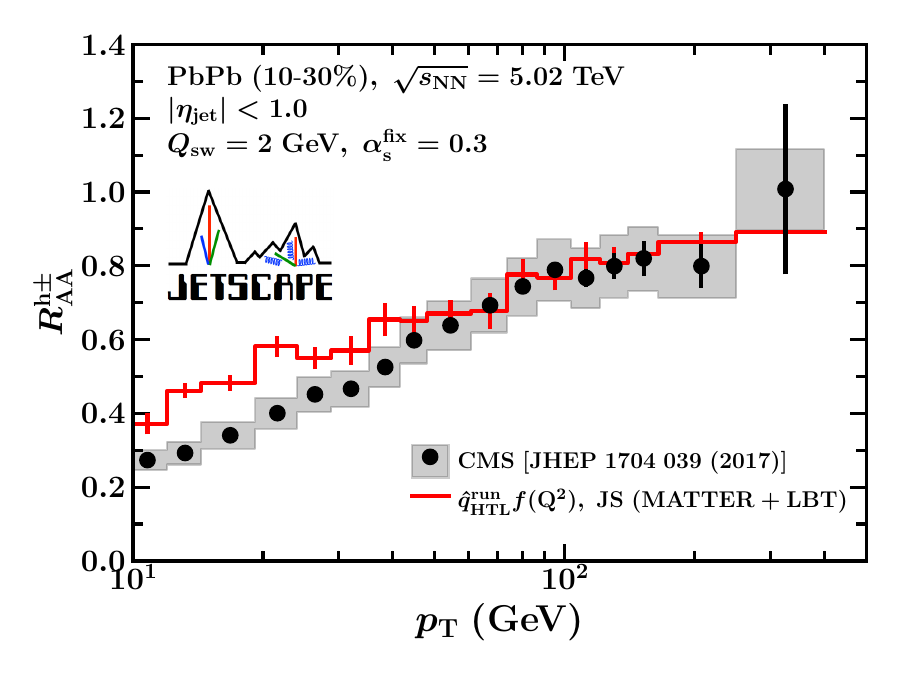}
  \includegraphics[width=.495\textwidth]{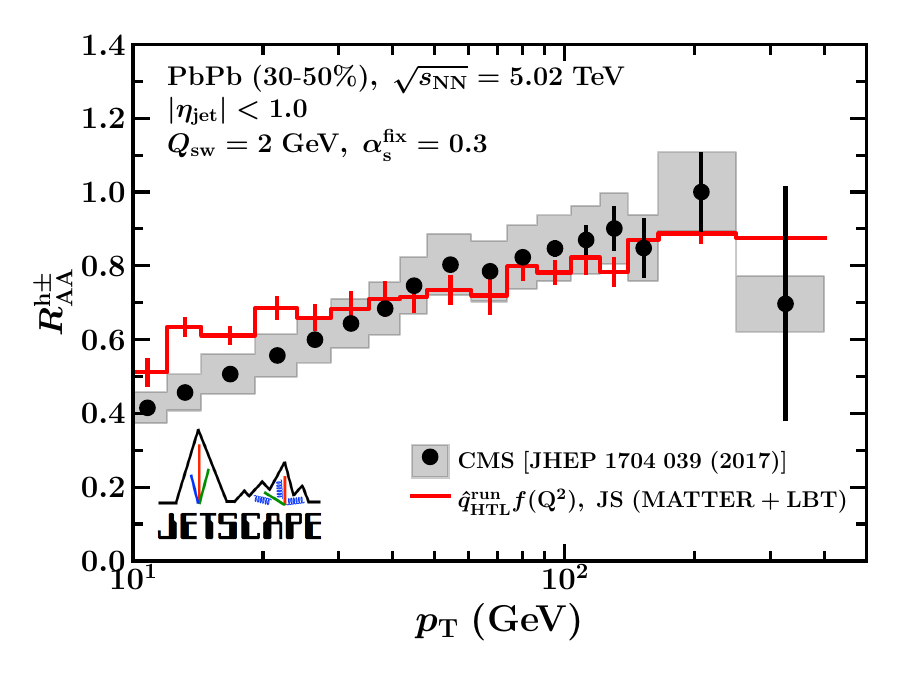}

\vspace{-1.25em}
  
  \caption{Nuclear modification factor of inclusive charged particle in Pb-Pb collisions at $\sqrt{s}=5.02$ ATeV for different centralities. 
  Here, results from the MATTER+LBT with the virtuality-dependent effective $\hat{q}$ are shown.   
  Experimental data are from CMS~\cite{CMS:2016xef}. 
  Figures from Ref.~\cite{JETSCAPE:2022hcb}. }
  \label{fig:cc_raa_5020TeV_cent}
\end{figure*}
\begin{figure*}[!h]
  \includegraphics[width=.495\textwidth]{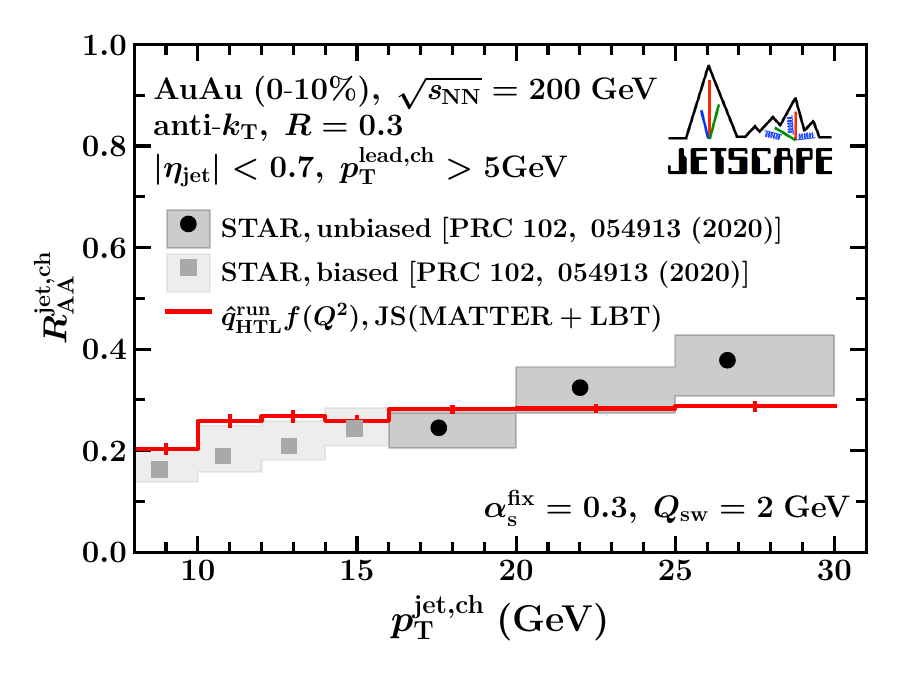}
  \includegraphics[width=.495\textwidth]{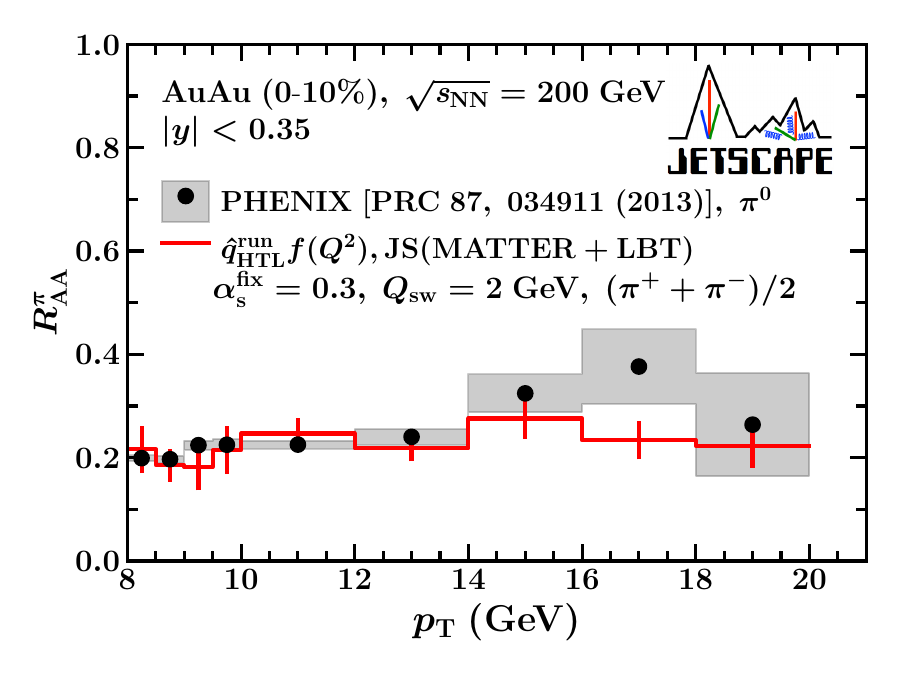}

\vspace{-1.25em}
  
  \caption{Nuclear modification factor of charged jet (left) and pion with (right) for $0$-$10\%$ Pb-Pb collisions at $\sqrt{s}=200$ AGeV. 
  Here, results from MATTER+LBT with the virtuality-dependent effective $\hat{q}$ are shown. 
  Experimental data are from STAR~\cite{STAR:2020xiv} for charged jet and from PHENIX~\cite{PHENIX:2012jha} for pion.
  Figures from 
  Ref.~\cite{JETSCAPE:2022hcb}}
  \label{fig:raa_200GeV}
\end{figure*}   
Figures~\ref{fig:jet_raa_5020TeV_cent} and \ref{fig:cc_raa_5020TeV_cent} show the results of the reconstructed jet $R_{\mathrm{AA}}$ and charged particle $R_{\mathrm{AA}}$ at $\sqrt{s}=5.02$ ATeV for various centrality classes. 
For both jets and charged particles, the two-stage jet energy loss model incorporating virtuality dependence successfully replicates centrality dependence. 
Furthermore, the excellent agreement observed in Fig.~\ref{fig:raa_200GeV}, 
which illustrates the charged jet and pion $R_{\mathrm{AA}}$ in central (0-10\%) Au-Au collisions at the top RHIC energy $\sqrt{s}=200$ ATeV, can also be confirmed.

Finally, we present the Soft Drop~\cite{Larkoski:2014wba} substructure observables characterizing hard splitting of jets. 
In the Soft Drop procedure, all components of a triggered jet are passed to the angular-ordered clustering process by the Cambridge-Aachen (C/A) algorithm. Then, the C/A jet is repeatedly deconstructed until a pair is found that satisfies the Soft Drop condition: 
\begin{align}
\label{eq:soft_drop}
\frac{\min\left(p_{T,1},
p_{T,2}\right)}{p_{T,1}+
p_{T,2}}>z_{\mathrm{cut}}\left(\frac{\Delta R_{12}}{R}\right)^{\beta}, 
\end{align}
where $p_{T,1}$ and $p_{T,2}$ are the transverse momenta of the prongs, 
$\Delta R_{12} = [\!\left(\eta_{1}-\eta_{2}\right)^{2}
+\left(\phi_{1}-\phi_{2}\right)^{2}]^{1/2}$ is their radial distance. 
Here, $z_{\mathrm{cut}}$ and $\beta$ are parameters that control the Soft Drop process. In the results presented below, $z_{\mathrm{cut}}=0.2$ and $\beta=0$ are adopted for all cases. 

\begin{figure*}[!h]
  \includegraphics[width=.495\textwidth]{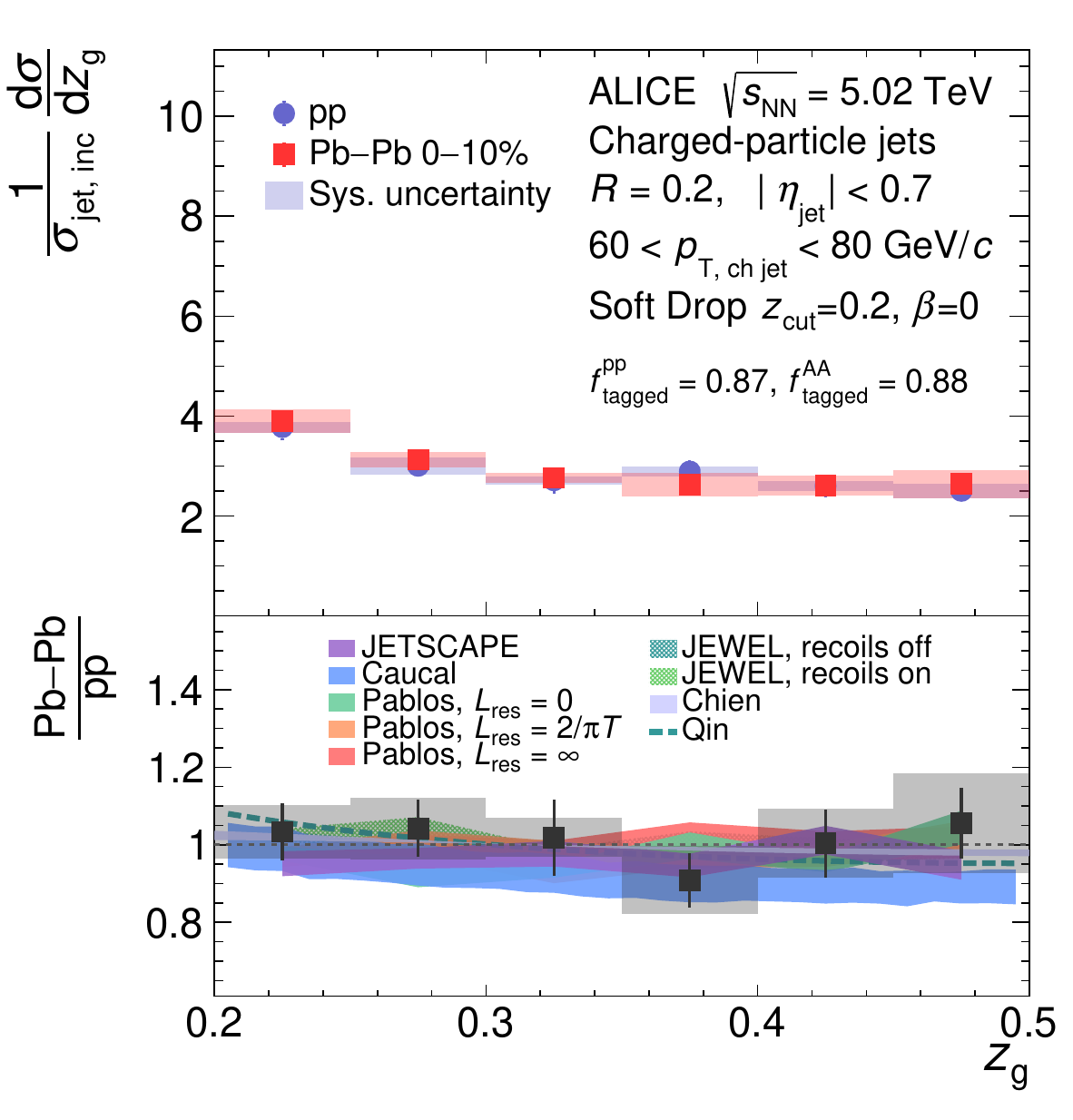}
  \includegraphics[width=.495\textwidth]{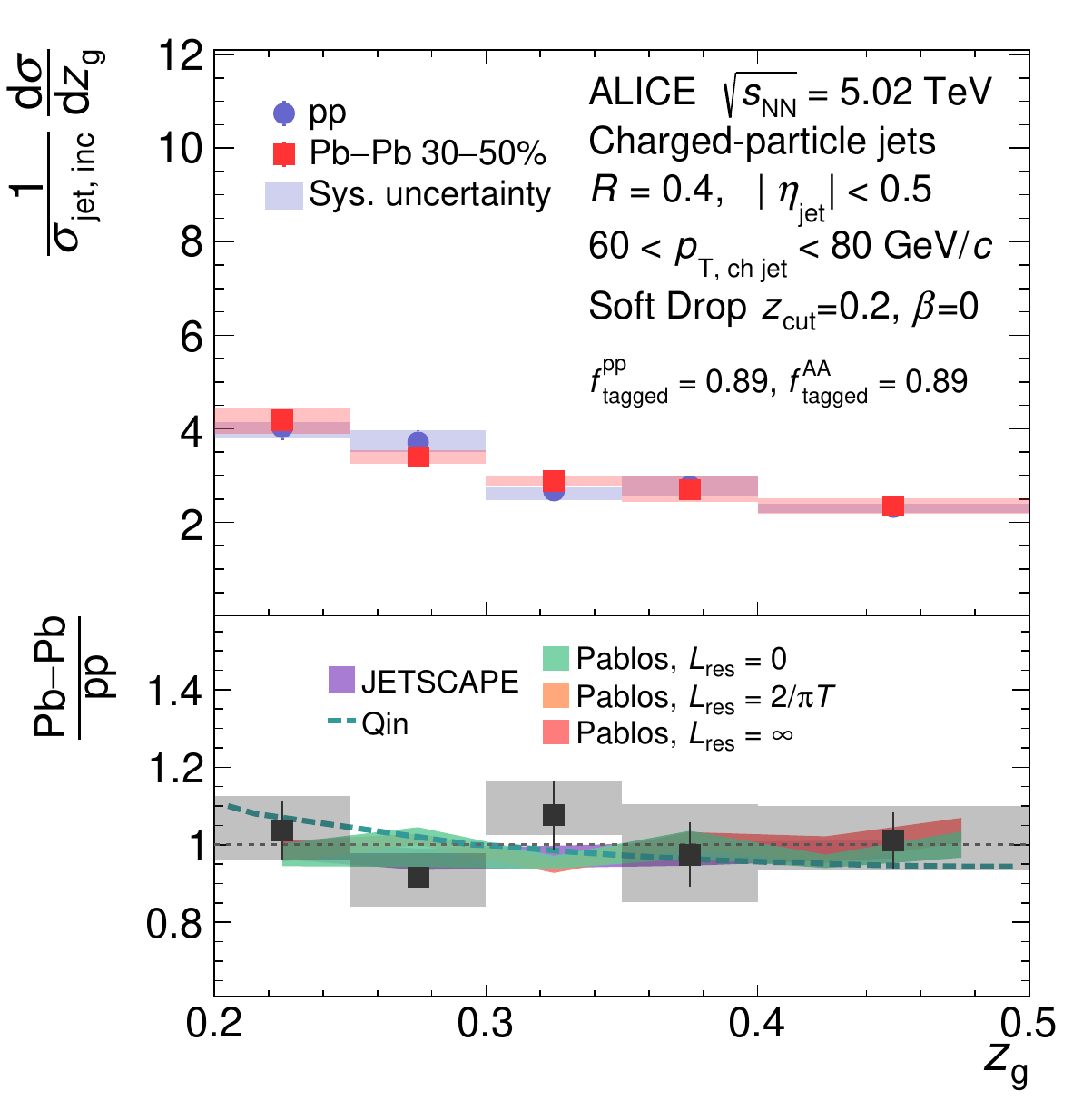}
  \includegraphics[width=.495\textwidth]{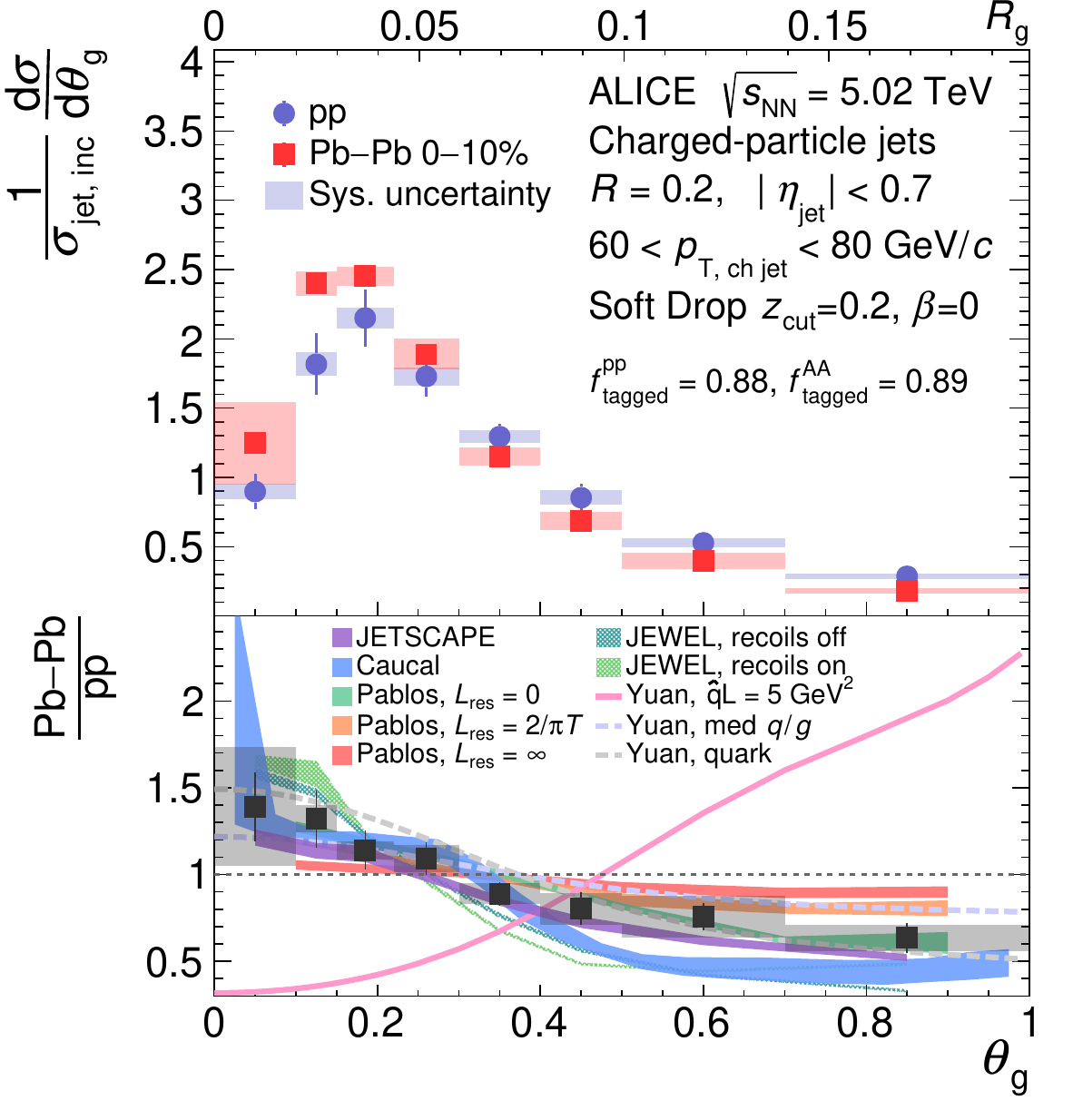}
  \includegraphics[width=.495\textwidth]{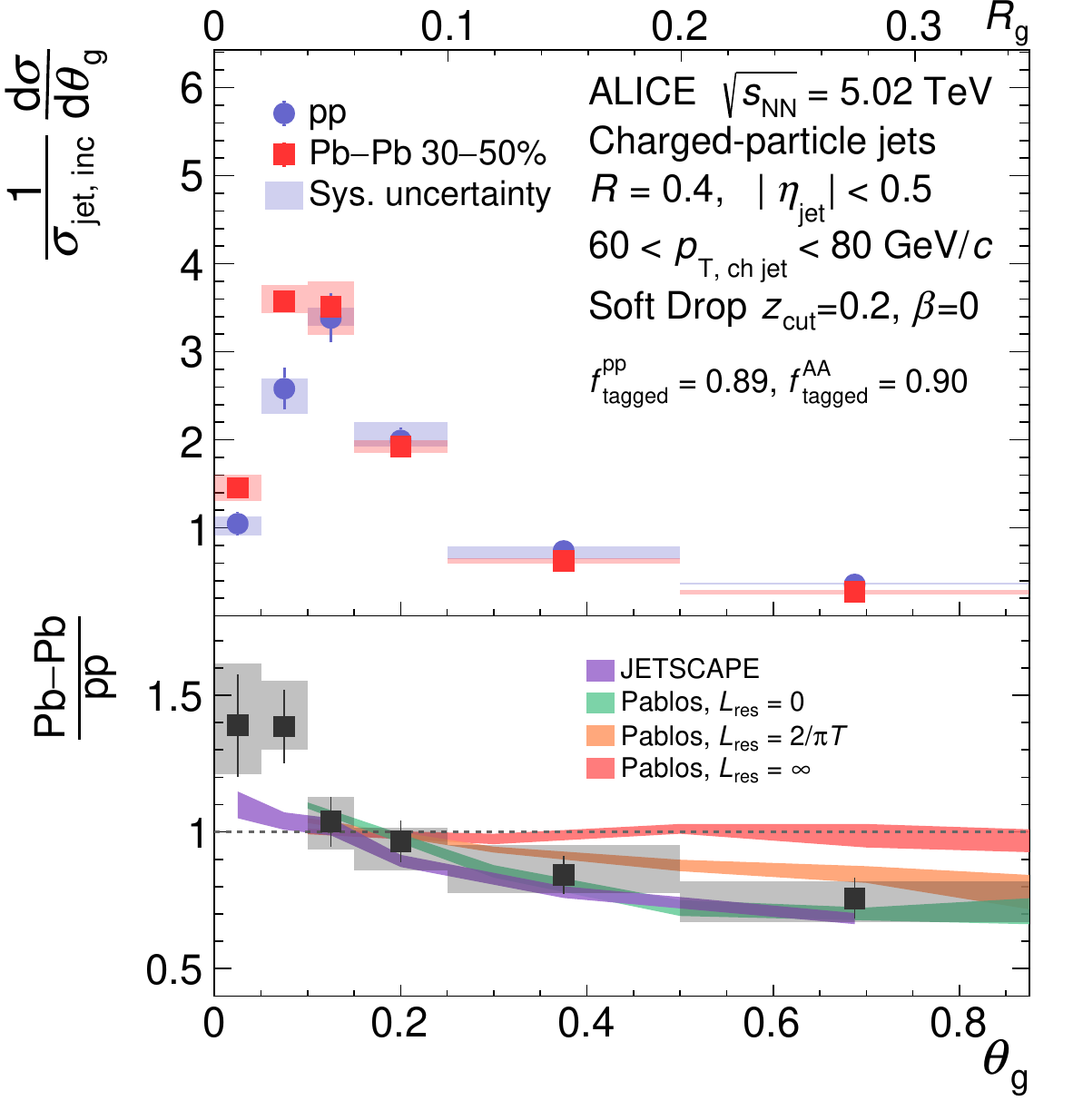}  

\vspace{-.75em}

  \caption{Distributions of $z_{g}$ (upper) and $\theta_{g}$ (lower) for charged jet in pp and Pb–Pb collisions at $\sqrt{s}=5.02$ ATeV. 
  Left: Results for 0-10\% collisions with $R=0.2$. 
  Right: Results for 30-50\% collisions with $R=0.4$. 
  Figures from ALICE~\cite{ALargeIonColliderExperiment:2021mqf}}
  \label{fig:sd_5020TeV_cent}
\end{figure*}

\begin{figure*}[!h]
  \includegraphics[width=.49\textwidth,height=0.45\textwidth]{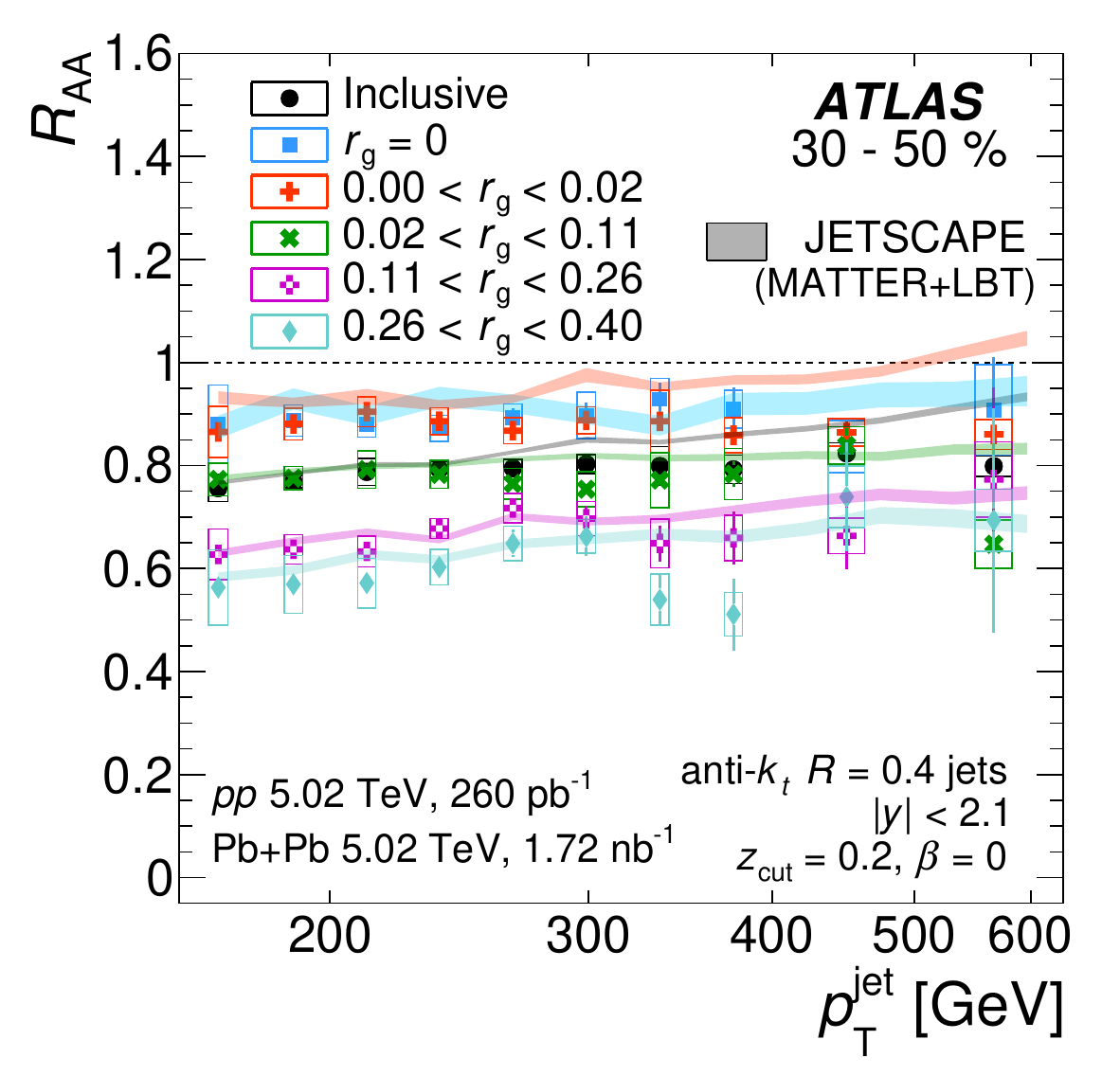}
  \includegraphics[width=.49\textwidth,height=0.45\textwidth]{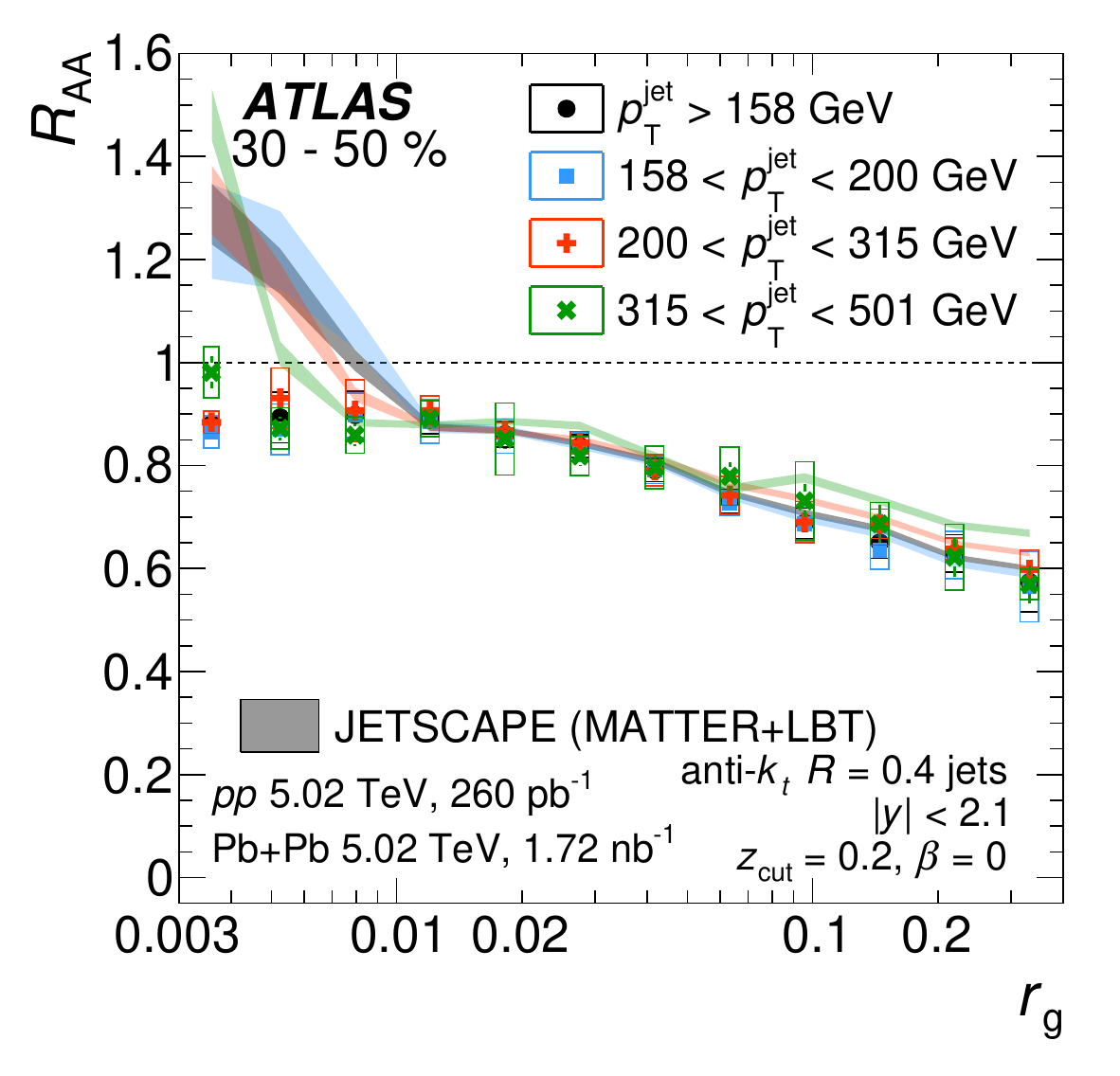}
  \includegraphics[width=.49\textwidth,height=0.45\textwidth]{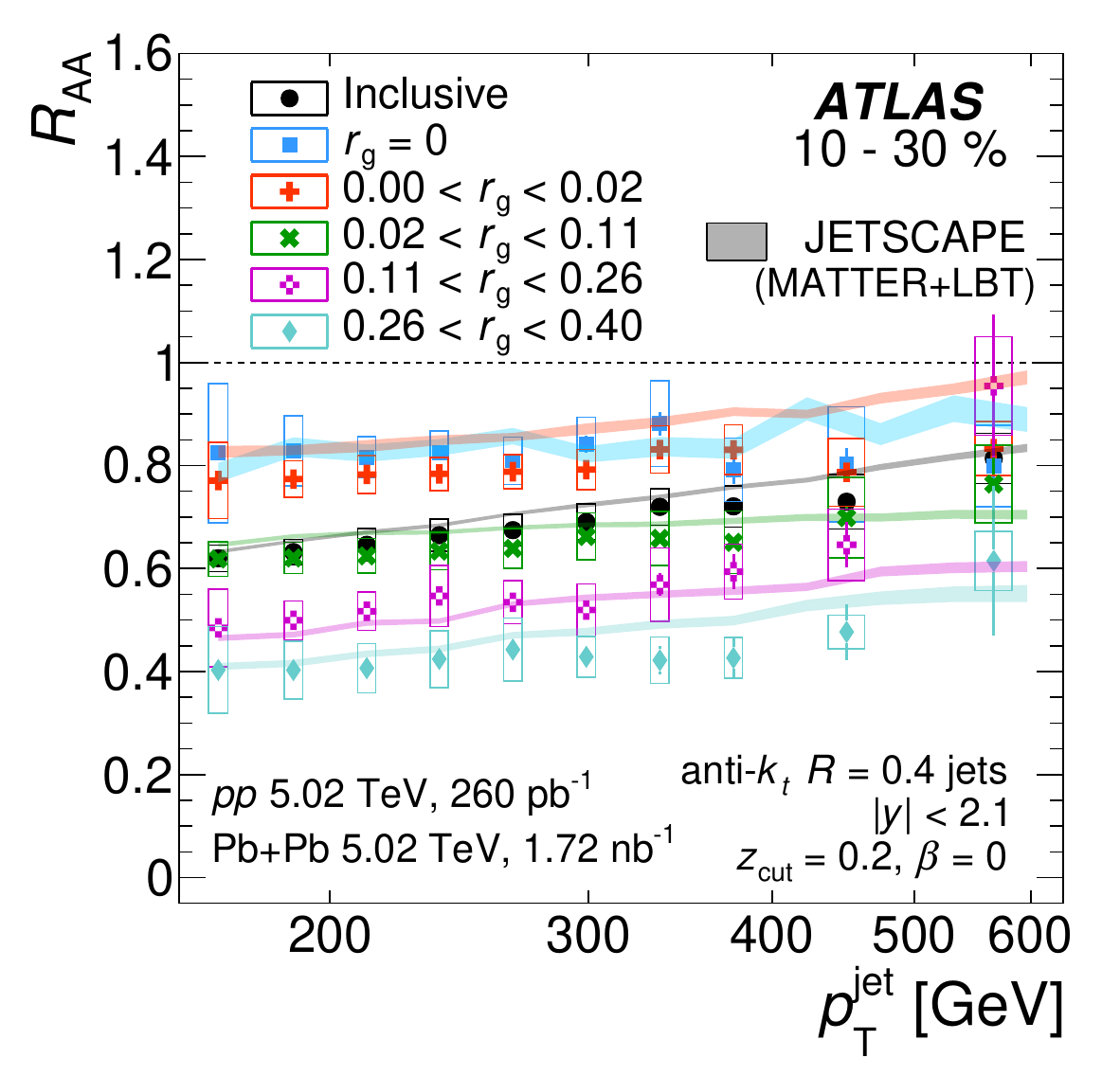}
  \includegraphics[width=.49\textwidth,height=0.45\textwidth]{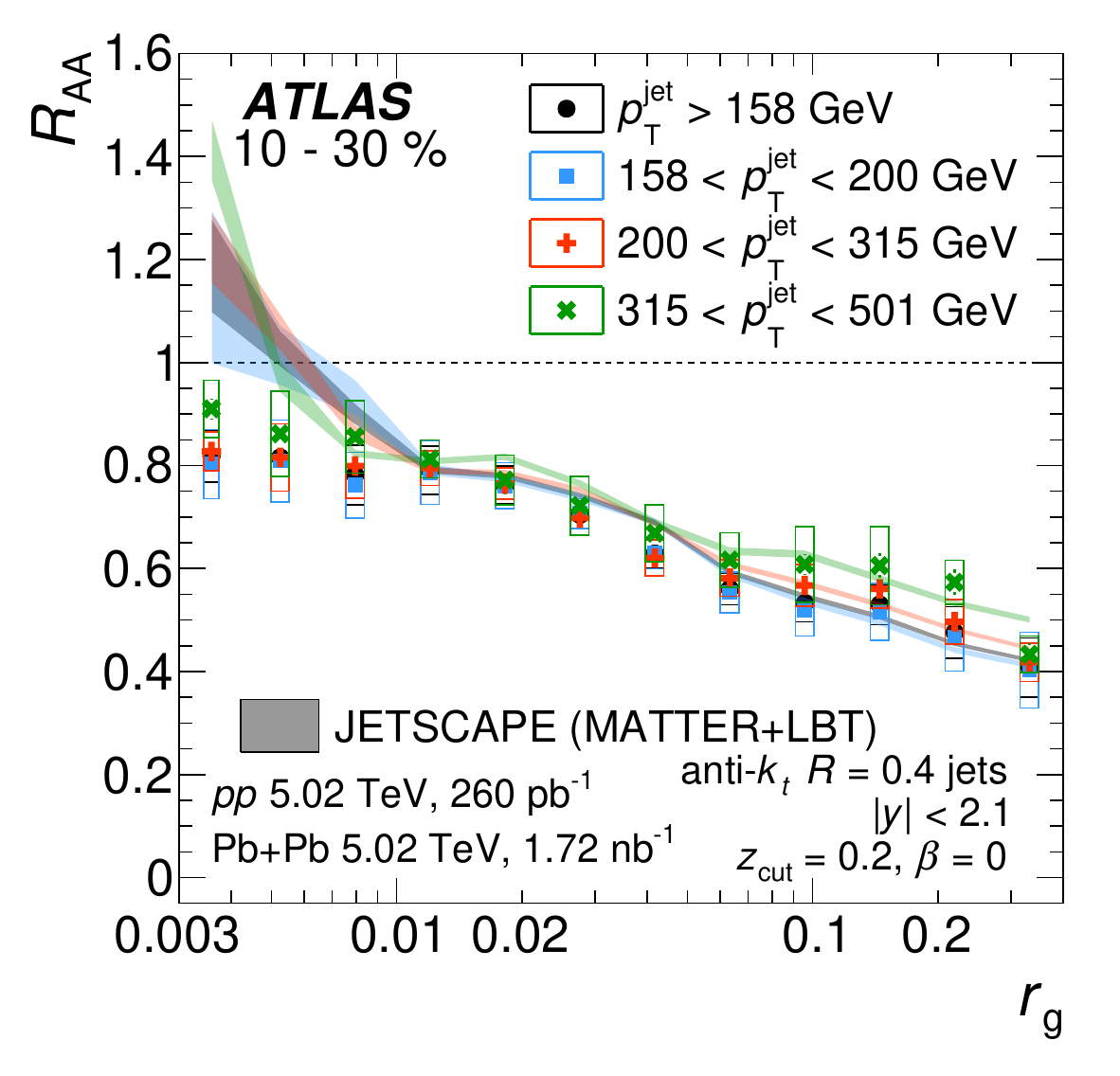}
  \includegraphics[width=.49\textwidth,height=0.45\textwidth]{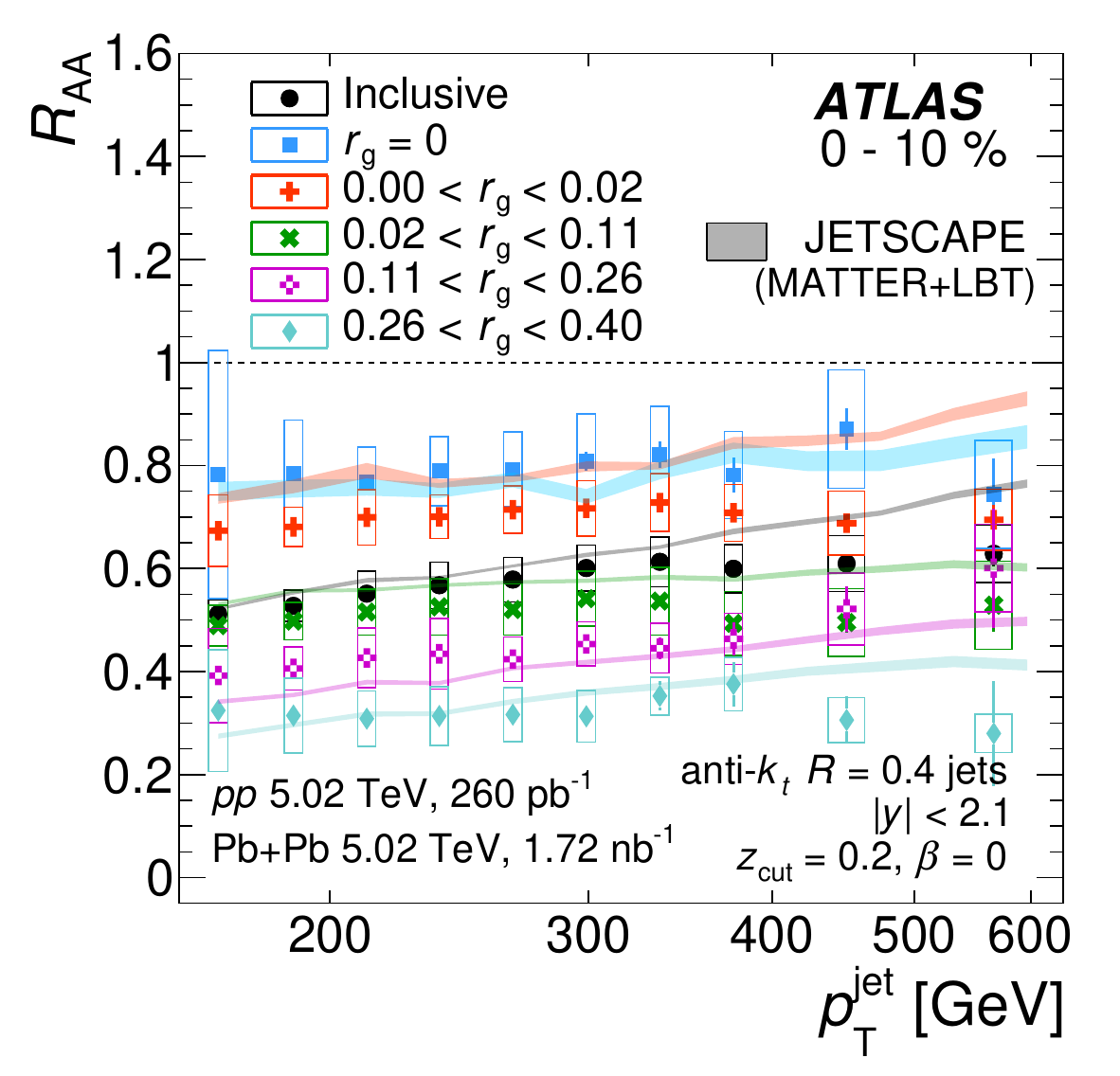}
  \includegraphics[width=.49\textwidth,height=0.45\textwidth]{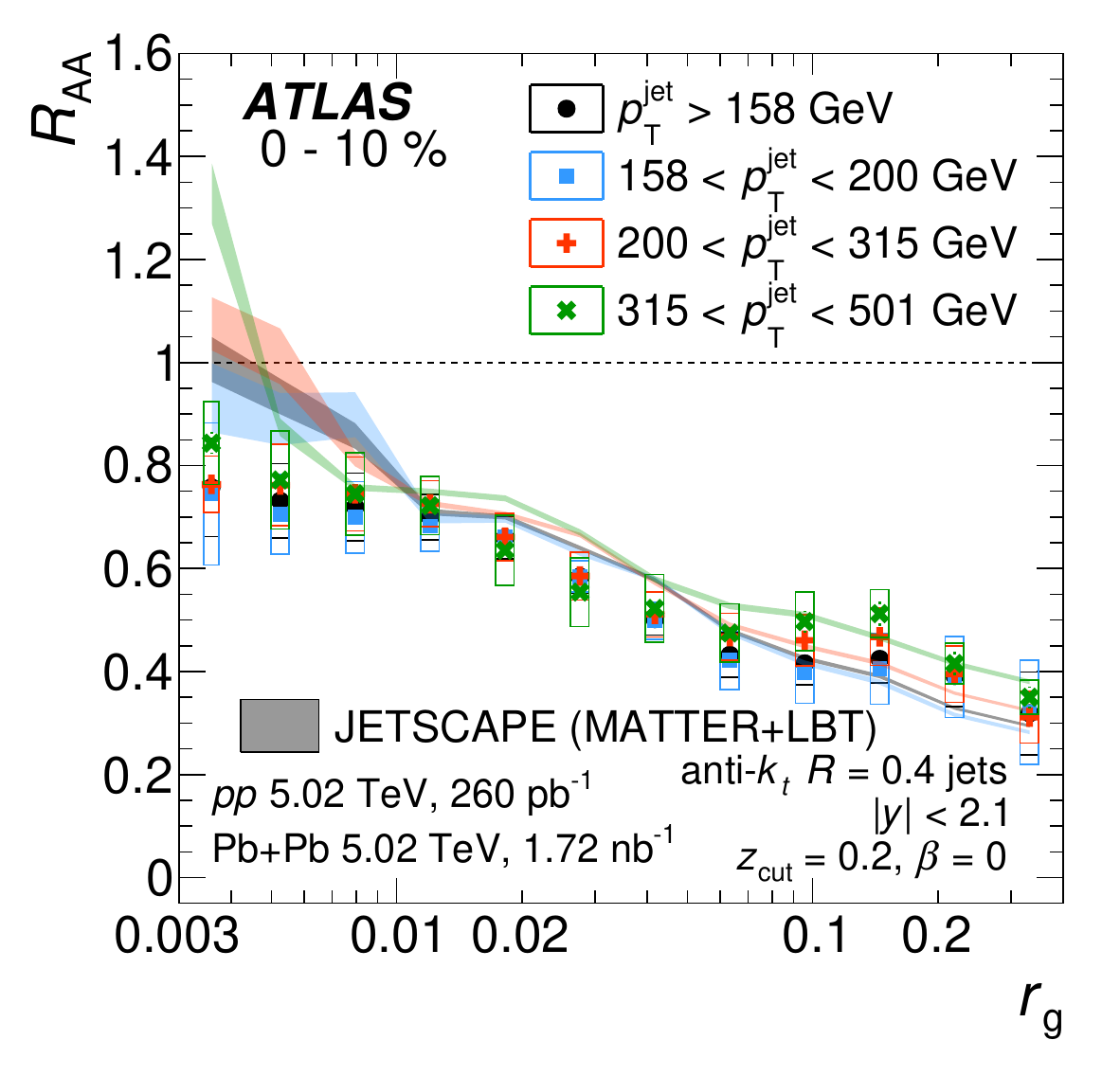}
  \vspace{-0.4cm}
  \caption{The $r_{g}$-dependence of the nuclear modification factor of jets in Pb-Pb collisions at $\sqrt{s}=5.02$ ATeV for different centralities.  
  Here, results from the MATTER+LBT with the virtuality-dependent effective $\hat{q}$~\cite{JETSCAPE:2023hqn} are compared to the ATLAS data~\cite{ATLAS:2022vii}. 
  Figures from Ref.~\cite{ATLAS:2022vii}. 
  }
  \label{fig:sd_raa_cent}
\end{figure*} 
Figure~\ref{fig:sd_5020TeV_cent} shows the distributions of Soft Drop observables 
\begin{align}
\label{eq:soft_drop_obs}
z_{g} = \frac{\min\left(p_{T,1},
p_{T,2}\right)}{p_{T,1}+
p_{T,2}},
\quad \theta_{g} = \frac{\Delta R_{12}}{R},
\end{align}
for charged jets $\sqrt{s}=5.02$ ATeV. 
Here, the JETSCAPE bands present results from MATTER+LBT with the virtuality-dependent effective $\hat{q}$~\cite{JETSCAPE:2023hqn}. 
Almost all the compared MC model calculations, including the MATTER+LBT setup of JETSCAPE, successfully capture the behavior indicated by the experimental results: almost no visible medium effects on the $z_{g}$ distribution and shifts towards smaller hard branching angles in $\theta_{g}$ distribution. 
One can see more details of the medium effects on the angle of jet hard splittings in Fig.~\ref{fig:sd_raa_cent}, which shows the dependence of full jet $R_{\mathrm{AA}}$ on $r_{g} = \Delta R_{12} = R \theta_{g}$. 
As $r_{g}$ of jets increases, there is an almost monotonic intensification of jet yield suppression.

        \subsection{Bayesian Analysis on jet and QGP properties} 
            One of the final goals of studying jets in heavy-ion collisions is to use them to constrain the properties of nuclear matter. These properties are usually embedded in the model parameters that need to be determined using experimental data. Due to the growing sophistication of jet-medium interaction models and the fast explosion of experimental data, traditional methods of extracting model parameters, like ``fit by eye" and least square ($\chi^2$), have become extremely time consuming or not very informative. The Bayesian statistical analysis method has been introduced in heavy-ion physics and shown to be a successful tool in constraining the bulk properties of the QGP, such as its shear and bulk viscosities~\cite{Novak:2013bqa,Bernhard:2019bmu,JETSCAPE:2020shq,Nijs:2020ors} and equation of state (EOS)~\cite{Pratt:2015zsa}. In this subsection, we focus on the performance of Bayesian analysis on hard probe observables.

The Bayesian analysis is based on the Bayes' theorem,
\begin{equation}
\label{eq:Bayes_theorem}
P\left( \boldsymbol\theta | \mathrm{data} \right) \propto P\left( \boldsymbol\theta \right) P\left( \mathrm{data}|\boldsymbol\theta \right),
\end{equation}
in which $P\left( \boldsymbol\theta | \mathrm{data} \right)$ on the left represents the posterior distribution of the model parameter vector $\boldsymbol\theta$ given the knowledge of experimental data, while $P\left( \boldsymbol\theta \right)$ on the right represents the prior distribution of $\boldsymbol\theta$ without the knowledge, and $P\left( \mathrm{data}|\boldsymbol\theta \right)$ measures the likelihood of a particular vector $\boldsymbol\theta$ by comparing its model output to data. Usually, a Gaussian form is taken for the likelihood function as
\begin{equation}
\label{eq:Gaussian_likelihood}
P\left( \mathrm{data}|\boldsymbol\theta \right) = \prod_i \frac{1}{\sqrt{2 \pi} \sigma_i} e^{- \left[y_i(\boldsymbol\theta) - y_i^\mathrm{exp}\right]^2 \big/ (2 \sigma_i^2)},
\end{equation}
where $y_i^\mathrm{exp}$ denotes the central value of the $i^\mathrm{th}$ data point, $y_i(\boldsymbol\theta)$ denotes the model output at this point, and $\sigma_i$ denotes the error that can be contributed by both experimental measurement and theoretical calculation. 

In reality, it is rather time consuming to calculate experimental observables using sophisticated Monte Carlo event generators, which makes it impossible to directly use these generators to scan through the multi-dimensional parameter space. For this reason, the Gaussian process emulator (GPE)~\cite{GPE1,GPE2} is introduced, which is first trained using the model outputs on limited sets of $\boldsymbol\theta$ and then used as a fast surrogate of the event generator during model-to-data comparison. One usual routine of Bayesian analysis is starting from a random location in the parameter space, and then performing the Markov Chain Monte Carlo (MCMC) random walks according to the probability given by Eq.~(\ref{eq:Bayes_theorem}) using the Metropolis-Hastings algorithm. After sufficient steps, the chain will reach equilibrium, and locations extracted from its further steps constituent the posterior distribution of $\boldsymbol\theta$.  

\begin{figure}[!h]
        \includegraphics[width=.45\textwidth]{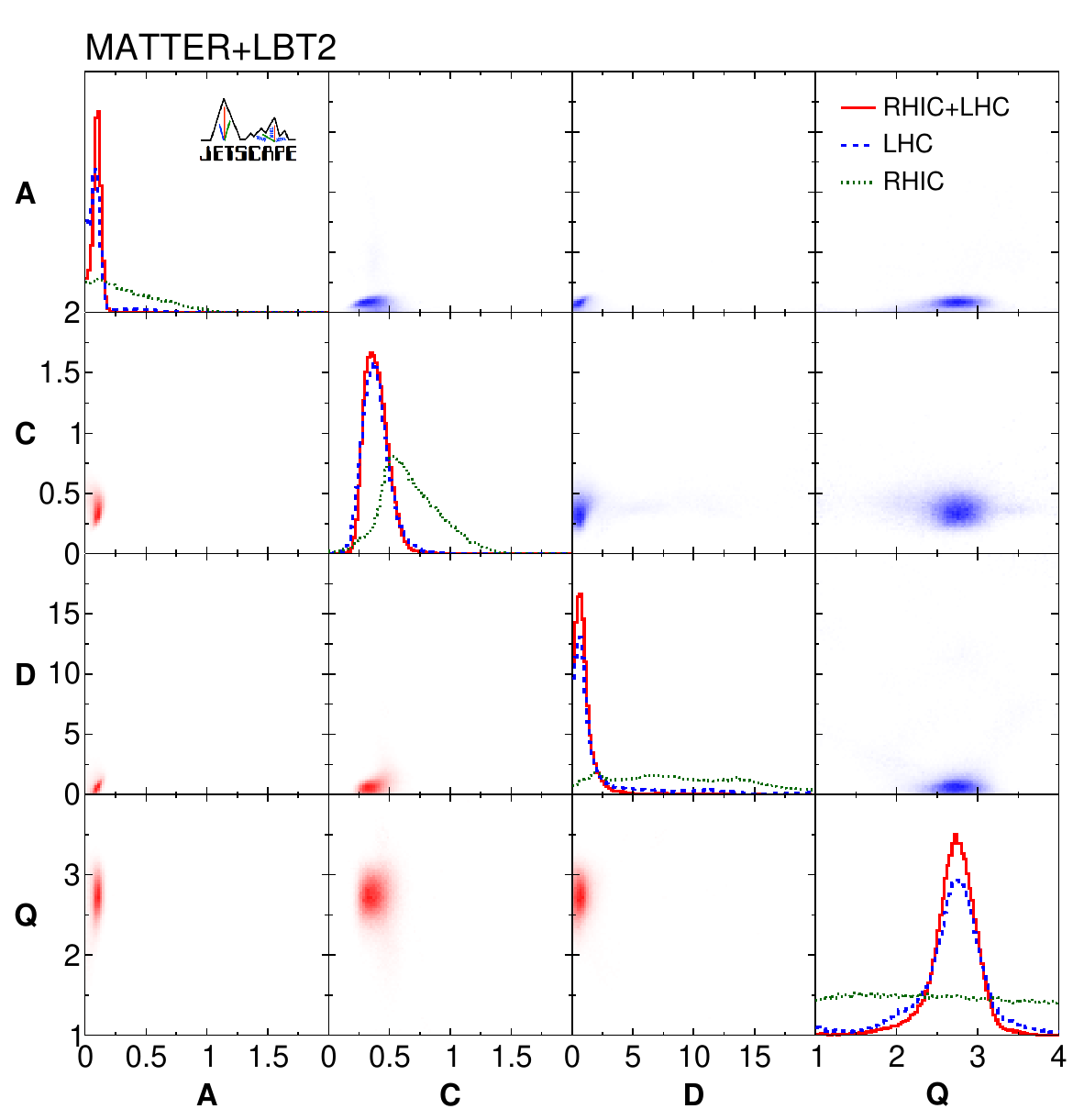}
        \hspace{0.02\textwidth}
        \includegraphics[width=.5\textwidth]{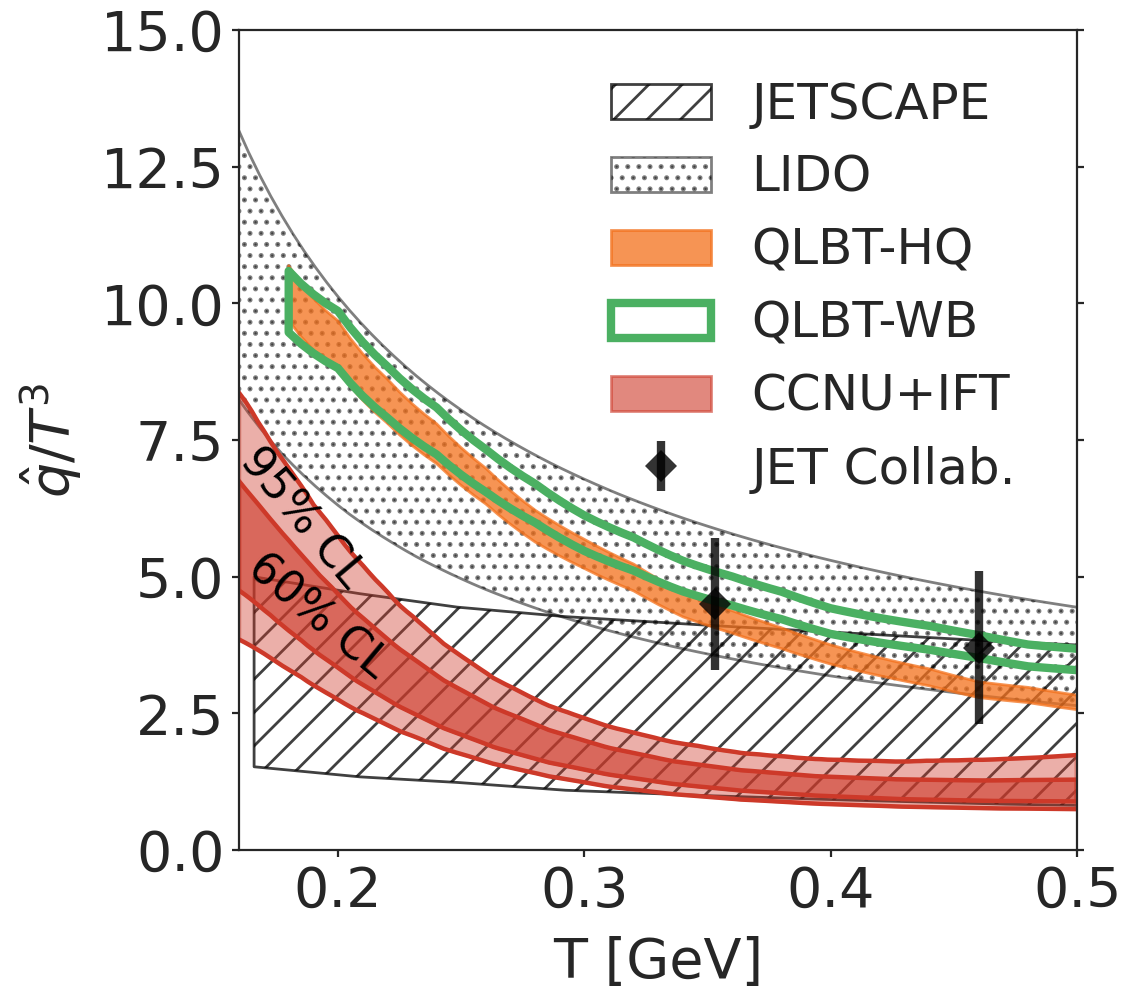}        
        \caption{Left: The model parameters of $\hat{q}$ extracted using the JETSCAPE framework (MATTER+LBT module) calculation. Right: Constraints on $\hat{q}$ inside the QGP from different model calculations. The figures are taken from Refs.~\cite{JETSCAPE:2021ehl,Xie:2022ght}.}
        \label{fig:Bayesian_qhat}
\end{figure}

For instance, in Ref.~\cite{JETSCAPE:2021ehl}, the JETSCAPE framework is used to calibrate the jet transport coefficient $\hat{q}$ by comparing the LBT+MATTER module to the data of single inclusive hadron $R_\mathrm{AA}$ at RHIC and LHC. In one of the model setups, $\hat{q}$ is parametrized as
\begin{align}
\label{eq:parametrization2}
\frac{\hat{q}\left(Q,E,T\right) |_{Q_0,A,C,D}}{T^3}=42C_R\frac{\zeta(3)}{\pi}&\left(\frac{4\pi}{9}\right)^2\\
\times\Bigg\{\frac{A\left[\ln\left(\frac{Q}{\Lambda}\right)-\ln\left(\frac{Q_0}{\Lambda}\right)\right]}{\left[\ln\left(\frac{Q}{\Lambda}\right)\right]^2}&\theta (Q-Q_0)+\frac{C\left[\ln\left(\frac{E}{T}\right)-\ln(D)\right]}{\left[\ln\left(\frac{ET}{\Lambda^2}\right)\right]^2}\Bigg\}\nonumber
\end{align}
with four model parameters $\boldsymbol\theta=(Q_0,A,C,D)$. When the parton virtuality is higher than the separation scale $Q_0$, the MATTER module is applied for the medium-modified parton shower, in which $\hat{q}$ is assumed to run with parton virtuality ($Q$), energy ($E$) and medium temperature ($T$). On the other hand, below $Q_0$, the LBT module is used, in which $\hat{q}$ only depends on $E$ and $T$. The form of the ansatz here is inspired by the perturbative calculation of $\hat{q}$ at the leading order. Through Bayesian calibration, one obtains the constraints on the four parameters as shown in the left panel of Fig.~\ref{fig:Bayesian_qhat}. In the figure, each diagonal sub-figure shows the posterior distribution of one parameter, while each off-diagonal one shows the correlation between two. Constraints from the RHIC data, LHC data and their combination are shown separately. 
In the end, a 90\% credible region (C.R.) of $\hat{q}$ can be obtained by evaluating it using $\boldsymbol\theta$ drawn from these posterior distributions, and discarding its values at both top and bottom 5 percents.

The advantage of physics-driven parametrization is the extracted model parameters can be easily mapped to well-known physics quantities, e.g., one can obtain the strong coupling constant $\alpha_\mathrm{s}$ from Eq.~(\ref{eq:parametrization2}). However, a parametrization inspired by a particular model may also limit the flexibility of this parametrization. In the example above, effects of non-perturbative interactions are hard to include in Eq.~(\ref{eq:parametrization2}), which limits the energy and temperature dependencies of $\hat{q}$ obtained in the end. To overcome this deficiency, an information field (IF) based Bayesian interference method is applied in a recent work~\cite{Xie:2022ght}. Instead of assuming a particular functional form of a field $F(x)$, $x=\ln(T/\mathrm{GeV})$ and $F(x)=\ln(\hat{q}/T^3)$ in this work, it starts with an average value $\langle F(x)\rangle=\mu(x)$ and a correlation function of deviations between different locations $\langle\delta F(x)\delta F(x')\rangle=C(x,x')$, with $\delta F(x)=F(x)-\mu(x)$. As a first trial, a constant $\mu=1.36$ is chosen based on the approximate value of $\hat{q}/T^3$ constrained in earlier studies, and a Gaussian form of the correlation function is taken as
\begin{equation}
\label{eq:IFcorrelation}
C(x,x')=\sigma^2\exp\left[-(x-x')^2/(2l^2)\right], 
\end{equation}
in which $\sigma=0.7$ estimates the strength of correlation and $l=\ln(2)$ estimates the correlation length. With this setup, one can then constrain the functional form of $F(x)$, or its Fourier coefficients in its frequency space, using the Bayesian calibration of jet models against experimental data. 

Shown in the right panel of Fig.~\ref{fig:Bayesian_qhat} is the jet quenching parameter $\hat{q}/T^3$ inside the QGP extracted using different methodologies. The two diamond points with error bars are taken from the JET Collaboration work~\cite{JET:2013cls}, in which the traditional $\chi^2$-fitting method is applied. The value at higher average temperature is extracted from the $R_\mathrm{AA}$ in central Pb+Pb collisions at $\sqrt{s_\mathrm{NN}}=2.76$~TeV, while the value at lower temperature is from central Au+Au collisions at $\sqrt{s_\mathrm{NN}}=200$~GeV. The error bars correspond to the systematical uncertainties from different jet models adopted in this analysis. The shaded band is the 90\% C.R. of $\hat{q}/T^3$ from the Bayesian calibration of the JETSCAPE calculation~\cite{JETSCAPE:2021ehl} as discussed above. The Bayesian method allows a simultaneous calibration against multiple sets of hadron $R_\mathrm{AA}$ data from RHIC and LHC, and therefore provides a functional form of $\hat{q}/T^3$ with respect to $T$. Similarly, the dotted band is the 90\% C.R. constrained from the LIDO model using the $R_\mathrm{AA}$ data of both single inclusive hadrons and jets~\cite{Ke:2020clc}. These two examples of Bayesian analysis both start with physics-motivated model parameters. To the contrary, the red bands are from the IF-based Bayesian interference method where the nuclear modification data of single inclusive hadrons, dihadrons and $\gamma$-hadrons are involved~\cite{Xie:2022ght}. This method can exclude the long-range correlation of $\hat{q}$ between different medium temperatures, which is inherited from a specific model assumption. In addition, $\hat{q}$ of charm quarks extracted from Bayesian analysis on the $D$ meson data~\cite{Liu:2021dpm} (shown by the orange and green bands) are also included for comparison.

\begin{figure}[!h]
        \includegraphics[width=.455\textwidth]{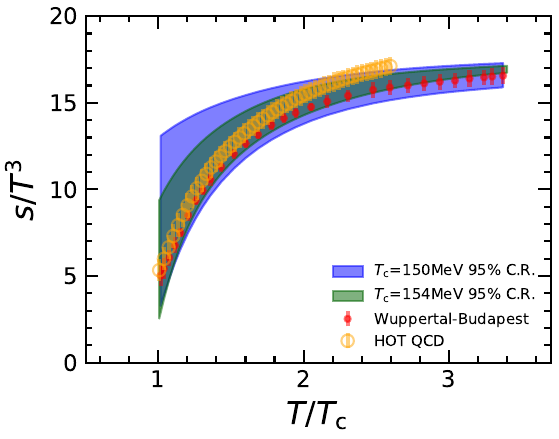}
        \hspace{0.02\textwidth}
        \includegraphics[width=.48\textwidth]{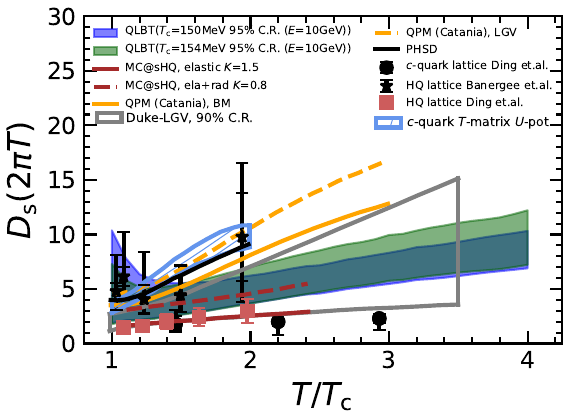}        
        \caption{The equation of state of the QGP (left) and the spatial coefficient of heavy quarks (right) simultaneously extracted from the QLBT model using the heavy flavor data. The figures are taken from Ref.~\cite{Liu:2021dpm}.}
        \label{fig:EOS}
\end{figure}

Apart from the jet quenching parameter, the Bayesian method has also been implemented to directly constrain the energy loss of jets and hadrons through the QGP~\cite{He:2018gks,Xing:2023ciw,Zhang:2023oid}. The average values of energy loss obtained from these analyses, including their flavor hierarchy $\Delta E_g > \Delta E_q \sim \Delta E_c > \Delta E_b$ between gluons ($g$), light ($q$), charm ($c$) and bottom ($b$) quarks, are shown to be consistent with QCD-based model calculations. Meanwhile, the distribution function of parton energy loss $W_\mathrm{AA}(x)$, with $x$ as the ratio of energy loss in one event to its average value, is also constrained in these studies, which can provide a more strict test of QCD calculations of parton interactions with dense nuclear matter. Additionally, a recent proof-of-principle study~\cite{Liu:2023rfi} has shown that hard probe observables can also be used to constrain the EOS of the QGP. In this work, heavy quark scattering with quasi-particles that constitute the QGP is introduced into the LBT model (named QLBT)~\cite{Liu:2021dpm}, where both the scattering rates and the quasi-particle masses depend on the strong coupling strength between partons. By calibrating QLBT using the $R_\mathrm{AA}$ and $v_2$ data of heavy mesons, the energy and temperature dependence of the coupling strength can be extracted, using which one can further calculate the thermal masses together with the EOS of the quasi-particle system. As shown in the left panel of Fig.~\ref{fig:EOS}, the 95\% C.R.'s of the EOS are in reasonable agreement with the lattice QCD data. The blue band, obtained using 150~MeV as the transition temperature ($T_\mathrm{c}$) between the QGP and the hadron state, coincides with the Wuppertal–Budapest (WB) lattice data~\cite{Borsanyi:2013bia} with the same $T_\mathrm{c}$; though the greed band, constrained with $T_{\mathrm{c}}=154$~MeV, slightly deviates from the HotQCD data~\cite{HotQCD:2014kol} with the same transition temperature. From the same analysis, the spatial diffusion coefficient of heavy quarks is simultaneously extracted, as shown in the right panel of Fig.~\ref{fig:EOS}, which agrees with both lattice QCD data and values from other model calculations, among which the grey band (Duke-LGV) is also obtained from a Bayesian analysis using a Langevin model of heavy quarks~\cite{Xu:2017obm}.

    \section{New developments}~\label{sec:new-stuff} 
    The modification of jets in heavy-ion collisions is now a well developed and sophisticated enterprise. This does not mean that the field is near completion, simply that the major phenomenological framework has been established, and mirrored with a simulation framework. The field is now in the midst of a detailed study on how much can be learnt from the myriad of current observables, and to what extent can simulation developments and new observables help in this regard. 

New phenomenology is yet to the established in the field of jets in small systems such as those in high multiplicity $p$-$p$ and $p$-$A$ collisions. Can the same set up for $A$-$A$ be straightforwardly extended to $p$-$A$ or will major modifications need to be made? The formulation of the answer is now ongoing, and we highlight some of the challenges faced below. 

Yet another direction that is now less than a decade out is that of jet modification in a confined nuclear environment, such as jets in an electron-ion collider. Very little is known about this topic with certainty given the limited amount of experimental data. 
        \subsection{Jets in small systems} 
            During heavy ion collisions, the initial elliptical shape of the overlapping nuclei leads to large pressure gradients.
Due to the low viscosity of QCD matter, the pressure gradients are converted into momentum anisotropies in the final state hadrons.
The transverse momentum distribution of produced particles can be decomposed into a Fourier series in the azimuthal angle $\phi$,
\begin{align}
        \frac{dN}{d\phi dp_T}
        =& \frac{dN}{dp_T} \left( 1 + \sum_{n=1}^{\infty} 2 v_n \cos(n(\phi-\Psi_n)) \right) \;,
\end{align}
where $\Psi_n$ is the event plane angle.
The second Fourier coefficient, $v_2$, is known as the elliptic flow coefficient.

\begin{figure}[h]
    \begin{center}
        \includegraphics[width=0.95\textwidth]{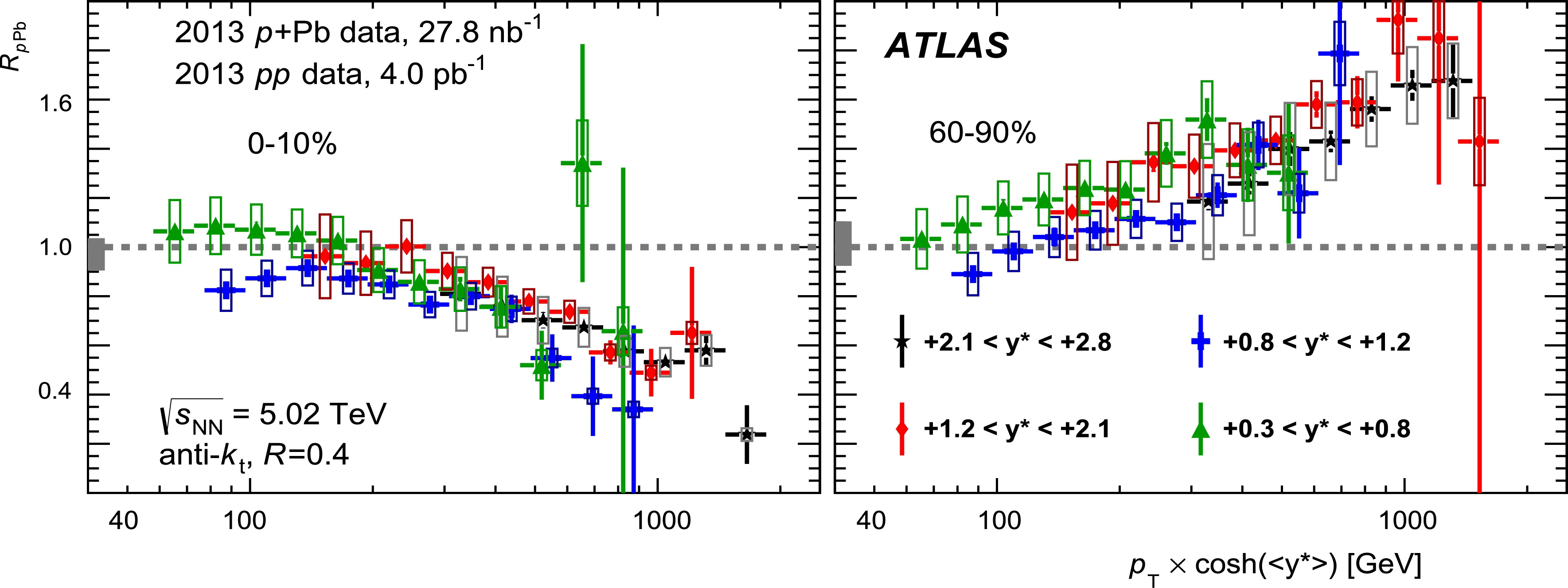}
        \vspace{-0.6cm}
    \end{center}
    \caption{
        Jet nuclear modification factor $R_{pPb}$ for jet as a function of $p_T \cosh(\langle y^*\rangle)$, where $\langle y^* \rangle$ is the midpoint of the different rapidity bins represented with different colors.
        The left panel shows the results for 0-10\% centrality and the right panel shows the results for 60-90\% centrality.
        Figure taken from Ref.~\cite{ATLAS:2014cpa}.
    }
    \label{fig:CenDep}
\end{figure}

\begin{figure}[h]
        \begin{center}
                \includegraphics[width=0.45\textwidth]{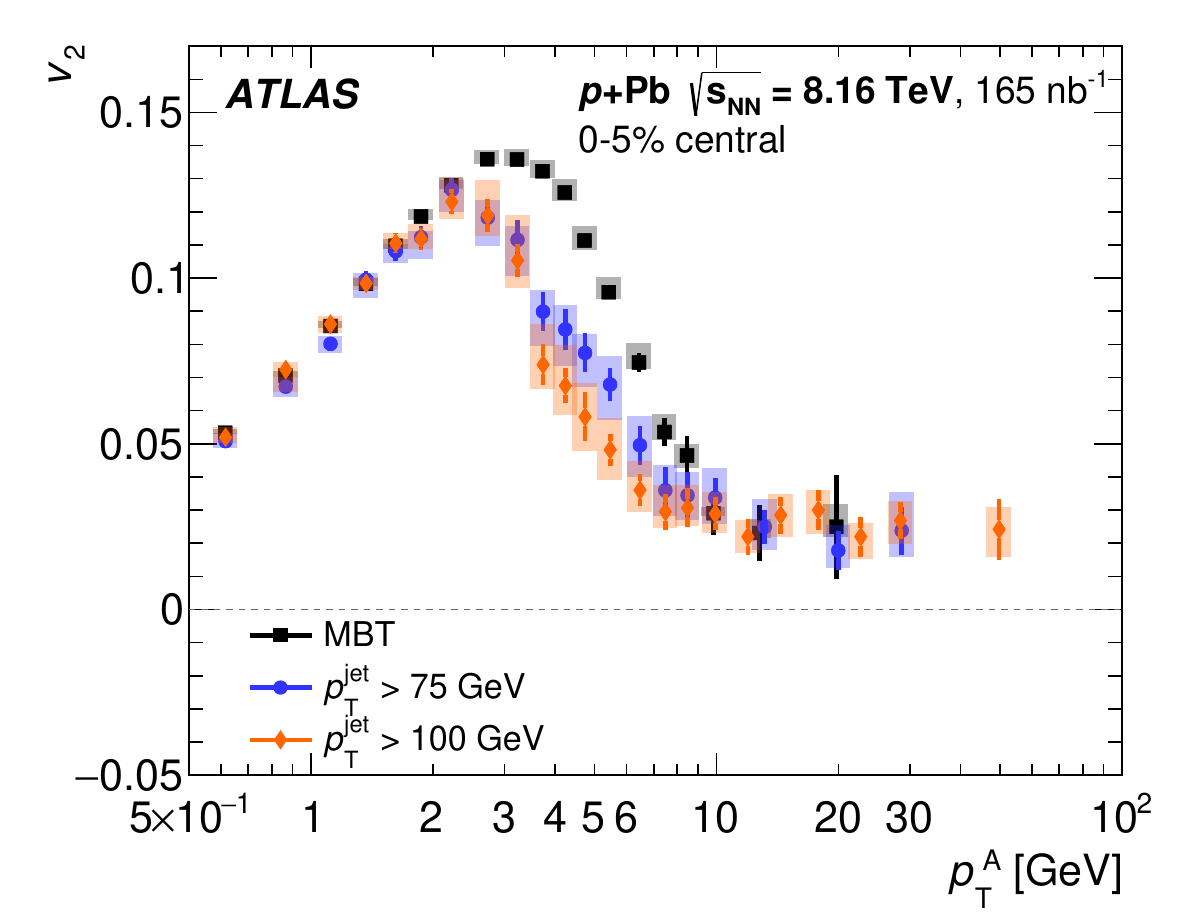}\includegraphics[width=0.45\textwidth]{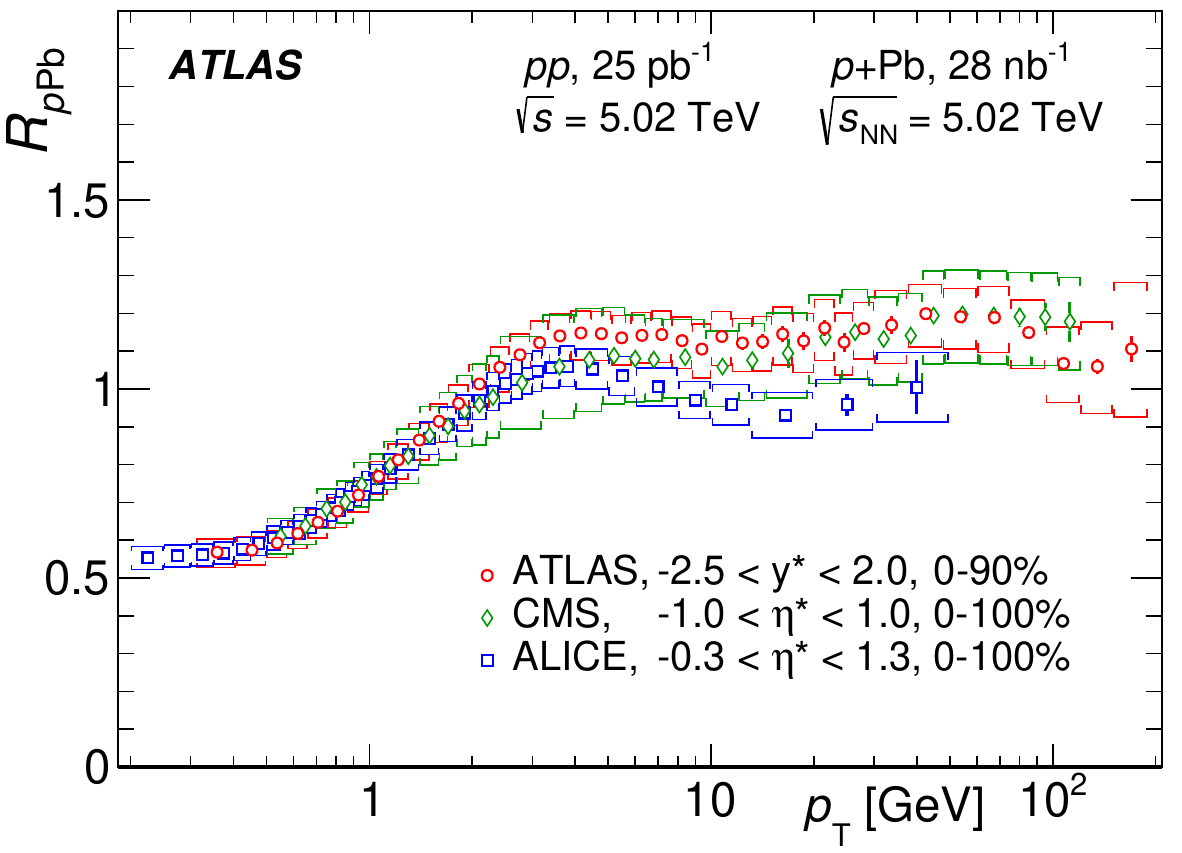}
                \vspace{-0.7cm}
        \end{center}
        \caption{
                Left: ATLAS measurement of charged particle elliptic flow coefficient $v_2$ as a function of $p_T$ for $p$-$Pb$ collisions at $\sqrt{s} = 8.16$~TeV. Figure from Ref~\cite{ATLAS:2019vcm}.
                Right: Charge-hadron nuclear suppression factor $R_{pPb}$ as a function of $p_T$ measured at ATLAS \cite{ATLAS:2022kqu}, CMS \cite{CMS:2016xef} and ALICE \cite{ALICE:2018vuu}. Figure from Ref.~\cite{ATLAS:2022kqu}.
        }\label{fig:v2RpPb}
\end{figure}

Observations of $v_2$ have been used as a strong signal of the formation of a QGP.
Recent experimental results have shown that even in small systems such as $p$-$p$ and $p$-$A$ collisions, there is a non-negligible $v_2$ (c.f. Fig.~\ref{fig:v2RpPb}).
These hints of collectivity in small systems have led to several theoretical investigations which have shown that these effects can be explained by hydrodynamic simulations coupled with novel initial state models \cite{ALICE:2015efi}.
Conversely, jet quenching has not been observed in small systems, the nuclear modification factor $R_{pPb}$ in $p$-$Pb$ collisions is consistent with unity for minimum bias events.
However, when measured as a function of multiplicities, one observes an enhancement $(R_{pPb}\gtrsim 1)$ for peripheral events and suppression $(R_{pPb} \lesssim 1)$ for central events (c.f. Fig.~\ref{fig:CenDep}).

When simulating heavy-ion collisions in the proceeding sections, we have assumed that the energy momentum of the hard partons is negligible compared to the energy momentum of the bulk medium.
Thus, the initial state of the hard partons is completely decoupled from the initial state of the bulk medium.
However, in small systems such as $p$-$p$ and $p$-$A$ collisions, the energy momentum of the jet production can be a substantial share of the total energy available in the initial state.
Due to energy-momentum conservation, the hard/soft parton production is necessarily correlated directly in the initial state \cite{Kordell:2016njg}.

Based on the JETSCAPE framework, the X-SCAPE framework has been released which makes use of energy-momentum conservation to introduce correlations between the hard scattering and the bulk medium \cite{JETSCAPE:2023xbc}.
Starting from a hard scattering sampling using PYTHIA, the initial partons involved in the scattering follow an initial state radiation (ISR) shower which evolves backward in time to obtain the initial partons coming from the two protons.
These initial partons' energy is then subtracted from the initial state of the bulk medium which is modeled using a three-dimensional Glauber model (3DGlauber) \cite{Shen:2017bsr,Shen:2022oyg,Zhao:2022ayk}.
This 3DGlauber samples hotspots inside the protons and create strings that are source of the hydrodynamic evolution done using MUSIC \cite{Schenke:2010nt}.

This model has obtained a good description of the hadronic and jet spectra in $p$-$p$ and $p$-$Pb$ collisions \cite{JETSCAPE:2023xbc}.
Moreover, the energy momentum conservation leads to a clear correlation between the event multiplicity and jet energies.
However, the initial hard partons subtracted are taken to be collinear with the proton momentum.
Although measurement of $v_2$ decreases with increasing $p_T$, we observe non-zero $v_2$ even at high $p_T\gtrsim 10$~GeV.
Correlations between the hard scattering and the bulk medium in transverse plane could be important to explain high-$p_T$ $v_2$.

A high-$p_T$ jet traveling along the event plane axis must traverse a longer path in the medium than a jet traveling perpendicular to the event plane.
Consequently, jet-medium interactions lead to correlations between the elliptic flow of the medium and the transverse directions of jets.
However, jet-medium interaction inevitably leads to energy loss, which is not observed in small systems as shown on the right panel of Fig.~\ref{fig:v2RpPb}.

To reconcile these observations, a new approach has been proposed \cite{Soudi:2023epi}, where the jet medium correlations are obtained from the initial state of the collision.
By introducing transverse momentum dependent (TMD) PDFs and FFs, the initial hard partons undergoing the hard scattering acquire an intrinsic transverse momentum in the center of mass reference frame of the hadronic scattering.
Subsequently, the center of mass of the hard partonic scattering is traveling with an overall transverse momentum $\bm{q}$ in the transverse plane of the hadronic scattering.
Due to momentum conservation, the remaining soft partons must also have a transverse momentum $-\bm{q}$ in the opposite direction.
A large portion of this transverse momentum is carried over to the soft partons in the final state, leading to a correlation between the hard-soft parton production.

Additionally, when one allows for the possibility of an intrinsic transverse momentum, new mechanisms can lead to a distribution of polarized partons inside the proton known as the Boer-Mulders' effect \cite{Boer:1997nt,Mulders:2000sh}, or a fragmentation of polarized partons into unpolarized hadrons known as the Collins' effect \cite{Collins:1992kk}.
The coupling of the Boer-Mulders' effect and the Collins' effect can lead to a new source of transverse momentum correlations.
Since the Boer-Mulders' PDFs and the Collins' FFs are not well constrained, this opens up a new approach to study TMD effects using azimuthal correlations in small systems.

Although $R_{pPb}$ is consistent with unity for minimum bias events, when measured at different centrality classes, one observes an enhancement $(R_{pPb}\!\gtrsim\! 1)$ for peripheral events and suppression $(R_{pPb}\!\lesssim\! 1)$ for central events.
If a QGP is formed in these small systems, the size of the droplet is expected to be small such that the energy of jets constitutes a significant fraction of the total energy.
To this end, a new phenomenological framework~\cite{JETSCAPE:2023xbc} has been developed that makes use of energy momentum conservation to introduce correlations between the QGP and hard scattering.

        \subsection{Jets in a confined medium} 
        The scattering of a quark within a nucleon in deep-inelastic scattering at $Q^2 \gg \Lambda^2_{QCD}$ leads to the production of a hard quark. If the process takes place within a nucleus, then the hard quark will evolve via radiation and scattering in the medium, leading to the production of a jet (see Fig.~\ref{fig:A-DIS-mult-scat}). 

The modification of this jet will take place via multiple scattering, radiation and hadronization in a confined medium~\cite{Accardi:2009qv}. At this time of writing there have been several attempts to understand the measured attenuation of the yield of high momentum hadrons~\cite{HERMES:2000ytc,HERMES:2003icw,Lehmann:2010zz}.
produced in the DIS on a large nucleus (A-DIS). Most of these have been using semi-analytic approaches~\cite{Deng:2009ncl,Majumder:2009zu,Arleo:2003jz}.
In all these approaches, cast before the advent of multi-stage frameworks, the same methodology as used for jets in deconfined matter, is simply extended to confined matter. Calculations have also been extended to include the modification of di-hadrons in dense confined matter and compared with experimental data~\cite{Majumder:2004pt}.
It would not be a surprise that the extracted values of $\hat{q}$ from the 3 theoretical approaches of Refs.~\cite{Deng:2009ncl,Majumder:2009zu,Arleo:2003jz}, are very different. The resolution of this is to extend the multi-stage framework to jets in cold nuclear matter. 

There exist at this time, a handful of simulators for jet propagation in a nuclear environment~\cite{Chang:2022hkt,Toll:2013gda}, however, these deal with different parameter ranges, have been constructed using different languages and with different observables in mind. Given the success of multi-stage frameworks in high energy heavy-ion collisions, it seems natural that nuclear DIS would also greatly benefit from such an approach. This along with jets in small systems represents the frontier of multi-stage generator frameworks.

    \section{Summary} 
    In this Chapter, we have described the physics behind multi-stage jet event generators. By its very nature, jet quenching is a multi-scale phenomenon. Jets start at large virtualities, smaller than, but comparable to, their energies, and undergo multiple emissions with few scatterings until the virtualities approach the medium induced scale $\mu^2 \sim \hat{q} \tau$. Partons which reach this scale maintain their virtuality via multiple scattering in the medium prior to each emission. Partons which constitute the shower arise not only from splits from jet partons, but may also include recoil partons knocked out of the medium by interaction with the jet. Some of these are hard enough that they retain their partonic nature (and can be described using pQCD). And some will lose energy, and thermalize with the dense medium. These will have to be described using non-perturbative methods. The deposited energy will lead to a source of energy momentum in the fluid medium.  

To describe the entirety of jet observables, all of the above aspects of jet modification, running on top of an evolving pre-calibrated fluid medium, have to be simulated. To include and test varying physics assumptions in extensive comparisons with data require modular frameworks with Bayesian statistical analysis. In this Chapter, we have given the reader an exposition of how this entire program works, the physics behind the multi-stage approach, a sampling of comparisons with data, and an introduction to Bayesian routines. 

While a majority of our results have been obtained using the publicly available JETSCAPE framework, we have also presented results from a smattering of other generators. 
We can state without qualification that any successful jet event generator will, in the end, contain within it some aspect of the multi-stage physics described in this chapter. Current frameworks such as JETSCAPE are now being extended to the realm of small systems and jets in cold nuclear matter. 

Experimental results for small systems, such as $p$-$p$, $p$-$A$ and $d$-$A$ collisions, have shown signs of collectivity similar to what is obtained in heavy-ion collisions.
However, investigations of jet quenching in small systems have yielded conflicting information.
In order to reconcile these observations, one needs to carefully model the interplay of energy-momentum conservation between the hard process and the bulk dynamics.
This new frontier of jet studies will open up new opportunities to test our understanding of jet evolution by simultaneously confronting observables from heavy-ion collision to small systems.

In this Chapter, we have attempted to give the reader an introduction into the sophisticated environment of multi-stage event generator frameworks, particularly applied to jets in heavy-ion collisions. The goal has been to draw a direct line from theory to phenomenology to simulation and data comparison. Also included was a brief introduction to Bayesian techniques which are now widely used as these extensive simulators are simultaneously compared with the widest range of observables. We concluded with current and upcoming challenges to this framework in small systems and electron ion collisions. We hope this chapter will contribute to and ease the introduction of new practitioners in this exciting scientific endeavor, as they use these frameworks to address novel observables.  

    \section{Acknowledgements}
    SC is supported by the National Natural Science Foundation of China (NSFC) under Grant Nos. 12175122 and 2021-867. IS is funded as a part of the European Research Council project ERC-2018-ADG-835105 YoctoLHC, and as a part of the Center of Excellence in Quark Matter of the Academy of Finland (project 346325). AM is supported by the US Department of Energy under Grant No. DE-SC0013460, and by the US National Science Foundation under Grant No. OAC-2004571, within the framework of the JETSCAPE collaboration (I.S. was also supported by these grants while at Wayne State University). RMY was (while at McGill University) supported by the Natural Sciences and Engineering Research Council of Canada and gratefully acknowledges the support of the Digital Research Alliance of Canada and Calcul Qu\'ebec for the computations performed on the B\'eluga, Graham and Narval computers.  
YT is supported by JSPS KAKENHI Grant No. 22K14041. 

    \bibliographystyle{ws-rv-van}
    \bibliography{refs}

\begin{thebibliography}{250}
\providecommand{\natexlab}[1]{#1}
\providecommand{\url}[1]{\texttt{#1}}
\expandafter\ifx\csname urlstyle\endcsname\relax
  \providecommand{\doi}[1]{doi: #1}\else
  \providecommand{\doi}{doi: \begingroup \urlstyle{rm}\Url}\fi

\bibitem{Field:1989uq}
R.~D. Field, {APPLICATIONS OF PERTURBATIVE QCD} Redwood City, USA:
  Addison-Wesley (1989) 366 p. (Frontiers in physics, 77).

\bibitem{Sjostrand:2014zea}
T.~Sj\"ostrand, S.~Ask, J.~R. Christiansen, R.~Corke, N.~Desai, P.~Ilten,
  S.~Mrenna, S.~Prestel, C.~O. Rasmussen, and P.~Z. Skands, {An introduction to
  PYTHIA 8.2}, \emph{Comput. Phys. Commun.} {\bf 191}, \penalty0 159--177
  (2015).
\newblock \doi{10.1016/j.cpc.2015.01.024}.

\bibitem{Bahr:2008pv}
M.~Bahr et~al., {Herwig++ Physics and Manual}, \emph{Eur. Phys. J. C}. {\bf
  58}, \penalty0 639--707  (2008).
\newblock \doi{10.1140/epjc/s10052-008-0798-9}.

\bibitem{Sherpa:2019gpd}
E.~Bothmann et~al., {Event Generation with Sherpa 2.2}, \emph{SciPost Phys.}
  {\bf 7}\penalty0 (3), \penalty0 034  (2019).
\newblock \doi{10.21468/SciPostPhys.7.3.034}.

\bibitem{Bjorken:1982qr}
J.~D. Bjorken, Highly relativistic nucleus-nucleus collisions: The central
  rapidity region, \emph{Phys. Rev.} {\bf D27}, \penalty0 140--151  (1983).

\bibitem{Kolb:2000sd}
P.~F. Kolb, J.~Sollfrank, and U.~W. Heinz, Anisotropic transverse flow and the
  quark-hadron phase transition, \emph{Phys. Rev.} {\bf C62}, \penalty0 054909
  (2000).

\bibitem{Kolb:2003dz}
P.~F. Kolb and U.~W. Heinz, Hydrodynamic description of ultrarelativistic
  heavy-ion collisions  (2003).

\bibitem{Teaney:2000cw}
D.~Teaney, J.~Lauret, and E.~V. Shuryak, Flow at the sps and rhic as a quark
  gluon plasma signature, \emph{Phys. Rev. Lett.} {\bf 86}, \penalty0
  4783--4786  (2001).

\bibitem{Gale:2013da}
C.~Gale, S.~Jeon, and B.~Schenke, {Hydrodynamic Modeling of Heavy-Ion
  Collisions}, \emph{Int. J. Mod. Phys. A}. {\bf 28}, \penalty0 1340011
  (2013).
\newblock \doi{10.1142/S0217751X13400113}.

\bibitem{Cooper:1974mv}
F.~Cooper and G.~Frye, {Comment on the Single Particle Distribution in the
  Hydrodynamic and Statistical Thermodynamic Models of Multiparticle
  Production}, \emph{Phys. Rev.} {\bf D10}, \penalty0 186  (1974).
\newblock \doi{10.1103/PhysRevD.10.186}.

\bibitem{Majumder:2010qh}
A.~Majumder and M.~Van~Leeuwen, {The Theory and Phenomenology of Perturbative
  QCD Based Jet Quenching}, \emph{Prog. Part. Nucl. Phys.} {\bf 66}, \penalty0
  41--92  (2011).
\newblock \doi{10.1016/j.ppnp.2010.09.001}.

\bibitem{Cao:2020wlm}
S.~Cao and X.-N. Wang, {Jet quenching and medium response in high-energy
  heavy-ion collisions: a review}  (2020).

\bibitem{Andersson:1983ia}
B.~Andersson, G.~Gustafson, G.~Ingelman, and T.~Sjostrand, Parton fragmentation
  and string dynamics, \emph{Phys. Rept.} {\bf 97}, \penalty0 31  (1983).

\bibitem{Fries:2003vb}
R.~J. Fries, B.~Muller, C.~Nonaka, and S.~A. Bass, Hadronization in heavy ion
  collisions: Recombination and fragmentation of partons, \emph{Phys. Rev.
  Lett.} {\bf 90}, \penalty0 202303  (2003).

\bibitem{Molnar:2003ff}
D.~Molnar and S.~A. Voloshin, Elliptic flow at large transverse momenta from
  quark coalescence, \emph{Phys. Rev. Lett.} {\bf 91}, \penalty0 092301
  (2003).

\bibitem{Greco:2003xt}
V.~Greco, C.~M. Ko, and P.~Levai, Parton coalescence and antiproton/pion
  anomaly at rhic, \emph{Phys. Rev. Lett.} {\bf 90}, \penalty0 202302  (2003).

\bibitem{Hwa:2002tu}
R.~C. Hwa and C.~B. Yang, Scaling behavior at high p(t) and the p/pi ratio,
  \emph{Phys. Rev.} {\bf C67}, \penalty0 034902  (2003).

\bibitem{Schenke:2012wb}
B.~Schenke, P.~Tribedy, and R.~Venugopalan, {Fluctuating Glasma initial
  conditions and flow in heavy ion collisions}, \emph{Phys. Rev. Lett.} {\bf
  108}, \penalty0 252301  (2012).
\newblock \doi{10.1103/PhysRevLett.108.252301}.

\bibitem{Moreland:2014oya}
J.~S. Moreland, J.~E. Bernhard, and S.~A. Bass, {Alternative ansatz to wounded
  nucleon and binary collision scaling in high-energy nuclear collisions},
  \emph{Phys. Rev. C}. {\bf 92}\penalty0 (1), \penalty0 011901  (2015).
\newblock \doi{10.1103/PhysRevC.92.011901}.

\bibitem{Putschke:2019yrg}
J.~H. Putschke et~al., {The JETSCAPE framework}  (2019).

\bibitem{Cao:2017zih}
S.~Cao et~al., {Multistage Monte-Carlo simulation of jet modification in a
  static medium}, \emph{Phys. Rev.} {\bf C96}\penalty0 (2), \penalty0 024909
  (2017).
\newblock \doi{10.1103/PhysRevC.96.024909}.

\bibitem{Collins:1985ue}
J.~C. Collins, D.~E. Soper, and G.~Sterman, {Factorization for Short Distance
  Hadron - Hadron Scattering}, \emph{Nucl. Phys.} {\bf B261}, \penalty0 104
  (1985).
\newblock \doi{10.1016/0550-3213(85)90565-6}.

\bibitem{Collins:1981uw}
J.~C. Collins and D.~E. Soper, Parton distribution and decay functions,
  \emph{Nucl. Phys.} {\bf B194}, \penalty0 445  (1982).

\bibitem{Collins:1988ig}
J.~C. Collins, D.~E. Soper, and G.~Sterman, {Soft Gluons and Factorization},
  \emph{Nucl. Phys.} {\bf B308}, \penalty0 833  (1988).
\newblock \doi{10.1016/0550-3213(88)90130-7}.

\bibitem{Collins:1989gx}
J.~C. Collins, D.~E. Soper, and G.~Sterman, {Factorization of Hard Processes in
  QCD}, \emph{Adv. Ser. Direct. High Energy Phys.} {\bf 5}, \penalty0 1--91
  (1988).

\bibitem{Salgado:2003gb}
C.~A. Salgado and U.~A. Wiedemann, {Calculating quenching weights}, \emph{Phys.
  Rev.} {\bf D68}, \penalty0 014008  (2003).
\newblock \doi{10.1103/PhysRevD.68.014008}.

\bibitem{He:2015pra}
Y.~He, T.~Luo, X.-N. Wang, and Y.~Zhu, {Linear Boltzmann Transport for Jet
  Propagation in the Quark-Gluon Plasma: Elastic Processes and Medium Recoil},
  \emph{Phys. Rev.} {\bf C91}, \penalty0 054908  (2015).
\newblock \doi{10.1103/PhysRevC.97.019902, 10.1103/PhysRevC.91.054908}.
\newblock [Erratum: Phys. Rev.C97,no.1,019902(2018)].

\bibitem{Schenke:2009gb}
B.~Schenke, C.~Gale, and S.~Jeon, {MARTINI: An Event generator for relativistic
  heavy-ion collisions}, \emph{Phys.Rev.} {\bf C80}, \penalty0 054913  (2009).
\newblock \doi{10.1103/PhysRevC.80.054913}.

\bibitem{Qin:2009bk}
G.-Y. Qin, J.~Ruppert, C.~Gale, S.~Jeon, and G.~D. Moore, {Jet energy loss,
  photon production, and photon-hadron correlations at RHIC}  (2009).

\bibitem{Majumder:2007iu}
A.~Majumder, {A comparative study of jet-quenching schemes}, \emph{J. Phys.}
  {\bf G34}, \penalty0 S377--388  (2007).
\newblock \doi{10.1088/0954-3899/34/8/S25}.

\bibitem{Hwa:2004ng}
R.~C. Hwa and C.~B. Yang, Recombination of shower partons at high p(t) in
  heavy-ion collisions, \emph{Phys. Rev.} {\bf C70}, \penalty0 024905  (2004).

\bibitem{Majumder:2005jy}
A.~Majumder, E.~Wang, and X.-N. Wang, {Modified fragmentation function from
  quark recombination}, \emph{Phys. Rev. C}. {\bf 73}, \penalty0 044901
  (2006).
\newblock \doi{10.1103/PhysRevC.73.044901}.

\bibitem{Majumder:2007hx}
A.~Majumder and B.~Muller, {Higher twist jet broadening and classical
  propagation}, \emph{Phys. Rev.} {\bf C77}, \penalty0 054903  (2008).
\newblock \doi{10.1103/PhysRevC.77.054903}.

\bibitem{Qin:2012fua}
G.-Y. Qin and A.~Majumder, {Parton Transport via Transverse and Longitudinal
  Scattering in Dense Media}, \emph{Phys.Rev.} {\bf C87}\penalty0 (2),
  \penalty0 024909  (2013).
\newblock \doi{10.1103/PhysRevC.87.024909}.

\bibitem{Majumder:2008zg}
A.~Majumder, {Elastic energy loss and longitudinal straggling of a hard jet},
  \emph{Phys. Rev.} {\bf C80}, \penalty0 031902  (2009).
\newblock \doi{10.1103/PhysRevC.80.031902}.

\bibitem{Baier:2002tc}
R.~Baier, {Jet quenching}, \emph{Nucl. Phys.} {\bf A715}, \penalty0 209--218
  (2003).
\newblock \doi{10.1016/S0375-9474(02)01429-X}.

\bibitem{Majumder:2012sh}
A.~Majumder, {Calculating the jet quenching parameter $\hat{q}$ in lattice
  gauge theory}, \emph{Phys. Rev.} {\bf C87}, \penalty0 034905  (2013).
\newblock \doi{10.1103/PhysRevC.87.034905}.

\bibitem{Kumar:2020pdl}
A.~Kumar, A.~Majumder, and J.~H. Weber.
\newblock {Lattice calculation of transport coefficient $\hat{q}$ in pure gluon
  plasma and (2+1)-flavor QCD plasma}.
\newblock In \emph{{10th International Conference on Hard and Electromagnetic
  Probes of High-Energy Nuclear Collisions}: {Hard Probes 2020~}}  (9, 2020).

\bibitem{Braaten:1989kk}
E.~Braaten and R.~D. Pisarski, Resummation and gauge invariance of the gluon
  damping rate in hot qcd, \emph{Phys. Rev. Lett.} {\bf 64}, \penalty0 1338
  (1990).

\bibitem{Braaten:1989mz}
E.~Braaten and R.~D. Pisarski, {Soft Amplitudes in Hot Gauge Theories: A
  General Analysis}, \emph{Nucl.Phys.} {\bf B337}, \penalty0 569  (1990).
\newblock \doi{10.1016/0550-3213(90)90508-B}.

\bibitem{Frenkel:1989br}
J.~Frenkel and J.~C. Taylor, {High Temperature Limit of Thermal QCD},
  \emph{Nucl. Phys.} {\bf B334}, \penalty0 199--216  (1990).
\newblock \doi{10.1016/0550-3213(90)90661-V}.

\bibitem{Dokshitzer:1977sg}
Y.~L. Dokshitzer, {Calculation of the Structure Functions for Deep Inelastic
  Scattering and e+ e- Annihilation by Perturbation Theory in Quantum
  Chromodynamics}, \emph{Sov. Phys. JETP}. {\bf 46}, \penalty0 641--653
  (1977).

\bibitem{Gribov:1972ri}
V.~N. Gribov and L.~N. Lipatov, {Deep inelastic e p scattering in perturbation
  theory}, \emph{Sov. J. Nucl. Phys.} {\bf 15}, \penalty0 438--450  (1972).

\bibitem{Gribov:1972rt}
V.~N. Gribov and L.~N. Lipatov, {e+ e- pair annihilation and deep inelastic e p
  scattering in perturbation theory}, \emph{Sov. J. Nucl. Phys.} {\bf 15},
  \penalty0 675--684  (1972).

\bibitem{Altarelli:1977zs}
G.~Altarelli and G.~Parisi, {Asymptotic Freedom in Parton Language},
  \emph{Nucl. Phys.} {\bf B126}, \penalty0 298  (1977).
\newblock \doi{10.1016/0550-3213(77)90384-4}.

\bibitem{Majumder:2013re}
A.~Majumder, {Incorporating Space-Time Within Medium-Modified Jet Event
  Generators}, \emph{Phys. Rev.} {\bf C88}, \penalty0 014909  (2013).
\newblock \doi{10.1103/PhysRevC.88.014909}.

\bibitem{Sirimanna:2021sqx}
C.~Sirimanna, S.~Cao, and A.~Majumder, {Final-state gluon emission in
  deep-inelastic scattering at next-to-leading twist}, \emph{Phys. Rev. C}.
  {\bf 105}\penalty0 (2), \penalty0 024908  (2022).
\newblock \doi{10.1103/PhysRevC.105.024908}.

\bibitem{Majumder:2014gda}
A.~Majumder and J.~Putschke, {Mass depletion: a new parameter for quantitative
  jet modification}, \emph{Phys. Rev. C}. {\bf 93}\penalty0 (5), \penalty0
  054909  (2016).
\newblock \doi{10.1103/PhysRevC.93.054909}.

\bibitem{Arnold:2008iy}
P.~B. Arnold, {Simple Formula for High-Energy Gluon Bremsstrahlung in a Finite,
  Expanding Medium}, \emph{Phys. Rev. D}. {\bf 79}, \penalty0 065025  (2009).
\newblock \doi{10.1103/PhysRevD.79.065025}.

\bibitem{Gyulassy:1993hr}
M.~Gyulassy and X.-N. Wang, Multiple collisions and induced gluon
  bremsstrahlung in qcd, \emph{Nucl. Phys.} {\bf B420}, \penalty0 583--614
  (1994).

\bibitem{Gyulassy:1999zd}
M.~Gyulassy, P.~Levai, and I.~Vitev, {Jet quenching in thin quark-gluon
  plasmas. I: Formalism}, \emph{Nucl. Phys.} {\bf B571}, \penalty0 197--233
  (2000).
\newblock \doi{10.1016/S0550-3213(99)00713-0}.

\bibitem{Gyulassy:2000er}
M.~Gyulassy, P.~Levai, and I.~Vitev, {Reaction operator approach to non-Abelian
  energy loss}, \emph{Nucl. Phys.} {\bf B594}, \penalty0 371--419  (2001).
\newblock \doi{10.1016/S0550-3213(00)00652-0}.

\bibitem{Wiedemann:2000za}
U.~A. Wiedemann, {Gluon radiation off hard quarks in a nuclear environment:
  Opacity expansion}, \emph{Nucl. Phys.} {\bf B588}, \penalty0 303--344
  (2000).
\newblock \doi{10.1016/S0550-3213(00)00457-0}.

\bibitem{Djordjevic:2003zk}
M.~Djordjevic and M.~Gyulassy, {Heavy quark radiative energy loss in QCD
  matter}, \emph{Nucl. Phys.} {\bf A733}, \penalty0 265--298  (2004).
\newblock \doi{10.1016/j.nuclphysa.2003.12.020}.

\bibitem{Djordjevic:2007at}
M.~Djordjevic and U.~Heinz, {Radiative heavy quark energy loss in a dynamical
  QCD medium}, \emph{Phys. Rev. C}. {\bf 77}, \penalty0 024905  (2008).
\newblock \doi{10.1103/PhysRevC.77.024905}.

\bibitem{Djordjevic:2008iz}
M.~Djordjevic and U.~W. Heinz, {Radiative energy loss in a finite dynamical QCD
  medium}, \emph{Phys. Rev. Lett.} {\bf 101}, \penalty0 022302  (2008).
\newblock \doi{10.1103/PhysRevLett.101.022302}.

\bibitem{Djordjevic:2009cr}
M.~Djordjevic, {Theoretical formalism of radiative jet energy loss in a finite
  size dynamical QCD medium}, \emph{Phys. Rev. C}. {\bf 80}, \penalty0 064909
  (2009).
\newblock \doi{10.1103/PhysRevC.80.064909}.

\bibitem{Xu:2014ica}
J.~Xu, A.~Buzzatti, and M.~Gyulassy, {Azimuthal jet flavor tomography with
  CUJET2.0 of nuclear collisions at RHIC and LHC}, \emph{JHEP}. {\bf 08},
  \penalty0 063  (2014).
\newblock \doi{10.1007/JHEP08(2014)063}.

\bibitem{Xu:2014wua}
J.~Xu, A.~Buzzatti, and M.~Gyulassy, {The tricky azimuthal dependence of jet
  quenching at RHIC and LHC via CUJET2.0}, \emph{Nucl. Phys. A}. {\bf 932},
  \penalty0 128--133  (2014).
\newblock \doi{10.1016/j.nuclphysa.2014.07.027}.

\bibitem{Xu:2015bbz}
J.~Xu, J.~Liao, and M.~Gyulassy, {Bridging Soft-Hard Transport Properties of
  Quark-Gluon Plasmas with CUJET3.0}, \emph{JHEP}. {\bf 02}, \penalty0 169
  (2016).
\newblock \doi{10.1007/JHEP02(2016)169}.

\bibitem{Xu:2014tda}
J.~Xu, J.~Liao, and M.~Gyulassy, {Consistency of Perfect Fluidity and Jet
  Quenching in semi-Quark-Gluon Monopole Plasmas}, \emph{Chin. Phys. Lett.}
  {\bf 32}\penalty0 (9), \penalty0 092501  (2015).
\newblock \doi{10.1088/0256-307X/32/9/092501}.

\bibitem{Shi:2022rja}
S.~Shi, R.~Modarresi~Yazdi, C.~Gale, and S.~Jeon, {Comparing the martini and
  cujet models for jet quenching: Medium modification of jets and jet
  substructure}, \emph{Phys. Rev. C}. {\bf 107}\penalty0 (3), \penalty0 034908
  (2023).
\newblock \doi{10.1103/PhysRevC.107.034908}.

\bibitem{Arnold:2002ja}
P.~Arnold, G.~D. Moore, and L.~G. Yaffe, Photon and gluon emission in
  relativistic plasmas, \emph{JHEP}. {\bf 06}, \penalty0 030  (2002).

\bibitem{Arnold:2001ba}
P.~Arnold, G.~D. Moore, and L.~G. Yaffe, Photon emission from ultrarelativistic
  plasmas, \emph{JHEP}. {\bf 11}, \penalty0 057  (2001).

\bibitem{Arnold:2001ms}
P.~Arnold, G.~D. Moore, and L.~G. Yaffe, {Photon emission from quark gluon
  plasma: Complete leading order results}, \emph{JHEP}. {\bf 12}, \penalty0 009
   (2001).

\bibitem{Jeon:2003gi}
S.~Jeon and G.~D. Moore, Energy loss of leading partons in a thermal qcd
  medium, \emph{Phys. Rev.} {\bf C71}, \penalty0 034901  (2005).

\bibitem{Arnold:2008vd}
P.~Arnold and W.~Xiao, {High-energy jet quenching in weakly-coupled quark-gluon
  plasmas}, \emph{Phys. Rev.} {\bf D78}, \penalty0 125008  (2008).
\newblock \doi{10.1103/PhysRevD.78.125008}.

\bibitem{Caron-Huot:2010qjx}
S.~Caron-Huot and C.~Gale, {Finite-size effects on the radiative energy loss of
  a fast parton in hot and dense strongly interacting matter}, \emph{Phys. Rev.
  C}. {\bf 82}, \penalty0 064902  (2010).
\newblock \doi{10.1103/PhysRevC.82.064902}.

\bibitem{Caron-Huot:2008zna}
S.~Caron-Huot, {O(g) plasma effects in jet quenching}, \emph{Phys. Rev. D}.
  {\bf 79}, \penalty0 065039  (2009).
\newblock \doi{10.1103/PhysRevD.79.065039}.

\bibitem{Ghiglieri:2013gia}
J.~Ghiglieri, J.~Hong, A.~Kurkela, E.~Lu, G.~D. Moore, and D.~Teaney,
  {Next-to-leading order thermal photon production in a weakly coupled
  quark-gluon plasma}, \emph{JHEP}. {\bf 05}, \penalty0 010  (2013).
\newblock \doi{10.1007/JHEP05(2013)010}.

\bibitem{Ghiglieri:2015ala}
J.~Ghiglieri, G.~D. Moore, and D.~Teaney, {Jet-Medium Interactions at NLO in a
  Weakly-Coupled Quark-Gluon Plasma}, \emph{JHEP}. {\bf 03}, \penalty0 095
  (2016).
\newblock \doi{10.1007/JHEP03(2016)095}.

\bibitem{Panero:2013pla}
M.~Panero, K.~Rummukainen, and A.~Sch\"afer, {Lattice Study of the Jet
  Quenching Parameter}, \emph{Phys. Rev. Lett.} {\bf 112}\penalty0 (16),
  \penalty0 162001  (2014).
\newblock \doi{10.1103/PhysRevLett.112.162001}.

\bibitem{Moore:2019lua}
G.~D. Moore and N.~Schlusser, {Full O(a) improvement in electrostatic QCD},
  \emph{Phys. Rev. D}. {\bf 100}\penalty0 (3), \penalty0 034510  (2019).
\newblock \doi{10.1103/PhysRevD.100.034510}.

\bibitem{Schlichting:2021idr}
S.~Schlichting and I.~Soudi, {Splitting rates in QCD plasmas from a
  nonperturbative determination of the momentum broadening kernel
  C(q\ensuremath{\perp})}, \emph{Phys. Rev. D}. {\bf 105}\penalty0 (7),
  \penalty0 076002  (2022).
\newblock \doi{10.1103/PhysRevD.105.076002}.

\bibitem{Yazdi:2022bru}
R.~M. Yazdi, S.~Shi, C.~Gale, and S.~Jeon, {Leading order, next-to-leading
  order, and nonperturbative parton collision kernels: Effects in static and
  evolving media}, \emph{Phys. Rev. C}. {\bf 106}\penalty0 (6), \penalty0
  064902  (2022).
\newblock \doi{10.1103/PhysRevC.106.064902}.

\bibitem{Kumar:2019uvu}
A.~Kumar, A.~Majumder, and C.~Shen, {Energy and scale dependence of $\hat{q}$
  and the \textquotedblleft{}JET puzzle\textquotedblright{}}, \emph{Phys. Rev.
  C}. {\bf 101}\penalty0 (3), \penalty0 034908  (2020).
\newblock \doi{10.1103/PhysRevC.101.034908}.

\bibitem{MehtarTani:2011tz}
Y.~Mehtar-Tani, C.~A. Salgado, and K.~Tywoniuk, {Jets in QCD Media: From Color
  Coherence to Decoherence}, \emph{Phys. Lett.} {\bf B707}, \penalty0 156--159
  (2012).
\newblock \doi{10.1016/j.physletb.2011.12.042}.

\bibitem{CasalderreySolana:2012ef}
J.~Casalderrey-Solana, Y.~Mehtar-Tani, C.~A. Salgado, and K.~Tywoniuk, {New
  picture of jet quenching dictated by color coherence}  (2012).

\bibitem{Bjorken:1982tu}
J.~D. Bjorken, {Energy Loss of Energetic Partons in Quark - Gluon Plasma:
  Possible Extinction of High p(t) Jets in Hadron - Hadron Collisions},
  \emph{Fermilab, Report No. FERMILAB-PUB-82-059-THY}  (unpublished).

\bibitem{Baier:1998yf}
R.~Baier, Y.~L. Dokshitzer, A.~H. Mueller, and D.~Schiff, {Radiative energy
  loss of high energy partons traversing an expanding {QCD} plasma},
  \emph{Phys. Rev.} {\bf C58}, \penalty0 1706--1713  (1998).
\newblock \doi{10.1103/PhysRevC.58.1706}.

\bibitem{Zakharov:1998sv}
B.~G. Zakharov, {Light-cone path integral approach to the Landau-
  Pomeranchuk-Migdal effect}, \emph{Phys. Atom. Nucl.} {\bf 61}, \penalty0
  838--854  (1998).

\bibitem{Baier:2001yt}
R.~Baier, Y.~L. Dokshitzer, A.~H. Mueller, and D.~Schiff, Quenching of hadron
  spectra in media, \emph{JHEP}. {\bf 09}, \penalty0 033  (2001).

\bibitem{Armesto:2011ir}
N.~Armesto, H.~Ma, Y.~Mehtar-Tani, C.~A. Salgado, and K.~Tywoniuk, {Coherence
  effects and broadening in medium-induced QCD radiation off a massive $q {\bar
  q}$ antenna}, \emph{JHEP}. {\bf 01}, \penalty0 109  (2012).
\newblock \doi{10.1007/JHEP01(2012)109}.

\bibitem{Mehtar-Tani:2017web}
Y.~Mehtar-Tani and K.~Tywoniuk, {Sudakov suppression of jets in QCD media},
  \emph{Phys. Rev. D}. {\bf 98}\penalty0 (5), \penalty0 051501  (2018).
\newblock \doi{10.1103/PhysRevD.98.051501}.

\bibitem{Mehtar-Tani:2014yea}
Y.~Mehtar-Tani and K.~Tywoniuk, {Jet (de)coherence in Pb\textendash{}Pb
  collisions at the LHC}, \emph{Phys. Lett. B}. {\bf 744}, \penalty0 284--287
  (2015).
\newblock \doi{10.1016/j.physletb.2015.03.041}.

\bibitem{Bass:2008rv}
S.~A. Bass, C.~Gale, A.~Majumder, C.~Nonaka, G.-Y. Qin, et~al., {Systematic
  Comparison of Jet Energy-Loss Schemes in a realistic hydrodynamic medium},
  \emph{Phys.Rev.} {\bf C79}, \penalty0 024901  (2009).
\newblock \doi{10.1103/PhysRevC.79.024901}.

\bibitem{Majumder:2011uk}
A.~Majumder and C.~Shen, {Suppression of the High $p_T$ Charged Hadron $R_{AA}$
  at the LHC}, \emph{Phys.Rev.Lett.} {\bf 109}, \penalty0 202301  (2012).
\newblock \doi{10.1103/PhysRevLett.109.202301}.

\bibitem{Mehtar-Tani:2018zba}
Y.~Mehtar-Tani and S.~Schlichting, {Universal quark to gluon ratio in
  medium-induced parton cascade}, \emph{JHEP}. {\bf 09}, \penalty0 144  (2018).
\newblock \doi{10.1007/JHEP09(2018)144}.

\bibitem{Schlichting:2020lef}
S.~Schlichting and I.~Soudi, {Medium-induced fragmentation and equilibration of
  highly energetic partons}, \emph{JHEP}. {\bf 07}, \penalty0 077  (2021).
\newblock \doi{10.1007/JHEP07(2021)077}.

\bibitem{Mehtar-Tani:2022zwf}
Y.~Mehtar-Tani, S.~Schlichting, and I.~Soudi, {Jet thermalization in QCD
  kinetic theory}, \emph{JHEP}. {\bf 05}, \penalty0 091  (2023).
\newblock \doi{10.1007/JHEP05(2023)091}.

\bibitem{Isaksen:2022pkj}
J.~H. Isaksen, A.~Takacs, and K.~Tywoniuk, {A unified picture of medium-induced
  radiation}, \emph{JHEP}. {\bf 02}, \penalty0 156  (2023).
\newblock \doi{10.1007/JHEP02(2023)156}.

\bibitem{Qin:2007rn}
G.-Y. Qin et~al., {Radiative and Collisional Jet Energy Loss in the Quark-
  Gluon Plasma at RHIC}, \emph{Phys. Rev. Lett.} {\bf 100}, \penalty0 072301
  (2008).
\newblock \doi{10.1103/PhysRevLett.100.072301}.

\bibitem{Qin:2009gw}
G.-Y. Qin and A.~Majumder, {A pQCD-based description of heavy and light flavor
  jet quenching}, \emph{Phys.Rev.Lett.} {\bf 105}, \penalty0 262301  (2010).
\newblock \doi{10.1103/PhysRevLett.105.262301}.

\bibitem{Arleo:2017ntr}
F.~Arleo, {Quenching of Hadron Spectra in Heavy Ion Collisions at the LHC},
  \emph{Phys. Rev. Lett.} {\bf 119}\penalty0 (6), \penalty0 062302  (2017).
\newblock \doi{10.1103/PhysRevLett.119.062302}.

\bibitem{Blaizot:2014jna}
J.-P. Blaizot, B.~Wu, and L.~Yan, {Quark production, Bose\textendash{}Einstein
  condensates and thermalization of the quark\textendash{}gluon plasma},
  \emph{Nucl. Phys. A}. {\bf 930}, \penalty0 139--162  (2014).
\newblock \doi{10.1016/j.nuclphysa.2014.07.041}.

\bibitem{Arnold:2000dr}
P.~Arnold, G.~D. Moore, and L.~G. Yaffe, Transport coefficients in high
  temperature gauge theories. i: Leading-log results, \emph{JHEP}. {\bf 11},
  \penalty0 001  (2000).

\bibitem{Arnold:2003rq}
P.~Arnold, J.~Lenaghan, and G.~D. Moore, Qcd plasma instabilities and bottom-up
  thermalization, \emph{JHEP}. {\bf 08}, \penalty0 002  (2003).

\bibitem{Adhya:2019qse}
S.~P. Adhya, C.~A. Salgado, M.~Spousta, and K.~Tywoniuk, {Medium-induced
  cascade in expanding media}, \emph{JHEP}. {\bf 07}, \penalty0 150  (2020).
\newblock \doi{10.1007/JHEP07(2020)150}.

\bibitem{Adhya:2021kws}
S.~P. Adhya, C.~A. Salgado, M.~Spousta, and K.~Tywoniuk, {Multi-partonic medium
  induced cascades in expanding media}, \emph{Eur. Phys. J. C}. {\bf
  82}\penalty0 (1), \penalty0 20  (2022).
\newblock \doi{10.1140/epjc/s10052-021-09950-8}.

\bibitem{Blaizot:2012fh}
J.-P. Blaizot, F.~Dominguez, E.~Iancu, and Y.~Mehtar-Tani, {Medium-induced
  gluon branching}, \emph{JHEP}. {\bf 01}, \penalty0 143  (2013).
\newblock \doi{10.1007/JHEP01(2013)143}.

\bibitem{nazarenko_2011}
S.~Nazarenko, \emph{{Wave turbulence}}. Springer  (2011).

\bibitem{zakharov2012kolmogorov}
V.~E. Zakharov, V.~S. L'vov, and G.~Falkovich, \emph{{Kolmogorov spectra of
  turbulence I: Wave turbulence}}. Springer Science \& Business Media  (2012).

\bibitem{Blaizot:2015jea}
J.-P. Blaizot and Y.~Mehtar-Tani, {Energy flow along the medium-induced parton
  cascade}, \emph{Annals Phys.} {\bf 368}, \penalty0 148--176  (2016).
\newblock \doi{10.1016/j.aop.2016.01.002}.

\bibitem{Majumder:2009zu}
A.~Majumder, {The in-medium scale evolution in jet modification}  (2009).

\bibitem{Kang:2014xsa}
Z.-B. Kang, R.~Lashof-Regas, G.~Ovanesyan, P.~Saad, and I.~Vitev, {Jet
  quenching phenomenology from soft-collinear effective theory with Glauber
  gluons}, \emph{Phys. Rev. Lett.} {\bf 114}\penalty0 (9), \penalty0 092002
  (2015).
\newblock \doi{10.1103/PhysRevLett.114.092002}.

\bibitem{Chien:2015vja}
Y.-T. Chien, A.~Emerman, Z.-B. Kang, G.~Ovanesyan, and I.~Vitev, {Jet Quenching
  from QCD Evolution}, \emph{Phys. Rev. D}. {\bf 93}\penalty0 (7), \penalty0
  074030  (2016).
\newblock \doi{10.1103/PhysRevD.93.074030}.

\bibitem{Sjostrand:1984}
T.~Sjöstrand, Jet fragmentation of multiparton configurations in a string
  framework, \emph{Nuclear Physics B}. {\bf 248}\penalty0 (2), \penalty0
  469--502  (1984).
\newblock ISSN 0550-3213.

\bibitem{Sjostrand:1985xi}
T.~Sjostrand, {A Model for Initial State Parton Showers}, \emph{Phys. Lett. B}.
  {\bf 157}, \penalty0 321--325  (1985).
\newblock \doi{10.1016/0370-2693(85)90674-4}.

\bibitem{Lipatov:1974qm}
L.~N. Lipatov, {The parton model and perturbation theory}, \emph{Yad. Fiz.}
  {\bf 20}, \penalty0 181--198  (1974).

\bibitem{Hoche:2014rga}
S.~H\"oche.
\newblock {Introduction to parton-shower event generators}.
\newblock In \emph{{Theoretical Advanced Study Institute in Elementary Particle
  Physics}: {Journeys Through the Precision Frontier: Amplitudes for
  Colliders}}, pp. 235--295  (2015).
\newblock \doi{10.1142/9789814678766_0005}.

\bibitem{Sjostrand:2006za}
T.~Sjostrand, S.~Mrenna, and P.~Z. Skands, {PYTHIA 6.4 Physics and Manual},
  \emph{JHEP}. {\bf 05}, \penalty0 026  (2006).
\newblock \doi{10.1088/1126-6708/2006/05/026}.

\bibitem{Cao:2017qpx}
S.~Cao and A.~Majumder, {Nuclear modification of leading hadrons and jets
  within a virtuality ordered parton shower}, \emph{Phys. Rev. C}. {\bf
  101}\penalty0 (2), \penalty0 024903  (2020).
\newblock \doi{10.1103/PhysRevC.101.024903}.

\bibitem{Bodek:1983qn}
A.~Bodek et~al., {Electron Scattering from Nuclear Targets and Quark
  Distributions in Nuclei}, \emph{Phys. Rev. Lett.} {\bf 50}, \penalty0 1431
  (1983).
\newblock \doi{10.1103/PhysRevLett.50.1431}.

\bibitem{EuropeanMuon:1988tpw}
M.~Arneodo et~al., {Shadowing in Deep Inelastic Muon Scattering from Nuclear
  Targets}, \emph{Phys. Lett. B}. {\bf 211}, \penalty0 493--499  (1988).
\newblock \doi{10.1016/0370-2693(88)91900-4}.

\bibitem{Eskola:2016oht}
K.~J. Eskola, P.~Paakkinen, H.~Paukkunen, and C.~A. Salgado, {EPPS16: Nuclear
  parton distributions with LHC data}, \emph{Eur. Phys. J. C}. {\bf
  77}\penalty0 (3), \penalty0 163  (2017).
\newblock \doi{10.1140/epjc/s10052-017-4725-9}.

\bibitem{Eskola:2009uj}
K.~J. Eskola, H.~Paukkunen, and C.~A. Salgado, {EPS09: A New Generation of NLO
  and LO Nuclear Parton Distribution Functions}, \emph{JHEP}. {\bf 04},
  \penalty0 065  (2009).
\newblock \doi{10.1088/1126-6708/2009/04/065}.

\bibitem{Li:2001xa}
S.-y. Li and X.-N. Wang, {Gluon shadowing and hadron production at RHIC},
  \emph{Phys. Lett. B}. {\bf 527}, \penalty0 85--91  (2002).
\newblock \doi{10.1016/S0370-2693(02)01179-6}.

\bibitem{Luo:2023nsi}
T.~Luo, Y.~He, S.~Cao, and X.-N. Wang, {Linear Boltzmann transport for jet
  propagation in the quark-gluon plasma: Inelastic processes and jet
  modification}  (6, 2023).

\bibitem{Auvinen:2009qm}
J.~Auvinen, K.~J. Eskola, and T.~Renk, {A Monte-Carlo model for elastic energy
  loss in a hydrodynamical background}, \emph{Phys. Rev. C}. {\bf 82},
  \penalty0 024906  (2010).
\newblock \doi{10.1103/PhysRevC.82.024906}.

\bibitem{Guo:2000nz}
X.-F. Guo and X.-N. Wang, {Multiple scattering, parton energy loss and modified
  fragmentation functions in deeply inelastic e A scattering}, \emph{Phys. Rev.
  Lett.} {\bf 85}, \penalty0 3591--3594  (2000).
\newblock \doi{10.1103/PhysRevLett.85.3591}.

\bibitem{Zhang:2003wk}
B.-W. Zhang, E.~Wang, and X.-N. Wang, {Heavy quark energy loss in nuclear
  medium}, \emph{Phys. Rev. Lett.} {\bf 93}, \penalty0 072301  (2004).
\newblock \doi{10.1103/PhysRevLett.93.072301}.

\bibitem{Majumder:2009ge}
A.~Majumder, {Hard collinear gluon radiation and multiple scattering in a
  medium}, \emph{Phys. Rev.} {\bf D85}, \penalty0 014023  (2012).
\newblock \doi{10.1103/PhysRevD.85.014023}.

\bibitem{Majumder:2007ae}
A.~Majumder, C.~Nonaka, and S.~A. Bass, {Jet modification in three dimensional
  fluid dynamics at next-to-leading twist}, \emph{Phys. Rev.} {\bf C76},
  \penalty0 041902  (2007).
\newblock \doi{10.1103/PhysRevC.76.041902}.

\bibitem{Luo:2018pto}
T.~Luo, S.~Cao, Y.~He, and X.-N. Wang, {Multiple jets and $\gamma$-jet
  correlation in high-energy heavy-ion collisions}, \emph{Phys. Lett.} {\bf
  B782}, \penalty0 707--716  (2018).
\newblock \doi{10.1016/j.physletb.2018.06.025}.

\bibitem{He:2018xjv}
Y.~He, S.~Cao, W.~Chen, T.~Luo, L.-G. Pang, and X.-N. Wang, {Interplaying
  mechanisms behind single inclusive jet suppression in heavy-ion collisions},
  \emph{Phys. Rev.} {\bf C99}\penalty0 (5), \penalty0 054911  (2019).
\newblock \doi{10.1103/PhysRevC.99.054911}.

\bibitem{He:2022evt}
Y.~He, W.~Chen, T.~Luo, S.~Cao, L.-G. Pang, and X.-N. Wang, {Event-by-event jet
  anisotropy and hard-soft tomography of the quark-gluon plasma}, \emph{Phys.
  Rev. C}. {\bf 106}\penalty0 (4), \penalty0 044904  (2022).
\newblock \doi{10.1103/PhysRevC.106.044904}.

\bibitem{Cao:2017hhk}
S.~Cao, T.~Luo, G.-Y. Qin, and X.-N. Wang, {Heavy and light flavor jet
  quenching at RHIC and LHC energies}, \emph{Phys. Lett. B}. {\bf 777},
  \penalty0 255--259  (2018).
\newblock \doi{10.1016/j.physletb.2017.12.023}.

\bibitem{Xing:2019xae}
W.-J. Xing, S.~Cao, G.-Y. Qin, and H.~Xing, {Flavor hierarchy of jet quenching
  in relativistic heavy-ion collisions}, \emph{Phys. Lett. B}. {\bf 805},
  \penalty0 135424  (2020).
\newblock \doi{10.1016/j.physletb.2020.135424}.

\bibitem{Xing:2021xwc}
W.-J. Xing, G.-Y. Qin, and S.~Cao, {Perturbative and non-perturbative
  interactions between heavy quarks and quark-gluon plasma within a unified
  approach}, \emph{Phys. Lett. B}. {\bf 838}, \penalty0 137733  (2023).
\newblock \doi{10.1016/j.physletb.2023.137733}.

\bibitem{Liu:2021dpm}
F.-L. Liu, W.-J. Xing, X.-Y. Wu, G.-Y. Qin, S.~Cao, and X.-N. Wang, {QLBT: a
  linear Boltzmann transport model for heavy quarks in a quark-gluon plasma of
  quasi-particles}, \emph{Eur. Phys. J. C}. {\bf 82}\penalty0 (4), \penalty0
  350  (2022).
\newblock \doi{10.1140/epjc/s10052-022-10308-x}.

\bibitem{Liu:2023rfi}
F.-L. Liu, X.-Y. Wu, S.~Cao, G.-Y. Qin, and X.-N. Wang, {Constraining the
  equation of state with heavy quarks in the quasi-particle model of QCD
  matter}, \emph{Phys. Lett. B}. {\bf 848}, \penalty0 138355  (2024).
\newblock \doi{10.1016/j.physletb.2023.138355}.

\bibitem{JETSCAPE:2019udz}
A.~Kumar et~al., {JETSCAPE framework: $p+p$ results}, \emph{Phys. Rev. C}. {\bf
  102}\penalty0 (5), \penalty0 054906  (2020).
\newblock \doi{10.1103/PhysRevC.102.054906}.

\bibitem{Park:2016jap}
C.~Park, C.~Shen, S.~Jeon, and C.~Gale, {Rapidity-dependent jet energy loss in
  small systems with finite-size effects and running coupling}, \emph{Nucl.
  Part. Phys. Proc.} {\bf 289-290}, \penalty0 289--292  (2017).
\newblock \doi{10.1016/j.nuclphysbps.2017.05.066}.

\bibitem{Park:2018acg}
C.~Park, S.~Jeon, and C.~Gale, {Jet modification with medium recoil in
  quark-gluon plasma}, \emph{Nucl. Phys. A}. {\bf 982}, \penalty0 643--646
  (2019).
\newblock \doi{10.1016/j.nuclphysa.2018.10.057}.

\bibitem{Park:2021yck}
C.~Park.
\newblock \emph{{Jet modification in strongly-coupled quark-gluon plasma}}.
\newblock PhD thesis, McGill U.  (2021).

\bibitem{Buzzatti:2011vt}
A.~Buzzatti and M.~Gyulassy, {Jet Flavor Tomography of Quark Gluon Plasmas at
  RHIC and LHC}, \emph{Phys. Rev. Lett.} {\bf 108}, \penalty0 022301  (2012).
\newblock \doi{10.1103/PhysRevLett.108.022301}.

\bibitem{Buzzatti:2013scw}
A.~Buzzatti.
\newblock \emph{{Jet quenching in Quark Gluon Plasma: flavor tomography at RHIC
  and LHC by the CUJET model}}.
\newblock PhD thesis, Columbia U. (main)  (2013).

\bibitem{Thoma:1990fm}
M.~H. Thoma and M.~Gyulassy, {Quark Damping and Energy Loss in the High
  Temperature {QCD}}, \emph{Nucl. Phys.} {\bf B351}, \penalty0 491--506
  (1991).
\newblock \doi{10.1016/S0550-3213(05)80031-8}.

\bibitem{Wicks:2005gt}
S.~Wicks, W.~Horowitz, M.~Djordjevic, and M.~Gyulassy, {Elastic, Inelastic, and
  Path Length Fluctuations in Jet Tomography}, \emph{Nucl. Phys.} {\bf A784},
  \penalty0 426--442  (2007).
\newblock \doi{10.1016/j.nuclphysa.2006.12.048}.

\bibitem{Qin:2009uh}
G.~Y. Qin, A.~Majumder, H.~Song, and U.~Heinz, {Energy and momentum deposited
  into a QCD medium by a jet shower}, \emph{Phys. Rev. Lett.} {\bf 103},
  \penalty0 152303  (2009).
\newblock \doi{10.1103/PhysRevLett.103.152303}.

\bibitem{Neufeld:2009ep}
R.~B. Neufeld and B.~Muller, {The sound produced by a fast parton in the
  quark-gluon plasma is a 'crescendo'}, \emph{Phys. Rev. Lett.} {\bf 103},
  \penalty0 042301  (2009).
\newblock \doi{10.1103/PhysRevLett.103.042301}.

\bibitem{Casalderrey-Solana:2016jvj}
J.~Casalderrey-Solana, D.~Gulhan, G.~Milhano, D.~Pablos, and K.~Rajagopal,
  {Angular Structure of Jet Quenching Within a Hybrid Strong/Weak Coupling
  Model}, \emph{JHEP}. {\bf 03}, \penalty0 135  (2017).
\newblock \doi{10.1007/JHEP03(2017)135}.

\bibitem{Tachibana:2017syd}
Y.~Tachibana, N.-B. Chang, and G.-Y. Qin, {Full jet in quark-gluon plasma with
  hydrodynamic medium response}, \emph{Phys. Rev.} {\bf C95}\penalty0 (4),
  \penalty0 044909  (2017).
\newblock \doi{10.1103/PhysRevC.95.044909}.

\bibitem{KunnawalkamElayavalli:2017hxo}
R.~Kunnawalkam~Elayavalli and K.~C. Zapp, {Medium response in JEWEL and its
  impact on jet shape observables in heavy ion collisions}, \emph{JHEP}. {\bf
  07}, \penalty0 141  (2017).
\newblock \doi{10.1007/JHEP07(2017)141}.

\bibitem{Chen:2017zte}
W.~Chen, S.~Cao, T.~Luo, L.-G. Pang, and X.-N. Wang, {Effects of jet-induced
  medium excitation in $\gamma$-hadron correlation in A+A collisions},
  \emph{Phys. Lett.} {\bf B777}, \penalty0 86--90  (2018).
\newblock \doi{10.1016/j.physletb.2017.12.015}.

\bibitem{Chang:2019sae}
N.-B. Chang, Y.~Tachibana, and G.-Y. Qin, {Nuclear modification of jet shape
  for inclusive jets and $\gamma$-jets at the LHC energies}, \emph{Phys. Lett.}
  {\bf B801}, \penalty0 135181  (2020).
\newblock \doi{10.1016/j.physletb.2019.135181}.

\bibitem{Cao:2022odi}
S.~Cao and G.-Y. Qin, {Medium Response and Jet-Hadron Correlations in
  Relativistic Heavy-Ion Collisions}, \emph{Ann. Rev. Nucl. Part. Sci.} {\bf
  73}, \penalty0 205--229  (2023).
\newblock \doi{10.1146/annurev-nucl-112822-031317}.

\bibitem{Schulc:2013kra}
M.~Schulc and B.~Tom\'a\v{s}ik, {Stimulation of static deconfined medium by
  multiple hard partons}, \emph{J. Phys. G}. {\bf 40}, \penalty0 125104
  (2013).
\newblock \doi{10.1088/0954-3899/40/12/125104}.

\bibitem{Floerchinger:2014yqa}
S.~Floerchinger and K.~C. Zapp, {Hydrodynamics and Jets in Dialogue},
  \emph{Eur. Phys. J. C}. {\bf 74}\penalty0 (12), \penalty0 3189  (2014).
\newblock \doi{10.1140/epjc/s10052-014-3189-4}.

\bibitem{Schulc:2014jma}
M.~Schulc and B.~Tom\'a\v{s}ik, {Anisotropic flow of the fireball fed by hard
  partons}, \emph{Phys. Rev. C}. {\bf 90}\penalty0 (6), \penalty0 064910
  (2014).
\newblock \doi{10.1103/PhysRevC.90.064910}.

\bibitem{Okai:2017ofp}
M.~Okai, K.~Kawaguchi, Y.~Tachibana, and T.~Hirano, {New approach to
  initializing hydrodynamic fields and mini-jet propagation in quark-gluon
  fluids}, \emph{Phys. Rev. C}. {\bf 95}\penalty0 (5), \penalty0 054914
  (2017).
\newblock \doi{10.1103/PhysRevC.95.054914}.

\bibitem{Pablos:2022piv}
D.~Pablos, M.~Singh, S.~Jeon, and C.~Gale, {Minijet quenching in a concurrent
  jet+hydro evolution and the nonequilibrium quark-gluon plasma}, \emph{Phys.
  Rev. C}. {\bf 106}\penalty0 (3), \penalty0 034901  (2022).
\newblock \doi{10.1103/PhysRevC.106.034901}.

\bibitem{Zapp:2012ak}
K.~C. Zapp, F.~Krauss, and U.~A. Wiedemann, {A perturbative framework for jet
  quenching}, \emph{JHEP}. {\bf 1303}, \penalty0 080  (2013).
\newblock \doi{10.1007/JHEP03(2013)080}.

\bibitem{Zapp:2013vla}
K.~C. Zapp, {JEWEL 2.0.0: directions for use}, \emph{Eur. Phys. J. C}. {\bf
  74}\penalty0 (2), \penalty0 2762  (2014).
\newblock \doi{10.1140/epjc/s10052-014-2762-1}.

\bibitem{Wang:2013cia}
X.-N. Wang and Y.~Zhu, {Medium Modification of $\gamma$-jets in High-energy
  Heavy-ion Collisions}  (2013).

\bibitem{JETSCAPE:2022jer}
A.~Kumar et~al., {Inclusive jet and hadron suppression in a multistage
  approach}, \emph{Phys. Rev. C}. {\bf 107}\penalty0 (3), \penalty0 034911
  (2023).
\newblock \doi{10.1103/PhysRevC.107.034911}.

\bibitem{JETSCAPE:2023hqn}
Y.~Tachibana et~al., {Hard Jet Substructure in a Multi-stage Approach}  (1,
  2023).

\bibitem{Tachibana:2020mtb}
Y.~Tachibana, C.~Shen, and A.~Majumder, {Bulk medium evolution has considerable
  effects on jet observables}, \emph{Phys. Rev. C}. {\bf 106}\penalty0 (2),
  \penalty0 L021902  (2022).
\newblock \doi{10.1103/PhysRevC.106.L021902}.

\bibitem{Stoecker:2004qu}
H.~Stoecker, {Collective flow signals the quark gluon plasma}, \emph{Nucl.
  Phys. A}. {\bf 750}, \penalty0 121--147  (2005).
\newblock \doi{10.1016/j.nuclphysa.2004.12.074}.

\bibitem{Casalderrey-Solana:2004fdk}
J.~Casalderrey-Solana, E.~V. Shuryak, and D.~Teaney, {Conical flow induced by
  quenched QCD jets}, \emph{J. Phys. Conf. Ser.} {\bf 27}, \penalty0 22--31
  (2005).
\newblock \doi{10.1088/1742-6596/27/1/003}.

\bibitem{Chaudhuri:2005vc}
A.~K. Chaudhuri and U.~Heinz, {Effect of jet quenching on the hydrodynamical
  evolution of QGP}, \emph{Phys. Rev. Lett.} {\bf 97}, \penalty0 062301
  (2006).
\newblock \doi{10.1103/PhysRevLett.97.062301}.

\bibitem{Betz:2008ka}
B.~Betz, J.~Noronha, G.~Torrieri, M.~Gyulassy, I.~Mishustin, and D.~H. Rischke,
  {Universality of the Diffusion Wake from Stopped and Punch-Through Jets in
  Heavy-Ion Collisions}, \emph{Phys. Rev. C}. {\bf 79}, \penalty0 034902
  (2009).
\newblock \doi{10.1103/PhysRevC.79.034902}.

\bibitem{Tachibana:2012sa}
Y.~Tachibana and T.~Hirano, {Emission of Low Momentum Particles at Large Angles
  from Jet}, \emph{Nucl. Phys. A}. {\bf 904-905}, \penalty0 1023c--1026c
  (2013).
\newblock \doi{10.1016/j.nuclphysa.2013.02.189}.

\bibitem{Tachibana:2014yai}
Y.~Tachibana.
\newblock \emph{{Hydrodynamic response to jet propagation in quark-gluon
  plasma}}.
\newblock PhD thesis, Tokyo U.  (2014).

\bibitem{Casalderrey-Solana:2006lmc}
J.~Casalderrey-Solana, E.~V. Shuryak, and D.~Teaney, {Hydrodynamic flow from
  fast particles}  (2, 2006).

\bibitem{Gubser:2007ga}
S.~S. Gubser, S.~S. Pufu, and A.~Yarom, {Sonic booms and diffusion wakes
  generated by a heavy quark in thermal AdS/CFT}, \emph{Phys. Rev. Lett.} {\bf
  100}, \penalty0 012301  (2008).
\newblock \doi{10.1103/PhysRevLett.100.012301}.

\bibitem{Casalderrey-Solana:2020rsj}
J.~Casalderrey-Solana, J.~G. Milhano, D.~Pablos, K.~Rajagopal, and X.~Yao, {Jet
  Wake from Linearized Hydrodynamics}, \emph{JHEP}. {\bf 05}, \penalty0 230
  (2021).
\newblock \doi{10.1007/JHEP05(2021)230}.

\bibitem{Aziz:2004qu}
M.~A. Aziz and S.~Gavin, {Causal diffusion and the survival of charge
  fluctuations in nuclear collisions}, \emph{Phys. Rev.} {\bf C70}, \penalty0
  034905  (2004).
\newblock \doi{10.1103/PhysRevC.70.034905}.

\bibitem{JETSCAPE:2020uew}
Y.~Tachibana et~al., {Hydrodynamic response to jets with a source based on
  causal diffusion}, \emph{Nucl. Phys. A}. {\bf 1005}, \penalty0 121920
  (2021).
\newblock \doi{10.1016/j.nuclphysa.2020.121920}.

\bibitem{Klasen:2017dsy}
M.~Klasen, C.~Klein-B\"osing, and H.~Poppenborg, {Prompt photon production and
  photon-jet correlations at the LHC}, \emph{JHEP}. {\bf 03}, \penalty0 081
  (2018).
\newblock \doi{10.1007/JHEP03(2018)081}.

\bibitem{andersson_1998}
B.~Andersson, \emph{The Lund Model}. Cambridge Monographs on Particle Physics,
  Nuclear Physics and Cosmology, Cambridge University Press  (1998).
\newblock \doi{10.1017/CBO9780511524363}.

\bibitem{Ferreres-Sole:2018vgo}
S.~Ferreres-Sol\'e and T.~Sj\"ostrand, {The space\textendash{}time structure of
  hadronization in the Lund model}, \emph{Eur. Phys. J. C}. {\bf 78}\penalty0
  (11), \penalty0 983  (2018).
\newblock \doi{10.1140/epjc/s10052-018-6459-8}.

\bibitem{Bali:1994de}
G.~S. Bali, K.~Schilling, and C.~Schlichter, {Observing long color flux tubes
  in SU(2) lattice gauge theory}, \emph{Phys. Rev. D}. {\bf 51}, \penalty0
  5165--5198  (1995).
\newblock \doi{10.1103/PhysRevD.51.5165}.

\bibitem{Webber:1983if}
B.~R. Webber, {A QCD Model for Jet Fragmentation Including Soft Gluon
  Interference}, \emph{Nucl. Phys. B}. {\bf 238}, \penalty0 492--528  (1984).
\newblock \doi{10.1016/0550-3213(84)90333-X}.

\bibitem{Bellm:2015jjp}
J.~Bellm et~al., {Herwig 7.0/Herwig++ 3.0 release note}, \emph{Eur. Phys. J.
  C}. {\bf 76}\penalty0 (4), \penalty0 196  (2016).
\newblock \doi{10.1140/epjc/s10052-016-4018-8}.

\bibitem{Han:2016uhh}
K.~C. Han, R.~J. Fries, and C.~M. Ko, {Jet Fragmentation via Recombination of
  Parton Showers}, \emph{Phys. Rev. C}. {\bf 93}\penalty0 (4), \penalty0 045207
   (2016).
\newblock \doi{10.1103/PhysRevC.93.045207}.

\bibitem{Kordell:2021prk}
M.~Kordell, II, R.~J. Fries, and C.~M. Ko, {Angular momentum eigenstates of the
  isotropic 3-D harmonic oscillator: Phase-space distributions and coalescence
  probabilities}, \emph{Annals Phys.} {\bf 443}, \penalty0 168960  (2022).
\newblock \doi{10.1016/j.aop.2022.168960}.

\bibitem{Fries:2003kq}
R.~J. Fries, B.~Muller, C.~Nonaka, and S.~A. Bass, Hadron production in heavy
  ion collisions: Fragmentation and recombination from a dense parton phase,
  \emph{Phys. Rev.} {\bf C68}, \penalty0 044902  (2003).

\bibitem{Kolb:2004gi}
P.~F. Kolb, L.-W. Chen, V.~Greco, and C.~M. Ko, Momentum anisotropies in the
  quark coalescence model, \emph{Phys. Rev.} {\bf C69}, \penalty0 051901
  (2004).

\bibitem{Wang:1991hta}
X.-N. Wang and M.~Gyulassy, {HIJING: A Monte Carlo model for multiple jet
  production in p p, p A and A A collisions}, \emph{Phys. Rev. D}. {\bf 44},
  \penalty0 3501--3516  (1991).
\newblock \doi{10.1103/PhysRevD.44.3501}.

\bibitem{Gyulassy:1994ew}
M.~Gyulassy and X.-N. Wang, {HIJING 1.0: A Monte Carlo program for parton and
  particle production in high-energy hadronic and nuclear collisions},
  \emph{Comput. Phys. Commun.} {\bf 83}, \penalty0 307  (1994).
\newblock \doi{10.1016/0010-4655(94)90057-4}.

\bibitem{Zapp:2008gi}
K.~Zapp, G.~Ingelman, J.~Rathsman, J.~Stachel, and U.~A. Wiedemann, {A Monte
  Carlo Model for 'Jet Quenching'}, \emph{Eur. Phys. J.} {\bf C60}, \penalty0
  617--632  (2009).
\newblock \doi{10.1140/epjc/s10052-009-0941-2}.

\bibitem{Pierog:2013ria}
T.~Pierog, I.~Karpenko, J.~M. Katzy, E.~Yatsenko, and K.~Werner, {EPOS LHC:
  Test of collective hadronization with data measured at the CERN Large Hadron
  Collider}, \emph{Phys. Rev. C}. {\bf 92}\penalty0 (3), \penalty0 034906
  (2015).
\newblock \doi{10.1103/PhysRevC.92.034906}.

\bibitem{Pierog:2009zt}
T.~Pierog and K.~Werner, {EPOS Model and Ultra High Energy Cosmic Rays},
  \emph{Nucl. Phys. B Proc. Suppl.} {\bf 196}, \penalty0 102--105  (2009).
\newblock \doi{10.1016/j.nuclphysbps.2009.09.017}.

\bibitem{Porteboeuf:2010um}
S.~Porteboeuf, T.~Pierog, and K.~Werner.
\newblock {Producing Hard Processes Regarding the Complete Event: The EPOS
  Event Generator}.
\newblock In \emph{{45th Rencontres de Moriond on QCD and High Energy
  Interactions}}, pp. 135--140, Gioi Publishers  (2010).

\bibitem{Casalderrey-Solana:2014bpa}
J.~Casalderrey-Solana, D.~C. Gulhan, J.~G. Milhano, D.~Pablos, and
  K.~Rajagopal, {A Hybrid Strong/Weak Coupling Approach to Jet Quenching},
  \emph{JHEP}. {\bf 10}, \penalty0 019  (2014).
\newblock \doi{10.1007/JHEP09(2015)175}.
\newblock [Erratum: JHEP 09, 175 (2015)].

\bibitem{Casalderrey-Solana:2015vaa}
J.~Casalderrey-Solana, D.~C. Gulhan, J.~G. Milhano, D.~Pablos, and
  K.~Rajagopal, {Predictions for Boson-Jet Observables and Fragmentation
  Function Ratios from a Hybrid Strong/Weak Coupling Model for Jet Quenching},
  \emph{JHEP}. {\bf 03}, \penalty0 053  (2016).
\newblock \doi{10.1007/JHEP03(2016)053}.

\bibitem{Park:2019sdn}
C.~Park, {Multi-stage jet evolution through QGP using the JETSCAPE framework:
  inclusive jets, correlations and leading hadrons}, \emph{PoS}. {\bf
  HardProbes2018}, \penalty0 072  (2019).
\newblock \doi{10.22323/1.345.0072}.

\bibitem{Chen:2020tbl}
W.~Chen, S.~Cao, T.~Luo, L.-G. Pang, and X.-N. Wang, {Medium modification of
  $\gamma$-jet fragmentation functions in Pb+Pb collisions at LHC}, \emph{Phys.
  Lett. B}. {\bf 810}, \penalty0 135783  (2020).
\newblock \doi{10.1016/j.physletb.2020.135783}.

\bibitem{JETSCAPE:2017eso}
S.~Cao et~al., {Multistage Monte-Carlo simulation of jet modification in a
  static medium}, \emph{Phys. Rev. C}. {\bf 96}\penalty0 (2), \penalty0 024909
  (2017).
\newblock \doi{10.1103/PhysRevC.96.024909}.

\bibitem{Bernhard:2019bmu}
J.~E. Bernhard, J.~S. Moreland, and S.~A. Bass, {Bayesian estimation of the
  specific shear and bulk viscosity of quark\textendash{}gluon plasma},
  \emph{Nature Phys.} {\bf 15}\penalty0 (11), \penalty0 1113--1117  (2019).
\newblock \doi{10.1038/s41567-019-0611-8}.

\bibitem{ATLAS:2015qmb}
G.~Aad et~al., {Measurement of charged-particle spectra in Pb+Pb collisions at
  $\sqrt{{s}_\mathsf{{NN}}} = 2.76$ TeV with the ATLAS detector at the LHC},
  \emph{JHEP}. {\bf 09}, \penalty0 050  (2015).
\newblock \doi{10.1007/JHEP09(2015)050}.

\bibitem{ALICE:2012aqc}
B.~Abelev et~al., {Centrality Dependence of Charged Particle Production at
  Large Transverse Momentum in Pb--Pb Collisions at $\sqrt{s_{\rm{NN}}} = 2.76$
  TeV}, \emph{Phys. Lett. B}. {\bf 720}, \penalty0 52--62  (2013).
\newblock \doi{10.1016/j.physletb.2013.01.051}.

\bibitem{CMS:2012aa}
S.~Chatrchyan et~al., {Study of High-pT Charged Particle Suppression in PbPb
  Compared to $pp$ Collisions at $\sqrt{s_{NN}}=2.76$ TeV}, \emph{Eur. Phys. J.
  C}. {\bf 72}, \penalty0 1945  (2012).
\newblock \doi{10.1140/epjc/s10052-012-1945-x}.

\bibitem{ATLAS:2014ipv}
G.~Aad et~al., {Measurements of the Nuclear Modification Factor for Jets in
  Pb+Pb Collisions at $\sqrt{s_{\mathrm{NN}}}=2.76$ TeV with the ATLAS
  Detector}, \emph{Phys. Rev. Lett.} {\bf 114}\penalty0 (7), \penalty0 072302
  (2015).
\newblock \doi{10.1103/PhysRevLett.114.072302}.

\bibitem{CMS:2016uxf}
V.~Khachatryan et~al., {Measurement of inclusive jet cross sections in $pp$ and
  PbPb collisions at $\sqrt{s_{NN}}=$ 2.76 TeV}, \emph{Phys. Rev. C}. {\bf
  96}\penalty0 (1), \penalty0 015202  (2017).
\newblock \doi{10.1103/PhysRevC.96.015202}.

\bibitem{Cacciari:2005hq}
M.~Cacciari and G.~P. Salam, {Dispelling the $N^{3}$ myth for the $k_t$
  jet-finder}, \emph{Phys. Lett. B}. {\bf 641}, \penalty0 57--61  (2006).
\newblock \doi{10.1016/j.physletb.2006.08.037}.

\bibitem{Cacciari:2011ma}
M.~Cacciari, G.~P. Salam, and G.~Soyez, {FastJet User Manual}, \emph{Eur. Phys.
  J. C}. {\bf 72}, \penalty0 1896  (2012).
\newblock \doi{10.1140/epjc/s10052-012-1896-2}.

\bibitem{CMS:2013lhm}
S.~Chatrchyan et~al., {Modification of Jet Shapes in PbPb Collisions at $\sqrt
  {s_{NN}} = 2.76$ TeV}, \emph{Phys. Lett. B}. {\bf 730}, \penalty0 243--263
  (2014).
\newblock \doi{10.1016/j.physletb.2014.01.042}.

\bibitem{ATLAS:2018gwx}
M.~Aaboud et~al., {Measurement of the nuclear modification factor for inclusive
  jets in Pb+Pb collisions at $\sqrt{s_\mathrm{NN}}=5.02$ TeV with the ATLAS
  detector}, \emph{Phys. Lett. B}. {\bf 790}, \penalty0 108--128  (2019).
\newblock \doi{10.1016/j.physletb.2018.10.076}.

\bibitem{CMS:2021vui}
A.~M. Sirunyan et~al., {First measurement of large area jet transverse momentum
  spectra in heavy-ion collisions}, \emph{JHEP}. {\bf 05}, \penalty0 284
  (2021).
\newblock \doi{10.1007/JHEP05(2021)284}.

\bibitem{CMS:2016xef}
V.~Khachatryan et~al., {Charged-particle nuclear modification factors in PbPb
  and pPb collisions at $ \sqrt{s_{\mathrm{N}\;\mathrm{N}}}=5.02 $ TeV},
  \emph{JHEP}. {\bf 04}, \penalty0 039  (2017).
\newblock \doi{10.1007/JHEP04(2017)039}.

\bibitem{CMS:2017qjw}
A.~M. Sirunyan et~al., {Nuclear modification factor of D$^0$ mesons in PbPb
  collisions at $\sqrt{s_\mathrm{NN}} = 5.02$ TeV}, \emph{Phys. Lett. B}. {\bf
  782}, \penalty0 474--496  (2018).
\newblock \doi{10.1016/j.physletb.2018.05.074}.

\bibitem{ALICE:2018lyv}
S.~Acharya et~al., {Measurement of D$^{0}$, D$^{+}$, D$^{*+}$ and D$_{s}^{+}$
  production in Pb-Pb collisions at $ \sqrt{{\mathrm{s}}_{\mathrm{NN}}}=5.02 $
  TeV}, \emph{JHEP}. {\bf 10}, \penalty0 174  (2018).
\newblock \doi{10.1007/JHEP10(2018)174}.

\bibitem{JETSCAPE:2022hcb}
W.~Fan et~al., {Multiscale evolution of charmed particles in a nuclear medium},
  \emph{Phys. Rev. C}. {\bf 107}\penalty0 (5), \penalty0 054901  (2023).
\newblock \doi{10.1103/PhysRevC.107.054901}.

\bibitem{STAR:2020xiv}
J.~Adam et~al., {Measurement of inclusive charged-particle jet production in Au
  + Au collisions at $\sqrt{s_{NN}}=$200 GeV}, \emph{Phys. Rev. C}. {\bf
  102}\penalty0 (5), \penalty0 054913  (2020).
\newblock \doi{10.1103/PhysRevC.102.054913}.

\bibitem{PHENIX:2012jha}
A.~Adare et~al., {Neutral pion production with respect to centrality and
  reaction plane in Au$+$Au collisions at $\sqrt{s_{NN}}$=200 GeV}, \emph{Phys.
  Rev. C}. {\bf 87}\penalty0 (3), \penalty0 034911  (2013).
\newblock \doi{10.1103/PhysRevC.87.034911}.

\bibitem{Larkoski:2014wba}
A.~J. Larkoski, S.~Marzani, G.~Soyez, and J.~Thaler, {Soft Drop}, \emph{JHEP}.
  {\bf 05}, \penalty0 146  (2014).
\newblock \doi{10.1007/JHEP05(2014)146}.

\bibitem{ALargeIonColliderExperiment:2021mqf}
S.~Acharya et~al., {Measurement of the groomed jet radius and momentum
  splitting fraction in pp and Pb$-$Pb collisions at $\sqrt{s_{NN}} = 5.02$
  TeV}, \emph{Phys. Rev. Lett.} {\bf 128}\penalty0 (10), \penalty0 102001
  (2022).
\newblock \doi{10.1103/PhysRevLett.128.102001}.

\bibitem{ATLAS:2022vii}
G.~Aad et~al., {Measurement of substructure-dependent jet suppression in Pb+Pb
  collisions at 5.02 TeV with the ATLAS detector}, \emph{Phys. Rev. C}. {\bf
  107}\penalty0 (5), \penalty0 054909  (2023).
\newblock \doi{10.1103/PhysRevC.107.054909}.

\bibitem{Novak:2013bqa}
J.~Novak, K.~Novak, S.~Pratt, J.~Vredevoogd, C.~Coleman-Smith, and R.~Wolpert,
  {Determining Fundamental Properties of Matter Created in Ultrarelativistic
  Heavy-Ion Collisions}, \emph{Phys. Rev. C}. {\bf 89}\penalty0 (3), \penalty0
  034917  (2014).
\newblock \doi{10.1103/PhysRevC.89.034917}.

\bibitem{JETSCAPE:2020shq}
D.~Everett et~al., {Phenomenological constraints on the transport properties of
  QCD matter with data-driven model averaging}, \emph{Phys. Rev. Lett.} {\bf
  126}\penalty0 (24), \penalty0 242301  (2021).
\newblock \doi{10.1103/PhysRevLett.126.242301}.

\bibitem{Nijs:2020ors}
G.~Nijs, W.~van~der Schee, U.~G\"ursoy, and R.~Snellings, {Transverse Momentum
  Differential Global Analysis of Heavy-Ion Collisions}, \emph{Phys. Rev.
  Lett.} {\bf 126}\penalty0 (20), \penalty0 202301  (2021).
\newblock \doi{10.1103/PhysRevLett.126.202301}.

\bibitem{Pratt:2015zsa}
S.~Pratt, E.~Sangaline, P.~Sorensen, and H.~Wang, {Constraining the Eq. of
  State of Super-Hadronic Matter from Heavy-Ion Collisions}, \emph{Phys. Rev.
  Lett.} {\bf 114}, \penalty0 202301  (2015).
\newblock \doi{10.1103/PhysRevLett.114.202301}.

\bibitem{GPE1}
J.~Sacks, W.~J. Welch, T.~J. Mitchell, and H.~P. Wynn, {Design and Analysis of
  Computer Experiments}, \emph{Statistical Science}. {\bf 4}\penalty0 (4),
  \penalty0 409 -- 423  (1989).
\newblock \doi{10.1214/ss/1177012413}.
\newblock URL \url{https://doi.org/10.1214/ss/1177012413}.

\bibitem{GPE2}
C.~E. Rasmussen and C.~K.~I. Williams, \emph{Gaussian process for machine
  learning}. The MIT Press  (2006).

\bibitem{JETSCAPE:2021ehl}
S.~Cao et~al., {Determining the jet transport coefficient qhat from inclusive
  hadron suppression measurements using Bayesian parameter estimation},
  \emph{Phys. Rev. C}. {\bf 104}\penalty0 (2), \penalty0 024905  (2021).
\newblock \doi{10.1103/PhysRevC.104.024905}.

\bibitem{Xie:2022ght}
M.~Xie, W.~Ke, H.~Zhang, and X.-N. Wang, {Information-field-based global
  Bayesian inference of the jet transport coefficient}, \emph{Phys. Rev. C}.
  {\bf 108}\penalty0 (1), \penalty0 L011901  (2023).
\newblock \doi{10.1103/PhysRevC.108.L011901}.

\bibitem{JET:2013cls}
K.~M. Burke et~al., {Extracting the jet transport coefficient from jet
  quenching in high-energy heavy-ion collisions}, \emph{Phys. Rev. C}. {\bf
  90}\penalty0 (1), \penalty0 014909  (2014).
\newblock \doi{10.1103/PhysRevC.90.014909}.

\bibitem{Ke:2020clc}
W.~Ke and X.-N. Wang, {QGP modification to single inclusive jets in a
  calibrated transport model}, \emph{JHEP}. {\bf 05}, \penalty0 041  (2021).
\newblock \doi{10.1007/JHEP05(2021)041}.

\bibitem{He:2018gks}
Y.~He, L.-G. Pang, and X.-N. Wang, {Bayesian extraction of jet energy loss
  distributions in heavy-ion collisions}, \emph{Phys. Rev. Lett.} {\bf
  122}\penalty0 (25), \penalty0 252302  (2019).
\newblock \doi{10.1103/PhysRevLett.122.252302}.

\bibitem{Xing:2023ciw}
W.-J. Xing, S.~Cao, and G.-Y. Qin, {Flavor hierarchy of parton energy loss in
  quark-gluon plasma from a Bayesian analysis}  (3, 2023).

\bibitem{Zhang:2023oid}
S.-L. Zhang, E.~Wang, H.~Xing, and B.-W. Zhang, {Flavor dependence of jet
  quenching in heavy-ion collisions}  (3, 2023).

\bibitem{Borsanyi:2013bia}
S.~Borsanyi, Z.~Fodor, C.~Hoelbling, S.~D. Katz, S.~Krieg, and K.~K. Szabo,
  {Full result for the QCD equation of state with 2+1 flavors}, \emph{Phys.
  Lett.} {\bf B730}, \penalty0 99--104  (2014).
\newblock \doi{10.1016/j.physletb.2014.01.007}.

\bibitem{HotQCD:2014kol}
A.~Bazavov et~al., {Equation of state in ( 2+1 )-flavor QCD}, \emph{Phys. Rev.
  D}. {\bf 90}, \penalty0 094503  (2014).
\newblock \doi{10.1103/PhysRevD.90.094503}.

\bibitem{Xu:2017obm}
Y.~Xu, J.~E. Bernhard, S.~A. Bass, M.~Nahrgang, and S.~Cao, {Data-driven
  analysis for the temperature and momentum dependence of the heavy-quark
  diffusion coefficient in relativistic heavy-ion collisions}, \emph{Phys. Rev.
  C}. {\bf 97}\penalty0 (1), \penalty0 014907  (2018).
\newblock \doi{10.1103/PhysRevC.97.014907}.

\bibitem{ATLAS:2014cpa}
G.~Aad et~al., {Centrality and rapidity dependence of inclusive jet production
  in $\sqrt{s_\mathrm{NN}} = 5.02$ TeV proton-lead collisions with the ATLAS
  detector}, \emph{Phys. Lett. B}. {\bf 748}, \penalty0 392--413  (2015).
\newblock \doi{10.1016/j.physletb.2015.07.023}.

\bibitem{ALICE:2015efi}
J.~Adam et~al., {Azimuthal anisotropy of charged jet production in
  $\sqrt{s_{\rm NN}}$ = 2.76 TeV Pb-Pb collisions}, \emph{Phys. Lett. B}. {\bf
  753}, \penalty0 511--525  (2016).
\newblock \doi{10.1016/j.physletb.2015.12.047}.

\bibitem{ATLAS:2019vcm}
G.~Aad et~al., {Transverse momentum and process dependent azimuthal
  anisotropies in $\sqrt{s_{\mathrm{NN}}}=8.16$ TeV $p$+Pb collisions with the
  ATLAS detector}, \emph{Eur. Phys. J. C}. {\bf 80}\penalty0 (1), \penalty0 73
  (2020).
\newblock \doi{10.1140/epjc/s10052-020-7624-4}.

\bibitem{ATLAS:2022kqu}
G.~Aad et~al., {Charged-hadron production in $pp$, $p$+Pb, Pb+Pb, and Xe+Xe
  collisions at $\sqrt{s_{_\text{NN}}}=5$ TeV with the ATLAS detector at the
  LHC}, \emph{JHEP}. {\bf 07}, \penalty0 074  (2023).
\newblock \doi{10.1007/JHEP07(2023)074}.

\bibitem{ALICE:2018vuu}
S.~Acharya et~al., {Transverse momentum spectra and nuclear modification
  factors of charged particles in pp, p-Pb and Pb-Pb collisions at the LHC},
  \emph{JHEP}. {\bf 11}, \penalty0 013  (2018).
\newblock \doi{10.1007/JHEP11(2018)013}.

\bibitem{Kordell:2016njg}
M.~Kordell and A.~Majumder, {Jets in d(p)-A Collisions: Color Transparency or
  Energy Conservation}, \emph{Phys. Rev.} {\bf C97}\penalty0 (5), \penalty0
  054904  (2018).
\newblock \doi{10.1103/PhysRevC.97.054904}.

\bibitem{JETSCAPE:2023xbc}
A.~Majumder et~al.
\newblock {A multistage framework for studying the evolution of jets and
  high-$p_T$ probes in small collision systems}.
\newblock In \emph{{11th International Conference on Hard and Electromagnetic
  Probes of High-Energy Nuclear Collisions}: {Hard Probes 2023}}  (8, 2023).

\bibitem{Shen:2017bsr}
C.~Shen and B.~Schenke, {Dynamical initial state model for relativistic
  heavy-ion collisions}, \emph{Phys. Rev. C}. {\bf 97}\penalty0 (2), \penalty0
  024907  (2018).
\newblock \doi{10.1103/PhysRevC.97.024907}.

\bibitem{Shen:2022oyg}
C.~Shen and B.~Schenke, {Longitudinal dynamics and particle production in
  relativistic nuclear collisions}, \emph{Phys. Rev. C}. {\bf 105}\penalty0
  (6), \penalty0 064905  (2022).
\newblock \doi{10.1103/PhysRevC.105.064905}.

\bibitem{Zhao:2022ayk}
W.~Zhao, C.~Shen, and B.~Schenke, {Collectivity in Ultraperipheral Pb+Pb
  Collisions at the Large Hadron Collider}, \emph{Phys. Rev. Lett.} {\bf
  129}\penalty0 (25), \penalty0 252302  (2022).
\newblock \doi{10.1103/PhysRevLett.129.252302}.

\bibitem{Schenke:2010nt}
B.~Schenke, S.~Jeon, and C.~Gale, {(3+1)D hydrodynamic simulation of
  relativistic heavy-ion collisions}, \emph{Phys. Rev. C}. {\bf 82}, \penalty0
  014903  (2010).
\newblock \doi{10.1103/PhysRevC.82.014903}.

\bibitem{Soudi:2023epi}
I.~Soudi and A.~Majumder, {Azimuthal Anisotropy at high transverse momentum in
  $p$-$p$ and $p$-$A$ collisions}  (8, 2023).

\bibitem{Boer:1997nt}
D.~Boer and P.~J. Mulders, {Time reversal odd distribution functions in
  leptoproduction}, \emph{Phys. Rev. D}. {\bf 57}, \penalty0 5780--5786
  (1998).
\newblock \doi{10.1103/PhysRevD.57.5780}.

\bibitem{Mulders:2000sh}
P.~J. Mulders and J.~Rodrigues, {Transverse momentum dependence in gluon
  distribution and fragmentation functions}, \emph{Phys. Rev. D}. {\bf 63},
  \penalty0 094021  (2001).
\newblock \doi{10.1103/PhysRevD.63.094021}.

\bibitem{Collins:1992kk}
J.~C. Collins, {Fragmentation of transversely polarized quarks probed in
  transverse momentum distributions}, \emph{Nucl. Phys. B}. {\bf 396},
  \penalty0 161--182  (1993).
\newblock \doi{10.1016/0550-3213(93)90262-N}.

\bibitem{Accardi:2009qv}
A.~Accardi, F.~Arleo, W.~K. Brooks, D.~D'Enterria, and V.~Muccifora, {Parton
  Propagation and Fragmentation in QCD Matter}, \emph{Riv. Nuovo Cim.} {\bf
  32}, \penalty0 439--553  (2010).
\newblock \doi{10.1393/ncr/i2009-10048-0}.

\bibitem{HERMES:2000ytc}
A.~Airapetian et~al., {Hadron formation in deep inelastic positron scattering
  in a nuclear environment}, \emph{Eur. Phys. J. C}. {\bf 20}, \penalty0
  479--486  (2001).
\newblock \doi{10.1007/s100520100697}.

\bibitem{HERMES:2003icw}
A.~Airapetian et~al., {Quark fragmentation to pi+-, pi0, K+-, p and anti-p in
  the nuclear environment}, \emph{Phys. Lett. B}. {\bf 577}, \penalty0 37--46
  (2003).
\newblock \doi{10.1016/j.physletb.2003.10.026}.

\bibitem{Lehmann:2010zz}
I.~Lehmann, {Nuclear attenuation: 2 dimensional dependences at HERMES},
  \emph{PoS}. {\bf DIS2010}, \penalty0 119  (2010).
\newblock \doi{10.22323/1.106.0119}.

\bibitem{Deng:2009ncl}
W.-t. Deng and X.-N. Wang, {Multiple Parton Scattering in Nuclei: Modified
  DGLAP Evolution for Fragmentation Functions}, \emph{Phys. Rev. C}. {\bf 81},
  \penalty0 024902  (2010).
\newblock \doi{10.1103/PhysRevC.81.024902}.

\bibitem{Arleo:2003jz}
F.~Arleo, {Quenching of hadron spectra in DIS on nuclear targets}, \emph{Eur.
  Phys. J. C}. {\bf 30}, \penalty0 213--221  (2003).
\newblock \doi{10.1140/epjc/s2003-01289-x}.

\bibitem{Majumder:2004pt}
A.~Majumder, E.~Wang, and X.-N. Wang, {Modified dihadron fragmentation
  functions in hot and nuclear matter}, \emph{Phys. Rev. Lett.} {\bf 99},
  \penalty0 152301  (2007).
\newblock \doi{10.1103/PhysRevLett.99.152301}.

\bibitem{Chang:2022hkt}
W.~Chang, E.-C. Aschenauer, M.~D. Baker, A.~Jentsch, J.-H. Lee, Z.~Tu, Z.~Yin,
  and L.~Zheng, {Benchmark eA generator for leptoproduction in high-energy
  lepton-nucleus collisions}, \emph{Phys. Rev. D}. {\bf 106}\penalty0 (1),
  \penalty0 012007  (2022).
\newblock \doi{10.1103/PhysRevD.106.012007}.

\bibitem{Toll:2013gda}
T.~Toll and T.~Ullrich, {The dipole model Monte Carlo generator Sar$t$re 1},
  \emph{Comput. Phys. Commun.} {\bf 185}, \penalty0 1835--1853  (2014).
\newblock \doi{10.1016/j.cpc.2014.03.010}.

\end{thebibliography}

\end{document}